\documentclass[a4paper,11pt]{article}
\usepackage{amsmath}

\usepackage{amssymb}
\newenvironment{psmallmatrix}
  {\bigl[\begin{smallmatrix}}
  {\end{smallmatrix}\bigr]}

\numberwithin{equation}{section}
\numberwithin{equation}{subsection}

\usepackage{amsfonts}
\usepackage{relsize} % to be used with \mathlarger{}
\usepackage{calligra} % lovely calligrapic fonts
\usepackage{graphicx}
\usepackage{slashed}
\usepackage{mathtools}

\begin{document}
\pagenumbering{gobble}

\title{{CONFORMAL GENERAL RELATIVITY} \\
{Unified Theory of the Standard Models of Elementary Particles and Modern Cosmology}}
\author{Renato Nobili \\ Senior scientist of the Department of Physics of Padova University\\
     {\normalsize renato.nobili@unipd.it}\\}
\date{9 February 2020}
\maketitle
\pagestyle{myheadings}
\markright{R.Nobili, Conformal General Relativity}

\renewcommand{\thesection}{\roman{section}} % set roman pagenumbering

\begin{abstract}
\noindent
This work may be defined as a modern philosophical approach to theoretical physics.
Since ancient times science and philosophy evolved in parallel, thus renewing from
time to time the epochal paradigms of human thought. We could not understand how
the scientists of the past could have achieved so many goals, if we neglect the
philosophical ideas that inspired their minds. Today, despite the spectacular
successes of the Standard Models of Elementary Particles (SMEP) and Modern Cosmology
(SMMC), theoretical physics seems to be run into a mess of contradictions that
preclude the access to higher views. We are still unable to explain why it is so
difficult to include gravitation into the SMEP, although General Relativity (GR)
works so well in the SMMC, why it is so difficult to get rid of all the divergences
of the SMEP, and ``why there is something rather than nothing''. This paper aims to
answer these and other questions by starting from a novel fundamental principle:
{\em the spontaneous breaking of conformal symmetry down to the metric symmetry
of GR}. This statement is very simple but its implementation is a little bit
complicated. To facilitate the reading, the paper is divided in a main sequence
of sections and subsections and a collection of Appendices. The first acting as a
sort of Ariadne's wire for guiding the reader through the labyrinth of specialized
topics that are necessary to understand the work.
\end{abstract}

\newpage
\markright{}

\quad

\centerline{This page intentionally left blank}
\newpage

\renewcommand{\thesection}{\roman{section}}
\pagenumbering{roman}% roman page numbers (and reset to 1)
\markright{R.Nobili, Conformal General Relativity -- Contents}
\tableofcontents
\newpage

\renewcommand{\thesection}{\arabic{section}}
\pagenumbering{arabic}

\markright{R.Nobili, Conformal General Relativity -- {\bf \ref{introduction}} Introduction}
\section{Introduction}
\vspace{-3mm}
\label{introduction}
The {\em metric symmetry} is the fundamental symmetry of General Relativity (GR). It
reflects the invariance of a total action of matter and geometry, $\cal A$,  under the
group of diffeomorphisms of spacetime parameters
$x\equiv\{x^0\!,x^1\!,x^2\!,x^3\}$ of the form $x^\mu\!\rightarrow\!
\bar x^\mu\!=\bar x^\mu(x)$.

Any one of these diffeomorphisms changes the metric tensor of the spacetime of $\cal A$,
$g_{\mu\nu}(x)$, into a gravitationally equivalent metric tensor $\bar g_{\mu\nu}(\bar x)$
satisfying equation
\begin{equation}
\label{MetricSymm}
\bar g_{\mu\nu}[\bar x(x)]\,d\bar x^\nu(x)\,d\bar x^\mu(x)
= g_{\rho\sigma}(x)\,d x^\rho d x^\sigma.
\end{equation}
As $ds^2$ has length--dimension 2 and $x^\mu$ are adimensional, $g_{\mu\nu}$ also
has length--dimension~$2$.

The {\em conformal symmetry} reflects the invariance of $\cal A$ under the
infinite {\em group of conformal diffeomorphisms}. These are obtained by
combining the metric diffeomorphisms  $x^\mu \rightarrow \bar x^\mu(x)$  with
{\em Weyl transformations}, which consist of multiplying each local quantity of
length--dimension $n$ by a scale factor $e^{\,n\,\beta(x)}$, where
$\beta(x)$ is any smooth function of $x$.

For consistency with GR, the Standard Model of Modern Cosmology (SMMC) depicts the
initial state of the universe as an infinite concentration of matter counterbalanced
by an infinite concentration of gravitational energy. In our view, the origin of the
universe must be instead ascribed to a spontaneous breaking of conformal symmetry down to metric
symmetry, which occurred in the vacuum state of a renormalizable quantum--field system.

We will show that this sort of decay opened up {\em ex nihilo} a conical
spacetime, first promoting in it a huge scale expansion ({\em inflation}),
and then a sudden transfer of energy from geometry to matter ({\em big bang})
via the materialization of a crowd of Higgs bosons.

In GR, energy transfer from geometry to matter is impossible because the
energy--momentum (EM) tensors of geometry, $\Theta^G_{\mu\nu}\! \equiv\!-
G_{\mu\nu}/\kappa$, and of matter, $\Theta^M_{\mu\nu}$, are separately
conserved. Here, $G_{\mu\nu}\! =\! R_{\mu\nu}\!-\!\frac{1}{2}\,g_{\mu\nu}
R^\rho_\rho$ is the gravitational tensor, $R_{\mu\nu}$ is the Ricci tensor
of a spacetime manifold equipped with a metric tensor $g_{\mu\nu}$ of
signature $\{+- -\, -\}$, and $\kappa\cong 1.686\times10^{-37}$GeV$^{-2}$
is the gravitational coupling constant (with GeV as natural unit). The
separate conservation follows from gravitational equation
$\Theta^{\,G}_{\mu\nu} +\Theta^M_{\mu\nu}=0$ and the second Bianchi
identity, which states the vanishing of the covariant divergence of
$G_{\mu\nu}$.

The energy transfer is instead possible in a suitable conformal--invariant
generalization of GR, here called {\em Conformal General Relativity} (CGR),
because in this case it is possible to construct a gravitational equation
that, although retaining the form $\Theta^{\,G}_{\mu\nu}+\Theta^M_{\mu\nu}=0$,
does not necessarily imply the separate conservation of $\Theta^{\,G}_{\mu\nu}$
and $\Theta^M_{\mu\nu}$.

The idea of CGR was born several years ago from a critique of the classical
gravitational equation of Einstein in regard to the problem of renormalizability.

For the sake of brevity, a classical theory will be called renormalizable if its
quantum theoretical implementation is renormalizable.

Let ${\cal L}^M(x)$,  $[g_{\mu\nu}(x)]$,  $g(x)$, $R(x)$ and $\kappa$ be respectively the
Lagrangian density of a matter field, the metric--matrix of the spacetime, its
determinant, the Ricci scalar and the gravitational coupling constant of GR. Then,
according to the Hilbert--Einstein view, the gravitational equation can be simply
obtained by requiring the invariance of the total action of matter and geometry,
${\cal A}\!=\!{\cal A}^M\! +\!{\cal A}^G$, where
\begin{equation}
\label{AG&AM}
{\cal A}^M= \int\!\!\sqrt{-g}\,{\cal L}^M(x)\,d^4x\,,\quad
{\cal A}^G= -\frac{1}{2\,\kappa}\int\!\!\sqrt{-g}\,R(x)\,d^4x\,,
\end{equation}
under infinitesimal variations of the contravariant metric tensor $g^{\mu\nu}(x)$.

Carrying out the functional derivatives, we obtain the  gravitational equation in
the form $T_{\mu\nu}(x)\equiv T^M_{\mu\nu}(x)+T^G_{\mu\nu}(x)=0$, where
\begin{eqnarray}
\label{TMmunuofx}
&& T^M_{\mu\nu}(x)=\frac{1}{\sqrt{-g(x)}}\frac{\delta {\cal A}^M}
{\delta g^{\mu\nu}(x)}=2\frac{\delta{\cal L}^M(x)}
{\delta g^{\mu\nu}(x)}- g_{\mu\nu}(x)\,{\cal L}^M(x)\,\,\,
\hbox{and}\\
\label{TGmunuofx}
&& T^{\,G}_{\mu\nu}(x)=\frac{1}{\sqrt{-g(x)}}\frac{\delta {\cal A}^G}
{\delta g^{\mu\nu}(x)} =-\frac{1}{\kappa}\Big[R_{\mu\nu}(x)-
\frac{1}{2}\,g_{\mu\nu}\,R(x)\Big]
\end{eqnarray}
are respectively the energy--momentum (EM) tensors of matter and geometry.

As is well--known, the big problem with this gravitational equation is that
$T^G_{\mu\nu}$ is non--renormalizable. So we would expect that if we take $R=0$
the matrix elements of $T^M_{\mu\nu}$ should be finite in every order
of the renormalized perturbation theory. (`Finite'' means independent of the
cut--off in the limit of large cut--off). Unfortunately, this is not always true.
The point is that, if ${\cal L}^M(x)$ includes a scalar field $\varphi(x)$ with
quartic self--interactions, $T^M_{\mu\nu}(x)$ does not satisfy the EM--tensor
conservation equation $\partial^\mu T^M_{\mu\nu}(x)=0$ and, moreover, its matrix
elements turn out to be cut--off dependent.

It is however possible to construct a new energy--momentum tensor,
\begin{equation}
\label{CCJ}
\Theta^M_{\mu\nu}(x) = T^M_{\mu\nu}(x)+ \frac{1}{6}\,\big[g_{\mu\nu}(x)\,
\square -\partial_\mu\partial_\nu\big]\varphi(x)^2\,,
\end{equation}
where $\square$ is the d'Alembert operator in the given metric, which defines
the same four--momentum, satisfies equation $\partial^\mu\Theta^M_{\mu\nu}(x)=0$,
as well as all the standard commutation relations of the algebra of currents,
and, further, has finite matrix elements.

This unexpected complication was discovered by Callan, Coleman and Jackiw in
1970, who called $\Theta^M_{\mu\nu}(x)$ the {\em improved EM tensor} \cite{CALLAN}.

At the end of their investigation, these authors also proved that a similar result
can be obtained if ${\cal A}^G$ is replaced with
\vspace{-1mm}
\begin{equation}
\label{barAG}
{\cal A}'^{\,G}=\int\!\sqrt{-g(x)}\,\frac{R(x)}{12}\,\varphi(x)^2\,d^4x\,.
\vspace{-1mm}
\end{equation}

In this case, the variation of ${\cal A}'^{\,G}$  with respect
to $g^{\mu\nu}(x)$ yields equation
\vspace{-1mm}
\begin{equation}
\label{barTGmunu}
T'^{\,G}_{\mu\nu}(x)=\frac{1}{\sqrt{-g}}\,\frac{\delta {\cal A}'^{\,G}}{\delta g^{\mu\nu}(x)}=
\frac{1}{6}\,\Big[R_{\mu\nu}(x)- \frac{1}{2}\,g_{\mu\nu}(x)\,R(x) +
g_{\mu\nu}\,D^2- D_\mu D_\nu\Big]\varphi(x)^2,
\vspace{-1mm}
\end{equation}
where $D^2=g^{\mu\nu} D_\mu D_\nu$ is the Beltrami--d'Alembert operator and $D_\mu$
are the covariant derivatives. To obtain this equation we have used Eqs (\ref{D2f})
(\ref{Rvariation}) of Appendix {\bf {\ref{BasFormApp}}}.

It is therefore evident that for $R \rightarrow 0$, tensor
$T'^{\,G}_{\mu\nu}+ T^M_{\mu\nu}$ converges exactly to the improved EM tensor
$\Theta^M_{\mu\nu}$ and, if the metric becomes Minkowskian, $D_\mu$ converges
to $\partial_\mu$.

An interesting implication of this result is that a total classical action of
matter and geometry of the form
\vspace{-1mm}
\begin{equation}
\label{totalcalAforvarphi}
{\cal A}'_\varphi = \int\!\sqrt{-g}\,\bigg[\frac{1}{2}\,g^{\mu\nu}(\partial_\mu \varphi)\,
\partial_\nu \varphi-\frac{\lambda}{4}\,\varphi^4+\varphi^2\, \frac{R}{12}\bigg]\,d^4x\,,
\quad (\lambda >0),
\vspace{-1mm}
\end{equation}
is invariant under conformal diffeomorphisms. To prove this we have only to carry out
in ${\cal A}'_\varphi$ the following Weyl transformations with arbitrary scale
factor $e^{\beta(x)}$:
\vspace{-1mm}
\begin{eqnarray}
\label{Weyltrans4}
&&\hspace{-8mm}\sqrt{-g(x)} \rightarrow e^{4\beta(x)}\!\sqrt{-g(x)}\,,\quad
g_{\mu\nu}(x) \rightarrow e^{2\beta(x)} g_{\mu\nu}(x)\,,\quad g^{\mu\nu}(x)
\rightarrow e^{-2\beta(x)} g^{\mu\nu}(x)\,,\nonumber \\
&&\hspace{-8mm}\varphi(x) \rightarrow e^{-\beta(x)} \varphi(x)\,, \quad R(x)
\rightarrow e^{-2\beta(x)}\big[R(x)-6\,e^{-\beta(x)}D^2 e^{\beta(x)}\big]\,,
\vspace{-1mm}
\end{eqnarray}
the latter of which is picked up from Eq (\ref{TildeR4}), and then verify that
under the action of these transformations ${\cal A}'_\varphi$ is transformed to
$\bar{\cal A}'_\varphi = {\cal A}'_\varphi +\Delta {\cal A}'_\varphi$, where
\vspace{-1mm}
\begin{equation}
\label{DeltaAprime}
\Delta{\cal A}'_\varphi\! =\! \int\!\sqrt{-g(x)}\,D_\mu \big[\,\varphi(x)^2
e^{-\beta(x)}\partial^\mu e^{\beta(x)}\big]d^4x\! \equiv\! \int\!
\partial_\mu\big[\,\varphi(x)^2e^{-\beta(x)}\partial^\mu e^{\beta(x)}\big] d^4x\,.
\vspace{-1mm}
\end{equation}

Since this difference is manifestly a surface term, we infer that ${\cal A}'_\varphi$
and $\bar{\cal A}'_\varphi$ are functionally equivalent; so, by carrying out the functional
variations with respect to  $g^{\mu\nu}(x)$ and $\varphi(x)$, we obtain the same gravitational
equation and motion equation for $\varphi$ as before.

Remarkably, this equivalence fails if the dimension of the spacetime is
different from  $2$ or $4$. The reader can easily verify this fact by
carrying out the analogous computations in an action integral defined
over an $n$--dimensional spacetime.

Note that, if the vacuum expectation value (VEV) of $\varphi^2$ were just
$1/\kappa$, the quantum implementation of ${\cal A}'_\varphi$ would provide
not only the improved EM tensor, but also a sort of gravitational equation.
In addition, since  ${\cal A}'_\varphi$ is free from dimensional constants,
this action would provide an excellent renormalizable approximation to GR.

And even if the conformal invariance of this classical action were destroyed by
the procedure of renormalization, we would nevertheless be left with the vivid
impression that conformal invariance and four--dimensionality of spacetime conspire
together to produce a sort of renormalizable theory of gravitation, with the hope
that the conformal invariance could be restored by suitable interactions with other
fields (as indeed happens in CGR).

But alas, the hope that ${\cal A}'_\varphi$ could represent a model
of renormalizable gravity is vain, because the positivity of the
$\varphi^2$--VEV would make gravitation to be repulsive.

One may have the idea of bypassing this difficulty by replacing ${\cal A}'_\varphi$
with
\begin{equation}
\label{totalcalAforsigma}
{\cal A}'_\sigma = \int\!\sqrt{-g}\,\bigg[\!-
\frac{1}{2}\,g^{\mu\nu}(\partial_\mu \sigma)\,
\partial_\nu \sigma-\frac{\bar\lambda}{4}\,\sigma^4
-\sigma^2\, \frac{R}{12}\bigg]\,d^4x\,,
\quad(\bar\lambda >0),
\end{equation}
where $\sigma(x)$ is a scalar ghost with nonzero VEV, which has
negative kinetic energy and  positive potential energy. Note
that the positivity of the self--interaction potential prevents
the total energy from going to $+\infty$ but not to $-\infty$.
In this case, in fact, the EM tensor of geometry described in
Eq (\ref{barTGmunu}) would be replaced by
\begin{equation}
\label{barTGmunusigma}
T{\,'}^G_{\mu\nu}=\frac{1}{\sqrt{-g}}\,\frac{\delta
{\cal A}'_\sigma}{\delta g_{\mu\nu}}=- \frac{1}{6}\,\sigma^2
\Big(R_{\mu\nu}- \frac{1}{2}\,g_{\mu\nu}\,R\Big) -\frac{1}{6}\,
\big(g_{\mu\nu}D^2- D_\mu D_\nu \big)\sigma^2,
\end{equation}
which has the right sign for gravity to be attractive.
But this idea also runs into troubles.

The problem arises from the amphibious nature of ghosts. It is
well--known that scalar ghosts play an important role in the
renormalization of Yang--Mills fields. In this case they
are harmless because they do not appear as asymptotic states
of the $S$--matrix.

Trying to domesticate them in other circumstances is almost
universally considered an inexcusable naivety. As a matter
of fact, almost all authors who in the second half of the
past century tried to domesticate them, had to abandon
their attempts. The author of this paper, who pursued the
same intent, being strongly disappointed with the theory of
strings, saved himself by working in biophysics for thirty years.

Before telling how the idea of the scalar ghost can be successfully
implemented in CGR, let us concisely explain what are the problems
with ghosts in classical and quantum field theories.

The catastrophic effect of a scalar ghost in a classical action is
that the total energy density of the system converges rapidly to
$-\infty$ over time.

However, if we consider the problem from the point of view of quantum
field theory (QFT), we have not to do with classical field amplitudes,
but rather with transition amplitudes and probabilities of physical
events. In this case, the problem with ghosts is that the norms of
their input and output states are negative, which entails the violation
of $S$--matrix unitarity \cite{CUTKOWSKY}.

The solution adopted in CGR is to sum up together actions ${\cal A}'_\varphi$
and ${\cal A}'_\sigma$ and a conformal--invariant interaction
term depending on $\varphi$ and $\sigma$ and arranged in such a way that
the total energy of the system remains bounded both from above and from
below.

An expedient of this sort has been proposed and exemplified in a simple
model by Ilhan and Kovner in 2013~\cite{ILHAN}.  A similar expedient can
be adopted in our case by introducing the conformal invariant action integral
\begin{equation}
\label{AsigmaAvarphi}
{\cal A} =\! \int\!\sqrt{-g}\,\bigg\{\frac{1}{2}\,g^{\mu\nu}
\Big[\big(\partial_\mu\varphi\big)\partial_\nu\varphi-
\big(\partial_\mu\sigma\big)\partial_\nu\sigma\Big] -
\frac{\lambda}{4}\big(\varphi^2- c^2\sigma^2\big)^2-\big(\sigma^2\!
-\varphi^2\big)\frac{R}{12}\bigg\}\,d^4x\,.
\end{equation}
This fulfils the bounded--energy condition provided that $c^2>1$ and the
initial values of $\varphi$ and $\sigma$ satisfy suitable initial conditions.
In particular, in order that gravity is always attractive, the VEV of
$\sigma^2- \varphi^2$ must always be positive. Another important requirement
is that potential energy density $U(\varphi,\sigma)\!=\!(\lambda/4)\,\big(\varphi^2
- c^2\sigma^2\big)^2$ converges to zero over time.

But in order that this really happens, the motion equations derived from $\cal A$
must contain a dissipative term. In the simplest case, it is sufficient
that the metric--matrix tensor has the form
\begin{equation}
\label{givencoord}
\hspace{12mm}\big[g_{\mu\nu}(\tau, \vec\rho\,)\big] =
\hbox{diag}\big[1, - \tau^2, - \tau^2 (\sinh\varrho)^2, -
\tau^2(\sinh\varrho^2 \sin\theta)^2\bigr], \,\,\,
\tau\in [0, \infty]\,,
\end{equation}
which entails that the spacetime is homogeneous, isotropic, flat and
originates at time $\tau=0$. This in turn entails that $R=0$ and $\varphi$,
$\sigma$ depend only on $\tau$.

In this case the motion equations derived from action (\ref{AsigmaAvarphi})
take the simple form
\begin{eqnarray}
\label{varphifrict}
&& \partial_\tau^2 \varphi(\tau) + \frac{3}{\tau}\,\partial_\tau \varphi(\tau) +
\lambda\,\big[\varphi(\tau)^2 - c^2\sigma(\tau)^2\big]\,\varphi(\tau) =0\,,\\
\label{sigmafrict}
&& \partial_\tau^2 \sigma(\tau) + \frac{3}{\tau}\,\partial_\tau\,\sigma(\tau) +
\lambda\,c^2\big[\varphi(\tau)^2 - c^2\sigma(\tau)^2\big]\,\sigma(\tau) =0\,,
\end{eqnarray}
which admit non--negative ab=nd finite solutions. Note that the frictional terms
proportional to $\partial_\tau \varphi(\tau)$ and $\partial_\tau \sigma(\tau)$
force $U(\varphi, \sigma)$ to converge to zero in the course of time.

\subsection{Remarkable properties of CGR}
\vspace{-1.5mm}
\begin{itemize}
\item[1)] The group  of conformal diffeomorphisms is the largest
group of coordinate transformations that preserve the causal order of
physical events in a generally curved spacetime (discussed and referenced
in Appendix {\bf \ref{ConfInvApp}}).
\vspace{-1.5mm}
\item[2)] CGR can be implemented only in a 4D spacetime. This is proven in
\S\,\ref{inclusCGR}.
\vspace{-1.5mm}
\item[3)] The entire history of the universe is confined to the interior of
a future cone $\bf{C}_\odot$.
\vspace{-1.5mm}
\item[4)] The spontaneous breakdown of conformal symmetry starts from the
inner boundary of ${\bf C}_\odot$ and its effects on the large scale propagate
homogeneously and isotropically all over the 3D hyperboloids that foliate the
interior of the cone.
\vspace{-1.5mm}
\item[5)] The action of CGR is free from dimensional constants. This is
admissible provided that all the dimensional constants of the theory
originate from the VEVs of suitable Nambu--Goldstone (NG) bosons. As will
be shown in the next, these VEVs depend on the temporal parameter $\tau$
that labels the hyperboloids of the conical spacetime.
\vspace{-1.5mm}
\item[6)] Since all coupling constants are adimensional, CGR is renormalizable
\cite{SCHWARTZ}.
\vspace{-1.5mm}
\item[7)] In CGR, the gravitational coupling constant of GR is replaced by a
quadratic function of two NG--boson VEVs, which decreases in time as the inverse
of a sigmoid (exemplified in Fig.s\,\ref{VacStFig4} and \ref{VacStFig5}
of Appendix {\bf \ref{VacDynApp}}).
\vspace{-1.5mm}
\item[8)]
\vspace{-1.5mm}
The time dependence of the gravitational attraction causes curious effects that
the SMMC tries to explain in other ways. For example: increased gravitational
redshift of distant stars, currently imputed to an accelerated expansion of
the universe \cite{RIESS}; anisotropy of cosmic microwave background, currently
ascribed to primordial quantum fluctuations surviving the superluminal inflationary
expansion of the universe \cite{MUKHANOV}; astronomic evidence of premature
formation of black holes \cite{6GIRLS} and demographic decrease of stars
\cite{SOBRAL},  both of which are still unexplained by the SMMC, etc.
\vspace{-1.5mm}
\item[9)] As will be clear in the following, and extensively in
\S\S\,\,\ref{unilargescale}, CGR works well also in the semi--classical
approximation. This may be surprising because it is commonly believed that
the conformal invariance of a classical action is destroyed by quantization.
If this were always the case, then CGR, as outlined in the previous subsection,
would be untenable. Fortunately, as it will be argued in the next four
subsections, there are particular circumstances in which conformal
invariance survives quantization.
\end{itemize}
\newpage

\subsection{Theoretical physics in between logic and dialectic}
\label{mainproblems}
Theoretical physics cannot properly be called {\em logical}, as if it were an axiomatic discipline.
It should rather be called {\em dialectical}, because, as a matter of fact, it is plagued with
tremendous contradictions. These, however, should not be regarded as incurable pathologies, but
rather as loci of fertility, from which new ideas may spring up and guide the physicist to higher
levels of comprehension. It seems therefore more appropriate to say that theoretical physics stands
in between mathematical logic and a sort of heuristic dialectic, in a limbic region where the first
pushes to reduce the uncertain margins of the second, with the aim of reaching a perfectly logical
and fully explanatory theory of everything.

The most disconcerting of all contradictions is perhaps the impressive success of GR in
cosmology and astrophysics, despite the evidence of its non--renormalizability. How can
it happen that the entire theoretical physics be held hostage by the length--dimension
$2$ of the gravitational coupling constant $\kappa$? In CGR, where $1/\kappa$ is
replaced by a biquadratic function of two NG--boson amplitudes, the problem does not
arise because the conformal invariance makes the theory renormalizable. The most
important consequence of this fact is the strong time--dependence of gravitational
attraction. But since this seems to account for certain unexplained phenomena, it
might be a bonus rather than a flaw.

Almost always, in the last sixty years, several contradictions emerged from the
bowels of QFT in the form of mathematical anomalies and singularities.

In the Standard Model of Elementary Particles (SMEP), this has happened to such
an extent that the winning strategy for invention, prediction and innovation seems
to have progressed mainly through the discovery of some good reasons for getting
rid of them; such is, for instance, the mutual cancelation of Adler--Bell--Jackiw's
triangle anomalies \cite{ABJ}.

Another contradiction comes from the dependence on momentum cutoff $\Lambda$ of the
zero--point energy densities (ZPEDs), which are positive for bosons and negative
for fermions.

As pointed out by Coleman in {\em Aspects of Symmetry} (1985, p.142), the
$\Lambda$--dependence of the Gaussian terms of a QFT path--integral cannot
simply be removed by the addition of mass and coupling--constant counterterms
to the Lagrangian density. Their presence in the 1--loop term of the effective
action is, in fact, a major problem for CGR. So the question arises whether
there are sufficient reasons for getting rid of them. Fortunately, as widely
argued in \S\,\ref{digressOnlnDet} of Appendix {\bf \ref{PathIntApp}}, and
hereafter summarized, the answer is positive.

\newpage

The 1--loop terms of the effective Lagrangian density that are proportional to
$\Lambda^4$ are harmless because their effect is to multiply the path--integral
by a phase factor that does not depend on mass parameters. This is proven in
\S\,\ref{digressOnlnDet}, near Eqs (\ref{calGm2}) (\ref{lnDetTerm2}).

For a single field of mass $m$, the mass dependence of the 1--loop term has the form
\vspace{-1mm}
\begin{equation}
\label{GenFullGamma}
G(m^2)  =  \hbar\,D\bigg(\frac{m^2\Lambda^2}{32\,\pi^2} - \frac{m^4}{64\,\pi^2}\ln\Lambda^2 +
\frac{m^4}{64\,\pi^2}\ln m^2\bigg),
\vspace{-1mm}
\end{equation}
where $D$ is the dimension of the field. To be precise, we have $D=1$ for a scalar
field, $D=3$ for a vector field, $D=-4$ for a Dirac field  and $D= -2$ for a Majorana
field. This is explained in detail in \S\,\ref{1looptermsgen} near Eq (\ref{GenFullGamma0}).
Thus, in particular, in a path integral with fields of different masses and spins,
the sum of the 1--loop terms proportional to $\Lambda^2$ is
\vspace{-1mm}
\begin{equation}
\label{VariuosSpinGamma}
{\mathbb S}_{\Lambda^2}=\frac{\hbar\,\Lambda^2}{32\,\pi^2}\,\Big(\sum m^2_S + 3\sum m^2_V -
 4\sum m^2_F - 2\sum m^2_M\Big).
\vspace{-1mm}
\end{equation}

In this regard, it is worth remembering the conjecture of Veltman (1981), according to which
this sum vanishes because of the mutual cancelation of the mass terms \cite{VELTMAN}.
Three decades later this conjecture had a role in predicting the mass of the Higgs boson
\cite{JONES}. Now, the question arises of whether also the other terms can vanish in
a similar way.

In which case, since the sum of all terms proportional $\log\Lambda$ is
\vspace{-1mm}
\begin{equation}
\label{LogLambdaGamma}
{\mathbb S}_{\,\ln \Lambda} =-\frac{\hbar\,\ln\Lambda}{16\,\pi^2}\,\Big(\sum m^4_S + 3\sum m^4_V -
 4\sum m^4_F - 2\sum m^4_M\Big),
 \vspace{-1mm}
\end{equation}
and that of all terms independent of $\Lambda$ is
\vspace{-1mm}
\begin{equation}
\label{NoLambdaGamma}
{\mathbb S}_0 =\frac{\hbar}{16\,\pi^2}\,\Big(\sum m^4_S\ln m_S
+ 3\sum m^4_V\ln m_V - 4\sum m^4_F\ln m_F - 2\sum m^4_M\ln m_M\Big),
%\vspace{-1mm}
\end{equation}
we should have the following conditions for the vanishing of the
sum of all 1--loop terms:
\vspace{-2mm}
\begin{eqnarray}
\label{squaredterms}
&&\hspace{-16mm} S^{(2)}= \sum m^2_S + 3\sum m^2_V - 4\sum m^2_F - 2\sum m^2_M  = 0; \\
\label{quarticterms}
&&\hspace{-16mm} S^{(4)} =\sum m_S^4 + 3\sum m_V^4 - 4\sum m^4_F - 2\sum m^4_M  = 0; \\
\label{quarticlogterms}
&&\hspace{-16mm} S^{(4*)} = \sum m_S^4\ln m_S + 3\sum m_V^4\ln m_V - 4\sum m^4_F\ln m_F
- 2\sum m^4_M\ln m_M  = 0.
\vspace{-3mm}
\end{eqnarray}

In principle, provided that the number of fields with different masses and spins is
sufficiently large and well--balanced, there is no reason why the above conditions could
not be simultaneously satisfied. An unexpected confirmation of this possibility has been
prospected by Alberghi, Kamenshchik {\em et al}. in 2008 \cite{ALBERGHI}, who proved that
the fields of the SMEP satisfy Eqs~(\ref{squaredterms})--(\ref{quarticlogterms})
provided that at least one massive fermion, even only of Majorana type, having mass
within specific ranges is added to list (see Appendix {\bf {\ref{DirMajorApp}}}).

But to corroborate this hypothesis we need a more robust theoretical
justification.

%no matters whether of Dirac

\subsection{Underlying conformal invariance and 1--loop--term cancelation}
\label{AsympConfInvinCGR}
Here we want to explain why the underlying conformal symmetry of CGR entails the
vanishing of the $1$--loop term of the effective action, so that the spontaneous
breakdown of conformal symmetry can produce a mass spectrum of the type described
by the SMEP.

To prove this, it is expedient to report the following important observation made
by Coleman (1985, \S\,6.3, p.138) in Ref. \cite{COLEMAN}, which regard the effects
produced by the spontaneous breakdown of a symmetry on the loop terms of an effective
Lagrangian density:

``{\em To renormalize the loop terms of order larger than one, we need to invoke
no more counterterms than would have been required if there had no spontaneous
symmetry breakdown; the ultraviolet divergences of the theory respect the symmetry
of the Lagrangian, even if the vacuum state does not; in other terms, the divergence
structure of a renormalizable field theory is not affected by the occurrence of
spontaneous symmetry breakdown. This is the secret of the renormalizability of
weak interactions}''.

Assume that a QFT has the conformal symmetry. If this symmetry is not spontaneously
broken, all NG bosons have zero VEVs and all Green functions are conformally invariant
Wightman functions \cite{NOBILI73}. If the fields have canonical dimensions, the
theory describes free massless fields, otherwise it describes fields with anomalous
dimensions \cite{TODOROV}. In both cases, the Green functions do not contain mass
constants.

Now assume that the symmetry is spontaneously broken, so that certain scalar fields
of the theory have nonzero VEVs. Hence, in accord with Coleman's observation, the
ultraviolet structure of the Green functions at high momenta is free from
dimensional parameters. Thus, passing from the Green--function representation
to the path--integral representation -- as described in Appendix
{\bf \ref{PathIntApp}}, especially in \S\,\ref{1loopgencase} -- the total
$1$--loop term of the effective action is free from mass terms. Which is
consistent with  Eqs (\ref{squaredterms})--(\ref{quarticlogterms}).

\subsection{The 1--loop term cancelation preserves the classical limit of a QFT}
\label{ShiftInvCGR}
At the beginning of the past century, Niels Bohr envisaged the guiding principle of the
nascent quantum mechanics in the so--called {\em correspondence principle}, which states
that, in the limit of large quantum numbers, the behavior of a quantum system approaches
that of a classical system. Unfortunately, this statement is not true, neither in
elementary quantum mechanics nor in QFT. In the first case, it fails with fermions
because these have no classical counterpart; in the second case it fails with GR
because this theory  cannot be quantized. As explained in \S\,\ref{introduction},
this problem does not arise in CGR because this theory is renormalizable and its
conformal symmetry breaks down to that of GR over time. But unfortunately, this
argument does not ensure the existence of the classical limit of CGR.

Among physicists, it is customary to assume that the VEV of the Higgs field coincides
with the field amplitude that minimizes the classical potential. This assumption is
simplistic, because the correct VEV must be retrieved by minimizing the potential of
the {\em effective action}, which in general differs from the classical one by a
non--negligible quantum correction. Nevertheless, in the SMEP, the assumption
works well. How is this possible?

In this regard, it is important to recall a result proved in Appendix
{\bf \ref{1looptermsgen}} near Eq~(\ref{mathbbG}): {\em If the sum of
all the 1--loop terms of the effective action vanishes, the VEVs of the
scalar fields coincide with those of the classical theory}. This point
deserves a further clarification.

Let us denote as  $\phi_c(x)$ the solution to the classical equation of a self--interacting scalar
field $\phi(x)$ of mass $m$, and as $\bar\phi(x)$ that of the corresponding quantum field. In
\S\,\ref{PathIntEval} it is shown that in general $\hat\phi_c(x)=\bar\phi(x)-\phi_c(x)$ is
nonzero for two important reasons:

1) The classical approximation of the effective action $\Gamma[\bar\phi]$ of $\phi(x)$ is not
the classical action ${\cal A}_{\hbox{\small cl}}[\phi_c]$, but coincides with the zero--loop
term of $\Gamma[\bar\phi]$, which is $\Gamma_0[\bar\phi] = {\cal A}_{\hbox{\small cl}}[\bar\phi]$.

2) The magnitude of $\hat\phi_c(x)$ is related to the one--loop term $\hbar\,\Gamma_1[\bar\phi]$
of $\Gamma[\bar\phi]$ and to the effective propagator of $\phi(x)$, $\Delta[\bar\phi;x,y]$,
by equation
\begin{equation}
\label{hat-phic-x}
\hat\phi_c(x) =\frac{i\hbar}{2}\!\int\frac{\delta \Gamma_1[\bar\phi\,]}{\delta \bar\phi(y)}
\,\Delta[\bar\phi;y,x]\,d^4y\,,
\end{equation}
which contains unremovable cut--off dependent terms that are present even if the loop
terms $\hbar^L\Gamma_L[\bar\phi]$ of order $L\ge 1$ are made finite by standard
renormalization procedures.

Thus, for instance, if $\phi(x)$ is the amplitude of a Higgs field, the
Higgs--bosom mass, $m(\bar\phi)$, is a function of $\bar\phi$. Correspondingly,
as described  in \S\,\ref{digressOnlnDet}, the potential of the classical Lagrangian
density is heavily distorted by the presence of the Gaussian term
\begin{equation}
\label{GomforHiggs}
G\big[m^2(\bar\phi)\big]  =  \hbar\,\bigg[\frac{m^2(\bar\phi)\,\Lambda^2}{32\,\pi^2} -
\frac{m^4(\bar\phi)}{64\,\pi^2}\ln\Lambda^2 +\frac{m^4(\bar\phi)\ln m^2(\bar\phi)}
{64\,\pi^2}\bigg]\,.\nonumber
\end{equation}
Thus, in order for $\Gamma_1[\bar\phi]$ to be zero, the Gaussian term described by
this equation must be canceled by the 1--loop terms of all other fields. If this
happen, we shall have $\hat\phi_c(x)=0$.

It is therefore evident that the correspondence principle of Bohr holds true
only if the effective actions describes a Higgs field interacting with other
massive fields.

Since, in accord with the SMEP, the Higgs boson gives mass of all other fields,
the masses appearing in Eqs (\ref{VariuosSpinGamma})--(\ref{NoLambdaGamma}) also
depend on $\bar\phi$. So, on account of Eqs (\ref{squaredterms})--(\ref{quarticlogterms}),
the 1--loop term of the effective action of the SMEP is expected to satisfy equation
$$
\Gamma_1[\bar\phi] =\Gamma^{\Lambda^2}_1[\bar\phi]+
\Gamma_1^{\,\ln \Lambda}[\bar\phi] +\Gamma^0_1[\bar\phi]=0\,.
$$
Using this in Eq (\ref{hat-phic-x}), we obtain $\hat\phi_c=0$, showing that the
underlying conformal invariance of the effective action entails the equality
$\bar\phi=\phi_c$. Since the conformal symmetry of CGR decays to metric
symmetry, the geometry of CGR is expected to evolve towards that of GR.
This means that particle accelerators can only unveil the last stage
of CGR.

\subsection{The dynamical rearrangement of conformal symmetry}
\label{DynRearrOfConfSymm}
Despite the simplicity of the founding principle of CGR, the description of the
spontaneous breakdown of CGR is rather complicated. The reason of this relies on
the following fact.

Any physical theory aims to establish a relation between two levels of description: one
{\em fundamental}, the other {\em phenomenological}, and tries to explain how the
second emerges from the first. In the framework of a non--relativistic QFT, we can take
as fundamental the level of the algebra of local fields $\Psi(x)$, called the
{\em fundamental fields}, or Heisenberg fields, in terms of which all the equations
of the theory can be expressed. As phenomenological level we can take the
representations of free physical fields in some Hilbert space $\mathbb{H}$,
namely the asymptotic fields $\psi^0(x)$. The relation between these two levels
of descriptions is called the {\em dynamical map} and is denoted by $\Psi\big(x;
\psi^0\big)$. This dual structure introduces a sophisticated mechanism for the
manifestation of symmetries.

Since this topic has been masterfully treated by Umezawa {\em et al.} in
Ref.\,\cite{UMEZAWA1}, here we limit ourselves to reporting a few important
concepts discussed by these authors.

Suppose that the basic equations of the theory are invariant under a group
${\cal G}$ of continuous transformations, $\Psi(x)\rightarrow \Psi'(x) =
T \Psi(x)\,T^{-1}$. It frequently happens that the fundamental state
$\vert\Omega\rangle$ of $\mathbb H$ does not manifest this symmetry. A
well--known example is that of a ferromagnet in which the spin--rotational
invariance is spontaneously broken. In this case, the original symmetry
is not simply lost, but gives rise to the spin--polarization of the
ferromagnet. This change in the manifestation of the symmetry is called
the {\em dynamical rearrangement of the symmetry}. This complicated
state of the things evidences the importance of distinguishing the
notion of {\em symmetry} from that of {\em invariance}.

A notable feature of the phenomenon of spontaneous breakdown of a
symmetry is that the degree of symmetry of the phenomenological
level can be lower than that of the fundamental level. This is
why we expect that the conformal symmetry of CGR can downgrade
to the metric symmetry of GR while preserving renormalizability.

Assume that the action $\cal A$ of a non--relativistic QFT is
invariant under a non--Abelian group $\cal G$ of continuous
transformations, and denote by $\vert\Omega\rangle$ the vacuum
state of the Hilbert space in which the asymptotic fields are
represented. The spontaneous breakdown of a symmetry divides
the elements of the Lie algebra of ${\cal G}$ in two classes:
(1) those that annihilate $\vert\Omega\rangle$, which belong
to a proper subgroup ${\cal S}\subset{\cal G}$ called the
{\em stability subgroup} of the theory; (2) the others which
instead multiply $\vert\Omega\rangle$ by an infinite constant. It
can easily be proved that such a partition is possible provided
that ${\cal S}$ is a {\em contraction} of ${\cal G}$
\cite{INONU&WIGNER}.

Since the theorem of N\"other identifies each element of the Lie
algebra as the charge $Q=\int\!j_0(x)\,d^3x$ of a conservative
current $j_\mu(x)$, we can envisage the charges of class (2) as
the manifestation of a condensate of massless bosons. And since
$\int \partial^\mu j_\mu(x)\,d^3x=0$, we can have $Q\vert\Omega\rangle\neq 0$
only if $\langle\Omega|j_\mu(x)j_\mu(0)|\Omega \rangle$ is singular at
$x^2=0$. This explains why the spontaneous breakdown of a symmetry
creates one ore more boson fields with nonzero VEVs and gapless energy
spectrum: namely, the boson fields of {\em Nambu--Goldstone} (NG).

If $\vert\Omega\rangle$ is invariant under spacetime translations, the
energy spectrum of the Hamiltonian exhibits zero--mass poles; meaning
that one or more NG fields represent massless particles with nonzero
VEVs; otherwise, the NG fields take the form of extended objects
depending on spacetime coordinates. The NG fields generated by
the spontaneous breakdown of conformal symmetry belong precisely
to the  second case (see Appendix {\bf \ref{BreakConfApp}}).

The application of these concepts to a relativistic QFT is far
from being obvious or straightforward. However, since we know that
non--relativistic theories are limiting cases of renormalizable
relativistic theories, there is no reason why the concepts here
introduced could not be extended to the relativistic case.

Further progress was achieved in 1964 by Jona--Lasinio, who
showed that the correct way to pose and solve the problem of the
spontaneous breaking of symmetries -- in relativistic theories
defined over a Minkowskian spacetime -- is one based on the
functional methods of the {\em effective action}. Appendix
{\bf \ref{PathIntApp}} is entirely devoted to this topic.

Unfortunately, using these methods to study the contraction of
the conformal symmetry of CGR into the dynamically rearranged
symmetry of GR, encounters additional difficulties that we have
not been able to overcome, mainly for the following reasons.

Firstly, in CGR isomorphic input and output fields do not exist
because the time parameter extends from $0$ to $+\infty$; secondly,
the spacetime is, on average, the interior of a future cone foliated
by a continuum of hyperboloidal 3D--surfaces; thirdly, the metric
tensor of the spacetime evolves continuously with a time--dependent
expansion factor. It is therefore clear that the construction of
a well--settled path--integral technique, capable of representing
adequately all these features, would take several years to be
accomplished.

All that we can do here is to indicate the sequence of steps through
which we attempted to individuate and logically connect the most
relevant tiles of this fascinating research.

The decay of conformal symmetry to metric symmetry involves the creation
of two NG bosons, $\varphi(x)$ and $\sigma(x)$, which must interact with
each other in such a way that the energy spectrum of CGR remains bounded
from above and below, as discussed in the end of \S\,\ref{introduction}.
Once evolved to GR, the ghost field $\sigma(x)$ and the scalar field
$\varphi(x)$ of CGR become respectively a constant and the Higgs field
of the SMEP, which in turn gives mass to almost all its decay products;
in practice, all other fields described by the SMEP.

In the last stage of CGR evolution, also the relation between the
original metric tensor of CGR described in Eq (\ref{givencoord})
and the standard metric tensor of GR, undergoes a structural change.
The system of coordinates used in \S\,\ref{introduction}, near Eq
(\ref{AsigmaAvarphi}), is replaced by that of the {\em proper--time}
coordinates of the comoving observers of the universe. Mathematically,
this transition is carried out through the intermediation of a
coordinate system that is reminiscent of the system of {\em conformal--time}
coordinates, as occurs in modern  cosmology. This rearrangement of
coordinates is described in detail in \S\S\,\ref{THREEWAYS}.

In these circumstances, the original vacuum state of CGR also undergoes
a {\em phase transition}, which can be described by a thermal Bogoliubov
transformation of the phenomenological fields (see Appendix
{\bf \ref{ThermVacApp}}). This transition occurs at the critical
big--bang time $\tau_B$, after which CGR takes the form of a
statistical QFT (see Section \ref{unilargescale}).

Let us point out that the evolution of CGR towards GR cannot be
described as a single physical process, but rather as a hierarchy
of physical and thermodynamical processes, which extends far beyond
the levels of NG--bosons, elementary particles and extended bodies,
up until the levels of indescribable complexity of celestial bodies
and living systems.

\newpage

\markright{R.Nobili, Conformal General Relativity -- {\bf \ref{futconegeom}}  Polar--hyperbolic  spacetimes}
\section{Polar--hyperbolic coordinates}
\label{futconegeom}
To implement CGR we must imagine the history of the universe confined to the
interior of a future cone, the simplest of which is a region of the Minkowskian
spacetime parameterized by polar--hyperbolic coordinates. In this case, the
Minkowskian parameters $\{x^0, x^1, x^2, x^3\}$ are related to the
polar--hyperbolic coordinates $\{\tau,\varrho, \theta, \phi\,\}$ by equations
$x^0 =\tau\,\cosh\varrho$, $x^1 = \tau\,\sinh\varrho\,\sin\theta \,\cos\phi$,
$x^2 = \tau\,\sinh\varrho\,\sin\theta\, \sin\phi$, and $x^3  = \tau\,
\sinh\varrho\,\cos\theta$. We will call $\tau =\sqrt{(x^0)^2 - (x^1)^2 -
(x^2)^2 - (x^3)^2}$ the {\em kinematic time}, and the parameters $\{\varrho,
\theta,\phi\}\equiv \vec\rho$ are the components of the {\em hyperbolic--Euler angle}.

The spacetime of CGR is an open future cone stemming from a point $V$ of a
pseudo--Riemannian manifold $\mathfrak M$ of signature $\{1, -1,-1, -1)$. The
outside of the cone can be assumed to be flat, but the interior is generally
curved because it contains the matter field. Presuming that the density of
matter near $V$ is zero, we can assume that the metric near $V$ is flat.
The worldlines stemming from $V$ are called {\em polar geodesics}.

\begin{figure}[!ht]
\centering
\mbox{%
\begin{minipage}{0.35\textwidth}
\includegraphics[scale=0.68]{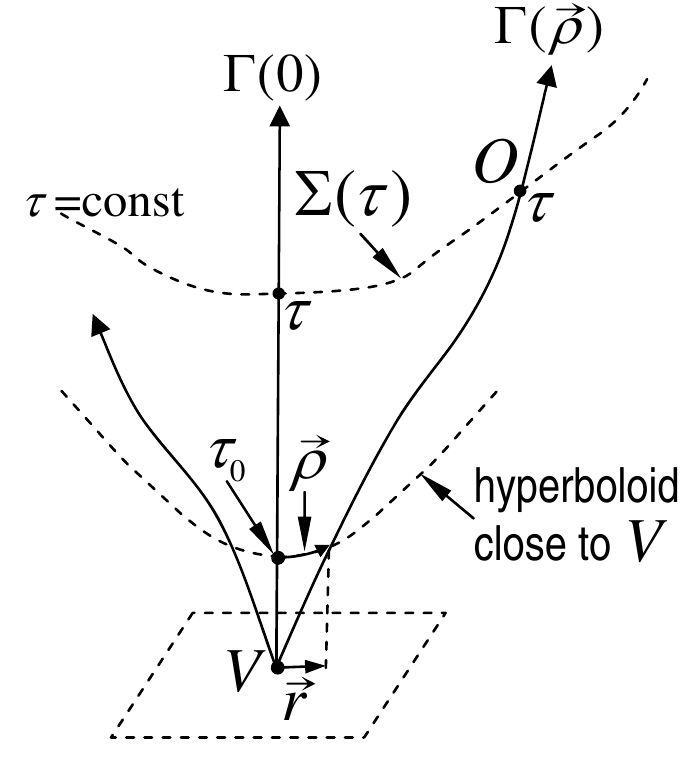}
\end{minipage}%
\begin{minipage}[c]{0.6\textwidth}
\caption{\small  Geodesics stemming from a point $V$ of a spacetime manifold $\mathfrak M$
and spanning the interior of a future cone of origin $V$ can be parameterized by
polar--hyperbolic coordinates $\{\tau, \vec\rho\,\}$. This is possible because any
geodesic of this type, $\Gamma(\vec\rho\,)$, is one to one with its direction
$\vec\rho\,=\{\varrho, \theta,\phi\}$ at~$V$. Kinematic time $\tau$ of an event
$O\!\in\!\Gamma(\vec\rho\,)$ can be defined as~the length of geodesic segment $VO$.
3D--surface $\Sigma(\tau)$ is the locus of all comoving observers synchronized at $\tau$.}
\label{geodesics}
\end{minipage}
}
\vspace{-2mm}
\end{figure}

In Fig.\,\ref{geodesics} is shown how a future cone of general type can be
parameterized by a system of polar--hyperbolic coordinates, provided that
each polar geodesic $\Gamma(\vec\rho\,)$ is one--to--one with its direction
$\vec\rho\,$ at~$V$. In this case, any polar geodesic -- but in general only
one -- can be transformed by a suitable diffeomorphism of the spacetime into
a straight line, identified as the axis of the future cone
$\Gamma(\vec\rho_0)\!\!\equiv\! \Gamma(0)$. We can therefore define the
{\em kinematic time} $\tau$ of an event $O\in\Gamma(\vec\rho\,)$ as the
length of geodesic segment $VO$; then the {\em hyperbolic angle} $\varrho\,$,
($0\le\varrho\le \infty$), as the derivative with respect to $\tau$ at
$\tau=0$ of the length of hyperboloidal arc between $\Gamma(0)$ and
$\Gamma(\vec\rho\,)$; lastly, we indicate by $\{\theta, \phi\}$ the Euler
angles of the projection $\vec r$ of $\Gamma(\vec\rho\,)$ onto the 3D--hyperplane
orthogonal to $\Gamma(0)$ at $V$. Since in the neighborhood of $V$ the metric is
Minkowskian, we can put $\vec\rho=\{\varrho,\theta,\phi\}$ and $\vec\rho_0=\{0, 0, 0\}$.
\vspace{-2mm}

Since along a polar geodesic we have $d\tau/ds =1$ and $\vec\rho= \hbox{constant}$, we can write
the squared line element of the conical spacetime as $ds^2 = d\tau^2 -\tau^2 \,\gamma_{ij}
(\tau, \vec\rho\,)\,d\rho^i\, d\rho^{\,j}$, where $i, j =1,2,3$, and impose the local--flatness
conditions near $V$:
\begin{eqnarray}
\label{initcond}
\!\!\!\!\lim_{\tau \rightarrow 0}\gamma_{11} = 1;\,\,\, \lim_{\tau \rightarrow 0}\gamma_{22}=
(\sinh\rho)^2;\,\,\,\lim_{\tau \rightarrow 0}\gamma_{33} =(\sinh\rho\,\sin\vartheta)^2;\,\,\,
\lim_{\tau \rightarrow 0}\gamma_{ij}=0\,\, (i\neq j).
\end{eqnarray}

Therefore, denoting the spacetime parameters $\{\tau, \vec\rho\,\}$ as $x$, we can write the
components

\noindent of the metric tensor as $g_{00}(x)=1$, $g_{0i}(x)=0$, $g_{ij}(x) = \gamma_{ij}(\tau, \vec\rho\,)$, and
the determinant of $\big[\gamma_{ij}(\tau, \vec\rho\,)\big]$ as $\gamma(\tau, \vec\rho\,)$. Hence, the volume
element is $\sqrt{-g(x)}\,d^4x \equiv \tau^3\!\sqrt{\gamma(\tau, \vec\rho)}\,d\varrho\,d\theta\,d\phi\,d\tau$.

Besides, denoting the inverse of matrix $\big[\gamma_{ij}(\tau, \vec\rho\,)\big]$ as
$\big[\gamma^{ij}(\tau, \vec\rho\,)\big]$ and the covariant derivatives with respect
to $x^\mu$ as $D_\mu$, we can write the squared gradient of a smooth scalar function
$f(\tau, \vec\rho\,)$ and the Beltrami--d'Alembert operator acting on $f(\tau, \vec\rho\,)$
respectively as
\begin{eqnarray}
\label{polarsqrgrad}
&&\hspace{-17mm}(D^\mu f)\, D_\nu f   = g^{\mu\nu}(\partial_\mu f)\, \partial_\nu f =
(\partial_\tau f)^2 -\frac{1}{\tau^2}\gamma^{ij}\!\big(\tau,\vec\rho\,\big)
(\partial_i f)\,\partial_j f\,; \\
\label{polarbeltdalemb}
&&\hspace{-17mm}D^2f  = \frac{1}{\sqrt{-g}}\,\partial_\mu\big(\sqrt{-g}
\,g^{\mu\nu}\partial_\nu f\big) = \partial_\tau^2 f + \partial_\tau\ln\big(\tau^3\!
\sqrt{\gamma\,}\,\big)\partial_\tau f -\frac{1}{\tau^2\sqrt{\gamma\,}}\,
\partial_i\big(\sqrt{\gamma\,}\gamma^{ij}\partial_j f\big).
\end{eqnarray}

We can easily verify that, if the metric--tensor matrix has the form
\begin{equation}
\label{gmunumatrix}
\big[g_{\mu\nu}(\tau, \vec\rho\,)\big] =\hbox{diag}\big[1, - \tau^2, -
\tau^2 (\sinh\varrho)^2,- \tau^2(\sinh\varrho^2 \sin\theta)^2\bigr]\,,
\end{equation}
then the squared line--element $ds^2=  (dx^0)^2- (dx^1)^2- (dx^2)^2- (dx^3)^2$ translated
to hyperbolic coordinates takes the form $ds^2 = d\tau^2 - \tau^2\big[d\varrho^2 +
(\sinh\varrho)^2 d\theta^2 +(\sinh\varrho\sin\theta)^2\big]$. Correspondingly, the
volume element $d^4x$ takes the form
$$
dV(\tau, \vec\rho\,)=\sqrt{-g(\tau, \vec\rho\,)}\,d\varrho\,d\theta\,d\phi\,d\tau
\equiv\tau^3 d\Omega(\vec\rho\,)\,d\tau;\,\,\,\hbox{with }\,d\Omega(\vec\rho\,)=
\big(\!\sinh\varrho\big)^2\!\sin\theta\,d\varrho\,d\theta\,d\phi\,,
$$
showing that the determinant of matrix $\big[g_{\mu\nu}(\tau, \vec\rho\,)\big]$
is $g(\tau, \vec\rho\,)=-\tau^6\big(\sinh\varrho\big)^4\!\sin\theta^2$ and the
volume element of the {\em unit hyperboloid} $\Omega$, i.e., the hyperboloid
at $\tau=1$, is $d\Omega(\vec\rho\, )$.

The squared gradient of a scalar function $f\equiv f(\tau, \vec\rho\,)$
consistent with this metric is
\begin{equation}
\label{hyperbsquaregrad} g^{\mu\nu}(x)(\partial_\mu f)\,\partial_\nu f =
(\partial_\tau f)^2 - \frac{1}{\tau^2}\bigg[(\partial_\rho f)^2 +
\frac{(\partial_\theta f)^2}{(\sinh \rho)^2} + \frac{(\partial_\phi
f)^2}{(\sinh\rho\, \sin \theta)^2}\bigg]\,,
\end{equation}
and the Beltrami--d'Alembert operator applied to $f(x)\equiv f(\tau, \vec\rho\,)$ is
\begin{equation}
\label{hyperbdalambert}
D^2 f(x)\equiv\frac{1}{\sqrt{-g(x)}}\,\partial_\mu\Bigl[\!\sqrt{-g(x)}\,
g^{\mu\nu}(x)\,\partial_\nu f(x)\Bigr]\!= \partial_\tau^2 f(x) +
\frac{3}{\tau}\,\partial_\tau f(x)- \Delta_\Omega f(x),
\end{equation}
where is evident that $(3/\tau)\,\partial_\tau f(x)$ works as a frictional term. In this
equation,
\begin{equation}
\label{unitlaplop}
\Delta_\Omega\,f \equiv \frac{1}{\tau^2\,(\sinh\varrho)^2} \bigg\{\partial_\varrho
\big[(\sinh\varrho)^2\partial_\rho f\big]+\frac{1}{\sin\theta}\,\partial_\theta
(\sin\theta\, \partial_\theta f) +\frac{1}{(\sin\theta)^2}\,\partial^2_\phi f\bigg\}
\end{equation}
represents the Laplacian operator applied to $f(\tau, \vec\rho\,)$.

\newpage

\subsection{Cylindrical and conical spacetimes in retrospect}
\label{spacetimes}
The SMMC encodes the properties of the expanding universe
on the large scale in the Robertson--Walker (RW) metric
\begin{equation}
\label{CylindrEq}
ds^2 = dt^2 -a(t)^2 \big(dr^2 + r^2\sin\theta\,
d\theta^2+r^2\sin\theta\,\cos\theta\,d\phi^2\big),\quad  t \ge 0.
\end{equation}
Here $t$ is the common proper time of ideal observers moving along worldlines
orthogonal to a starting 3D--hyperplane $\Sigma_0$, and $\vec r =\{r, \theta, \phi\}$
are standard polar coordinates. The  expansion rate of the universe is determined by
the Hubble law $H(t)= \dot a(t)/a(t)$, where $\dot a(t)$ is time derivative.
Note that if $a(t)$ is multiplied by a constant $c$, $H(t)$ remains the same; it
is customary to choose $c$ so that $a(t)=1$ today. $H(t)$ and the energy density
of the matter field $\rho(t)$ are related by the gravitational equation $3\,H(t)^2
= \kappa\,\rho(t)$.

It is known that, if the an expanding universe is seen to be isotropic by all
comoving observers, then it is homogeneous \cite{PEACOCK}. Since the SMMC
assumes that the state of the universe on the large scale is homogeneous in
each spacelike hyperplane, the spacetime foliates into a set of parallel
3D--hyperplanes. For this reason, it can be called {\em cylindrical}.

Instead, CGR encodes the properties of the universe on the large scale in
the metric
\begin{equation}
\label{ConicalEq}
ds^2 = d\tau^2 -\tau^2a(\tau)\,\big(d\varrho^2 + \sinh\theta\, d\theta^2+
\sinh\theta\,\cos\theta\,d\phi^2\big),\quad  \tau \ge 0.
\end{equation}
Here $x= \{\tau, \varrho, \theta, \phi\}$ are the polar--hyperbolic coordinates
described in the previous section and the expansion factor $a(\tau)$ is the same
in each spacelike hyperboloid. Since the history of the universe is confined to
a future cone and the state of the matter field on the large scale is homogeneous
in each spacelike hyperboloid, the spacetime can be defined {\em conical}.

However, in CGR the expansion rate of the universe is not only determined by
the energy density of the matter field, as in the SMMC, but also by the
dynamics of the vacuum state. These additional properties will be treated
in \S\,\ref{THREEWAYS} and \S\,\ref{CGRafterBB}.

\begin{figure}[!h]
\centering
\includegraphics[scale=1, trim=0 3mm 0 4mm, clip]{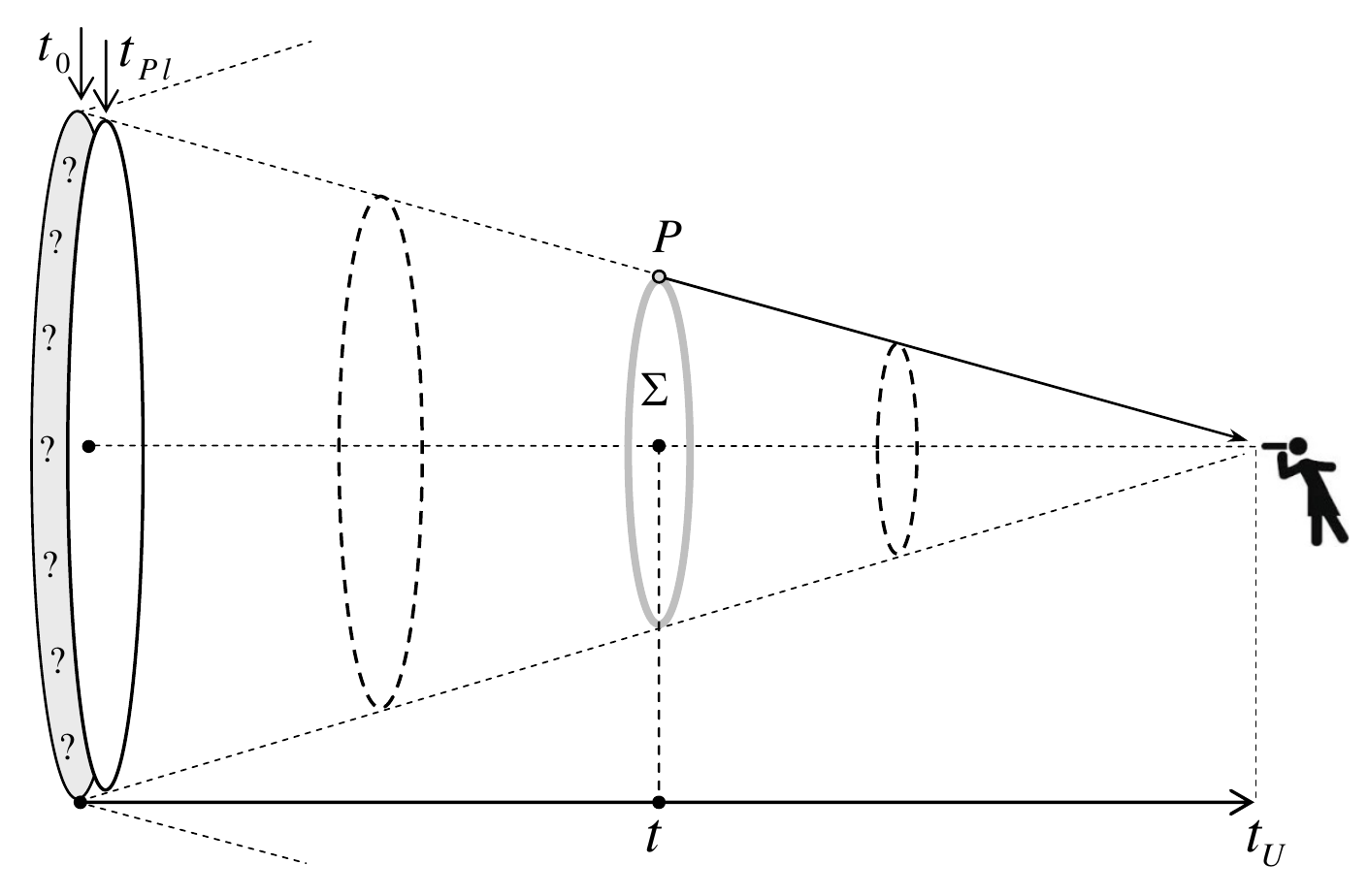}
\caption{\small{\em The flat cylindrical spacetime}. According to the
SMMC, the initial energy of the universe is uniformly concentrated
in a thin, infinitely extended spacetime layer orthogonal to the time
axis at the mythic Planck time $t_0\equiv t_{Pl}$. What happens in this
layer is unknown. If the initial distribution of matter were not
uniform, the state of the universe would change unpredictably in the
course of time. At the present universe age $t_U$, a comoving observer
can only see a luminous body $P$ if this lies in the intersection of its
own past light--cone and the hyperplane $\Sigma$ orthogonal to its own
worldline at time $t$ (thick gray circle). The time taken by a light
ray to travel from $P$ to the observer is equal to $t_U\!-t$. Therefore,
the physical structure of the universe can only be inferred by observing
the celestial bodies that have existed on the past light cone.}
\label{cylSpacetime}
\end{figure}

\begin{figure}[!ht]
\centering
\includegraphics[scale=1, trim=0 0 0 1mm, clip]{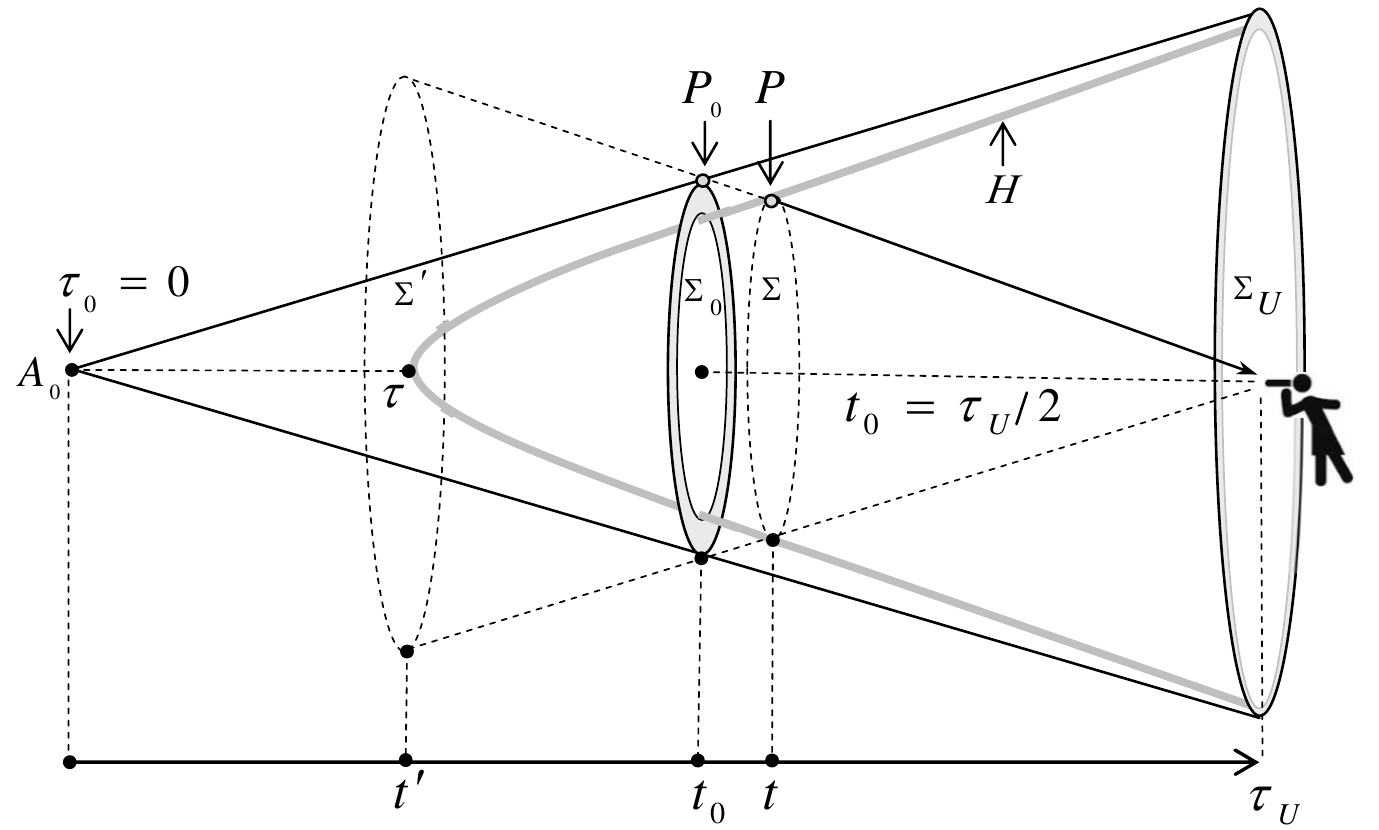}
\caption{\small {\em The flat conical spacetime ${\cal C}_\odot$}. The
spacelike surfaces are hyperboloids with vertices lying on the worldline
of an observer living today at universe age $\tau_U$. A star located
at a point $P$ of a 3D--hyperplane $\Sigma$, which emits light at proper
time $t$, belongs to the intersection of the light cone stemming from apex
$A_0$ and a hyperboloid $H$ (thick gray line), which intercepts the
observer's worldline at a time $\tau$, even long before $t$. An observer
who interprets the data as in the SMMC believes that the light emitted by
$P$ reflects the state of the matter at a point $P_0$ existing in the
intersection of its own past light cone with the 3D--hyperplane $\Sigma_0$
at time $t_0=\tau_U/2$. Actually, the star that it really sees belongs
instead to the intersection of $H$ with the light cone of ${\cal C}_\odot$.
Similarly, an event occurred on its own worldline at $\tau$, which is tangent
to the 3D--hyperplane $\Sigma'$ at time $t'$, is believed to reflect the
state of the matter in a 3D--hyperplane $\Sigma$ at time $t$. So, it
can hardly realize that the spacetime is conical. The images captured by
its telescope are actually anamorphic projections of those really occurred
in the interior of ${\cal C}_\odot$.}
\label{conSpacetime}
\end{figure}

The importance of distinguishing between cylindrical and conical spacetimes is
that the age of the universe is differently evaluated in the two cases. In the
first, it is the common length of the worldline of ideal comoving synchronized
observers which start from the hyperplane~$\Sigma_0$, cross orthogonally the
set of parallel hyperplanes and reaches the hyperplane $\Sigma_U$ at the present
universe age $t_U$ (Fig.\,\ref{cylSpacetime}). In the second, it is the length
of the worldline stemming from the apex of the future cone at $\tau=0$,
crosses the vertices of the set of 3D--hyperboloids and reaches the
3D--hyperboloid at universe age $\tau_U$ (Fig.\,\ref{conSpacetime}).

Fig.\,\ref{cylSpacetime} describes a non--expanding (flat) cylindrical
spacetime in retrospect.

Fig.\,\ref{conSpacetime} describes a flat conical spacetime ${\cal C}_\odot$
in retrospect. In this case, the spacelike surfaces are 3D--hyperboloids
starting from the degenerate hyperboloid at $\tau_0=0$, i.e., the light--cone
of ${\cal C}_\odot$, which expand and flatten more and more along the
polar--hyperbolic geodesic of the observer to that with the vertex at the
present universe age $\tau_U$. The geometrical properties of ${\cal C}_\odot$,
in particular the expansion factor $a(\tau)$, depend on the energy density
of the universe on the large scale, which is negligible if ${\cal C}_\odot$
is flat as the spacetime represented in Fig.\,\ref{conSpacetime}; in this
case, we can set $a(\tau)=1$. This means that this figure provides only
a qualitative representation of the topological structure of CGR's
spacetime.

In CGR, as in the SMMC, the expansion factor depends on the Hubble law. So,
unless the Hubble parameter is negligible, Fig.s \ref{cylSpacetime} and
\ref{conSpacetime} do not adequately represent the spacetimes described
by Eqs (\ref{CylindrEq}) and (\ref{ConicalEq}). However, Eq (\ref{ConicalEq})
is inadequate in any case because in CGR the strength of the gravitational
attraction depends strongly on the scale factor of vacuum dynamics, the
distorting effects of which are totally ignored in Fig.\,\ref{conSpacetime}.
An adequate representation of CGR's spacetime is provided by Fig.\,\ref{Figure06}
of \S\,\ref{Kin&PropST}.

\markright{R.Nobili, Conformal General Relativity -- {\bf \ref{mainreasons}} The breakdown of conformal symmetry}
\section{The spontaneous breakdown of conformal symmetry}
\label{mainreasons}
Assume that the action of a quantum field system is invariant under a continuous
group of symmetries. The standard procedure for investigating the possibility
of a spontaneous breakdown of symmetry is to look for a vacuum state that is
invariant under a subgroup of the full group, called the {\em stability
subgroup} of the symmetry. The theorem of Goldstone then ensures that the
broken part of the symmetry is not simply lost, but materializes into a
vacuum excitation consisting of one or more boson fields with gapless
energy spectrum, which are called the Nambu--Goldstone (NG) bosons of
the broken symmetry.

The idea that the universe may have originated from a spontaneous breakdown
of the conformal symmetry was advanced in 1976 by Fubini \cite{FUBINI}, who
proved that such an event can occur in three different ways corresponding to
the three possible stability subgroups of the conformal group $O(2, 4)$: the
{\em Poincar\'e group} $O(1,3)$, the {\em deSitter group} $O(2,3)$ and the
{\em anti--deSitter group} $O(1, 4)$, as proven in Appendix {\bf \ref{BreakConfApp}}.
So, the number of NG--bosons is three. Two of these, $\phi(x)$ and $\varphi(x)$,
respectively associated to the first two stability subgroups, are physical
fields, but the third, $\sigma(x)$, is a scalar ghost provided with geometric
meaning. The VEV of $\phi$ is a constant, but the VEVs of $\varphi(x)$ and $\sigma(x)$
depend on $x$.

In Fubini's paper, the action integral is invariant under the group of global
conformal transformations, in which $O(2,3)$ is the stability subgroup and
$\varphi(x)$  is the NG--boson. Instead, in CGR the symmetry breaking does
not choose either $\varphi(x)$ or $\sigma(x)$, but a conformal invariant
function of these. As anticipated in \S\,\ref{introduction} and clarified
in the next subsection, these fields interact in such a way that $\varphi(x)$
behaves as a Higgs field with variable mass, while $\sigma(x)$ acts as the
promoter of spacetime inflation. For our purposes, there will be no need to
solve the motion equations of these fields, but only determine the
kinematic--time course of their VEVs from the instant of the spontaneous
breakdown of conformal symmetry to the moment at which a thermodynamic phase
transition will cause the big bang and start of the history of the universe,
as described in Appendices {\bf \ref{VacDynApp}} and {\bf \ref{ThermVacApp}}.

The inclusion of the NG--boson ghost might rise objections because it is
generally believed that such an unphysical field violates $S$--matrix
unitarity. But in a theory in which ghost modes and physical modes interact
in such a way that the total energy is bounded from below, the violation
does not occur (Ihlan \& Kowner, 2013). This indeed happens provided that
the interaction potential of $\sigma(x)$ and $\varphi(x)$ satisfies
suitable conditions.

\newpage

\subsection{The evolving vacuum of CGR and the conditions for its stability}
\label{Evolvingvac}
To investigate whether certain scalar fields $\phi_i(x)$ of a given QFT
are the NG--bosons of a spontaneously broken symmetry, we must study the
VEVs $\langle 0|\phi_i(x)|0\rangle\equiv\bar\phi_{i}(x)$. Provided that
the one--loop term of its effective action vanishes, the QFT  admits the
classical limit (see \S\,\ref{ShiftInvCGR}). Then, putting $\phi_i(x)
=\bar\phi_{i}(x)+\hat\phi_{i}(x)$ in the classical Lagrangian density
${\cal L}(x)$ of the theory, where $\hat\phi_{i}(x)$ represents the
deviation from VEV $\bar\phi_{i}(x)$, we can determine $\bar\phi_{i}(x)$
by solving the {\em vacuum stability equations}
\begin{equation}
\label{FUBINIEQ}
\partial_\mu \frac{\delta {\cal L}(x)}{\delta [\partial_\mu \hat\phi_{i}(x)]} -
\frac{\delta {\cal L}(x)}{\delta \hat\phi_{i}(x)}\bigg|_{\hat\phi_{i}=0} =0\,,
\end{equation}
The specific $x$--dependencies of $\bar\phi_{i}(x)$ tells us which part
of the symmetry group is broken.

In principle, any solution to Eq (\ref{FUBINIEQ}) can be accepted. However,
if we presume that the vacuum state is homogeneous and isotropic, the
$x$--dependencies must be coherent with this assumption. For example, if
the spacetime is Minkowskian, all $\bar\phi_{i}$ are constant because
the vacuum state is Lorentz--invariant. In the conventional approach, we
``choose'' what subgroup of the symmetry group should survive, but in the
mechanism of spontaneous symmetry breaking it is ${\cal L}(x)$, together
with the symmetry conditions for the vacuum state, which decides,
through Eq (\ref{FUBINIEQ}), in which way it wants to be broken.

Consider, for example, the action integral ${\cal A}^H =\int\!
{\cal L}^H(x)\,dx^4$, where
\begin{equation}
\label{LHvarphi}
{\cal L}^H(x) = \frac{1}{2}\,\big[\partial^\mu \phi(x)\big]\partial_\mu
\phi(x)-\frac{\lambda}{4}\Big[\phi(x)^2 - \frac{\mu^2_H}{2\lambda} \Big]^2
\end{equation}
is the classical Lagrangian density of a Higgs field $\phi(x)$ of mass
$\mu_H$ and self--coupling constant $\lambda$. Denoting as $\bar\phi(x)$
a possible VEV of field $\phi(x)$ and as $\hat\phi(x)$ the deviation from
$\hat\phi(x)$ , we can put $\phi(x)= \bar\phi(x) + \hat \phi(x)$ and
determine $\bar\phi(x)$ by solving the stability equation
\begin{equation}
\label{HiggsEq}
-\frac{\delta {\cal A}^H}{\delta\,\hat\phi(x)}\bigg|_{\hat \phi =0} =
\,\square\,\bar\phi(x)-\lambda\,\bigg[\bar\phi(x)^2 -
\frac{\mu^2_H}{2\lambda} \bigg]\bar\phi(x) =0.
\end{equation}
However, since the vacuum is Lorentz--invariant, the correct solution is
$\bar\phi(x) = \mu_H/\sqrt{2\lambda}$, i.e., the minimum of the potential
energy density. It is therefore evident that non--trivial $x$--dependencies
may occur only if the spacetime is not Minkowskian.

This is just the case of CGR. In this theory, in fact, the history of
the universe is confined to the interior of a future cone $C_\odot$
parameterized by {\em polar--hyperbolic} coordinates $x=\{\tau, \vec\rho\,\}$
and equipped with a polar--hyperbolic metric tensor $g_{\mu\nu}(x)
\equiv g_{\mu\nu}(\tau, \vec\rho\,)$, with $g_{00}(\tau, \vec\rho\,)=1$ and
$g_{0i}(\tau, \vec\rho\,)=0$, $i=1,2,3$, as described in
\S\,\ref{futconegeom}.

If the conical spacetime is flat, the metric tensor and its volume element
simplify to
\begin{eqnarray}
\label{flatgmunu}
&&  \eta_{\mu\nu}(x)\equiv \eta_{\mu\nu}(\tau, \vec\rho\,)=\hbox{diag}
\bigl[1, -\tau^2,-\tau^2 (\sinh\varrho)^2, -\tau^2(\sinh\varrho\,
\sin\theta)^2\bigr]\,; \nonumber\\
\vspace{-3mm}
\label{flatg}
&& \sqrt{-\eta(x)}\,d^4x \equiv \sqrt{-\eta(\tau, \vec\rho\,)}\,d^4x =
\tau^3 (\sinh\varrho)^2\sin\theta d\varrho\,d\theta\,d\phi\,d\tau\,.
\vspace{-3mm}
\end{eqnarray}

Now consider the classical action
\begin{equation}
\label{actintA0}
{\cal A}_0 = \int_{C_\odot}\sqrt{-\eta(x)}\,{\cal L}_0(x)\,dx^4 \equiv
\int_\Omega\int_0^{\infty}\!\!\tau^3 {\cal L}_0(\tau, \vec\rho\,)\,d\tau\,d\Omega(\vec\rho\,) \,,
\end{equation}
where $C_\odot$ denotes a conical flat spacetime equipped with metric tensor
(\ref{flatgmunu}), $\eta(x)$ is the determinant of matrix $\big[\eta_{\mu\nu}(x)\big]$,
$d\Omega(\vec\rho\,) =(\sinh\varrho)^2\sin\theta\,d\varrho\,d\theta\,d\phi$ is
the 3D--volume element of unit hyperboloid $\Omega$ (see \S\,\ref{futconegeom}),
and assume that the Lagrangian density has the form
\begin{equation}
\label{L0ofx}
{\cal L}_0 = \frac{1}{2}\,\eta^{\mu\nu}\Big[\big(\partial_\mu\varphi
\big)\partial_\nu\varphi-\big(\partial_\mu\sigma\big) \partial_\nu\sigma\Big]
-\frac{\lambda}{4}\big(\varphi^2 -c^2\sigma^2\big)^2\,,
\end{equation}
where $c$ is an adimensional constant.

The interaction potential density in the right--hand side of Eq (\ref{L0ofx}),
$$
U(\varphi, \sigma) = \frac{\lambda}{4}\big(\varphi^2 -c^2\sigma^2\big)^2\,,
$$
is chosen in such a way that the total energy density be bounded from below
for suitable initial conditions of $\sigma(x)$ and $\varphi(x)$, as
discussed at the ends \S\,\ref{introduction} and of this subsection.

The condition for the vacuum state to be homogenous and isotropic is expressed
by the strict $\tau$--dependence of the VEVs of $\varphi(x)$ and $\sigma(x)$.
Without fear of confusion, we can denote these VEVs respectively as $\varphi(\tau)$
and $\sigma(\tau)$, and write $\varphi(x) = \varphi(\tau) +\hat\varphi(x)$,
$\sigma(x) = \sigma(\tau) +\hat\sigma(x)$, where $\hat\varphi(x)$ and
$\hat\sigma(x)$ represent the quantum excitations of the NG fields as
variations from $\varphi(\tau)$ and $\sigma(\tau)$. Of course, the VEVs
of $\hat\varphi(x)$ and $\hat\sigma(x)$ are assumed to be zero. Therefore,
the {\em vacuum--stability equations} are obtained from ${\cal A}_0$ as follows
\begin{eqnarray}
\label{deltaAvarphi}
&& \hspace{-14mm}\frac{-1}{\sqrt{-\eta(x)}}\,\frac{\delta {\cal A}_0}{\delta\,
\hat\varphi(x)}\bigg|_{\genfrac{}{}{0pt}{1}{\widehat{\varphi} =0 }
{\widehat{\sigma}=0}}= \ddot\varphi(\tau)
+ 3\frac{\dot\varphi(\tau)}{\tau}+ \lambda\big[\,\varphi(\tau)^2 -
c^2\sigma(\tau)^2\big]\varphi(\tau) =0\,,\\
\label{deltaAsigma}
&&\hspace{-14mm}\frac{1}{\sqrt{-\eta(x)}}\,\frac{\delta {\cal A}_0}{\delta\,
\hat\sigma(x)}\bigg|_{\genfrac{}{}{0pt}{1}{\widehat{\varphi} =0 }{\widehat{\sigma}=0}}
= \ddot\sigma(\tau) +3\frac{\dot\sigma(\tau)}{\tau} +\lambda\, c^2 \big[\,\varphi(\tau)^2 -
c^2\,\sigma(\tau)^2 \big]\sigma(\tau) =0\,,
\end{eqnarray}
where dot superscripts stand for $\tau$--derivatives.

The same equations can also be obtained from the variations with respect
to $\varphi(\tau)$ and $\sigma(\tau)$ of classical action
\begin{eqnarray}
\label{deltabarA0}
{\cal A}_{\hbox{\tiny cl}} \equiv {\cal A}_0\Big|_{\genfrac{}{}{0pt}{1}{\widehat{\varphi}
=0}{\widehat{\sigma}=0}} &=&\int_{C_\odot}\sqrt{-\eta(x)}\,{\cal L}_0\{\varphi(\tau),
\sigma(\tau)\}\,d^4x \equiv\nonumber\\ && \Omega\int_0^{\infty} \frac{\tau^3}{2}
\,\Big\{\dot\varphi(\tau)^2-\dot\sigma(\tau)^2-\frac{\lambda}{2} \big[\varphi(\tau)^2
- c^2\sigma(\tau)^2\big]^2\Big\}\,d\tau\,,
\end{eqnarray}
where $\Omega$ is the (infinite) volume of the unit hyperboloid of $C_\odot$.

These equations clearly show that, if $\sigma(\tau)$ evolves to a constant value
$\sigma_0$, Eq (\ref{deltaAvarphi}) describes the evolution of a Higg's field
of vacuum expectation $\varphi_0 = \lambda\, c\,\sigma_0$ and mass
$\mu_H = 2\lambda\, c\,\sigma_0$, homogeneously filling the hyperboloids
of $C_\odot$. This  would give $c = \mu_H/2\lambda\,\sigma_0$.

Of course, the integration of Eqs (\ref{deltaAvarphi}) and (\ref{deltaAsigma})
needs appropriate initial conditions for $\varphi(\tau)$ and $\sigma(\tau)$.
As for $\dot \varphi(\tau)$ and $\dot\sigma(\tau)$, these must vanish at
$\tau=0$, since otherwise the frictional terms in the right--hand sides
of the above equations would be initially infinite. These conditions are
necessary to control the time course of $\varphi(\tau)$ and $\sigma(\tau)$
and the excursions of their respective amplitude ranges. Note that the
frictional terms $3\,\dot \varphi(\tau)/\varphi(\tau)$ and $3\,\dot
\sigma(\tau)/\sigma(\tau)$ in the right--hand sides of the equations
play an important role in the dynamics of the vacuum state, because they
force the potential energy density
$$
\bar U(\tau) =\frac{\lambda}{4} \big[\varphi(\tau)^2 -c^2\sigma(\tau)^2\big]^2\,,
$$
to reach the minimum at $\tau\rightarrow\infty$, thus making $\varphi(\tau)-c\,
\sigma(\tau)$ converge to zero over time.

Therefore, for suitable initial values of $\varphi(0)$, $\sigma(0)$, with
$0<\varphi(0)<c\,\sigma(0)$ and $\dot \varphi(0)=\dot\sigma(0)=0$, for
$\tau\rightarrow \infty$, $\varphi(\tau)$ and $\sigma(\tau)$ converge
respectively to constant values $\varphi_0=\varphi(\infty)$ and
$\sigma_0=\sigma(\infty)$, such that $\varphi_0 = c\,\sigma_0$.

Putting $c=\mu_H/\sigma_0\sqrt{2\lambda}$ and $\alpha(\tau) =
\sigma(\tau)/\sigma(\infty)$,  $U(\tau)$
can be written as
\vspace{-1mm}
\begin{equation}
\label{VacPotEnDens}
\bar U(\tau) = \frac{\lambda}{4} \bigg[\varphi(\tau)^2 -
\frac{\mu_H^2}{2\lambda}\,\alpha(\tau)^2\bigg]^2\!.
\vspace{-1mm}
\end{equation}

Thus, provided that $\sigma(\infty)$ is finite, for $\tau\rightarrow \infty$,
$\alpha(\tau)$ and $U(\tau)$ converge respectively to 1 and to the potential
energy density of the standard Higgs field described by Eq (\ref{LHvarphi}).

Correspondingly, vacuum--stability equations (\ref{deltaAvarphi})
and (\ref{deltaAsigma}) become
\begin{eqnarray}
\label{varphieqB}
&& \ddot\varphi(\tau) + 3\frac{\dot\varphi(\tau)}{\tau} =
\lambda \bigg[\frac{\mu_H^2}{2\lambda}\,\alpha(\tau)^2-\varphi(\tau)^2\bigg]
\varphi(\tau)\,,\quad 0< \varphi(\tau) \le \frac{\mu_H}{\sqrt{2\lambda}}\,;\\
\label{sigmaeqB}
&& \ddot\alpha(\tau) +3 \frac{\dot\alpha(\tau)}{\tau} =
\frac{\mu_H^2}{2\,\sigma^2_0} \bigg[\frac{\mu_H^2}{2\lambda}\,
\,\alpha(\tau)^2-\varphi(\tau)^2\bigg]\alpha(\tau)\,,\quad 0< \alpha(\tau) \le 1\,.
\end{eqnarray}
The solutions to these equations and their discussion are deferred to Appendix
{\bf \ref{VacDynApp}}.

\subsection{Three different ways of implementing the vacuum stability equations}
\label{THREEWAYS}
In the previous subsection, in order to fulfil the conditions for the spontaneous
breakdown of conformal symmetry, the stability equations of the vacuum state are
derived from an action (\ref{deltabarA0}) integrated over a flat conical spacetime
equipped with polar--hyperbolic metric tensor
$$
\eta_{\mu\nu}(\tau, \vec\rho\,)=\hbox{diag}\bigl[1, -\tau^2, -\tau^2
(\sinh\varrho)^2,-\tau^2(\sinh\varrho\,\sin\theta)^2\bigr].
$$

The squared line--element of which is then
$$
ds^2= d\tau^2 - \tau^2\big(d\varrho^2 +\sinh\varrho^2d\theta^2+
\sinh\varrho^2\sin\theta^2 d\phi^2\big).
$$
This is called the {\em kinematic--time representation} because it is the
analog of the namesake representation introduced by Brout {\em et al.} in
1978, who were the first to introduce a polar--hyperbolic metric and a
ghost scalar field as the promoter of spacetime inflation.

An equivalent representation is obtained by carrying out on $\eta_{\mu\nu}
(\tau, \vec\rho\,)$, and on all other local quantities of the theory, a general
Weyl transformation with scale factor $e^{\beta(\tau)}\equiv \alpha(\tau)$, where
$\alpha(\tau)$ is the scale factor appearing in Eq (\ref{sigmaeqB}). Therefore,
since $\eta_{\mu\nu}(\tau, \vec\rho\,)$ has length--dimension, $2$ we have
$\widehat{\eta}_{\mu\nu}(\tau, \vec\rho\,) =\alpha(\tau)^2\,\eta_{\mu\nu}
(\tau, \vec\rho\,)$ or, in detail,
\begin{equation}
\label{hatetamunu}
\widehat{\eta}_{00}(\tau, \vec\rho\,) =
\alpha(\tau)^2,\,\,\, \widehat{\eta}_{0i}(\tau, \vec\rho\,)=0,\,\,\,
\widehat{\eta}_{ij}(\tau, \vec\rho\,)=\alpha(\tau)^2 \eta_{ij}
(\tau, \vec\rho\,)\,.
\end{equation}
and, since $\eta^{\mu\nu}(x)$ has length--dimension $-2$, we shall have
$\widehat{\eta}^{\,\,\mu\nu}(x)=\alpha(\tau)^{-2}\eta^{\mu\nu}(x)$.

It is therefore evident that coordinate system (\ref{hatetamunu}) is not
polar--hyperbolic. In the following, all the quantities transformed in this
way will be superscripted by a hat. Thus, for example, the squared
line--element constructed with $\widehat{\eta}_{\,\,\mu\nu}(\tau, \vec\rho\,)$
shall be written as $d\widehat{s}^{\,\,2}=\alpha(\tau)^2 ds^2$, the conical
spacetime as $\widehat{C}_\odot$, and we shall write $\widehat{\varphi}
(\tau)=\varphi(\tau)/\alpha(\tau)$ and $\widehat{\sigma}(\tau) =
\sigma(\tau) \alpha(\tau)\equiv\sigma_0$, because $\varphi(\tau)$ and
$\sigma(\tau)$ have length--dimension $-1$. This representation is the
analog of the {\em conformal--time representation} used in the SMMC in
alternative to the so--called {\em proper--time representation} (Peacock,
1999; Mukhanov, 2005). Recall that in GR {\em proper time} means the time
measured by (ideal) comoving observers equipped with synchronized clocks.

In CGR, the analog of the proper--time representation is obtained by
modifying the conformal--time representation by defining the
{\em proper--time} $\widetilde{\tau}$ and its differential
$d\widetilde{\tau}$.

First, we perform a Weyl transformation with the scale factor
$\alpha(\tau)$ provided by Eq (\ref{sigmaeqB}), then we define the
{\em proper--time} $\widetilde{\tau}$ and its differential
$d\widetilde{\tau}$ as
\begin{equation}
\label{tau2tilddtau}
\widetilde{\tau\,}(\tau) = \int_0^{\tau} \alpha(\tau')\,d\tau'\,,
\quad d\widetilde{\tau} = \alpha(\tau)\,d\tau\,,
\end{equation}
the inverse of which are defined by
\begin{equation}
\label{ttau2ctau}
\tau(\widetilde{\tau}) = \int_0^{\tilde\tau}
\frac{d\widetilde{\tau\,}'}{\widetilde{\alpha}
(\widetilde{\tau\,}')}\,,\quad d\tau =
\frac{d\widetilde{\tau}}{\widetilde{\alpha}(\widetilde{\tau})}\,.
\end{equation}
This operation can be carried out without problems because function
$\alpha(\tau)$ is monotonic. As shown in Figs.\,\ref{VacStFig5}A and
\ref{VacStFig5}B of Appendix {\bf \ref{VacDynApp}}, $\alpha(\tau)$ has
a pronounced sigmoidal profile, therefore, compared to $\tau$, the
initial tract of the proper--time scale is strongly compressed.

We can express $\tau$ and $\alpha(\tau)$ as functions of $\widetilde{\tau}$
by writing $\tau=\tau(\widetilde{\tau\,})$ and $\alpha(\tau)=
\alpha[\tau(\widetilde{\tau\,})] \equiv\tilde \alpha(\widetilde{\tau,})$.
More generally, we can express spacetime parameters $x\equiv\{\tau,\vec\rho\,\}$
as functions of proper--time parameters $\widetilde{x} =\{\widetilde{\tau},
\vec\rho\,\}$, by writing $x = x(\widetilde{x\,})$. In particular, we can
express any adimensional function $f(x)$ as a function of $\widetilde{x}$,
by writing $\widetilde{f}(\widetilde{x\,})\equiv f[x(\widetilde{x\,})]$.

We can easily prove that the derivative of $\widetilde{f}(\widetilde{\tau\,})$
with respect to $\tau$ is related to that of $f(\tau)$ with respect to
$\widetilde{\tau}$ by equation chain
\begin{equation}
\label{dertildef}
\partial_\tau \widetilde{f}(\widetilde{\tau\,}) = \partial_{\widetilde{\tau}}
\widetilde{f}(\widetilde{\tau\,})\,\frac{d\widetilde{\tau}(\tau)}{d\tau} =
\alpha(\tau)\,\partial_{\widetilde{\tau}}\widetilde{f}(\widetilde{\tau\,})\equiv
\widetilde{\alpha\,}[\tau(\widetilde{\tau\,})]\,\partial_{\widetilde{\tau}}
\widetilde{f}(\widetilde{\tau\,})\equiv \widetilde{\alpha}(\widetilde{\tau\,})\,
\partial_{\widetilde{\tau}}\widetilde{f}(\widetilde{\tau\,})\,.
\end{equation}

In the following all local quantities transformed in this way will be
superscripted by a tilde. Thus, for example, the proper--time representation
of the metric tensor will be written as $\widetilde{\eta}_{\mu\nu}(\tilde x)
\equiv \widetilde{\eta}_{\mu\nu}(\tilde \tau, \vec\rho\,)$ or, in detail,
\begin{equation}
\label{tildegmunu}
\widetilde{\eta}_{00}(\tilde x)\equiv \widetilde{\eta}_{00}
(\tilde \tau, \vec\rho\,)= 1\quad\hbox{and}\quad \widetilde{\eta}_{ij}
(\tilde x)\equiv \widetilde{\eta}_{ij}(\tilde \tau, \vec\rho\,)
= \tilde\alpha(\tilde \tau)^2 \eta_{ij}(\tilde \tau, \vec\rho\,)\,,
\end{equation}
clearly showing that the coordinate system is polar--hyperbolic. Therefore,
the proper--time representation of the squared line element, has the form
\begin{equation}
\label{dwidetildede2}
d\widetilde{s}^{\,\,2}(\widetilde{\tau}, \vec\rho\,) = d\widetilde{\tau}^2-
\widetilde{\eta}_{ij}(\widetilde{\tau}, \vec\rho\,)\,dx^i dx^j\,.
\end{equation}

It is therefore evident that in CGR the conformal--time representation is
a sort of bridge between two different polar--hyperbolic representations
of the spacetime, i.e., the kinematic--time and proper--time representations.

By applying this transformations to Lagrangian density (\ref{L0ofx}), we obtain
\begin{equation}
\label{TILDESMEPLAGR}
\widetilde{\cal L}_0(\tilde \tau)\equiv \widetilde{\cal L}_0
\big\{\widetilde{\varphi}(\widetilde{\tau\,}), \sigma_0\big\}\,,
\vspace{-1mm}
\end{equation}
because $\widetilde{\varphi}(\widetilde{\tau\,})=\varphi(\widetilde{\tau\,})
/\widetilde{\alpha}(\widetilde{\tau\,})$ and $\tilde \sigma(\tilde \tau)=
\sigma(\tilde \tau)/\tilde\alpha(\tilde\tau)= \sigma_0$.  The symbol of
Lagrangian density also is superscripted by a tilde to indicate that the
metric tensor is $\widetilde{\eta}_{\mu\nu}(\tilde x)$.

To obtain the vacuum--stability equations in the proper--time representation,
we can proceed directly by changing Eqs (\ref{varphieqB}) and (\ref{sigmaeqB})
as follows:

(1) put $\varphi(\tau)= \widehat{\varphi}(\tau)\,\alpha(\tau)$ in both
equations and combine the results, so as to obtain
\begin{equation}
\partial^2_\tau\, \widehat{\varphi}(\tau) +\bigg[\frac{3}{\tau} +
2\,\frac{\partial_\tau \alpha(\tau)}{\alpha(\tau)}\bigg]
\partial_\tau \widehat{\varphi}(\tau)=\alpha(\tau)^2\bigg(\lambda -
\frac{\mu_H^2}{2\,\sigma^2_0}\bigg)\Bigg[\frac{\mu_H^2}{2\lambda}
-\widehat{\varphi}(\tau)^2\Bigg]\widehat{\varphi}(\tau)=0\,;
\end{equation}
(2) put $\partial_\tau =\alpha(\tau)\,\partial_{\widetilde{\tau}}
\equiv\widetilde{\alpha}(\widetilde{\tau\,})\,\partial_{\widetilde{\tau}}$,
because $d\widetilde{\tau}/d\tau =\alpha(\tau)\equiv\widetilde{\alpha}
(\widetilde{\tau\,})$,  and replace everywhere $\tau$ with
$\tau(\widetilde{\tau\,})$; (3) put $\widehat{\varphi\,}
[\tau(\widetilde{\tau\,})]\equiv\widetilde{\varphi}
(\widetilde{\tau\,})$ and simplify the result, so as to obtain
\begin{equation}
\label{proptimevaceq}
\partial^2_{\widetilde{\tau}}\,\widetilde{\varphi\,}(\widetilde{\tau\,})+
3\bigg[\frac{1}{\tau(\widetilde{\tau\,})}+ \frac{\partial_{\widetilde{\tau}\,}
\widetilde{\alpha\,}(\widetilde{\tau\,})}{\widetilde{\alpha\,}(\widetilde{\tau\,})}\bigg]
\partial_{\widetilde{\tau}\,}\widetilde{\varphi\,}(\widetilde{\tau\,})=
\bigg(\lambda -\frac{\mu_H^2}{2\,\sigma^2_0}\bigg)\Bigg[\frac{\mu_H^2}{2\lambda} -
\widetilde{\varphi\,}^2(\widetilde{\tau\,})\Bigg]\widetilde{\varphi}
(\widetilde{\tau\,})\,.
\end{equation}

We see that, in passing from the conformal--time representation to the
proper--time representation, Eqs (\ref{varphieqB}) and (\ref{sigmaeqB})
together provide the motion equation of a homogeneous Higgs field of mass
$\mu_H$ and self--coupling constant $\lambda$, with an additional frictional
term $3\,(\partial_{\widetilde{\,\tau}}\ln\widetilde{\alpha\,})\,
\partial_{\widetilde{\tau}}\widetilde{\varphi\,}$.

Assuming $\mu_H\cong 126$ GeV and $\lambda\cong 0.1291$ -- in agreement
with the values provided by the SMEP -- then writing the gravitational
coupling constant as $\kappa = 1/M^2_{rP}$, where $M_{rP}\cong 2.4328\times
10^{18}$GeV is the reduced mass of Planck and, lastly, putting
$\sigma_0 = \sqrt{6/\kappa} = \sqrt{6}\, M_{rP}$, we obtain
$\mu_H^2/2\,\sigma^2_0\approx 10^{-36}$. It is therefore evident that
by replacing $\lambda-\mu_H^2/2\,\sigma^2_0$ with $\lambda$ we do not
make any appreciable error.

In Appendix {\bf \ref{VacDynApp}}, the stability equations of the dynamical
vacuum are solved numerically and graphically exemplified in both the
conformal--time and proper--time representations.

We can formalize the direct transition from the kinematic--time to the
proper--time representation as follows. Let ${\mathcal P}$ be an operator
that performs this transition: for any quantity $Q_n(x)$ or constant of
length--dimension $n$, we have ${\mathcal P}\, Q_n(x)= \widetilde{\alpha}
(\widetilde{\tau})^n \widetilde{Q}_n(\widetilde{x})$.

Therefore, since scalar fields $\sigma(\tau)$,  $\varphi(\tau)$ and constant
$\sigma_0$ have length--dimension $-1$, while $\alpha(\tau)$, $\tau$ and
$\kappa$ have respectively length--dimensions  0, 1 and $2$, we obtain
\begin{eqnarray}
\label{CalP}
&& {\mathcal P}\vspace{0.5mm}\sigma(\tau) = \sigma_0;\quad
{\mathcal P}\vspace{0.5mm}\varphi(\tau)= \widetilde{\varphi}
(\widetilde{\tau})\,\widetilde{\alpha}(\widetilde{\tau})^{-1};\quad
{\mathcal P}\sigma_0=\sigma_0\,\widetilde{\alpha}(\widetilde{\tau})^{-1}\,;
\nonumber\\
&&{\mathcal P}\alpha(\tau)=\widetilde{\alpha}(\widetilde{\tau})\,;
\quad{\mathcal P}\vspace{0.5mm}\tau= \tau(\tilde\tau)\,
\widetilde{\alpha}(\widetilde{\tau})\,;\quad{\mathcal P}\vspace{0.5mm}
\kappa =\kappa\,\widetilde{\alpha}(\widetilde{\tau})^2\,.
\end{eqnarray}
So, the first and third in sequence give ${\mathcal P}\!\big[{\mathcal P}
\vspace{0.5mm}\sigma(\tau)\big] = {\mathcal P}\vspace{0.5mm}\sigma_0 =
\sigma_0/\widetilde{\alpha}(\widetilde{\tau})\equiv \sigma_0/\alpha(\tau)
=\sigma(\tau)$.

\newpage

\subsection{Conical and goblet--shaped representations of CGR spacetime}
\label{Kin&PropST}
\begin{figure}[!ht]
\centering
\mbox{%
\begin{minipage}{0.42\textwidth}
\includegraphics[scale=0.4]{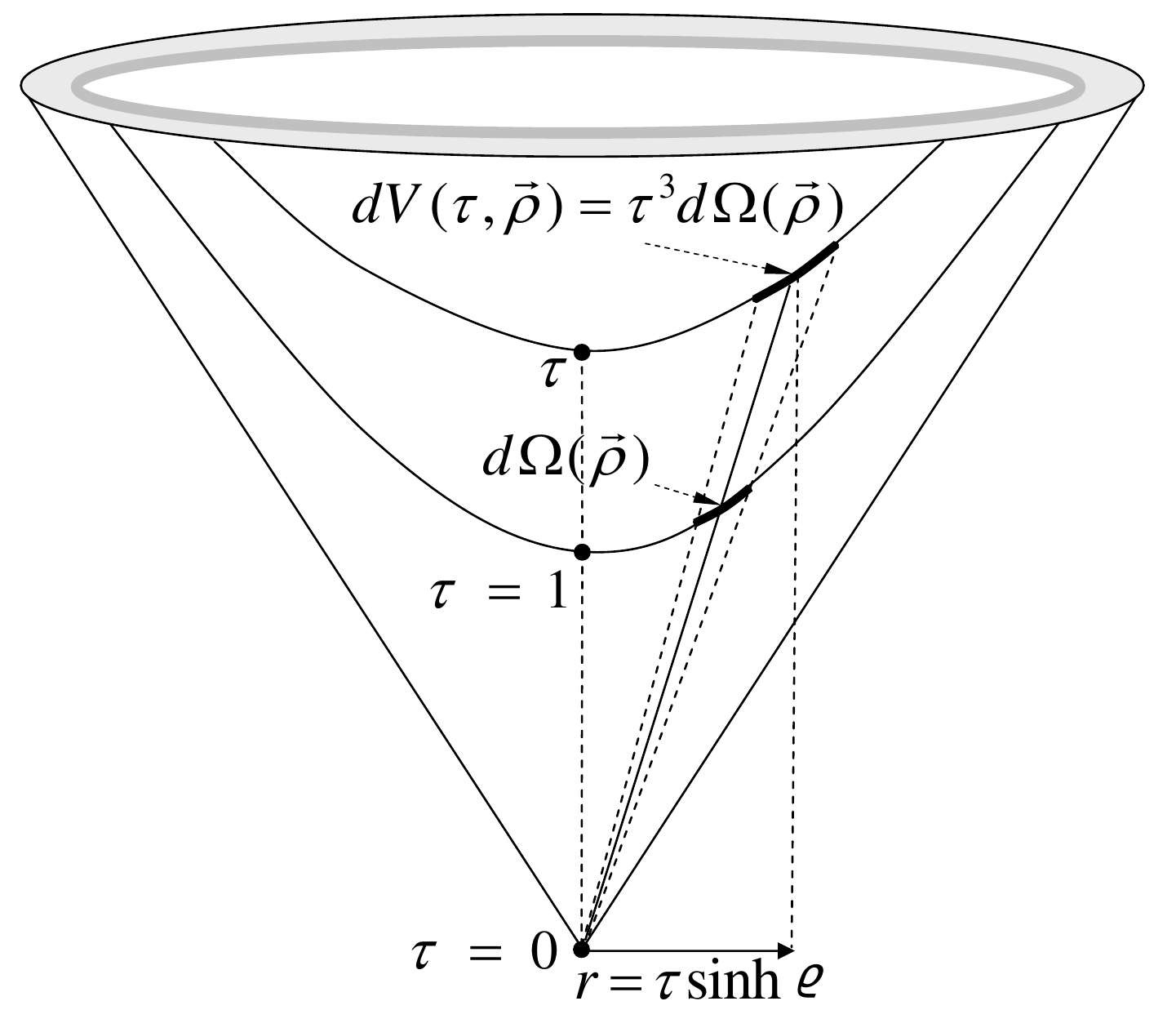}
\end{minipage}
\begin{minipage}[c]{0.5\textwidth}
\caption{\small Kinematic structure of future cone $C_\odot$ in polar--hyperbolic
coordinates: $\tau$ = kinematic time; $\vec\rho =\{\varrho, \theta, \phi\} = $
hyperbolic--Euler angles; $d\Omega(\vec\rho\,)$ = element of hyperbolic--Euler--angles
at $\tau=1$; $dV(\tau,\vec\rho\,)=\tau^3 d\Omega(\vec\rho\,)\,d\tau$ = volume--element
of the spacelike hyperboloid at $\tau$; $r$ = radial position of volume element
$dV(\tau, \vec\rho\,)$. Note ratio $dV(\tau_2,\vec\rho\,)/ dV(\tau_1,\vec\rho\,)$
evolving in time as $(\tau_2/\tau_1)^3$.
}
\label{BeforeBBFig1}
\end{minipage}
}
\end{figure}

The flat conical spacetime $C_\odot$ represented in Fig.\,\ref{conSpacetime} of
\S\,\ref{spacetimes} reflects the structure of metric--tensor matrix
\begin{equation}
\label{FRWmetmetrictens}
\big[\eta_{\mu\nu}(\tau, \vec\rho\,)\big] = \mbox{diag}\big[1,  -\tau^2,
-\tau^2\bigl(\sinh\varrho)^2,-\tau^2\bigl(\sinh\varrho\,
\sin\theta\bigr)^2\bigr]\,,
\end{equation}
where $\{\tau, \varrho,\theta,\phi\}$ are the hyperbolic--polar coordinates introduced
in \S\,\ref{futconegeom}. The details of its structure are shown in Fig.\ref{BeforeBBFig1}.

Denoting by $\sqrt{|\eta(\tau, \vec\rho\,)|\,}  = \tau^3\bigl(\sinh\varrho\big)^2\!\sin\theta$
the squared root of the matrix determinant, we can write the 3D--volume element at $\tau=1$ as
$d\Omega(\vec\rho\,) =(\sinh\varrho)^2\sin\theta\,d\varrho\,d\theta\,d\phi$, and the
4--D volume at any place and time as $d^4x= d\Omega(\vec\rho\,)\,d\tau$.

Unfortunately, this representation does not take into account that the physical
events should be ideally referred  to observers provided with their own rulers
and clocks. In the absence of any sort of matter field, it is impossible to
tell what is length--unit and how the observers could synchronize their clocks.

In the SMMC it is customary to define the reference frame of observers comoving
with the expanding universe and endowed with synchronized clocks that mark a
common proper time. We can do this because the gravitational equation of GR
allows us to establish a precise relation between the time scale and the
parameters of the matter field by equation $G_{00}(x) = 3\,H(t)^2 =\kappa\,\rho(t)$,
where $a(t)$ is the expansion factor of the universe, $\rho(t)$ is the density of
energy and  $H(t) =\dot a(t)/a(t)$ is the Hubble parameter.

In CGR, the relation between time scale and matter--field parameters cannot
be established as easily because the conformal invariant gravitational equation
is radically different.

However, to introduce the proper--time representation we can translate the metric matrix described by Eq (\ref{CalP})
directly to the metric matrix in the proper time representation by applying the operator $\cal P$ defined in the end
of the previous subsection:
\vspace{-1mm}
\begin{equation}
\label{tildemetrictensor} {\cal P} g_{\mu\nu}(\tau,\vec\rho\,) = \widetilde{g}_{\mu\nu}(\widetilde{\tau},\vec\rho\,) =
\hbox{diag}\Bigl[1, - \widetilde{c}(\widetilde{\tau\,})^2,-\widetilde{c} (\widetilde{\tau\,})^2\!\sinh\varrho^2,-
\widetilde{c}(\widetilde{\tau\,})^2\!\sinh\varrho^2 \sin\theta^2\Bigr],
\vspace{-1mm}
\end{equation}
where $\widetilde{c}(\widetilde{\tau\,}) = \tau(\widetilde{\tau\,})\, \widetilde{\alpha}
(\widetilde{\tau\,})$, $\tau(\widetilde{\tau\,})$ is the kinematic time as a function
of proper time $\widetilde{\tau\,}$ and $\widetilde{\alpha}(\widetilde{\tau\,})
\equiv \alpha[\tau(\widetilde{\tau\,})]$.

Therefore, denoting by $\sqrt{|\widetilde{g}
(\widetilde{\tau\,},\vec\rho\,)|} = \big[\tau(\widetilde{\tau\,})\,\widetilde{\alpha}
(\widetilde{\tau\,})\big]^3(\sinh\varrho)^2\sin\theta$ the squared root of the
metric--matrix determinant, we can write the $\cal P$--transform of the proper--time
spacetime as $\widetilde{C}_\odot$, the 3D--volume element at $\widetilde{\tau}$ as $\sqrt{|\widetilde{g}(\widetilde{\tau\,},\vec\rho\,)|}\,(\sinh\varrho)^2\sin\theta\,
d\varrho\,d\theta\,d\phi$, and the 4--D volume--element at any point of $\widetilde{C}_\odot$
as $d^4\widetilde{x}\equiv\big[\tau(\widetilde{\tau\,})\,\widetilde{\alpha}
(\widetilde{\tau\,})\big]^3 d\Omega(\vec\rho)\,d\,\widetilde{\tau}$.

\begin{figure}[!ht]
\centering
\mbox{%
\begin{minipage}{0.40\textwidth}
\includegraphics[scale=0.37]{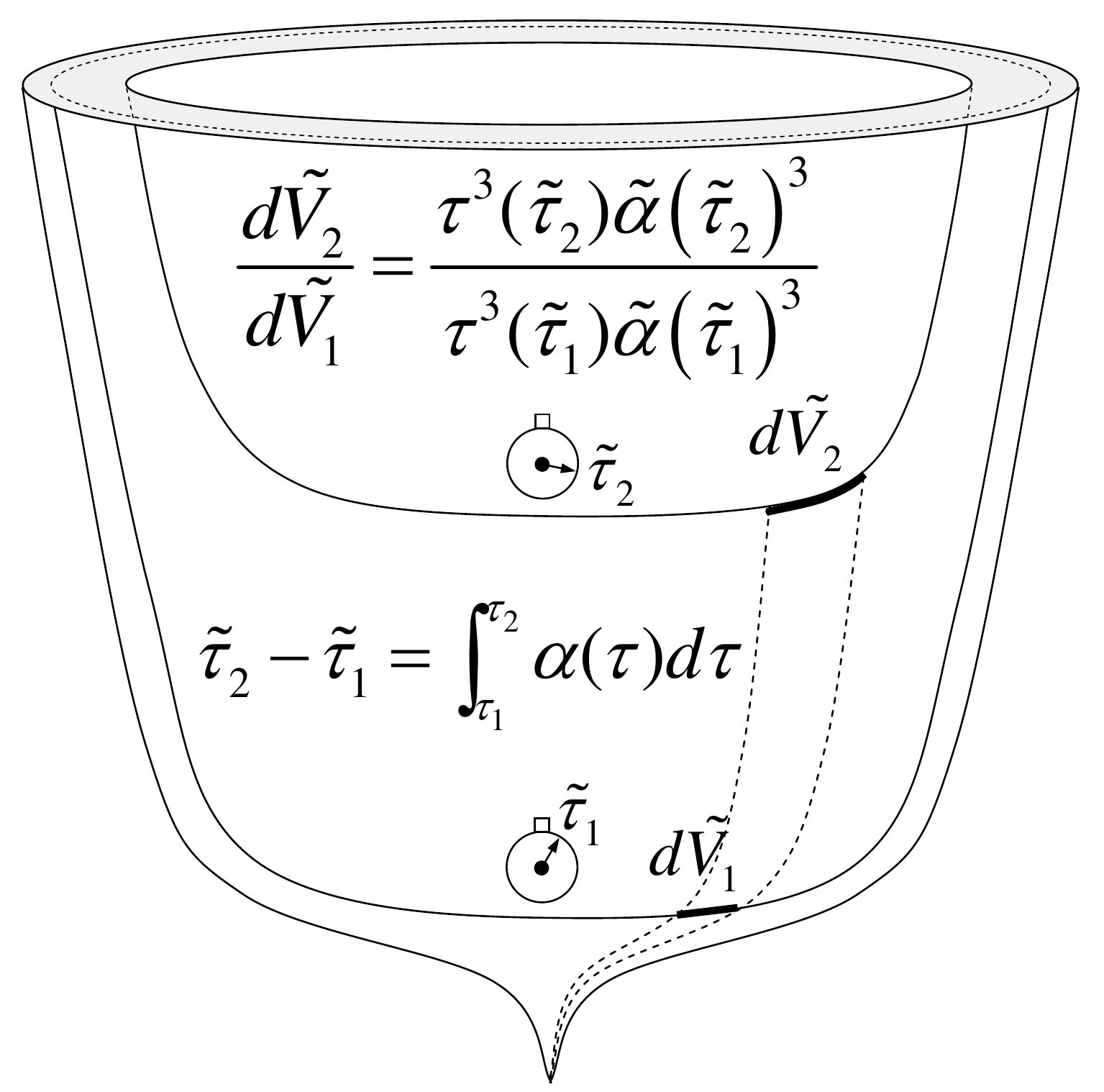}
\end{minipage}
\begin{minipage}[c]{0.57\textwidth}
\caption{\small Qualitative features of the goblet--shaped future cone $\widetilde{C}_\odot$ in proper--time
coordinates $\{\widetilde{\tau}, \vec\rho\,\}$; $\widetilde{\alpha}(\widetilde{\tau\,})= $\!scale factor of the
evolving vacuum state; $\tau(\widetilde{\tau\,})\,\tilde\alpha(\tilde\tau)$ = scale factor of metric tensor
$\widetilde{g}_{\mu\nu}(\widetilde{\tau},\vec\rho\,)$; $d\widetilde{V}_i=\widetilde{\alpha}(\widetilde{\tau}_i)^3
d\Omega(\vec\rho\,)$ $(i=1,2,3)$ = 3D--sections of a polar--geodesic tube of worldlines stemming from the future--cone
origin (dotted lines); $\big(\widetilde{V}_1/\widetilde{V}_2\big)^{1/3} p =
\tau(\widetilde{\tau\,}_1)\,\tilde\alpha(\tilde\tau_1)/\tau(\widetilde{\tau\,}_2) \,\tilde\alpha(\tilde\tau_2)$ =
linear expansion factor. Note flattening of spacelike surfaces in early epoch. } \label{BeforeBBFig2}
\end{minipage}
}
\end{figure}

Since the scale factor of vacuum dynamics causes a strong compression of the initial
tract of the time scale, $\widetilde{\alpha}(\widetilde{\tau\,})$ has the profile of
a sigmoid. Therefore, the future cone in the proper time representation has
qualitatively the goblet--shape form represented in Fig.\ref{BeforeBBFig2}.
\vspace{-1mm}
\vspace{-1mm}
Here we see very clearly that, independently of the gravitational equation, what
decides the structure of the spacetime is the relation between proper time and
kinematic time.

The partition of the spacetime into tubes of worldlines has important implications for
the propagation of conservative quantities. This holds for flows of electrical charge
and baryon number, but may not hold for flows of EM of matter because in CGR these
can be modified by the transfer of EM flows of geometry, as explained in
\S\,\ref{introduction}.

However, if the matter field in $\widetilde{C}_\odot$ is in thermodynamic equilibrium,
no work or heat can be exchanged among adjacent tubes. This means that the expansion
of the universe is adiabatic and that the entropy of the universe on the large scale
is conserved.

\newpage
Now consider the polar--hyperbolic metric tensor of a universe expanding with scale factor
$a(\tau, \vec\rho)$, $\mathbf{g}_{\mu\nu}(\tau,\vec\rho\,) = \hbox{diag}\big[1, - \tau^2
a(\tau, \vec\rho)^2, -\tau^2 a(\tau,\vec\rho)^2\sinh\varrho^2,-\tau^2 a(\tau,\vec\rho)^2
\sinh\varrho^2\sin\theta^2\big]$. Its ${\cal P}$--transform $\widetilde{\mathbf g}_{\mu\nu}
(\widetilde{\tau},\vec\rho\,) = \hbox{diag}\big\{1,-\tau(\widetilde{\tau})^2\,\widetilde{a}
(\widetilde{\tau\,},\vec\rho)^2\,\widetilde{\alpha}(\widetilde{\tau\,})^2 \big[1,
\sinh\varrho^2,\sinh\varrho^2\sin\theta^2\big]\!\bigr\}$ depends on both the scale
factor of the dynamical vacuum $\widetilde{\alpha}(\widetilde{\tau\,})$ and the expansion
factor of the universe, $\widetilde{a}(\widetilde{\tau\,},\vec\rho)$ as a function of
proper--time coordinates $\widetilde{x}\equiv\{\widetilde{\tau\,}, \vec\rho\}$.

\begin{figure}[!h]
\centering
\includegraphics[scale=0.5, trim=0 2mm 0 0, clip]{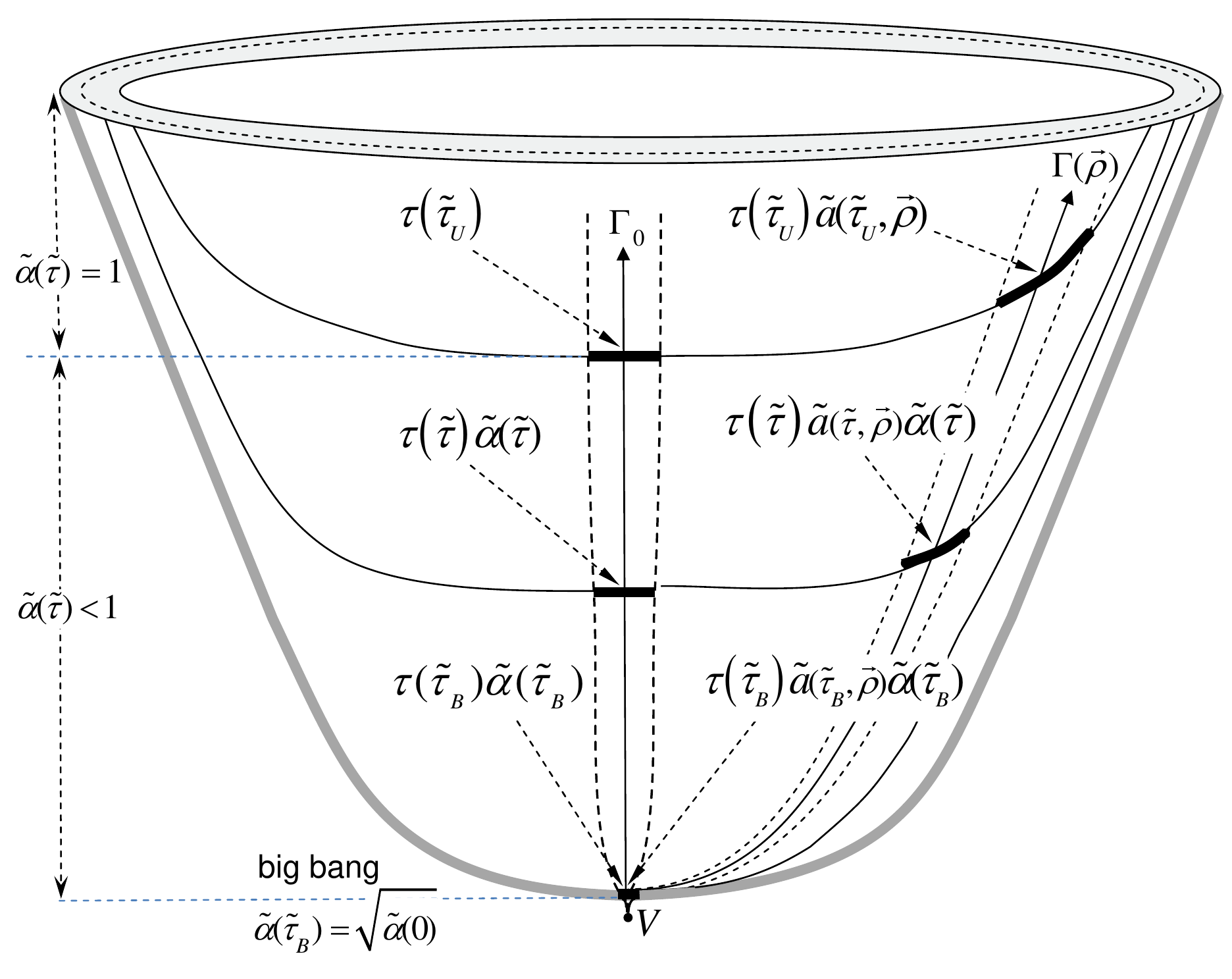} % trim=left bottom right top
\vspace{-2mm}
\caption{\small Qualitative features of the universe after big bang under the combined
action of the expansion factor of the universe $\widetilde{a}(\widetilde{\tau\,},\vec\rho)$
and of the scale factor of vacuum dynamics $\widetilde{\alpha}(\widetilde{\tau\,})$.
Proper time $\widetilde{\tau}$ ranges from big--bang time $\widetilde{\tau}_B$ (on the
bottom) to age--of--universe $\widetilde{\tau}_U$ (on the top). Unit vector $\vec\rho$
is the initial direction of a worldline stemming from vertex $V$ (arrowed solid lines).
Dashed lines flanking axial worldline $\Gamma_0$ represents a co--expanding tube of
nearby worldlines depending only on $\widetilde{\alpha}(\widetilde{\tau\,})$; those
flanking non--axial worldline $\Gamma(\vec\rho)$ denote a co--expanding tube of
nearby worldlines depending also on $\widetilde{a}(\widetilde{\tau},\vec\rho)$. Thick
black lines pointed to by diagonal dashed arrows denote the diameters of the tubes
as functions of proper time $\widetilde{\tau}$. Those of the tube wrapped around
$\Gamma_0$ vary in time as $\tau(\widetilde{\tau\,})\,\widetilde{\alpha}
(\widetilde{\tau})$, those of the tube wrapped around $\Gamma(\vec\rho)$ vary in
time as $\tau(\widetilde{\tau\,})\,\widetilde{a}(\widetilde{\tau}, \vec\rho)
\,\widetilde{\alpha}(\widetilde{\tau})$.}
\label{Figure06}
\vspace{-3mm}
\end{figure}

Therefore, differently from the spacetime if the SMMC, where the expansion factor depend
only on time, in CGR it depends also on the direction of the geodesics $\Gamma(\vec \rho)$
stemming from the origin of the conical spacetime $V$, as shown in Fig.\,\ref{Figure06}.
\vspace{-2mm}

The fact that the diameter of the tube around $\Gamma_0$ does not depend on
the expansion factor of the universe is of paramount importance for
determining the behaviors of conservative quantities in co--moving reference
frames. Let us clarify this point.

The most important fact regarding the SMMC is the relation between the
Hubble parameter $H(t)$ and the isotropy--homogeneity of energy density
$\rho(t)$ and pressure $p(t)$.

Determining the gravitational equation from the spatially flat Robertson--Walker
(RW) metric
$$
ds^2 = dt^2-a(t)^2 (dx^2 + dy^2+ dz^2)\,,
$$
we obtain the zero--zero components of gravitational tensor $G_{\mu\nu}$ and Ricci
tensor $R_{\mu\nu}$,
\begin{equation}
\label{G00&R00}
G_{00}(t) = 3 \frac{\dot a(t)^2}{a(t)^2} = \kappa\,\rho(t)\,;\quad
R_{00}(t) = -3\,\frac{\ddot a(t)}{a(t)}=\frac{\kappa}{2}\,\big[\rho(t)+3\,p(t)\big]\,,
\end{equation}
which are manifestly invariant under scale transformation $a(t)\rightarrow C\,a(t)$.
It is customary to take the constant $C$ that makes $a(t_U)=1$ at the present age of
the universe $t_U$.

Defining the Hubble parameter as $H(t)=\dot a(t)/a(t)$ and putting $L(t) = L_0\, a(t)$,
where $L_0$ is the distance between any two point of the 3D--space orthogonal to the
time axis, we obtain the Hubble law and the length--acceleration law,
\begin{equation}
\label{H&Loft}
H(t) = \frac{\dot a(t)}{a(t)} = \sqrt{\frac{\kappa\,\rho(t)}{3}}\,;\quad
\ddot L(t) = - \frac{k}{6}\, L(t)\big[\rho(t)+3\,p(t)\big]\,.
\end{equation}
If $L_0$ is the distance from a point of the time axis, we see that $L(t)=0$ is a
stagnation point of universe expansion. This means that all the worldlines sufficiently
close to the time axis do not sense the Hubble expansion. Since by a suitable gauge
transformation of the RW metric (see  \S\,\ref{GaugeInvGmunu} of Appendix
{\bf \ref{GravGaugInvApp}}) we can make any worldline of the universe to be a
time axis, each point of the spacetime is a stagnation point of universe expansion.
This means that the expansion of the universe can only be tested by astronomical
observations.

By transferring these concepts to the spacetime represented in Fig.\,\ref{Figure06},
we can easily realize that the distance of worldlines close to $\Gamma_0$ do not
depend appreciably on $\widetilde{a}(\widetilde{\tau}, \vec\rho)$, as described in
the caption. Of course, if we carry out a suitable gauge transformation of the CGR
metric, we can transform the worldline $\Gamma(\vec\rho)$, and all the worldlines
in the co--expanding tube around it, to the axial co--expanding tube directed by
$\Gamma_0$. Actually, finding the equivalent of Eqs (\ref{G00&R00}) and (\ref{H&Loft})
for CGR is a complicated task that goes beyond the goals of this paper. This point
will be further discussed in Section \ref{MachEinstPrinc}.

For the remaining part of our paper, we shall limit ourselves to use the arguments
here discussed to study how the scale factor of vacuum dynamics constraint the
entropy conservation and several important implications for the structure of CGR.

These topics will be widely described and discussed in \S\S\,\ref{EntrConserv} and
\ref{crossroad}.

\newpage

\subsection{Quantum corrections to early vacuum dynamics}
\label{QCorrToEVac}
The dynamics of the vacuum state described sofar provides only the classical
background of a quantum--theoretical scenario. To complete the picture, we must
also describe the effects of quantization. Let us recall that in
\S\,\ref{Evolvingvac}, near Eqs (\ref{deltaAvarphi})--(\ref{deltaAsigma}),
and in \S\,\ref{CGR&SMEP}, we have denoted the VEVs of the NG--boson fields
as $\widetilde{\varphi}\,(\widetilde{\tau})$ and $\widetilde{\sigma}\,
(\widetilde{\tau})\equiv\sigma_0$.  To carry out the quantization, of the
dynamical vacuum, we must evaluate the  effective Lagrangian density
${\cal L}_{\hbox{\tiny eff}}(x)$ that describes the interaction of the
quantum excitations $\hat\varphi(\widetilde{x\,})$ and $\hat\sigma(\widetilde{x\,})$
of the NG--boson fields $\widetilde{\varphi}\,(\widetilde{x\,})$ and
$\widetilde{\sigma}\,(\widetilde{x\,})$, respectively regarded as deviations
from the NG--boson VEVs with all other fields entering into play after the
decay of the Higgs bosons. This task is greatly facilitated if
${\cal L}_{\hbox{\tiny eff}}(x)$ can be well--approximated by the classical
Lagrangian density of CGR ${\cal L}_{\hbox{\tiny cl}}(x)$; which is indeed
possible for the reasons explained in \S\S\,\ref{AsympConfInvinCGR}--\ref{ShiftInvCGR}.

The effects of quantum excitations can then be evaluated by expanding
$\hat\varphi(\widetilde{x\,})$ and $\hat\sigma(\widetilde{x\,})$ in series
of a creation--annihilation operators, so that the quantum excitations are
treated as free fields. The effects of quantization can then be evaluated
by applying the methods of {\em adiabatic} and {\em sudden} approximations
described in Ref.~\cite{MESSIAH}.

Let us summarize the most significant aspects of this methods. As shown in
Fig.\,\ref{VacStFig4}B of Appendix {\bf \ref{VacDynApp}}, the Higgs--field
VEV $\widetilde{\varphi}(\widetilde{\tau\,})$ changes very slowly and
smoothly during the time interval from $\widetilde{\tau}=0$ to big--bang
time $\widetilde{\tau}_B$. In these circumstances, the temperature of the
evolving vacuum can be assumed to be nearly zero. The effects of quantization
can then be calculated in the {\em adiabatic approximation}, so that the
quantum--field amplitude of the Higgs field is well--approximated by
a simple phase factor depending on $\widetilde{\tau}$ \cite{BERRY}.

By contrast, in a small time interval across $\widetilde{\tau}_B$, the
unitary operator that acts on $\widetilde{\varphi}\,(\widetilde{x\,})$
does not depart appreciably from 1. So, the amplitude remains equal to
$\widetilde{\varphi}(\widetilde{\tau\,})$ all over the 3D--hyperboloid
$\Sigma_B$; the {\em sudden approximation} is just this. Once reached
the peak of amplitude exemplified in Fig.\ref{VacStFig4}B, the state
of the Higgs field becomes a ``democratic'' superposition of
$\hat\varphi(\widetilde{x\,})$ and $\hat\sigma(\widetilde{x\,})$
amplitudes with arbitrary phase, which collapses quickly into a gas
of Higgs bosons in thermal equilibrium at a certain temperature $T_B$.

Soon after, all Higgs bosons lying in $\Sigma_B$ decay into the SMEP inventory,
which evolves adiabatically during the expansion of the universe. The best
approximation to this stage of CGR is a thermodynamic expansion of the
matter field, which remains in equilibrium at a temperature decreasing
nearly uniformly in each evolving hyperboloid (see \S\,\ref{unilargescale}).

\newpage

% =================================================================================================

\markright{R.Nobili, Conformal General Relativity -- {\bf \ref{CGR&SMEP}} SMEP and gravity in CGR}
\section{How to include SMEP and gravity in CGR}
\label{CGR&SMEP}
In \S\,\ref{introduction}, CGR has been introduced as a theory in which the
action remains invariant under the group of conformal diffeomorphisms. Except
for the requirement that its general form be determined by the spontaneous
decay of conformal symmetry to metric symmetry, no other attempt has
been done to understand its architecture.

In \S\,\ref{Evolvingvac}, we focused on the invariance of CGR under the subgroup
of  {\em global conformal  transformations} $O(2, 4)$. The properties of this are
described in Appendix {\bf \ref{BreakConfApp}}. The global character of the subgroup
is implicit in the fact that the classical action ${\cal A}_0$ described in Eq
(\ref{actintA0}) is defined over a flat conical spacetime. This means that ${\cal A}_0$
represents the interaction of two classical NG bosons dissolved in an empty spacetime:
a physical scalar field $\varphi$ associated with the {\em deSitter subgroup}$O(2,3)$ -- to
be identified as a classical Higgs field -- and a ghost scalar field $\sigma$ associated
with the {\em anti--deSitter subgroup} $O(1, 4)$ -- to be identified as the agent of
spacetime inflation. The motion equations of these fields, described by Eqs
(\ref{deltaAvarphi}) and (\ref{deltaAsigma}), represent the conditions for
the stability of the dynamical vacuum of CGR.

In this section, we want to enrich this theoretical background by studying a
way to include the SMEP and the gravitational field in CGR. We will do this
in the kinematic--time representation, being it clear that the inclusion of
the SMEP is possible only in the latter stage of the evolution of CGR towards GR.

Differently form the SMEP, where the spacetime is Minkowskian, hence cylindrical,
and the vacuum state $|\Omega\rangle$ is independent of time, the spacetime of
CGR is conical and the vacuum state depends on kinematic time $\tau$, so we shall
denote it as $|\Omega(\tau)\rangle$. For these reasons, the NG--boson VEVs of
$\varphi(x)$ and $\sigma(x)$, respectively $\varphi(\tau)\equiv \langle\Omega(\tau)|
\,\varphi(x)|\Omega(\tau)\rangle$ and $\sigma(\tau)\equiv \langle\Omega(\tau)|
\,\sigma(x)|\Omega(\tau)\rangle$, evolve in time as described by the
vacuum--stability equations.

As explained in \S\,\ref{ShiftInvCGR}, in virtue of the underlying conformal
invariance of the theory, $\varphi(\tau)$ and $\sigma(\tau)$ coincide with the
classical limits of the two fields. It is therefore convenient to put $\varphi(x)=
\varphi(\tau)+ \hat\varphi(x)$ and $\sigma(x)=\sigma(\tau)+\hat\sigma(x)$,
where $\hat\varphi(x)$ and $\hat\sigma(x)$ represent the quantum excitations
of the two NG--boson fields. Of course, we must assume $\langle\Omega(\tau)|\hat\varphi(x)
|\Omega(\tau)\rangle=0$ and $\langle \Omega(\tau)|\hat\sigma(x)|\Omega(\tau)\rangle=0$.

As shown in Appendix {\bf \ref{VacDynApp}}, the classical solutions to this model
explain fairly well the transfer of energy from geometry to matter through the
materialization of a multitude of Higgs bosons in a thin hyperboloidal layer at
a critical time $\tau_B$ (the big--bang time).

The SMEP is a great triumph of modern physics, but is also the locus of several
unsolved problems: gravitation is excluded; right--handed neutrinos are out of
chart; dark energy and dark matter are unexplained; the cosmological constant
problem is unsolved.

To correct these shortcomings, let us briefly describe how the SMEP, or a suitable
completion of it, may be included in CGR. An evident difficulty with this a idea is
that the Higgs boson of CGR does not match that of the SMEP, where the homologous
field, $\varphi(x)$, is instead introduced as the norm of a complex isoscalar
multiplet
$$
%pmatrix round par
{\boldsymbol \varphi}(x) =
\begin{bmatrix}
\varphi^+(x)\\
\varphi^0(x)
\end{bmatrix}\!=\frac{1}{\sqrt{2}}
\begin{bmatrix}
\varphi_1(x)+i\,\varphi_2\\
\varphi_3(x)+i\,\varphi_4
\end{bmatrix}\! \equiv \varphi(x)\, {\boldsymbol e(x)}\,, \quad \hbox{where }\,
{\boldsymbol e(x)}=
\frac{1}{\sqrt{2}}
\begin{bmatrix}
e^{\textstyle i\theta_1(x)}\\
e^{\textstyle i\theta_2(x)}
\end{bmatrix}\!.
$$

This difficulty can be overcome by identifying ${\boldsymbol \varphi}^\dag
\!\cdot{\boldsymbol\varphi}= \varphi^2$ as the squared amplitude of the
Higgs boson field of CGR. The NG--boson fields ${\boldsymbol\varphi}(x)
= \varphi(x)\,{\boldsymbol e(x)}$ can then be used to give mass to a subset
of the additional massless fields ${\boldsymbol\Psi(x)}$ of the CGR.

The SMEP completion of CGR can be obtained by replacing the Lagrangian
density ${\cal L}_0(x)={\cal L}_0\big\{\sigma(x),\varphi(x)\big\}$
described by Eq (\ref{L0ofx}), with a SMEP--inclusive Lagrangian density
of the form
\vspace{-2mm}
\begin{equation}
\label{SMEPLAGR}
{\cal L}(x) \equiv {\cal L}\big\{\varphi(x)\,{\boldsymbol e(x)},\sigma(x),
{\boldsymbol\Psi}(x)\big\}= {\cal L}_0(x)+ {\cal L}_I(x)+{\cal L}_R(x)\,,
\vspace{-2mm}
\end{equation}
where curly brackets indicate the inclusion of partial spacetime--derivatives.
The classical action of ${\cal L}(x)$ shall then be written as
\vspace{-2mm}
\begin{equation}
\label{SMEPA}
{\cal A}= \int_{{\bf C}_\odot} \sqrt{-g(x)}\,{\cal L}(x)\, dx^4,
\vspace{-1mm}
\end{equation}
where ${\bf C}_\odot$ is the conical spacetime of the SMEP--inclusive CGR
and $g(x)$ is the determinant of matrix $\big[g_{\mu\nu}(x)\big]$, where
$g_{\mu\nu}(x)$ is the polar--hyperbolic metric tensor
of ${\bf C}_\odot$.

The Lagrangian density
\vspace{-2mm}
\begin{equation}
\label{L_Iofx}
{\cal L}_I(x) = {\cal L}_I\big\{{\boldsymbol \varphi(x)},\sigma(x),
{\boldsymbol\Psi}(x)\big\}
\vspace{-1mm}
\end{equation}
represents a conformal--invariant interaction of the isospin doublet
with a subset of the massless fields ${\boldsymbol\Psi}(x)$, where
${\boldsymbol\varphi(x)}$ and $\sigma(x)$ play the role of mass donors.

The additional term
\vspace{-1mm}
\begin{equation}
\label{L_Rofx}
{\cal L}_R(x)=-\big[ \sigma(x)^2-\varphi(x)^2\big]\,\frac{R(x)}{12}\,,
\vspace{-1mm}
\end{equation}
where $R(x)$ is the Ricci scalar constructed out of the polar--hyperbolic
metric tensor $g_{\mu\nu}(x)$, represents the gravitational interaction.
As we shall prove in \S\,\ref{inclusCGR}, this term is absolutely
necessary to ensure the conformal invariance of ${\cal A}$.

The total EM--tensor of matter and geometry can be obtained by the
variational equation of Hilbert--Einstein
\begin{equation}
\label{CGRThetamunu}
\Theta_{\mu\nu}(x) =  \frac{2}{\sqrt{-g(x)}}\frac{\delta{\cal A}}
{\delta g^{\mu\nu}(x)}\,.
\end{equation}

Let us put ${\cal A}= {\cal A}^M+{\cal A}^G$, where ${\cal A}^M=
\int_{{\bf C}\odot}\!\!\sqrt{-g(x)}\,\big[{\cal L}_0(x)+
{\cal L}_I(x)\big] dx^4$ is the action of the matter field  and
${\cal A}^G=\int_{{\bf C}\odot}\!\!\sqrt{-g(x)}\,{\cal L}_R(x)\,dx^4$
that of the geometry.  Although in CGR the separate conservation of
the EM--tensors of matter and geometry is impossible, we can
nevertheless re--write Eq (\ref{CGRThetamunu}) in the form
$\Theta_{\mu\nu}(x)=T^{\,M}_{\mu\nu}(x)+T^{\,\,G}_{\mu\nu}(x)$,
where
$$
T^{\,M}_{\mu\nu}(x)\! =\! \frac{2}{\sqrt{-g(x)}}\frac{\delta{\cal A}^M}
{\delta g^{\mu\nu}(x)}\,,\quad T^{\,\,G}_{\mu\nu}(x)\!=\!\frac{2}{\sqrt{-g(x)}}
\frac{\delta{\cal A}^G}{\delta g^{\mu\nu}(x)}\,.
$$
Since the gravitational equation is simply $\Theta_{\mu\nu}(x)=0$, we can
write it in the form
\begin{equation}
\label{CGRGRAVEQ}
T^M_{\mu\nu}(x)  + \frac{1}{6}\big[g_{\mu\nu}(x)D^2-D_\mu D_\nu\big]
\big[\varphi^2(x)-\sigma^2(x)\big] =\frac{\sigma^2(x)\!-\!\varphi^2(x)}{6}
\,G_{\mu\nu}(x) =0\,,
\end{equation}
where $G_{\mu\nu}(x)=R_{\mu\nu}(x)-\frac{1}{2}\,g_{\mu\nu}(x)\,R(x)$
is the gravitational tensor and $D_\mu$ are the covariant derivatives
constructed from $g_{\mu\nu}$.

To obtain this equation, the formulas of tensor calculus,
\vspace{-2mm}
\begin{eqnarray}
\label{standardforms}
&& D_\mu \big[g_{\rho\sigma}(x)\,F_{\lambda\dots}(x)\big]=
g_{\rho\sigma}(x)\,D_\mu F_{\lambda\dots}(x);\quad R(x)=
R_{\mu\nu}(x)\,g^{\mu\nu}(x)\,; \\
&&\delta R(x) = R_{\mu\nu}(x)\, \delta g^{\mu\nu}(x)+
\frac{1}{2}\,\big[g_{\mu\nu}(x) D^2-D_\mu D_\nu\big]\delta g^{\mu\nu}(x)\,;
\quad \hbox{and, consequently, } \nonumber \\
&&\frac{1}{\sqrt{-g(x)}} \frac{\delta}{\delta g^{\mu\nu}(x)}\int_{{\bf C}_\odot}\!\!
\sqrt{-g(x)}\,f(x)\,R(x)\,d^nx = f(x)\,G_{\mu\nu}(x)+\bigl(g_{\mu\nu}D^2 -
D_\mu D_\nu \bigr)f(x)\,,\nonumber
\vspace{-4mm}
\end{eqnarray}
where $C_\odot$ is a conical spacetime of CGR, must be used.

All these equations are proven in Appendix {\bf\ref{BasFormApp}} near
Eqs (\ref{RicciTensor})--(\ref{useful formula}) with $n=4$.

The reader can immediately realize that the right--hand side of
Eq (\ref{CGRGRAVEQ}) is just the improved EM--tensor $\Theta^M_{\mu\nu}(x)$
described in \S\,\ref{introduction}. Therefore, the total EM--tensors of
matter and geometry should rather be identified respectively as
\begin{eqnarray}
\label{TRUETMmunu}
&& \Theta^M_{\mu\nu}(x) =T^M_{\mu\nu}(x) + \frac{1}{6}\,\big[g_{\mu\nu}(x)
D^2-D_\mu D_\nu\big]\big[\varphi^2(x)-\sigma^2(x)\big]\,;\\
\label{TRUETGmunu}
&& \Theta^G_{\mu\nu}(x)= \frac{1}{6}\,\big[\sigma^2(x)-\varphi^2(x)\big]
\,G_{\mu\nu}(x)\,.
\end{eqnarray}

\newpage

We can therefore re--write Eq (\ref{CGRGRAVEQ}) in the compact form
\begin{equation}
\label{CGRGRAVEQ2}
\Theta^M_{\mu\nu}(x) =\frac{\sigma^2(x)\!-\!\varphi^2(x)}{6}\,G_{\mu\nu}(x)\,.
\end{equation}

This equation plays in CGR a role similar to that of the gravitational field
equation of standard GR, $\Theta^M_{\mu\nu}(x)  = (1/\kappa)\,G_{\mu\nu}(x)$,
where $\kappa$ is the gravitational coupling constant.

This means that in CGR $1/\kappa$ must be replaced by $\frac{1}{6}\,
\big[\sigma(x)^2-\varphi(x)^2\big]\approx\frac{1}{6}\,\sigma(\tau)^2$.
This approximation is valid because $\sigma(x)$ and $\varphi(x)$ are
dominated by their VEVs, $\sigma(\tau)$ and $\varphi(\tau)$, and
$\varphi(\tau)$ is negligible with respect to $\sigma(\tau)$ (proven
and exemplified in Appendix {\bf \ref{VacDynApp}}).

\subsection{How to prove the conformal invariance of the SMEP--inclusive CGR}
\label{inclusCGR}
In order for action (\ref{SMEPA}) to be invariant under conformal
diffeomorphisms, the following additional conditions must be satisfied:
\begin{itemize}
\item[1.] The Ricci scalar $R(x)$ must be constructed from $g_{\mu\nu}(x)$
as in GR.
\item[2.] The Dirac matrices $\gamma^\mu$, appearing in the kinetic--energy
terms of spinor fields in Minkowski spacetime, must be replaced by $\gamma^\mu(x)
= \gamma^a e^\mu_a(x)$, where $\gamma^a$ are standard Dirac matrices and
$e^\mu_a(x)$ are Einstein's {\em vierbein} depending on gravitation as in GR.
\item[3.] The kinetic Lagrangian densities of these fields must have the form
\begin{equation}
\label{SpinorLagDensOf}
{\cal L}^F(x) = \frac{i}{2}\big[D_\mu \bar\psi(x)]\, \gamma^\mu(x)\, \psi(x)-
\frac{i}{2} \bar\psi(x)\, \gamma^\mu(x) D_\mu \psi(x)\,,
\end{equation}
where $D_\mu = \partial_\mu + \Gamma_\mu(x)$ are the covariant derivatives
for spinors \cite{BIRREL}.
\item[4.] Similar expressions must hold also for Majorana spinors.
\end{itemize}

To make sure that ${\cal A}$ is conformal invariant, not simply a scale
invariant, we must verify whether it is invariant under Weil transformations
with scale factor $e^{\beta(x)}$, where $\beta(x)$ is any smooth real
function of spacetime parameters. To accomplish this, we must multiply
each quantity of length--dimension $n$, appearing in ${\cal A}$, by
$e^{n\beta(x)}$, carry out possible derivatives and verify whether
we reobtain ${\cal A}$ possibly up to a surface term.

Recall that scalar fields have length--dimension $-1$; spinor fields
have length--dimension $-3/2$; spacetime parameters, partial derivatives
$\partial_\mu$ and covariant gauge fields, have length--dimension $0$;
metric--tensor components of $g_{\mu\nu}(x)$ have length--dimension 2;
those of $g^{\mu\nu}(x)$ have length--dimension $-2$; $R(x)$ has
length--dimension $-2$.

Denote any quantity of length--dimension $n$ as $Q_n(x)$ and mark all
Weyl--transformed quantities with hat--superscript. In particular, the
Weyl transformations which act on the quantities that appear in the
action integral of CGR produce the following results:
\begin{eqnarray}
\label{WEYLTRANSF}
&&\hspace{-8mm} \widehat{Q}_n(x)=e^{n\beta(x)}Q_n(x)\,;\quad\widehat{g}_{\mu\nu}(x)
= e^{2\beta(x)}g_{\mu\nu}(x)\,;\quad \widehat{g}^{\,\,\mu\nu}(x)=
 e^{-2\beta(x)}g^{\mu\nu}(x)\,;\nonumber\\
&&\hspace{-8mm} \sqrt{-\widehat{g\,}(x)} = e^{4\beta(x)} \sqrt{-g(x)}\,;
\quad e^\mu_a(x)\rightarrow\widehat{e}^{\,\,\mu}_a(x) =
e^{-\beta(x)}e^\mu_a(x)\,;\nonumber\\
&&\hspace{-8mm}\widehat{L}^F(x) = e^{-4\beta(x)} L^F(x)\,;
\quad\widehat{L}_I(x) =e^{-4\beta(x)} L_I(x)\,; \nonumber \\
&&\hspace{-8mm}\widehat{\Gamma}^{\,\rho}_{\mu\nu}(x)=
\Gamma^\rho_{\mu\nu}(x) +\delta^\rho_\nu\,\partial_\mu\beta(x)
+\delta^\rho_\mu\,\partial_\nu\beta(x) -g_{\mu\nu}(x)
\,\partial^\rho\beta(x)\,; \nonumber\\
&&\hspace{-8mm}\widehat{R}(x) =  e^{-2\beta(x)}\big[R(x) -
6\,e^{-\beta(x)}D^2 e^{\beta(x)}\big]\,;\nonumber\\
&&\hspace{-8mm}\widehat{R}_{\mu\nu}(x)= R_{\mu\nu}(x)+
e^{-2\beta(x)}\big\{4\,\big[\partial_\mu  e^{\beta(x)}\big]
\,\partial_\nu  e^{\beta(x)}- g_{\mu\nu}(x)\,\big[\partial^\rho
e^{\beta(x)}\big]\,\partial_\rho e^{\beta(x)}\big\}-\nonumber \\
&&\hspace{10mm}e^{-\beta(x)}\big[2\,D_\mu\partial_\nu
\,e^{\beta(x)}+g_{\mu\nu}(x)\,D^2 e^{\beta(x)}\big]\,; \\
&&\hspace{-8mm} \widehat{G}_{\mu\nu}(x) = R_{\mu\nu}(x) -
\frac{1}{2}\,g_{\mu\nu}(x)R(x) + e^{-2\beta(x)}\big\{4\,\big[\partial_\mu
e^{\beta(x)}\big]\,\partial_\nu e^{\beta(x)}-\nonumber\\
&& \hspace{10mm} g_{\mu\nu}(x)\big[\partial^\rho e^{\beta(x)}\big]\,
\partial_\rho e^{\beta(x)}\big\}+2 e^{-\beta(x)} g_{\mu\nu}(x)(D^2-
D_\mu\partial_\nu)\, e^{\beta(x)}\,.\nonumber
\end{eqnarray}
Here, $\big[g^{\mu\nu}(x)\big]$ is the inverse of matrix $\big[g_{\mu\nu}(x)\big]$,
$g(x)$ is the determinant of $\big[g^{\mu\nu}(x)\big]$, $\delta^\mu_\nu$ is the
Kronecker delta, $\partial^\mu =g^{\mu\nu}(x)\,\partial_\nu$, and $D^2 f(x) =
\big[\sqrt{-g(x)}\,\big]^{-1} \partial_\mu \big[\sqrt{-g(x)}\,\partial^\mu f(x)\big]$
is the Beltrami--d'Alembert operator acting on a smooth scalar function $f(x)$.
Detailed explanations are found in Appendix~{\bf \ref{BasFormApp}} near
Eqs~(\ref{Gammavariations}) and (\ref{TildeR4}).

Let us prove that the action integral of the SMEP--inclusive CGR constructed
in this way is conformal invariant. To simplify the subject, let us denote
the three parts of Lagrangian density ${\cal L}(x)$, described by
Eq (\ref{SMEPLAGR}), as follows
\begin{equation}
{\cal L}_0 = \frac{g^{\mu\nu}\!}{2}\big[(\partial_\mu\varphi)
\partial_\nu\varphi\!-\!(\partial_\mu\sigma)\partial_\nu\sigma\big]\!-\!
\frac{\lambda}{4}\big(\varphi^2- c^2\sigma^2\big)^2\!;\,\,\, {\cal L}_R
= \big(\varphi^2\! -\! \sigma^2\big)\frac{R}{12};\,\,\,{\cal L}_I=
{\cal L}_I\!\big\{\sigma, {\boldsymbol\varphi}, {\boldsymbol\Psi}\big\}.
\nonumber
\end{equation}

Carrying out the Weyl transformations of all the terms of these functions, we obtain
\begin{eqnarray}
\label{3Weyltransf}
& & \hspace{-14mm}{\cal L}_0\rightarrow  e^{-4\beta} {\cal L}_0+ \Delta {\cal L}_0\,,
\quad {\cal L}_R\rightarrow e^{-4\beta} L_R+ \Delta {\cal L}_R\,,\quad
{\cal L}_I\rightarrow e^{-4\beta} {\cal L}_I\,,\quad \hbox{where}\\
& & \hspace{-14mm}\Delta {\cal L}_0=  (\varphi^2- \sigma^2)(\partial_\mu\beta)
\,\partial^\mu\beta -(\partial_\mu \beta)\,\partial^\mu(\varphi^2-
\sigma^2)\,\,\,\hbox{and }\,\Delta {\cal L}_R= -(\varphi^2 -\sigma^2)
\, e^{-\beta}D^2 e^{\beta}\,;\nonumber
\end{eqnarray}
clearly showing that neither ${\cal L}_0$ nor ${\cal L}_R$ separately
considered are conformal invariant.

Instead, the Lagrangian density ${\cal L}^F$ given by Eq (\ref{SpinorLagDensOf}),
is conformal invariant because the Weyl--transformed terms of the antisymmetric
spacetime derivatives of massless spinors cancel exactly those  produced by the
Weyl--transformed terms of the gauge fields.

We assert this without providing the cumbersome proof.
\newpage

Using identity $\varphi^2 e^{-\beta}D^2 e^{\beta} \equiv
D_\mu\big(\varphi^2 e^{-\beta}\partial^\mu e^{\beta}\big)+\varphi^2
\big(\partial_\mu\beta\big)\,\partial^\mu\beta -(\partial_\mu\beta)\,
\partial^\mu \varphi^2$, where $D_\mu$ are the covariant derivatives
constructed from $g_{\mu\nu}$, together with the similar identity with
$\sigma^2$ in place of $\varphi^2$, we get
\begin{equation}
\label{violation}
\Delta {\cal L}_0 + \Delta {\cal L}_R = - D_\mu \big[(\varphi^2 -
\sigma^2)\,e^{-\beta}\partial^\mu e^{\beta}\big]\equiv\frac{1}
{\sqrt{-g}}\,\partial_\mu\big[\sqrt{-g}\,\big(\varphi^2 -
\sigma^2)\,e^{-\beta}\partial^\mu e^{\beta}\big]\,,
\end{equation}
showing that the conformal invariance of the action integral of ${\cal L}_0 +
{\cal L}_R$ is violated.

Fortunately, however, this violation is harmless. Note, in fact, that, on account
of the conformal invariance of the action integrals of ${\cal L}^F$ and ${\cal L}_I$,
the action $\cal A$ of the SMEP--inclusive CGR, introduce in Eq (\ref{SMEPA}),
undergoes the Weyl transformation
\vspace{-1mm}
$$
\int_{{\bf C}_\odot}\sqrt{-g(x)}\,{\cal L}(x)\,dx^4 \rightarrow \int_{{\bf C}_\odot}
\sqrt{-g(x)}\,{\cal L}(x)\,dx^4 +\Delta {\cal A}\,,
\vspace{-1mm}
$$
where
\begin{equation}
\vspace{-2mm}
\label{DeltaAisSurTerm}
\Delta {\cal A} = \int_{{\bf C}_\odot}\partial_\mu\big\{\sqrt{-g(x)}\,\big[\varphi(x)^2 -
\sigma(x)^2\big]\,e^{-\beta(x)}\partial^\mu e^{\beta(x)}\big\}\,d^4x.
\end{equation}

Since this is a surface term, $\Delta {\cal A}$ is functionally equivalent
to zero, which proves the conformal invariance of $\cal A$.

Now, let us put $\varphi(x) = \varphi(\tau) + \hat\varphi(x)$ and  $\sigma(x) =
\sigma(\tau) + \hat\sigma(x)$, where $\hat\varphi(x)$ and $\hat\sigma(x)$
represent the deviations from the classical solutions $\varphi(\tau)$
and $\sigma(\tau)$, and denote the total Lagrangian density ${\cal L}(x)$
and its components as
\vspace{-2mm}
\begin{eqnarray}
\label{SMEPLAGR2}
&& {\cal L}(x) \equiv {\cal L}\{[\varphi(\tau)+ \hat\varphi(x)]
\,{\boldsymbol e(x)}, \sigma(\tau)+\hat\sigma(x),
{\boldsymbol\Psi}(x)\};\\
&& {\cal L}_0(x)\equiv {\cal L}_0\{\varphi(\tau) +\hat\varphi(x),
\sigma(\tau)+\hat\sigma(x)\};\nonumber\\
&& {\cal L}_I(x) \equiv {\cal L}_I\big\{[ \varphi(\tau) +
\hat\varphi(x)]\,{\boldsymbol e(x)},\sigma(\tau)
+\hat\sigma(x),{\boldsymbol\Psi}(x)\big\};
\nonumber \\
&& {\cal L}_R(x) \equiv -\big[\sigma(\tau) +\hat\sigma(x)\big]^2-
\big[\varphi(\tau)+\hat\varphi(x)\big]^2\big\}\,\frac{R(x)}{12}\,;
\nonumber\\
&& \label{hatA}
{\cal A}(\hat\varphi, \hat\sigma) =
\int_{{\bf C}_\odot}\sqrt{-g(x)}\,{\cal L}(x)\,d^4x\,.
\vspace{-1mm}
\end{eqnarray}

One may think that, by generalizing Eqs (\ref{WJ2calA}) and (\ref{Gamma2WJ})
of Appendix {\bf \ref{PathIntApp}}, it is possible to construct the path
integral over variations $\hat\varphi(x)$ and $\hat\sigma(x)$, so as to
obtain the effective action $\Gamma[\bar\varphi(\tau), \bar\sigma(\tau)\,]$
of CGR, satisfying equation
\begin{equation}
\label{EffActofCGRMT}
\hspace{4mm}e^{\textstyle\,\frac{i}{\hbar}\,\Gamma[\bar\varphi(\tau),
\bar\sigma(\tau)\,]} = e^{\textstyle\,\frac{i}{\hbar}\,
\big\{W[J_{\bar\varphi}, J_{\bar\sigma}\,]-\int[\bar\phi(x)J_{\bar\varphi}(x)
+\bar\sigma(x)J_{\bar\sigma}(x)]\,d^4x \big\}},
\end{equation}
where $W[J_{\bar\varphi}, J_{\bar\sigma}]$ is the generator of the Green
functions of CGR, and $J_{\bar\varphi}(x)$, $J_{\bar\sigma}(x)$ are the
external currents coupled to $\bar\varphi(x)$ and  $\bar\sigma(x)$. For
details see \S\,\ref{PathInTAndGRFunx} near Eqs (\ref{GRNSFUNCTS}).

Unfortunately, the construction of the path integral of CGR is very difficult.

\newpage
\markright{R.Nobili, Conformal General Relativity -- {\bf \ref{CGRafterBB}} CGR dynamics after big bang}
\section{CGR dynamics after big bang}
\label{CGRafterBB}
During the decay of CGR to GR, there is a proper time $\widetilde{\tau}_B$,
the big-bang time, at which the vacuum dynamics gives way to the history
of the universe.

This happens when Higgs--field VEV $\widetilde{\varphi}(\widetilde{\tau})$,
once reached the absolute maximum at~$\widetilde{\tau}_B$, enters a regime of
damped oscillations at the Compton frequency of the Higgs--boson mass. This
behavior is numerically simulated in  \S\,\ref{proptimesolutions} of Appendix
{\bf \ref{VacDynApp}} and exemplified in Fig.\ref{VacStFig4}. Always at
$\widetilde{\tau}_B$, the inflationary expansion of the spacetime stretches
the system so violently to determine a sudden transfer of energy from geometry
to matter through the materialization of a crowd of Higgs bosons on the
spacelike hyperboloid at $\widetilde{\tau}_B$. At the same time, the scale
factor of vacuum dynamics, $\widetilde{\alpha} (\widetilde{\tau})$, passes
from a state of accelerated increase to one of decelerated increase that
leads it to converge asymptotically to one.

Thus, in parallel with the evolution of the vacuum state, another history
takes place that makes CGR very similar to the SMMC: the rise and evolution
of the universe as a thermodynamic process. We would rather say that the
dynamics of the vacuum state is the natural prehistory of CGR, during which
the inflationary expansion of spacetime and the occurrence of the big bang
find their theoretical reasons.

In contrast with this scenario, the universe described by the SMMC has a
history but not a prehistory. In fact, here to explain how the universe could
have emerged from the mythical age of Planck, the cosmologist must invoke
the creation and decay of a primordial scalar field with the incredible
mass of about $10^{13}$GeV \cite{LINDE}, \cite{LYTH}, (Mukhanov, 2005).

In \S\,\ref{THREEWAYS}, we introduced three equivalent ways of describing
the vacuum--stability equations: the kinematic--time, the conformal--time
and the proper--time representations. If we use these representations to
describe the temporal course of CGR after big bang, we put ourselves in a
position to understand a little better their physical significance:
\begin{itemize}
\item[I.] The {kinematic--time representation} provides a description of the
universe from the point of view of the comoving observers today. Thinking
about the far past in the light of the kinematic time representation, they
are led to describe all natural events as subject to the inflationary power
of the ghost scalar field $\sigma(x)$. In accord with this interpretation,
they ascribe all the adimensional constants of the theory to the NG--boson
fields, which therefore appear as the universal donors of mass.
\vspace{-1mm}
\item[II.] The {conformal--time} representation provides instead a
description of the universe as might have been seen by ideal observers
comoving and co--expanding with the universe. Since in the reference
frames of these observers all rulers and clocks also co--expand,
these observers cannot actually detect any change of scale in the
magnitude of geometrical and physical quantities. As in the namesake
case of the SMMC, this representation produces an anamorphic deformation
of the spacetime geometry that hides the effects of universe expansion.
In CGR, however, it works as a mathematical bridge between the
kinematic--time and proper-time representations.
\vspace{-1mm}
\item[III.] After big bang, as CGR tends to evolve toward GR, the
proper--time representation allows us to describe the time course of the
universe as might have been seen by coeval observers equipped with fixed
rulers and synchronized clocks. Since in this representation the spacelike
terms of the metric tensors undergo a quadratic change of scale -- while the
timelike term does not -- all bodies appear to move along underwent a strong
compression in the initial tract. This occurs to such an extent that the
kinematic--time interval taken by the evolution of the vacuum state before
big bang seems to have shrunk to a point. In these circumstances the
description of the matter field becomes so complicated that the evolution
of CGR  can only be described as a thermodynamic process (see \S\,\ref{unilargescale}).
For all these reasons, the proper--time representation after big bang is not
an option but a necessity.
\end{itemize}

Here are the most important facts occurring after $\widetilde{\tau}_B$:

1) Both in the SMMC and in CGR the expansion factor of the universe depends
on the energy density of the cosmic background and of the matter field. But
in CGR it also depends on the spatial curvature of the spacelike hyperboloids
(proven in \S\,\ref{CBGravecCGR}).

2) In the SMMC the gravitational coupling is constant.  But in CGR it increases
by a factor of $\widetilde{\alpha}(\widetilde{\tau})^{-2}$. This factor is in
the order of magnitude of $10^{25}$ at big bang, but decreases very rapidly
and converges asymptotically to one in the course of time (proven in the
next subsection). This is the way how CGR converges to GR.

3) Soon after their sudden creation, the Higgs bosons decay progressively into
the inventory of elementary particles of the SMEP. The energy delivered by this
process increases the temperature from nearly zero to about the equivalent of
the Higgs--boson mass (proven in Section \ref{unilargescale}). The thermodynamics
of this process is discussed in Appendix {\bf \ref{ThermVacApp}}.

\newpage

\subsection{The proper--time representation of CGR's gravitational equation}
\label{PhenafterBB}
In \S\,\ref{CGR&SMEP} is described how  gravitation and the SMEP can be
included into CGR by forming an action $\cal A$ that comprises
the NG fields  $\varphi(x)$, $\sigma(x)$ interacting with the decay products
of the Higgs bosons, ${\boldsymbol{\Psi}}(x)$. Then, by variation of $\cal A$
with respect to metric tensor $g^{\mu\nu}(x)$ have obtained the
gravitational equation in the kinematic--time representation described
by Eq~(\ref{CGRGRAVEQ2}), in which replace for notational convenience
$\Theta^M_{\mu\nu}(x)$ with $\mathbb{T}_{\mu\nu}(x)$.

For the reasons mentioned in point III of the previous subsection, it is
opportune to have the gravitational equation in the proper--time
representation, so that in place of the above mentioned fields we have
their counterparts $\widetilde{\varphi}(\widetilde{x\,})$,
$\widetilde{\sigma}(\widetilde{x\,})\equiv\sigma_0$ and
$\widetilde{\boldsymbol{\Psi}}(\widetilde{x})$.

However, to do this, there is no need to start from the action $\cal A$
rewritten  in the proper time representation, but simply apply to
Eq~(\ref{CGRGRAVEQ2}) the operator ${\mathcal P}$ introduced at the
end of \S\,\ref{THREEWAYS} near Eqs (\ref{CalP}), which acts on a
local operator $Q_n(x)$ as ${\mathcal P}\, Q_n(x)= \widetilde{\alpha}
(\widetilde{\tau})^n\widetilde{Q}_n(\widetilde{x})$.

Since the mixed--index EM--tensor of matter $\mathbb{T}^\mu_\nu$,
the mixed--index gravitational tensor $G^\mu_\nu$ and scalar
field $\varphi$ have respectively length--dimension $-2$, $2$
and $-1$, we have
\begin{equation}
\label{PThetaM_G}
{\mathcal P}\Big[\mathbb{T}^\mu_\nu(x)\!=\frac{\sigma(x)^2\!-\!\varphi(x)^2}{6}
\,G^\mu_\nu(x)\Big]\!=\frac{1}{\widetilde{\alpha}(\widetilde{\tau})^2}
\bigg[\widetilde{\mathbb{T}}^\mu_\nu(\widetilde{x})=\frac{\sigma_0^2\!-\!
\widetilde{\varphi}(\widetilde{x})^2}{6}\,\widetilde{\alpha}(\widetilde{\tau})^2
\widetilde{G}^\mu_\nu(\widetilde{x})\bigg]\,,
\end{equation}
which leads us to establish the gravitational equation in the proper--time
representation
\begin{equation}
\label{TetaBGB}
\widetilde{\mathbb{T}}^\mu_\nu(\widetilde{x}) =
\frac{\sigma^2_0 - \widetilde{\varphi}^2(\widetilde{x\,})}{6}\,
\widetilde{\alpha}(\widetilde{\tau})^2\widetilde{G\,}^\mu_\nu(\widetilde{x\,})\cong
\frac{\widetilde{\alpha}(\widetilde{\tau})^2}{\kappa}\,\widetilde{G\,}^\mu_\nu
(\widetilde{x\,});
\end{equation}
showing that the gravitational coupling constant of CGR is divided by
$\widetilde{\alpha}(\widetilde{\tau\,})^2$. In particular, the
very--well--approximated $00$ component of the gravitational equation is
\begin{equation}
\label{TetaBGBII}
\widetilde{G}^0_0(\widetilde{x\,}) \cong \frac{\kappa}{\widetilde{\alpha}
(\widetilde{\tau})^2}\,\widetilde{\mathbb{T}}^0_0(\widetilde{x})
\equiv \frac{\kappa}{\widetilde{\alpha}(\widetilde{\tau})^2}\,
\widetilde{\rho}(\widetilde{x\,})\,.
\end{equation}

The symbol of very good approximation ($\cong$) is justified because
$\sigma^2_0\equiv6/\kappa\cong 3.551\!\times\! 10^{37}$GeV$^{2}$$\gg
\widetilde{\varphi\,}^2(\widetilde{x\,})$. Since at $\widetilde{\tau}_B$
it is $\widetilde{\alpha}(\widetilde{\tau}_B\,)^{-2} \approx 10^{17}$, while
to day it is about 1, we see that at big bang the gravitational attraction
is enormously larger than today.

To Eq (\ref{TetaBGBII}), we add for completion the trace reversed equation
\begin{equation}
\label{TReverseTetaBGBII}
\widetilde{R}^0_0(\widetilde{x\,}) \cong \frac{\kappa\,
\big[\,\widetilde{\rho}(\widetilde{x\,})+3\,\widetilde{p}
(\widetilde{x\,})\big]}{2\,\widetilde{\alpha}(\widetilde{\tau})^2}\,.
\end{equation}

For details, see Eq (\ref{TraceRev}) of Appendix {\bf \ref{GravGaugInvApp}}
and Eq (\ref{R00Append}) of Appendix {\bf\ref{HomIsotSptApp}}.

\newpage

\markright{R.Nobili, Conformal General Relativity -- {\bf \ref{MachEinstPrinc}} Mach principle, Hubble law, dark energy}
\section{Mach principle, Hubble law and dark energy in CGR}
\label{MachEinstPrinc}
According to the Mach--Einstein doctrine, here referred to as the {\em Mach principle},
in the universe there is an inertial frame that is globally determined by the distant
bodies. It was traditionally called the reference frame of ``fixed stars", but today it
should rather be called the frame of {\em galaxy clusters}, because the galaxies move
slowly with different speeds. The existence of such a frame is evident in the observed
simplicity of the universe on the large scale, but how this may happen in a universe
ruled by GR is still a mystery.

The SMMC replaces the Mach principle with the {\em Hubble law}: basing on the
astronomic evidence of sky isotropy and universe expansion, and on the
{\em Copernican principle} which states that humans are not privileged
observers of the  universe, we are led to infer that the universe on the large
scale is homogeneous and parameterized by an absolute time~\cite{PEACOCK2}.

In CGR, the Mach principle and Hubble law follow primarily from the conical
structure of the spacetime, which imposes the dynamical expansion sketched
in Fig.\,\ref{Figure06}, and secondarily on the gravitational equation, which
imposes the dependence of the expansion rate on the energy density of the matter
field, as proven and discussed in  Appendix {\bf \ref{HomIsotSptApp}}.

In this case, however, to pose well the problem we must distinguish between the
cosmological structure of the universe on the  large scale -- let us call it
the {\em  cosmic background}~-- and the gravitational effects caused by the
celestial bodies and their peculiar motions, because all statements related to
the Principle of Mach involve this separation,

We can do this by splitting the metric tensor of CGR in the form
\vspace{-2mm}
\begin{equation}
\label{G=GM+H}
\widetilde{g}_{\mu\nu}(\widetilde{\tau},\vec\rho\,; \boldsymbol{\zeta})=
\widetilde{g}^{\,B}_{\mu\nu}(\widetilde{\tau},\vec\rho\,)+\widetilde{h}_{\mu\nu}
(\widetilde{\tau},\vec\rho\,; \boldsymbol{\zeta})\,,
\vspace{-2mm}
\end{equation}
where $\widetilde{g}^{\,B}_{\mu\nu}$ is metric of the cosmic background as
a function of the proper--time coordinates $\widetilde{x} = \{\widetilde{\tau},
\vec r\,\}$, described by Eq~(\ref{tildemetrictensor}), and $\widetilde{h}_{\mu\nu}$
represents the deviation from $\widetilde{g}^{\,B}_{\mu\nu}$ (where $\boldsymbol{\zeta}$
is the set of variables that are necessary to describe the peculiar motions). Of course,
we must take care of not confusing $\widetilde{g}^{\,B}_{\mu\nu}$ with the metric tensor of
a Minkowskian background.

Denoting as $\widetilde{D}^\mu$ the contravariant derivatives constructed from
total metric tensor $\widetilde{g}_{\mu\nu}$ and as $\widetilde{\cal D}^\mu$
those constructed from metric tensor $\widetilde{g\,}^B_{\mu\nu}$, the obvious
identities $\widetilde{D}^\mu \widetilde{g}_{\mu\nu}=0$ and
$\widetilde{\cal D}^\mu\widetilde{g}^B_{\mu\nu}=0$ shall then hold.

If $\widetilde{h}_{\mu\nu}$ can be regarded as a slight perturbation of $\widetilde{g\,}^B_{\mu\nu}$,
we can put $\widetilde{D}^\mu= {\widetilde{\cal D}}^\mu +\Delta\widetilde{\cal D}^\mu$, with
$\Delta\widetilde{\cal D}^\mu$ is negligible relative to ${\widetilde{\cal D}}^\mu$. So, in
summary, we have $\widetilde{D}^\mu\widetilde{h}_{\mu\nu}\cong {\widetilde{\cal D}}^\mu
\widetilde{h}_{\mu\nu}=0$, which can be regarded as the Lorentz--gauge condition for
$\widetilde{h}_{\mu\nu}$ (cf. \S\,\ref{GaugeInvGmunu} of Appendix {\bf \ref{GravGaugInvApp}}).

\subsection{The gravitational equations of the cosmic background in GR}
\label{CBGravTensInCHR}
The SMMC represents the cosmic background as a cylindrical spacetime with metric tensor
$g_{\mu\nu}(t) = \hbox{diag}\big[1, a^2(t),  a^2(t),  a^2(t)\big]$  where $a(t)$ is the
expansion factor of the universe.

As shown in \S\,\ref{CylGraveq} of Appendix {\bf \ref{HomIsotSptApp}}, the temporal
curvature $R^B_{00}(t)$, the Hubble parameter $H(t)$ and the zero--zero component
of the gravitational tensor $G^B_{00}(t)$, satisfy equations
\begin{equation}
\label{RWHubbPar}
R^B_{00}(t)= -3\,\frac{\ddot a(t)}{a(t)}\,,\quad
H(t)=\frac{\dot a(t)}{a(t)},\quad G^B_{00}(t)\equiv 3\,H(t)^2= \kappa\,\rho(t)\,,
\end{equation}
where $\kappa$ is the gravitational coupling constant of GR and $\rho(t)$ is
the energy density of the cosmic background as a functions of absolute time $t$.

In \S\,\ref{CylGravPerts}, near Eq (\ref{CylGravPotEq}), it is shown that in the
presence of celestial bodies, the metric tensor $g_{\mu\nu}(t)$ changes to
$\bar g_{\mu\nu}(x) =g_{\mu\nu}(t) + h_{\mu\nu}(x)$, where $h_{\mu\nu}(x)$ is
related to the energy density of the bodies, $\delta\rho(x)$, by equation
$\delta G_{00}(x)= \kappa \,\delta\rho(x)$. Thus, in summary, the total
zero--zero component of the gravitational equation of the universe satisfies
equation
\begin{equation}
\label{RWHubbPar2}
G_{00}(x)= 3\, H(t)^2+ \kappa\, \delta\rho^H(x)\,.
\end{equation}

If the celestial bodies move slowly compared to the speed of light, and their
gravitational effects are sufficiently weak and independent of time,
Eq (\ref{RWHubbPar2}) can be further simplified by expressing the gravitational
field as a Newtonian potential $\Phi(x)$, in which case we find $h_{\mu\nu}(x)=
2\Phi(x)\, \delta_{\mu\nu}$, where $\delta_{\mu\nu}$ is the Kronecker delta in 4D.
The mathematical reason of this strange equation is explained in detail
in \S\,\ref{CylGravPerts}.

In these circumstances, as extensively described by Eq (\ref{ds2hmunu})], the
squared--line element of metric tensor $\bar g_{\mu\nu}(x)$  can be cast
in the form
\begin{equation}
\label{ds2ofh00}
d\bar s^2 =  dt^2\big[1+2\Phi(t, \vec r\,)\big]-a(t)^2\,\big[1-2\Phi(t,
\vec r\,)\big]\big(dr^2+r^2 d\theta^2+r^2\sin\theta^2 d\phi\big)\,.
\end{equation}
Here we have put $x=\{t, \vec r\,\}$ and denoted, as usual, the Euler angles
of radius vector $\vec r$ as $\theta$ and $\phi$. Since the determinant of
$\bar g_{\mu\nu}(x)$ differs from that of $g_{\mu\nu}(t)$ by a term proportional
to $\Phi(t, \vec r\,)^2$ -- hence to $\kappa^2$ -- we  see that the volume
element of the expanding universe is practically unaffected by the presence of the
celestial bodies.

Note that, if $a(t)$ is multiplied by a constant factor $C$, the Hubble parameter
$H(t)$ remains unvaried, we can choose $C$ so that $a(t)=1$ just at age of universe
$t=t_U$. In which case the squared--line element of the spacetime today takes the
form
\begin{equation}
\label{ds2ofh00tU}
d\bar s^2 =  dt^2\big[1+2\Phi(t_U, \vec r\,)\big]-\big[1-2\Phi(t_U, \vec r\,)
\big]\big(dr^2+r^2 d\theta^2+r^2\sin\theta^2 d\phi\big)\,.
\end{equation}

\subsection{The gravitational equations of the cosmic background in CGR}
\label{CBGravecCGR}
The main difference between the cosmic background of the SMMC and that of CGR, is that
the temporal curvature of the first, $R^B_{00}(t)$, differs from zero, whereas that
of the second, $\widetilde{R}^B_{00}(\widetilde\tau)$, is zero. Let us briefly
clarify this point.

To describe the cosmic background of CGR, we have introduced in \S\,\ref{ConGraveq}
of Appendix {\bf \ref{HomIsotSptApp}} a polar--hyperbolic metric, depending on
kinematic--time coordinates $x=\{\tau,\varrho, \theta, \phi\}$, which we write
in this context as
\begin{equation}
\label{FRWmetMach}
\hspace{-10mm}g^B_{\mu\nu}(x) = \hbox{diag}\big[1, -c(\tau)^2,
-c(\tau)^2\sinh\!\varrho^2,- c(\tau)^2(\sinh\!\varrho\,\sin\theta)^2\big]\,,
\end{equation}
where $c(\tau)= a(\tau)\,\tau$,  in which $a(\tau)$ is the expansion factor of the
cosmic background of CGR. The determinant of this metric is easily found to be
$\sqrt{-g(x)}= c(\tau)^3\,(\sinh \varrho)^2\sin\theta$.

Note that the frictional term $(3\,\dot c/c)\,\partial_\tau f= 3\,\big(1/\tau +
\dot a/a\big)\partial_\tau f$ depends on the Hubble parameter of CGR, $H(\tau)=
\dot a(\tau)/a(\tau)$, and, if $a(\tau)=1$, we have $c(\tau)=\tau$, implying that
the spacetime is flat.

In \S\,\ref{ConGraveq}, it is also proven that the temporal curvature and the
total curvature of the spacetime depend respectively on $a(\tau)$, which we
rewrite in this context as
\begin{eqnarray}
\vspace{-4mm}
\label{R00termMach}
&& \hspace{-18mm}  R^B_{00}(\tau)=-3\frac{\ddot c(\tau)}{c(\tau)} =
-3\bigg[\frac{\ddot a(\tau)}{a(\tau)}
+2 \frac{\dot a(\tau)}{a(\tau)}\bigg];\\
\label{R&RtermMach}
&& \hspace{-18mm} R^B(\tau)= -6\bigg[\frac{\ddot c(\tau)}{c(\tau)}+\frac{\dot c(\tau)^2-1}
{c(\tau)^2}\bigg]=-6\bigg[\frac{\ddot a(\tau)}{a(\tau)}+2\frac{\dot a(\tau)}{a(\tau)}
+2\frac{\dot c(\tau)^2-1}{c(\tau)^2}\bigg].
\end{eqnarray}

Imposing the conditions $R^B_{00}(\tau)=0$ and $R(\tau)\neq 0$, we can easily prove
the expansion factor must have the general form $a(\tau)= A\,\big(1 - \tau_B/\tau \big)$,
where $A$ is an arbitrary positive constant and $\tau_B$ is the origin of the kinematic
time. It comes natural to identify it with the big--bang time. This expression of $a(\tau)$
clearly implies that the cosmic background of CGR has necessarily the topology of a
truncated cone. This geometrical mismatch does not occur in the SMMC, where
$R^B_{00}(x)=0$ entails $R^B(x)=0$.

To understand the physical relevance of this point, let us consider the zero--zero
component of the trace reversed gravitational equation,
\begin{equation}
\label{R00Mach}
R^B_{00}(\tau) = \kappa\, \big[\mathbb{T}_{00}(\tau) - \frac{1}{2}
\mathbb{T}^\lambda_\lambda(\tau)\big] \equiv \frac{\kappa}{2}\,\big[\rho^B(\tau)
+ 3\,p^B(\tau)\big]\,,
\end{equation}
where $\rho^B(\tau)$ and $p^B(\tau)$ are the energy density and pressure of the
cosmic background, as defined with other notation in \S\,\ref{TraceRev} of
Appendix {\bf \ref{GravGaugInvApp}}. In this equation, the dependence of
$\mathbb{T}_{\mu\nu}(\tau)$ on the scale factor of vacuum dynamics,
$\alpha(\tau)$, is provisionally ignored.

Eq (\ref{R00Mach}) shows that $R^0_0(\tau) =0$ is possible only if $\rho(\tau)+
3p(\tau)=0$, in which case the zero--zero component of the gravitational tensor
satisfies equation
\begin{equation}
\label{GmunutermMach}
G^B_{00}(\tau) = R^B_{00}(\tau)-\frac{1}{2}\, R^B(\tau)=-\frac{1}{2}\, R^B(\tau)=
3\,\frac{\dot c(\tau)^2-1}{c(\tau)^2} =
\kappa\, \mathbb{T}_{00}(\tau) \equiv \kappa\,\rho^B(\tau)\,.
\end{equation}
we see that the Ricci scalar is related to $\rho(\tau)$ by equation $R(\tau)=
-2\,\kappa\,\rho(\tau)$, which is consistent with the fact that the curvature
of the hyperboloidal surfaces of a the truncated conical spacetime is negative.
Since,  as explained in \S\,\ref{RicciTensSign} of Appendix {\bf \ref{BasFormApp}},
the vanishing of $R^B_{00}(\tau)$ means that the curvature is purely spatial,
we infer that $R^B(\tau)$ is not the curvature of the cosmic background but that
of the hyperboloidal surfaces of the conical spacetime.

Considering that one of the most important discoveries of the SMMC is the
{\em dark~energy}, which is estimated to be about three times greater than
that of the matter field, and noting that the energy density and the pressure
of the truncated conical background are related equation $\rho^B(\tau)+
3\,p^B(\tau)=0$, we are led to identify quite naturally the density of dark
energy with $\rho^B(\tau)$ and that of the matter field as the product of
the work done by the gradient of pressure $p^B(\tau)= - \rho^B(\tau)/3$
between adjacent hyperboloids.

Putting $c(\tau)= a(\tau)\,\tau$ in the last two steps of equation
Eq (\ref{GmunutermMach}), and identifying $H(\tau)=\dot a(\tau)/a(\tau)$
with the Hubble parameter of the truncated conical spacetime, we can
rearrange the equation in the form,
\begin{equation}
\label{G00HubbleParMach}
H(\tau) = \sqrt{\frac{\kappa\,\rho^B(\tau)}{3}+\frac{1}{a(\tau)^2\,\tau^2}}
-\frac{1}{\tau}\,.
\end{equation}

For $\tau\rightarrow \infty$ the hyperboloids of the conical spacetime flatten and
$H(\tau)$ approaches the Hubble parameter of the cylindrical spacetime described
by the second of Eqs (\ref{RWHubbPar}).

Since $H(\tau)$ remains unvaried if $a(\tau)$ is multiplied by a constant, it
is customary to choose this constant so that the expansion factor equals 1
just today, and that the value of $H(\tau)$ just coincides with the value
of Hubble constant $H_0$ provided by astronomic observations of nearest
celestial bodies, so that $H(\tau_U) = H_0$, where $\tau_U$ is
the age of the universe. In formulas, by putting
\begin{equation}
\label{G00HubbleParUMach}
a(\tau) = \frac{1-\tau_B/\tau}{1-\tau_B/\tau_U}\quad\hbox{and }\,H(\tau_U) =
\sqrt{\frac{\kappa\,\rho^B(\tau_U)}{3}+\frac{1}{\tau^2_U}}-\frac{1}{\tau_U}=H_0\,,
\end{equation}
we obtain the best approximation to the analogous relation of the SMMC.

Backdating the kinematic time parameter to a value $\tau < \tau_U$, we
obtain instead
\begin{equation}
\label{G00HubbleParDMach}
H(\tau) = \sqrt{\frac{\kappa\,\rho^B(\tau)}{3}+\bigg(\frac{\tau_U-\tau_B}
{\tau-\tau_B}\bigg)^2\frac{1}{\tau^2_U}}-\frac{1}{\tau}\,.
\end{equation}

This relation between the Hubble law and the energy of cosmic background
stored in the curvature of the expanding hyperboloids has no analog in
the SMMC.

Proceeding as in \S\,\ref{CBGravTensInCHR}, we can determine the analog
of Eq (\ref{ds2ofh00}) in the polar--hyperbolic coordinate system, for
$\tau> \tau_B$, of the truncated conical background
\begin{eqnarray}
\label{KTConicalHarmGaugeMach}
&&\hspace{-16mm}d\bar s^2(x) = d\tau^2\big[1\!+2\Phi(x)\big]-\nonumber\\
&&a(\tau)^2 \big[1\!-2\Phi(x)\big]\tau^2\big(d\varrho^2 +
\sinh\varrho^2\,d\theta^2+\sinh\varrho^2\,\sin\theta^2 d\phi^2\big)\,.
\end{eqnarray}
and the analog of Eq (\ref{ds2ofh00tU})
\begin{eqnarray}
\label{KTConicalHarmGaugeMachtU}
&&\hspace{-16mm}d\bar s^2(x_U) = d\tau^2\big[1\!+2\Phi(x_U)\big]-\nonumber\\
&&\big[1\!-2\Phi(x_U)\big]\tau^2_U\big(d\varrho^2 +
\sinh\varrho^2\,d\theta^2+\sinh\varrho^2\,\sin\theta^2 d\phi^2\big)\,,
\end{eqnarray}
where $x_U =\{\tau_U, \varrho, \theta, \phi\}$.

To complete the picture, we must insert into the gravitational equation
the dependence on the scale factor of vacuum dynamics, $\alpha(\tau)$, and
rewrite all the equation in proper time coordinates. To carry out this
further step, we must convert Eq~(\ref{TetaBGBII}) to the form
$$
\widetilde{G}^B_{00}(\widetilde{x\,}) \cong  \frac{\kappa}{\widetilde{\alpha}
(\widetilde{\tau})^2}\,\widetilde{\rho}(\widetilde{x\,})\,.
$$

Carrying out the same operations in Eq (\ref{G00HubbleParMach}), we obtain
\begin{equation}
\label{HubbleParDMach}
\widetilde{H}(\widetilde{\tau}) =
\sqrt{\frac{\kappa\,\widetilde{\rho\,}^B\!(\widetilde{\tau})}{3}
+\frac{1}{\widetilde{a}(\widetilde{\tau})^2\,\tau(\widetilde{\tau})^2}}-
\frac{1}{\tau(\widetilde{\tau})}\,.
\end{equation}
where
$$
\widetilde{a}(\widetilde{\tau}) = \frac{1-\tau_B/\tau(\widetilde{\tau})}{1-\tau_B/\tau_U}.
$$
and $\tau(\widetilde{\tau})$ is the kinematic time as a function of the proper time
[see \S\,\ref{THREEWAYS} near Eq (\ref{ttau2ctau})].

The important point regarding the Hubble law in CGR is that the Hubble parameter depends explicitly
on the expansion factor of the cosmic background.

This circumstance rises the question of whether the difference between Eq (\ref{HubbleParDMach})
and $H(t) = \dot a(t)/a(t)$ may be detected by astronomical observations \cite{WENDY}; an
eventuality which is even more interesting if the dynamic of the universe has a significant
change after the age of photon decoupling.

\newpage

\markright{R.Nobili, Conformal General Relativity -- {\bf \ref{unilargescale}} Big bang as thermodynamic process}
\section{The big bang as a thermodynamic process}
\label{unilargescale}
Assume that in each
hyperboloidal section of the spacetime the matter field is uniform and in thermal
equilibrium at a temperature $T$, and denote its energy density as $\epsilon_*(T)$,
its pressure as $p_*(T)$ and its entropy density as $s_*(T)$
(with Boltzmann constant $k_B=1$).

The second law of thermodynamics states that any adiabatic change  of the matter
field in a volume $V$ produces a change in entropy
\begin{equation}
\label{entropychange}
d\big[s_*(T)\,V\big] = \frac{d\big[\epsilon_*(T)\,V\big]+ p_*(T)\,dV}{T}\,.
\end{equation}

By equating  the coefficients of $V dT$ we obtain the first law of thermodynamic
\begin{equation}
\label{Joule}
\frac{d\epsilon_*(T)}{dT}=\frac{ds_*(T)}{dT} \equiv c_V(T)\,,
\end{equation}
where $c_V(T)$ is the specific heat at constant volume, and, by equating the
coefficients of $dV$, we obtain the formula of entropy density
\begin{equation}
\label{sstarofT}
s_*(T) = \frac{\epsilon_*(T)+ p_*(T)}{T}\,.
\end{equation}

In general, if the matter field is a gas of particles of rest mass $m$ and
degeneracy factor $g$ (number of spin components) in thermal equilibrium
at temperature $T$, we can determine energy density $\epsilon(T)$, pressure
$p(T)$, entropy density $s(T)$ and particle density $n(T)$ by carrying out
the integrations
\begin{eqnarray}
\label{energydens}
\epsilon(T) &=& \frac{g}{2\pi^2}\int_0^\infty
\frac{E(m,p)\,p^2}{e^{[E(m,p)-\mu]/T} \pm 1}\,dp \equiv g\,a_{\epsilon}(m/T)
\,T^4\,;\\
\label{pressure}
p(T) &=& \frac{g}{6\pi^2}\int_0^\infty \frac{p^4}{E(m,p)
\big\{e^{[E(m,p)-\mu]/T}\pm 1\big\}}\,dp \equiv g\,a_p(m/T)\,T^4 \,;\\
\label{entropydens}
s(T) &=& g\,\frac{\epsilon(T) + p(T)}{T}\equiv  g\,a_s(m/T)\,T^3\,;\\
\label{particledens}
n(T) &=& \frac{g}{2\pi^2}\int \frac{p^2}{e^{[E(m,p)-\mu]/T} \pm 1}\,dp
\equiv  g\,a_n(m/T)\,T^3\,;
\end{eqnarray}
where, $m$, $p$ and $E(m, p) = \sqrt{m^2 + p^2}$ and $\mu$ are respectively
the mass, the momentum, the energy of the particle and the chemical potential.
This latter is zero for massless particles, and can be neglected if $T$
is sufficiently large (in which case we can safely put $\mu=0$). Signs
$\pm$ in the denominator refer respectively to the case of fermions or
bosons.

The reader can easily verify that by replacing the integration differential
$dp$ with $dx=dp/T$, the integrals take just the forms shown on the right.

\newpage

For $m=0$ or $m\ll T$, we can replace $E(m, p)$ by its relativistic limit $p$.
Carrying out the integrations over $p$,  Eqs (\ref{energydens})--(\ref{particledens})
simplify to
\begin{eqnarray}
\label{nonrelInts}
&& \hspace{-18mm}a_\epsilon(0) = \frac{\pi^2}{30}\,\, \hbox{for bosons};
\quad  a_\epsilon(0) =(7/8)(\pi^2/30)\,\, \hbox{for fermions};\nonumber\\
&& \hspace{-18mm}a_p (0)= \frac{1}{3}\,a_\epsilon;\quad a_s(0) =
(4/3)\,a_\epsilon(0) \quad a_n(0)= (3/4)\,\zeta(3)\,a_\epsilon;
\end{eqnarray}
where $\zeta(3)\cong 1.20206\dots$ is the Riemann zeta--function of 3.

If instead $m\gg T$, we can replace  $E(m, p)$ by its non--relativistic
limit $m+ p^2/2m$ and function $e^{[E(m,p)-\mu]/T}$ by $e^{m/T }
e^{(p^2/2m-\mu)/T}$. In this case, Eqs (\ref{energydens})--(\ref{particledens})
converge to
\begin{eqnarray}
\label{NRenergydens}
\epsilon_0(T) &=& \frac{g\,m}{2\pi^2}\, e^{-(m-\mu)/T}\!\!\int_0^\infty\! e^{-p^2/2mT}p^2 dp
= g\,m\bigg(\frac{m\,T}{2\pi}\bigg)^{3/2}\!\!e^{-(m-\mu)/T}\,;\\
\label{NRpressure}
p_0(T) &=& \frac{g\,m}{6\pi^2}\, e^{-(m-\mu)/T}\!\!\int_0^\infty\! e^{-p^2/2mT}p^4 dp
 \equiv g\,T\bigg(\frac{m\,T}{2\pi}\bigg)^{3/2}\!\!e^{-(m-\mu)/T} \,;\\
\label{NRentropydens}
s_0(T) &=& \frac{g}{T}\big [\epsilon_0(T) + p_0(T)\big]= g\,\frac{m}{T}
\bigg(\frac{m\,T}{2\pi}\bigg)^{3/2}\!\!e^{-(m-\mu)/T}\,;\\
\label{NRparticledens}
n_0(T) &=& \frac{g}{2\pi^2}\, e^{-(m-\mu)/T}\!\!\int\! e^{-p^2/2mT}p^2 dp =
g\,\bigg(\frac{m\,T}{2\pi}\bigg)^{3/2}\!\! e^{-(m-\mu)/T}\,;
\end{eqnarray}
so that at this limit the entropy density is replaced by $s_0(T)= m\,n_0(T)/T$.

If there are several species of particles, Eqs (\ref{energydens})--(\ref{particledens})
and (\ref{NRenergydens})--(\ref{NRparticledens}) must be replaced by the sum of similar
expressions over all species \cite{WALD}.

If the species were in thermodynamical equilibrium at temperature $T$, we would have
\begin{eqnarray}
\label{stareqs}
&& \epsilon_*(T)= T^4 \sum_i g_i a_\epsilon(m_i/T);\quad p_*(T)\!=
\!T^4 \sum_i g_i a_p(m_i/T);\nonumber\\
&& s_*(T)= T^3 \sum_i g_i a_s(m_i/T);\quad n_*(T)= T^3 \sum_i g_i a_n(m_i/T)\,.
\end{eqnarray}

By comparing the relativistic limit ($T\gg m$) and the non--relativistic limit
($T \ll m$), we see that the distribution functions are suppressed by the
factor $e^{-m/T}$. As the temperature drops below particle's mass, particles
and anti--particles tend to annihilate into photons or lighter particles
until the density and pressure of the primordial plasma gets dominated by
photons and neutrinos, although partially restored by particle--antiparticle
pair production. At higher energies these annihilations also occur. At low
temperatures, the thermal energies of the particles are not sufficient for
pair production.

If this equilibrium had persisted until today, the universe would mostly
be photons. To understand the present state of the universe, it is crucial
to understand the deviations from equilibrium. As long as the temperature of
the universe is greater than the rest mass of an electron, $0.511$ MeV,
pair--creation continues; but when the universe cools down below $0.511$ MeV,
the electrons remain bounded to protons, the mean life time of photons become
comparable to the age of the universe and pair--creation no longer occurs.

In these circumstances, some stable particles of species $i$ with sufficiently
small mass (Dirac neutrinos and perhaps sterile Majorana neutrinos), decouple
from the matter field at temperatures $T_i$ greater than $T$ and remain
``freezed out'' in this state with distribution functions $\epsilon_i(T_i)>
\epsilon_i(T)$, $p_i(T_i)> p_i(T)$, $s_i(T_i)> s_i(T)$ and  $n_i(T_i)> n_i(T)$
\cite{DODELSON} \cite{KOMATSU} \cite{BAUMANN}.

We are interested in the exploiting the consequences of the entropy
conservation, we shall determine this conservation law for entropy
densities in the co--expanding tubes of nearby worldlines shown in
Fig.\,\ref{Figure06} of \S\,\ref{Kin&PropST}. As there discussed,
it will be sufficient to consider the conservation property in the
axial tube, whose diameter depends only on the scale factor of
vacuum dynamics, $\widetilde{\alpha}(\widetilde{\tau})$, but not
on the expansion factor of the universe, $\widetilde{a}
(\widetilde{\tau})$, because the Hubble expansion is stagnant
along the worldline of the comoving observer.

The same considerations can be extended to the co--expanding
tube directed by any other worldline $\Gamma(\vec\rho)$, because
the  worldline of any comoving observer can be transformed to
that of any other comoving observer by a suitable gauge transformation
of the metric tensor. However, for our purposes we only need to
determine the entropy density in a small volume element
$dV(\widetilde{\tau}_B$, in the beginning of the axial tube, at big--bang
temperature $T_B$, and that in the corresponding co--expanded volume
$dV(\widetilde{\tau}_U)$ at the present universe age $\widetilde{\tau}_U$,
at the present background temperature $T_{BK}$.

While $T_{BK}$ is know by direct measurements of cosmic radiation,
we need only to determine  $T_B$, which is just the temperature
of Higgs field at big bang.

In Appendix {\bf \ref{VacDynApp}} near Eq (\ref{HiggsEdens}), we
have shown that the energy density of the Higgs field in the
hyperboloidal section of the spacetime, at big bang temperature
$\tilde\tau_B$, is
\begin{equation}
\label{HiggsEdensQ}
\widetilde{U} (\tilde\tau_B)=\frac{\mu_H^4}{16\,\lambda}
\cong  1.186\times 10^8 \,\mbox{GeV$^4$}\,.
\end{equation}

Since we presume that at big--bang time $\widetilde{\tau}_B$ the Higgs bosons soon
created are in thermal equilibrium as a gas of free particles non yet decayed, we
can infer the temperature of the Higgs  bosons by solving equation $\epsilon_*(T_B) =
\mu_H^4/16\lambda$ for $T_B$  by numerical methods.

To obtain $T_B$ we must compute Eq (\ref{energydens}), with $g=1$ and $m=\mu_H=125.1$
GeV, for dense sets of $T$--values, and find the value that minimizes $\epsilon_*(T)
-\mu_H^4/16\lambda$. Then, using Eq (\ref{pressure}) and (\ref{entropydens}) with
$T=T_B$, we can calculate Higgs--fluid pressure $p_*(T_B)$, entropy density $s_*(T_B)$
and boson density $n_*(T_B)$. Numerical computations give:
\vspace*{1mm}
\begin{eqnarray}
\label{bigbangtemp}
&&\hspace{-10mm}T_B \cong 141.03\,\hbox{GeV}\quad\hbox{big bang temperature}\,; \\
\label{epsstar}
&& \hspace{-10mm}\epsilon_*(T_B) = \frac{\mu_H^4}{16\,\lambda} \cong
1.186\times 10^8\,\hbox{GeV}^4
\quad\hbox{energy density at big bang}\,;\\
\label{gepsstar}
&&\hspace{-10mm}g_{\epsilon*}(T_B) = \frac{30}{\pi^2}\,\frac{\epsilon_*(T_B)}{T^4_B}
\cong 0.9112\quad\hbox{effective degeneracy of }\,\epsilon_*(T_B)\,;\\
\label{higgspress}
&&\hspace{-10mm}p_*(T_B) \cong 3.554\times 10^7 \, \hbox{GeV}^4 \quad\hbox{pressure at big bang}\,;\\
\label{gpstar}
&&\hspace{-10mm}g_{p*}(T_B) = \frac{30}{\pi^2}\,\frac{p_*(T_B)}{T^4_B} \cong 0.2731
\quad\hbox{effective degeneracy of }\,p_*(T_B)\,;\\
\label{pstaronestar}
&&\hspace{-10mm}\frac{p_*(T_B)}{\epsilon_*(T_B)} = \frac{g_{p*}(T_B)}{g_{\epsilon*}(T_B)}
\cong  0.2997\quad(\hbox{relativistic limit = }1/3)\,; \\
\label{sstar}
&&\hspace{-10mm}s_*(T_B) = \frac{\epsilon_*(T_B)\!+\!p_*(T_B)}{T_B} \cong 1.093\times 10^6
\hbox{GeV}^3\,\,\hbox{entropy density at big bang}\,;\\
\label{gsstar}
&&\hspace{-10mm}g_{s*}(T_B) = \frac{45}{2\pi^2}\frac{s_*(T_B)}{ T_B^3} \cong
0.8883\quad\hbox{effective degeneracy of }\,s_*(T_B)\,;\\
\label{gestarder}
&&\hspace{-10mm}\frac{d\epsilon_*(T_B)}{d T_B} \cong 3.513\times 10^6\,\hbox{GeV}^3\,;\quad
\frac{d g_{\epsilon*}(T_B)}{d T_B} \cong 1.142\times 10^{-3}\,\hbox{GeV}^{-1}\,;\\
\label{geptarder}
&&\hspace{-10mm}\frac{d p_*(T_B)}{d T_B} \cong 1.093\times 10^6\,\hbox{GeV}^3\,;\quad
\frac{d g_{p*}(T_B)}{d T_B} \cong 6.514\times 10^{-4}\,\hbox{GeV}^{-1}\,;\\
\label{sstarder}
&&\hspace{-10mm}\frac{d s_*(T_B)}{d T_B}\cong 2.491\times 10^4\,\hbox{GeV}^2;\quad
\frac{d g_{s*}(T_B)}{d T_B} \cong 1.345\times 10^{-3}\hbox{GeV}^{-1};\\
\label{higgdens}
&&\hspace{-10mm}n_*(T_B) \cong 2.655 \times 10^5\,\hbox{GeV}^3\quad \hbox{Higgs--boson density at big bang\,.}\\
&&\hbox{\small Table 1. Magnitudes of most significant thermodynamic quantities at big bang.}\nonumber\\
%\vspace{-4mm}
&&\hspace{-17mm}\hbox{These may be compared with the cosmic--background data observed today:}\nonumber\\
\label{TBK}
&&\hspace{-10mm} T_{BK}\cong 2.726 \, \mbox{$^{\mbox{\tiny o}}$K} \cong 2.350\times 10^{-13}\,
\hbox{GeV}\,\,\,\hbox{(temperature of cosmic background)}\,;\\
\label{BKepstardata}
&&\hspace{-10mm} g_{*\epsilon}(T_{BK}) \cong 3.738\,; \quad \epsilon_*(T_{BK}) =
\frac{\pi^2}{30}\,g_{*\epsilon}(T_{BK})\,T^4_{BK}\cong 3.750\times 10^{-51}\,\hbox{GeV}^4\,;\\
\label{BKsstardata}
&&\hspace{-10mm} g_{*s}(T_{BK})\cong  4.725\,;  \quad s_*(T_{BK})=\frac{2\pi^2}{45}\,g_{*s}(T_{BK})\,T_{BK}^3
\cong 2.69\times 10^{-38}\,\hbox{GeV}^3\,.\\
&&\hbox{\small  Table 2. Magnitudes of most significant thermodynamic quantities today.}\nonumber
\vspace{-10mm}
\end{eqnarray}

Here, $g_{*\epsilon}(T_{BK})$, $s_*(T_{BK})$ are respectively the energy density and entropy density of
photons and neutrinos in the cosmic background, and $g_{*\epsilon}(T_{BK})$, $g_{*s}(T_{BK})$ are their respective
degeneracy factors (other possible contributions are ignored)  \cite{KOLB} \cite{EGAN} \cite{MANGANO} \cite{FULLER}.

\subsection{The time course of entropy density after big bang}
\label{EntropyTimeCourse}
The dynamics of the vacuum state described in Appendix {\bf \ref{VacDynApp}} provides a good description of the
universe after the proper time of big bang, $\widetilde{\tau}_B$. A computation, which will be carried out
in \S\,\ref{crossroad} near Eq~(\ref{tildeTauB}), yields $\widetilde{\tau}_B \approx 7.6 \times 10^{-9}$s,
which can be set equal to zero, because it is absolutely negligible with respect to the age of the universe
$\tilde\tau_U\cong 4.358\times 10^{17}$s.

\begin{figure}[!h]
\centering
\includegraphics[scale=0.82, trim = 6mm 0mm 0 0mm, clip]{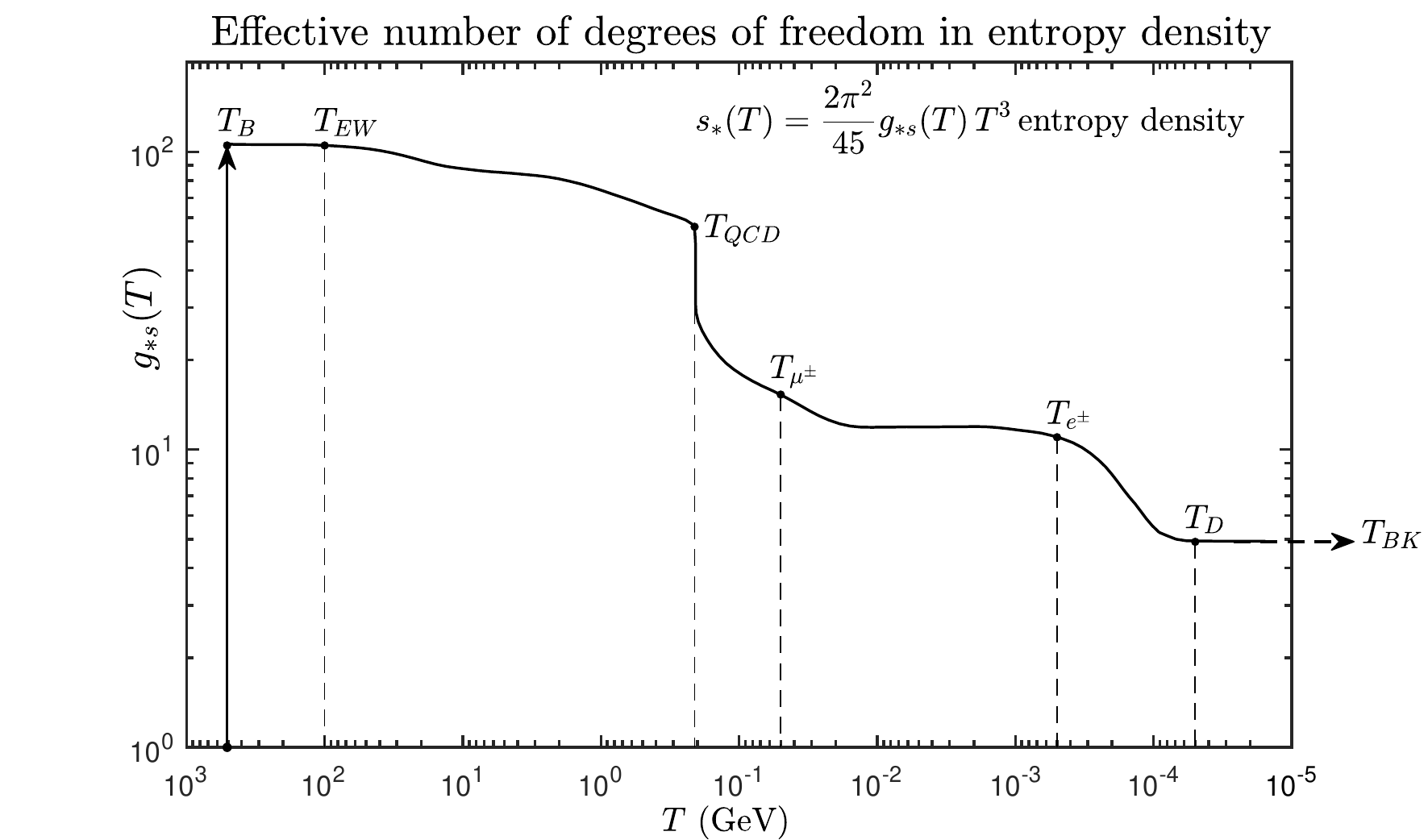}
\caption{\small {\bf Reinterpretation of the Kolb--Turner diagram}.
The vertical arrow on the left side represents the effect of the phase transition of the vacuum
state at big bang: the Higgs boson crowd that fill the critical hyperboloid at the moment of
big bang warms up suddenly to a thermodynamic state at temperature $T_B\!\cong\! 141$\,GeV. Soon
after this moment, the Higgs bosons decay in cascade into the inventory of SMEP through an
entropy--conserving process. {\bf On the top}: relation between entropy density $s_*(T)$ and
effective degrees of freedom $g_{*s}(T)$. {\bf Notable temperatures}: electroweak unification
at $T_{EW}\!\approx\!100$\,GeV; QCD phase--transition $T_{QCD}\!\cong\!200$\,MeV; $\mu^\pm$
annihilation $T_{\mu^\pm}\!\approx\!0.5$\,MeV; $e^\pm$ annihilation $T_{e^\pm}\!\approx\!0.5$\,MeV;
photon decoupling $T_D\!\cong\!5\times 10^{-5}$GeV; cosmic background $T_{BK}\cong 2.350\times
10^{-13}$GeV $\cong 2.726\,\mbox{$^{\mbox{\tiny o}}$K}$. ({\em Adapted from Fig.3.5 of
Ref.~\cite{KOLB}, pp. 65--67}).}
\label{kolbturner}
\end{figure}

The thermodynamic state of the system after $\widetilde{\tau}_B$ lasts a short time because the
Higgs bosons decay very rapidly into a complicate mix of particles, which can only be described
as a thermodynamic system. Fig.\,\ref{kolbturner} represents the thermal history of the universe
according to the SMMC: a sequence of thermodynamical stages with different entropy densities.

\subsection{Entropy conservation after big bang}
\label{EntrConserv}
From \S\,\ref{Kin&PropST} near Fig.\,\ref{Figure06} we know that the density of a conservative
quantity from big--bang time to today has evolved within the geodesic tube wrapped around the
worldline of the co--moving reference frame is independent of the universe expansion.

Since any two adjacent tubes, although intersecting in different ways across the infinite
array spacelike hyperboloids,  are nevertheless cosmological equivalent because the universe
is homogeneous and isotropic and have the same temperature $T$ no exchange of heat can take
place between each other. Possible deviations from this regime can only be caused by thermal
fluctuations of energy density at the age of big bang.

Since in these conditions the evolution of the entire system is almost adiabatic, and in expansion, we
infer that the entropy densities within each tube is almost exactly conserved.

Denoting by $s_*(T_1)$
and $s_*(T_{2})$ the entropy at two different position of the worldline, with temperatures $T_1$
and $T_2$, respectively measured at proper times $\widetilde{\tau}_1$ and $\widetilde{\tau}_2$,
are related to the the scale factor of vacuum dynamics, $\widetilde{\alpha}(\widetilde{\tau})$, by equation
%\vspace{-1mm}
\begin{equation}
\label{generalentropyratio}
\bigg[\frac{s_*(T_1)}{s_*(T_2)}\bigg]^{1/3}\!\!\! =\frac{\tau(\widetilde{\tau}_2)
\, \widetilde{\alpha}(\widetilde{\tau}_2)}{\tau(\widetilde{\tau}_1)
\, \widetilde{\alpha}(\widetilde{\tau}_1)}= \bigg[\frac{g_{*s}(T_1)}
{g_{*s}(T_2)}\bigg]^{1/3}\frac{T_1}{T_2}\,.
%\vspace{-1mm}
\end{equation}

In particular, for entropy densities $s_*(T_B)$ and $s_*(T_{BK})$, respectively at proper
big--bang time $\widetilde{\tau}_B$ and proper universe age $\widetilde{\tau}_U$,
using the data of Tables 1 and 2, we obtain
%\vspace{-1mm}
\begin{equation}
\label{entropyratio}
\frac{\tau(\widetilde{\tau}_B)\,\sqrt{\alpha(0)}}{\tau(\widetilde{\tau}_U)\,
\widetilde{\alpha}(\widetilde{\tau}_U)} =\bigg[\frac{s_*(T_{BK})}{s_*(T_B)}\bigg]^{1/3}\!\!\! =
\bigg[\frac{g_{*s}(T_{BK})}{g_{*s}(T_B)}\bigg]^{1/3}\frac{T_{BK}}{T_B}\cong 2.91
\times 10^{-15}\equiv A\,,
\end{equation}
from which  we derive
\begin{equation}
\label{entropyratio2}
\frac{\tau(\widetilde{\tau}_U)}{\tau(\widetilde{\tau}_B)}
\frac{T_{BK}}{T_B}\bigg[\frac{g_{*s}(T_{BK})} {g_{*s}(T_B)}\bigg]^{1/3}\!\!=
\frac{\tau(\widetilde{\tau}_U)}{\tau(\widetilde{\tau}_B)}\,\frac{A}{\sqrt{\alpha(0)}}\,. % A/sqrta0 = 0.0074
\end{equation}

Using the the first and the second of Eqs (\ref{s0tosj}) of Appendix {\bf \ref{VacDynApp}}, we obtain
the scale factors at kinematic times $\tau_B=\tau(\widetilde{\tau}_B)$ and
$\tau_U=\tau(\widetilde{\tau}_U) > \tau_B$:
\begin{eqnarray}
\label{sqrtalpha0}
&&\alpha(\tau_B)\equiv  \widetilde{\alpha}(\widetilde{\tau}_B) =
1-\frac{\tau^2_B}{\tau_c^2}\cong \sqrt{\alpha(0)}\,,\\
\label{tauUEqs}
&&\alpha(\tau_U)\equiv  \widetilde{\alpha}(\widetilde{\tau}_U) =  1- \frac{\tau_B^4}{\tau_c^2 \tau_U^2}
\cong  1- \frac{\tau_B^2}{\tau_U^2}=1- \frac{\tau_B^2}{\tau(\widetilde{\tau}_U)^2}\,,
\end{eqnarray}
where we have put $\tau_B^2/\tau_c^2 \cong 1$ in the last step of the second equation.

\newpage

\subsection{The prodigious melting pot of CGR}
\label{crossroad}
We have finally arrived to the central crossroad of our investigations at which all physical
constants, theoretical constrains and logical implications imposed by the fundamental
principle of CGR converge together to provide a spectacular series of predictions.

Since the topic is a little bit complicated and articulated, we ask the reader to consider the
order of the topics, steps and methods of our computations. The most important results here
presented have been achieved by graphical methods, which are made possible by extraordinary
computational power of MATLAB programming \cite{MATLAB} through a sort of continuous dialectic
between routine--compilations and command--line operations.

To facilitate the computations, let us start from determining the kinematic time $\tau$ corresponding
to a proper time $\widetilde{\tau}$ of the deceleration era. First of all, we approximate the scale
factor of the dynamical vacuum $\alpha(\tau)$ by joining smoothly its initial and final branches, as
described in \S\,\ref{predictions} of Appendix {\bf \ref{VacDynApp}}, i.e., respectively,
$$
\alpha_i(\tau)= \frac{\alpha(0)}{1 -\tau^2/\tau_c^2}\,,\quad
\alpha_f(\tau)= 1 -\frac{\tau_B^4}{\tau_c^2{\tau}^2}\,,
$$
where $\tau_c > \tau_B$, with $\tau_c \cong \tau_B$, is the critical time at which the
spacetime would blow up.

Then, we carry out on these branches the following integrations:
\begin{eqnarray}
\label{integraltildetauc}
&&\hspace{-17mm}\widetilde{\tau} =\int_0^{\tau}\!\!\alpha_i(\tau')\,d\tau'=\frac{\alpha(0)\,\tau_c}{2}
\ln \frac{\tau_c+\tau}{\tau_c-\tau} \cong\alpha(0)\bigg(\!\tau+\frac{\tau^3}{3\,\tau_c}+
\cdots\!\!\bigg),\quad (0\leq \tau\leq \tau_B)\,;\\
\label{tildetau2tilde}
&&\hspace{-17mm}\widetilde{\tau} - \widetilde{\tau}_B = \int_{\tau_B}^{\tau}\!\!\alpha_f(\tau')\,d\tau' =
\tau-\tau_B+\frac{\tau_B^4}{\tau_c^2} \bigg(\frac{1}{\tau}-\frac{1}{\tau_B}\bigg),\quad
(\tau\geq\tau_B)\,;
\end{eqnarray}
where $\widetilde{\tau}_B$ is the proper time corresponding to $\tau_B$.
Putting $\tau=\tau_B$ in Eq (\ref{integraltildetauc}), we obtain
\begin{equation}
\label{tildeTauB}
\widetilde{\tau}_B=\frac{\alpha(0)\,\tau_c}{2}\ln\frac{(1+\tau_B/\tau_c)^2}{1-\tau^2_B/\tau_c^2}\cong
\frac{\alpha(0)\,\tau_c}{2}\ln\!\frac{4}{\sqrt{\alpha(0)}}\approx 7.6\times 10^{-9}\hbox{sec}\,,
\end{equation}
which is very small compared compared to any significant age of the universe.

In the last step of Eq (\ref{tildeTauB}), we have used two fundamental relations of vacuum dynamics that
hold almost exactly in the extreme boundary conditions prescribed by CGR:  the value of big bang
time $\tau_B\cong \tau_c$ and equation $\alpha(\tau_B)\cong \sqrt{\alpha(0)}$, respectively provided
by Eqs (\ref{s0oftau}) and (\ref{s0stauj}) of Appendix {\bf \ref{VacDynApp}},
\begin{equation}
\label{TheB}
\tau_B \cong \frac{\sqrt{8\,\lambda}\,\sigma_0}{\alpha(0)\,\mu^2}\cong \frac{5.093\times 10^{-10}}{
\alpha(\tau_B)^2}\,\hbox{sec}\,.
\end{equation}
where $\sigma_0 \cong 5.959\times 10^{18}$ GeV is related to the gravitational coupling constant
of GR $\kappa$ by equation $\sigma_0 =\sqrt{6/\kappa}$, $\mu\cong 88.47$\,GeV is related to
Higgs--boson mass $\mu_H \cong 125.1$\,GeV by equation $\mu=\mu_H/\sqrt{2}$ and
$\lambda\cong 0.1291$ is the self--coupling constant of the Higgs field. In the following
the numerator appearing in the second part of Eq (\ref{TheB}) will be denoted as
\begin{equation}
\label{Bnumb}
B =5.093\times 10^{-10}\,\hbox{sec}\,.
\end{equation}

Therefore, since $\tau_B^2/\tau_c^2\cong 1$ and $\widetilde{\tau}_B$ is negligible,
Eq (\ref{tildetau2tilde}) will be simplified to
\begin{equation}
\label{propkinrel}
\widetilde{\tau} - \widetilde{\tau}_B \cong \tau-\tau_B+\frac{\tau^{\,4}_B}{\tau_c^2}
\bigg(\frac{1}{\tau}-\frac{1}{\tau_B}\bigg)\cong  \tau-2 \tau_B + \frac{\tau_B^2}{\tau}\,.
\end{equation}

Solving this equation for positive values of $\tau$, and approximating $\widetilde{\tau}
-\widetilde{\tau}_B$ to $\widetilde{\tau}$, we obtain from Eq (\ref{propkinrel}) the
kinematic time as a function of proper time,
\vspace{-2mm}
\begin{equation}
\label{proptime2kintime}
\tau(\widetilde{\tau}) \cong \tau_B + \frac{\widetilde{\tau}-\widetilde{\tau}_B}{2}
\Bigg(1+\sqrt{1+ \frac{2\,\tau_B}{\widetilde{\tau}-\widetilde{\tau}_B}}\,\,\Bigg) \cong
\tau_B + \frac{\widetilde{\tau}}{2} \Bigg(1+\sqrt{1+ \frac{2\,\tau_B}
{\widetilde{\tau}}}\,\,\Bigg)\,.
\vspace{-2mm}
\end{equation}
Note that the term $\widetilde{\tau}_B$ in the last step can be safely omitted because
$\widetilde{\tau}_B/\widetilde{\tau}_U \cong 1.75\times 10^{-26}$.

Since the age of the universe evaluated by the cosmologists in several independent ways, insists on
$\tilde\tau_U\cong 13.82\times 10^9 \hbox{Gyr} \cong 4.36\times 10^{17}\,\hbox{sec}$, we obtain for
the well--approximated age of the universe in kinematic--time units the expression
\vspace{-2mm}
\begin{equation}
\label{tildetau2tau}
\tau_U \cong \tau_B +  \frac{\widetilde{\tau}_U}{2} \Bigg(1+\sqrt{1+ \frac{2\,\tau_B}
{\widetilde{\tau}_U}}\,\Bigg)\,.
\end{equation}

Now we want to show that the same quantities $\tau_U$, $\tau_B$ and $\widetilde{\tau}_U$ are mutually
related by a second equation which is totally different from Eq (\ref{tildetau2tau}). So, by combining
the two equations, we will ba able to determine a fix value $\tau_B$, thus unlocking the numerical
values of all the significant parameters of CGR.

To achieve this result, let us first combine Eq (\ref{entropyratio2}) with Eq (\ref{tauUEqs}),
so as to obtain
\begin{equation}
\label{alphaArel}
\alpha(\tau_B) = \bigg(1- \frac{\tau_B^2}{\tau_U^2}\bigg)\,\frac{\tau_U}{\tau_B}\,A
= \bigg(\frac{\tau_U}{\tau_B} - \frac{\tau_B}{\tau_U}\bigg)\,A\,.
\end{equation}
where for clarity the symbol $\cong$ has been replaced by that of equality.

Then combine Eqs (\ref{alphaArel}) and (\ref{TheB}), so as to eliminate variable
$\alpha(\tau_B)$ and obtain
\begin{equation}
\label{relBA}
\sqrt{\frac{B}{\tau_B}} = \bigg(\frac{\tau_U}{\tau_B} - \frac{\tau_B}{\tau_U}\bigg)\,A\,,
\end{equation}
where the numerical parameter $B$ defined in Eq (\ref{Bnumb}) has been used. Rearranging
the terms of this equation, we obtain the algebraic equation of second order in $\tau_U$
$$
\tau_U^2- \tau_U \frac{\sqrt{B}}{A}\,\sqrt{\tau_B} - \tau_B^2 =0.
$$

Finally, solving  this equation for positive values of $\tau_U$, we obtain,
\vspace{-1mm}
\begin{equation}
\label{solvedforTauU}
\tau_U \cong\frac{\sqrt{\tau_B}}{2}\Bigg(1+\sqrt{1+\frac{4\,\tau_B}{C^2}}\Bigg)\,C,
\,\,\,\hbox{where } C =  \frac{\sqrt{B}}{A} \cong 7.731\times 10^9\, \hbox{sec}^{1/2},
\vspace{-1mm}
\end{equation}

\begin{figure}[!h]
\centering
\includegraphics[scale=0.85, trim = 10mm 0mm 0 0, clip]{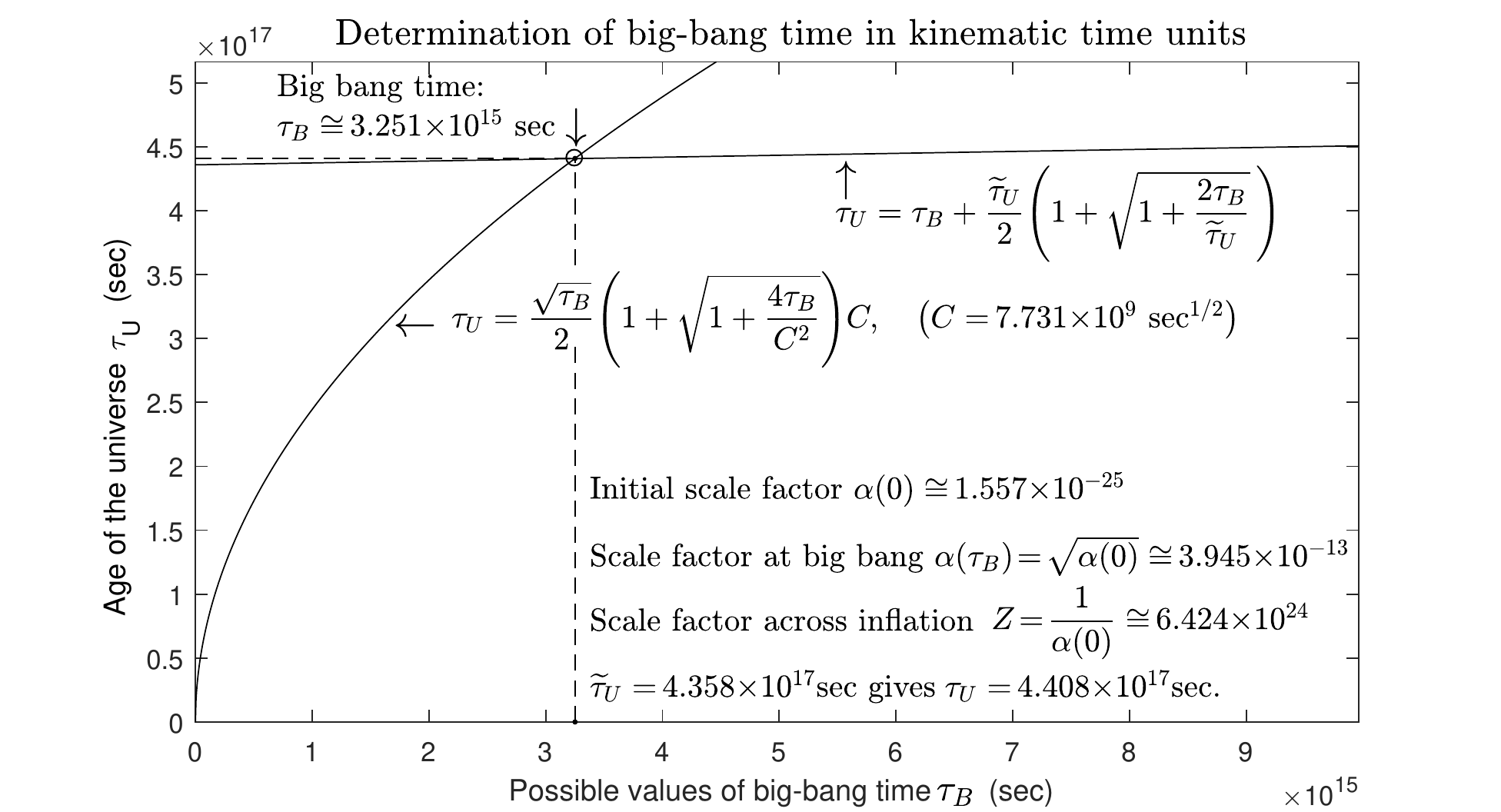}
\caption{\small The value of the big bang time in kinematic time units $\tau_B$
(downward arrow), is obtained by intersecting two different curves
respectively representing the age of the universe, $\tau_U$, as functions
of $\tau_B$. The first curve (upward arrow) is described by Eq (\ref{tildetau2tau}),
the second (leftward arrow) by Eq (\ref{solvedforTauU}). Once determined $\tau_B$,
we are in a position to calculate $\tau_U$, the initial value of scale
factor $\alpha(0)$, that at big bang, is $\alpha(\tau_B)=\sqrt{\alpha(0)}$
and the expansion factor of spacetime over time, $Z=1/\alpha(0)$.
\label{crit_time_det}
\vspace{-6mm}
}
\end{figure}

As shown in Fig.\,\ref{crit_time_det}, the curves described by Eqs (\ref{tildetau2tau})
and (\ref{solvedforTauU}) intersect at  proper time $\tau_B=3.251\times 10^{15}$, thus
determining the values of big--bang time $\tau_B$ and of $\alpha(\tau_B)$.

\newpage

Let us show how the doubling of the linear expansion factor of the
entropy variation from big bang to today alters the fundamental parameters of CGR.
\begin{figure}[!h]
\centering
\includegraphics[scale=0.85, trim = 8mm 0mm 0 0, clip]{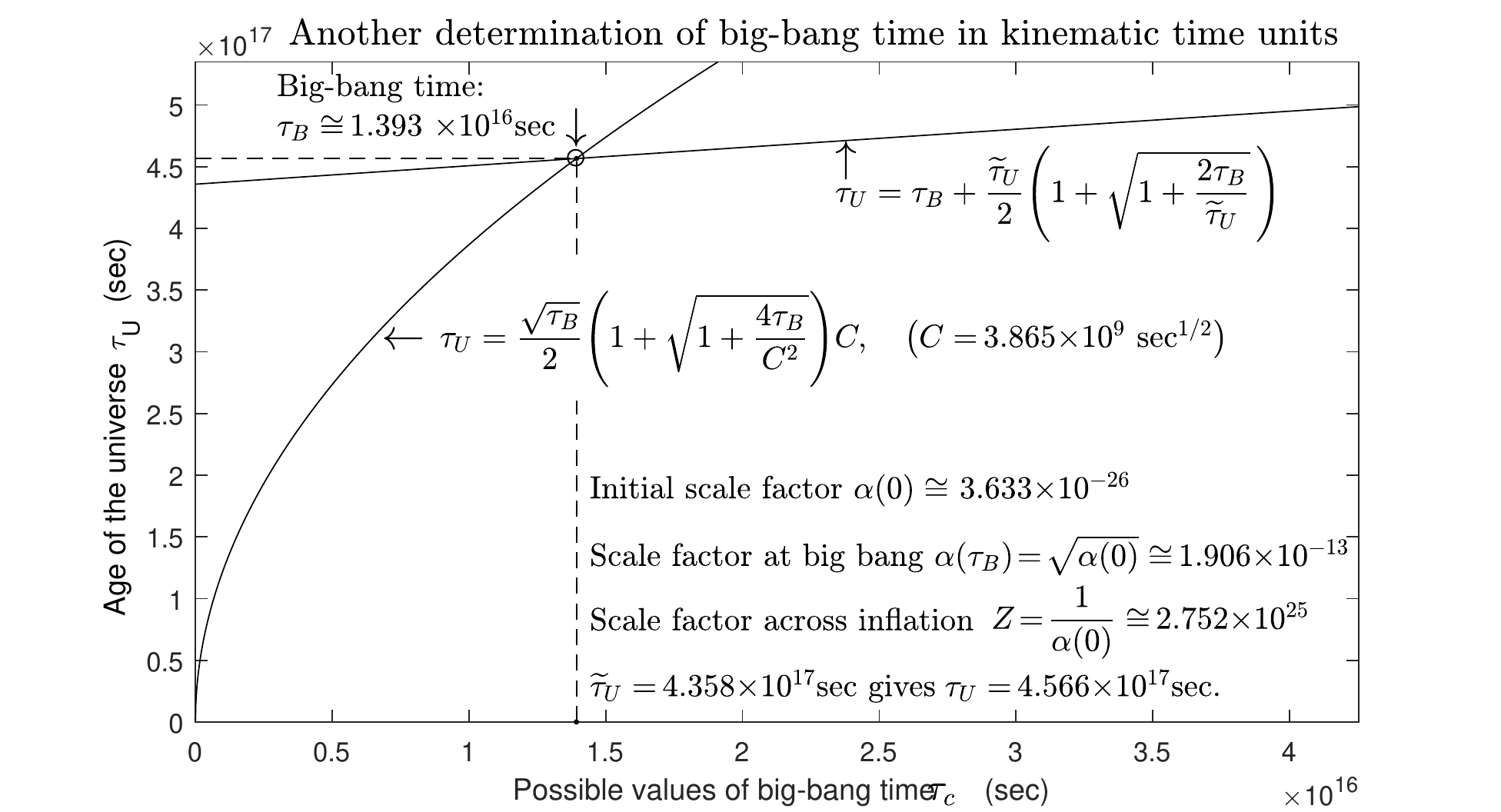}
\caption{\small (For comparison with Fig.\,\ref{crit_time_det}). By
increasing the entropy density of the universe today by a factor of two,
the value of $C$ halves, the big--bang time $\tau_B \cong 1.393\times
10^{16}\,$sec (downward arrow) has increased by a factor of $\cong 4.28$.
Correspondingly, the initial value of scale factor $\alpha(0)$ decreases by
a factor of
$\cong 0.233$, so that its total variation across inflation, $Z=1/\alpha(0)$,
increases by a factor of $\cong 4.286$. Since these predictions are so sensitive
to the variation of the linear expansion factor $A$, introduced in Eq
(\ref{entropyratio}), which is proportional to the present entropy density
of the universe, the question arises of whether the doubling of entropy
density may be due to right--handed sterile neutrinos, possibly thermalized
by interactions with the standard left--handed ones.}
\label{scfactonentr}
\end{figure}

The possibility that the entropy of the cosmic background is larger than that
predicted by the SMEP has been advanced by several authors. The hypothesis that
the existence of sterile neutrinos enhance the entropy by a factor of two or
tree has been advanced by Egan and Lineweaver in 2010 and by Fuller {\em et al.}
in 2011.

The factor might be even higher if hybrid Dirac--Majorana neutrinos of
the types described in Appendix {\bf \ref{DirMajorApp}} should exist. However,
since this argument is merely speculative, we avoid discussing further about it.
\newpage

\begin{figure}[!h]
\centering
\includegraphics[scale=0.77, trim = 18mm 0 0 0, clip]{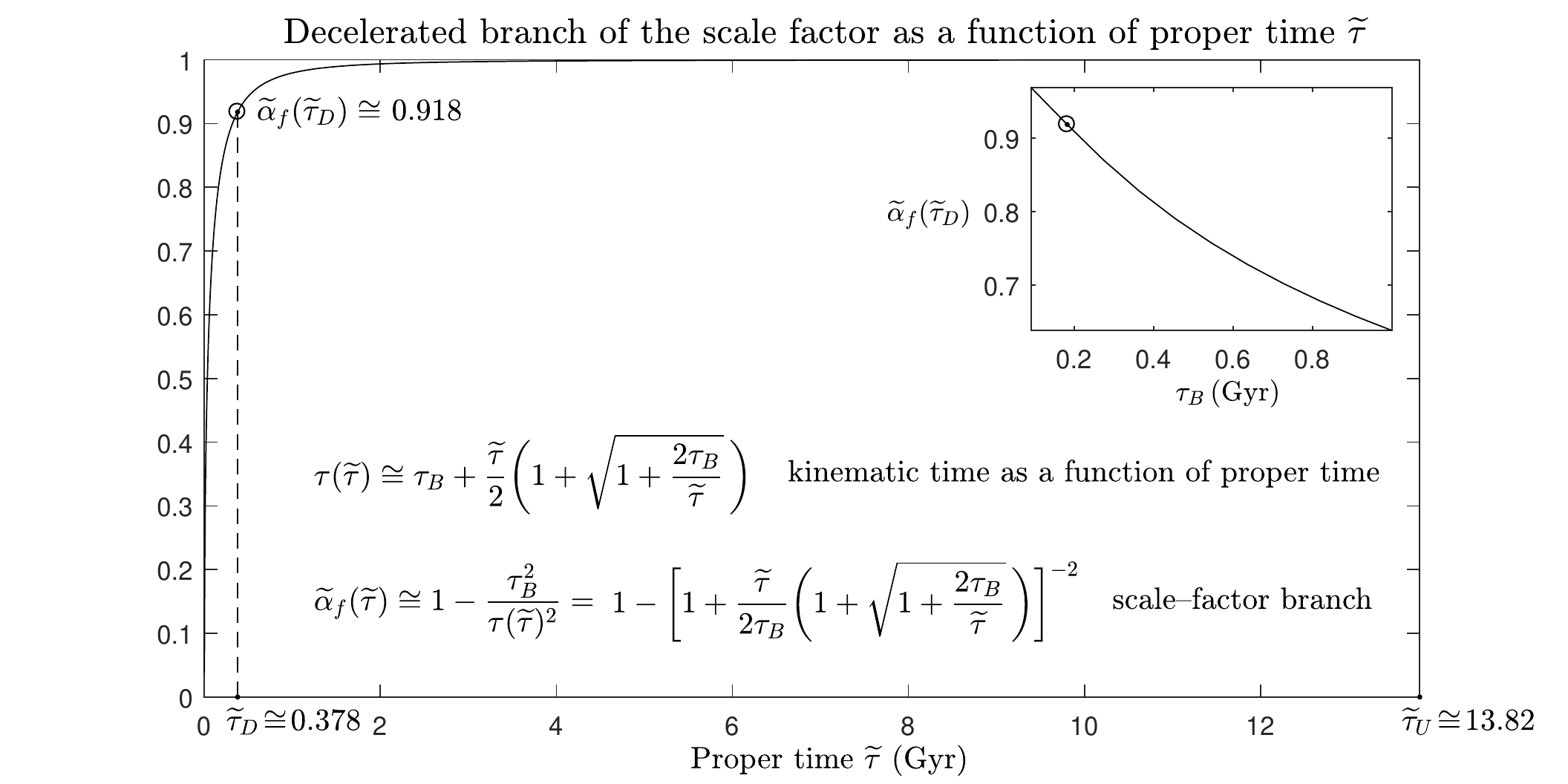}
\caption{\small Profile of the scale factor of vacuum dynamics as a
function of proper time $\widetilde{\tau}$ (in Gyr) for a given value
of big--bang time, $\tau_B$. The profile is originally defined by the
second of Eqs (\ref{s0tosj}), in Appendix {\bf \ref{VacDynApp}}, as
the accelerated branch of the scale factor $\alpha_f(\tau)\cong 1
-\tau_B^2/\tau^2$, which is a function of kinematic time $\tau$. To obtain
the scale factor as a function of the proper time we reported from Eq
(\ref{proptime2kintime}) the expression for $\tau(\widetilde{\tau})$ shown
in the figure. The proper time ranges from the big--bang time
$\widetilde{\tau}_0\cong 0$, to the present age of the universe
$\widetilde{\tau}_U\cong 13.82$ Gyr. Marked on the time axis, is
also the photon--decoupling time, $\widetilde{\tau}_D\cong 0.378$ Gyr,
and the value of the scale factor at $\widetilde{\tau}_D$,
$\widetilde{\alpha}({\widetilde{\tau}_D})$. The inset on the top--right
of the figure shows how the values of $\widetilde{\alpha}
({\widetilde{\tau}_D})$ vary with $\tau_B$.}
\label{HubbleParEntr}
\end{figure}

For $\tau(\widetilde{\tau})\ge \tau_B$, the scale factor
$\widetilde{\alpha}_f(\widetilde{\tau})$, has the expression
shown in Fig.\,\ref{HubbleParEntr},
\vspace{-3mm}

This figure shows very clearly that, at photon--decoupling time, the scale
factor of vacuum dynamics differs appreciably, if not considerably, from
its asymptotic value 1. This fact appears even more relevant if we consider
that the strength of the gravitational attraction is proportional to
$1/\widetilde{\alpha}({\widetilde{\tau}})^2$, as proven in
\S\,\ref{PhenafterBB} near Eq (\ref{TetaBGB}).

This curious effect leads us to predict that the astronomic observations of
events occurred soon after $\widetilde{\tau}_D$ should unveil remarkable deviations
from the predictions of the SMMC: in particular, increased gravitational redshift
of distant stars, currently imputed to accelerated expansion of the universe
(Riess {\em et al.}, 1998), formation of supermassive black holes (Pezzulli
{\em et al.}, 2016); Baados {\em et al.}, 2018), demographic decrease of stars
(Sobral {\em et al.}, 2012), and other unexpected phenomena that we will
describe in Sec.\,\ref{CMBanysot}.

\newpage

\markright{R.Nobili, Conformal General Relativity -- {\bf \ref{CMBanysot}} Anisotropies of the cosmic background}
\section{The lower bound of cosmic background anisotropies}
\label{CMBanysot}
The anisotropies of the cosmic microwave background (CMB), detected by spatial or
terrestrial infrared--sensitive telescopes, can be expressed as a sum of terms, called
multi--poles, characterized by progressively finer angular features. The multi--pole
expansion of the CMB is a mathematical series of spherical harmonics of degrees ranging
from $\,\ell =$ 2 to 10000, whose power spectrum extends from $\cong 35 \mu$K$^2$ to
$6\times 10^3 \mu$K$^2$ \cite{HINSHOW} \cite{WRIGHT} \cite{SHIROKOFF}.

The SMMC explains the CMB as the delayed manifestation of very strong quantum fluctuations
occurred in causally disconnected regions of the primordial spacetime, which survived the
expansion of the universe at superluminal speed during the acute stage of inflation (Mukhanov,
Ch.5, 2005). Unfortunately, the QFT does not explain how a superpositions of virtual quanta
can evolve unitarily to thermal fluctuations.

CGR instead explains the CMB anisotropies as thermal fluctuations of the Higgs field
at the big--bang temperature of about 141 GeV, which favored the gravitational collapse
of the Higgs boson gas and its decay products into clumps of various sizes and shapes.
This happened because at big bang the gravitational attraction is enhanced by a factor
of $\alpha(0)^{-1}$, i.e., about $2.409\times 10^{25}$ times stronger than today, as
described in Fig.\,\ref{scfactonentr} of \S\,\ref{crossroad}.

In GR, the mechanism of gravitational collapse was investigated in 1902 by Jeans \cite{JEANS},
who showed that a homogeneous sphere of non--relativistic gravitating fluid becomes unstable
as its radius exceeds a critical value $R_J$, known as the {\em radius of Jeans} \cite{WEINBERG2}.

Presuming that at big bang the Higgs boson gas behaves as an adiabatic fluid at a constant pressure,
we infer that the sum of gravitational energy $U_G$ and thermal energy $U_T$ of a Jeans sphere
are initially in equilibrium. Therefore, the simplest way to determine $R_J$ is by requiring
that the derivative of $U_G+U_T$ with respect to radius is zero.

In the Newtonian approximation, the gravitational potential $\Phi$ is related to matter density $\rho$
by equation $\nabla^2\Phi = 4\pi G\rho$, where $\nabla^2$ is the operator of Laplace and
$G\equiv\kappa/8\pi$ is the gravitational coupling constant of Newton. Therefore, at the surface
of a sphere of radius $R$ and mass $M=4\pi\rho R^3/3$, we have $\Phi(R)\!=\!GM/R$.

Since the contribution to $U_G$ exerted by the spherical shell of radius $R$ and thickness $dR$ is
$dU_G= -\Phi(R)\,4\pi\rho R^2 dR$, we obtain by integration,
\begin{equation}
\label{U_G}
U_G = - \frac{16\,\pi^2 G\,\rho^2\, R^5}{15} = - \frac{3\,G\,\rho^2\, V^2}{5\,R}\,,\quad\mbox{where }\,
V=\frac{4}{3}\,\pi R^3\,.
\end{equation}

To determine $U_T$, we must know the temperature $T$ and the specific heat capacity at
constant pressure $c_P$ of the matter inside the sphere. Presuming that the matter field
is initially homogeneous and isotropic, we find immediately the heat capacity of the sphere
\vspace{-1mm}
\begin{equation}
\label{U_T}
U_T = c_P\,V\,T\,.
\vspace{-1mm}
\end{equation}
Therefore, by imposing the initial equilibrium condition $d(U_T+U_G)/dR=0$, we obtain
\begin{equation}
\label{R_Ggas}
R_J=\sqrt{\frac{3\,c_P \,T}{4\,\pi\, G\,\rho^2}}.
\end{equation}

To translate these concepts to CGR, we proceed as follows:

1) Dive in the representation of the truncated conical universe, described in Fig.\,\ref{Figure06}
of \S\,\ref{Kin&PropST}, and focus on the worldline--tube wrapped around the axial worldline $\Gamma^0$
of the comoving observer at proper time $\widetilde{\tau}_B$. In this way, we can neglect all
the aspects of CGR dynamics concerning the behavior of the matter in worldline--tubes wrapped
around worldlines $\Gamma(\vec \rho\,)$ stemming from the base of this cosmological representation
with other directions $\vec \rho$.

2) Replace in Eq (\ref{R_Ggas}) $\rho^2$ with $\epsilon_*(T_B)^2$, where $\epsilon_*(T_B)=
1.186\times 10^8\,\hbox{GeV}^4$ is the energy density at big bang provided by
Eq (\ref{epsstar}) of \S\,\ref{unilargescale}.

3) Since in standard thermodynamics $c_P$ is related to the specific heath capacity
at constant volume $c_V$ by equation $c_P=c_V\gamma$, where $\gamma$ is the adiabatic
factor, replace $c_P$  appearing in the Eq~(\ref{R_Ggas}), with $c_P(\tau_B)=
c_V(\tau_B)\,\gamma_*(T_B)$, where
$$
c_V(T_B)\equiv T_B\,\frac{d s_*(T_B)}{dT_B} \cong 3.51\times 10^6\,\hbox{GeV}^3\,,
$$
is the specific heath capacity at constant volume of the Higgs boson gas calculated
using Eqs (\ref{bigbangtemp}) and (\ref{sstarder}) provided in Table~1 of
\S\,\ref{unilargescale}. The adiabatic factor
$$
\gamma_*(T_B) \equiv 1 + \frac{p_*(T_B)}{\epsilon_*(T_B)} \cong 1.2997
$$
is determined as the ratio between the enthalpy density $\epsilon_*(T_B) + p_*(T_B)$ and
energy density $\epsilon_*(T_B)$ of the Higgs boson gas at big bang. Of note, the reason why
$\gamma_*(T_B)$ differs so much from the adiabatic factor $\gamma = 5/3\cong 1.6667$ of a perfect
gas of neutral particles is that the energy density of the Higgs boson gas at big bang is nearly
relativistic.

4) Replace $G$ with $G/\alpha(\tau_B)^2$, where $\alpha(\tau)$ is the scale factor as a function
of kinematic time $\tau$. This is equivalent to multiplying $R_J$ by $\alpha(\tau)$.

We obtain, thereby, the critical radius of the Jeans--sphere at kinematic time $\tau_B$:
\begin{equation}
\label{RadOfJeans}
R_J(\tau_B)=a(\tau_B)\, R_0(\tau_B),\,\, \hbox{where } R_0(\tau_B)=
\sqrt{\frac{3\,c_P(T_B)\,T_B}{4\pi G\,\epsilon_*(T_B)^2}}\cong  25.18 \hbox{\,cm}\,.
\end{equation}

We derive from this equation the additional quantities:
\begin{eqnarray}
\label{addquants}
&&\hspace{-10mm} V_J(T_B) =(4/3)\,\pi\,R_J(\tau_B)^3\quad \hbox{volume of the Jeans sphere
at big bang,} \nonumber\\
&& \hspace{-10mm}N_J(T_B) =n_*(T_B)V_J(\tau_B) \quad\,\,\,\hbox{mean number of Higgs bosons in }
V_J(T_B),\nonumber\\
&& \hspace{-10mm}\Delta N_J(T_B) \equiv \sqrt{N_J(\tau_B)}\!\!\qquad\quad\hbox{standard deviation of
$N_J(T_B)$,}\nonumber\\
&& \hspace{-10mm}\frac{\Delta N_J(T_B)}{N_J(T_B)}\equiv \frac{1}{\sqrt{N_J(T_B)}}
\qquad\hbox{entropy fluctuation of Higgs boson number,}
\end{eqnarray}
where $n_*(T_B) = 2.655\times 10^5$GeV$^3$ is the density of Higgs bosons at big bang
as given by Eq (\ref{higgdens}) of Table 1 in \S\,\ref{unilargescale}.

Exploiting the entropy conservation property stated in \S\,\ref{EntrConserv}, and using
the first of Eqs~(\ref{BKsstardata}) listed in Table 2 of \S\,\ref{unilargescale},
we can relate Eq (\ref{addquants}) with the thermal fluctuation of the Higgs sphere
of radius $R_J(\tau_B)$ resurfacing today through the photon decoupling era,
$$
\frac{\Delta N_J(T_B)}{N_J(T_B)}\equiv\frac{1}{\sqrt{N_J(\tau_B)}} =
\Delta \ln\big[g_{s*}(T_{BK})\,T_{BK}^3\big]\cong
3\,\frac{\Delta T_{BK}}{T_{BK}}\,,
$$
where $\Delta$ is regarded as a discrete differential. The last step of this
equation--chain reflects the fact that $g_{s*}(T_{BK})$ does not vary
appreciably with $T_{BK}$. Therefore, equation
\begin{equation}
\label{MinCMB}
W_{\hbox{\small min}} = \Delta\,T_{BK}^2 = \frac{T_{BK}^2}{9 N_J(T_B)}
\end{equation}
is the spectral power of CMB anisotropies caused by the thermal
fluctuation of the Jeans spheres collapsed soon after the big bang.
Presuming that these spheres are regions of minimum size warmed up by the
violence of the gravitational collapse, we advance the hypothesis that
Eq (\ref{MinCMB}) provides the lower bound of CMB anisotropies.

Here is the table of the results for tree different values of the scale factor:

\vspace{3mm}

\centerline{
\begin{tabular}{|l|l|l|l|}
\hline Variables & Fig.\,\ref{crit_time_det} of \S\,\ref{crossroad} &Fig.\,\ref{scfactonentr} of
\S\,\ref{crossroad}&Fig.\,\ref{anisotr} of next page\\
\hline
$\alpha(\tau_B)$\quad & $\cong 3.90\times 10^{-13}$  & $\cong 1.90\times 10^{-13}$  & $\cong 2.05\times 10^{-13}$\\
\hline
$R_J(\tau_B)$\quad & $\cong 98.2$ fm & $\cong  48.0$ fm & $\cong 51.6$ fm \\
%\hline
%$R_J(\tau_B)$\quad & $\cong 497$ GeV$^{-1}$& $\cong 243$ GeV$^{-1}$ & $\cong 202$ GeV$^{-1}$\\
\hline
$W_{\hbox{\small min}}$ & $\cong 54.2\times 10^{-13}\mu$K$^2$ & $\cong 46.4\times 10^{-12}\mu$K$^2$ &
$\cong37.3\times 10^{-12}\mu$K$^2$\\
\hline
\end{tabular}}
\vspace{3mm}

Since $R_J(\tau_B)$ ranges in the order of tens of femto--meters, we
may say that the gravitational collapse at big bang is {\em femto--granular}.

\newpage

\begin{figure}[!h]
\begin{center}
\scalebox{0.31}{\includegraphics{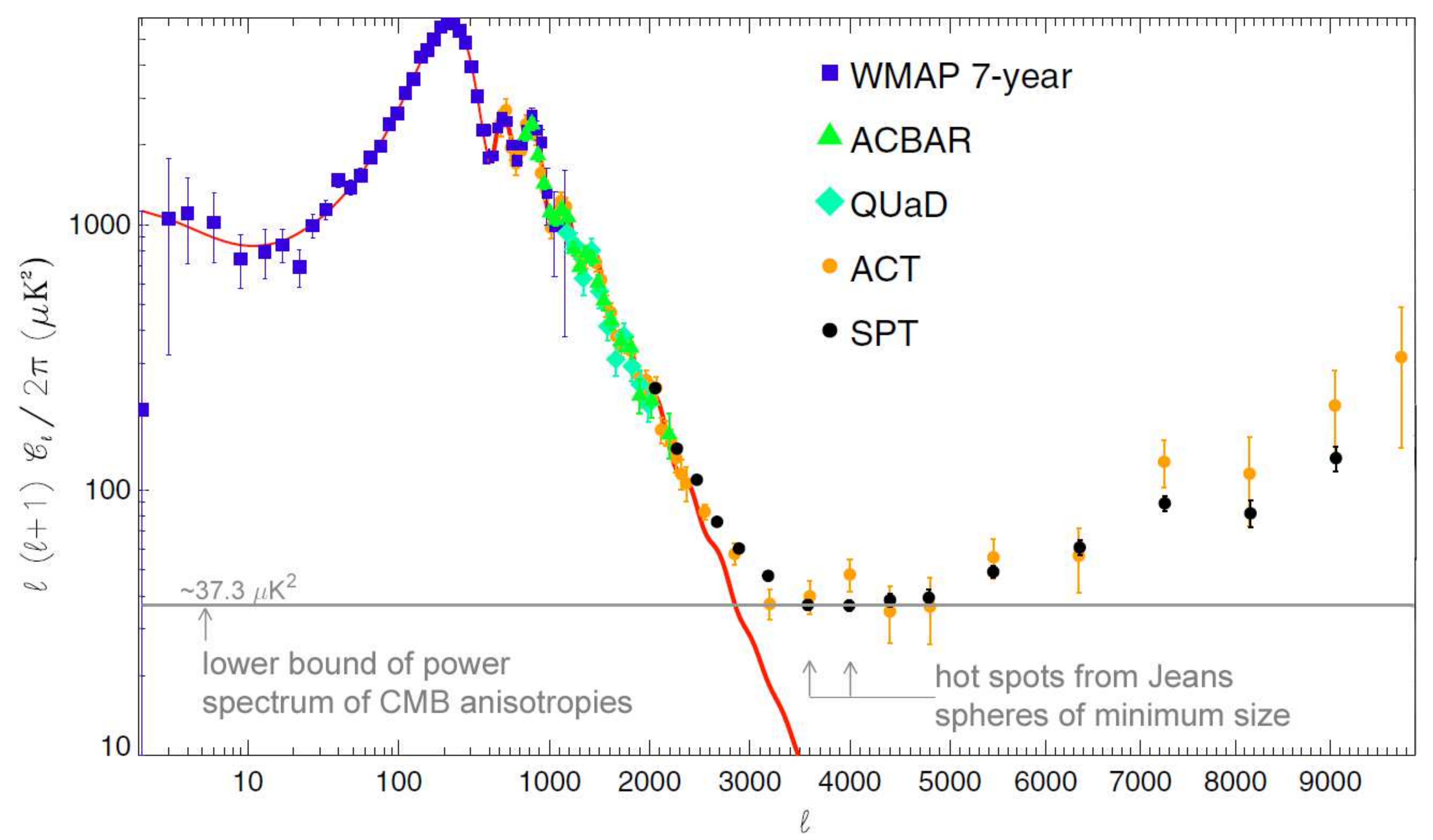}}
\end{center}
\vspace{-10mm}
\caption{\small Predicted level of CMB anisotropies of minimum size in $\mu$K$^2$
compared with data from five astronomical missions: WMAP (Wilkinson Microwave Probe
Telescope, 2001--2008); ACBAR (Arcminute Cosmology Bolometer Array Receiver,
2002--2006); QUaD (Q\&U Extragalactic Survey Telescope + Degree Angular Scale
Interferometer, 2003); ACT (Atacama Cosmology Telescope, 2014); STP (South Pole
Telescope, 2007--2011). SPT data with 3.5\% calibration error from figure~4 of
the paper of Shirokoff {\em et al.} (2011).
\label{anisotr}
}
\end{figure}

Fig.\,\ref{anisotr} shows that the hot spots of minimum spectral power are those
provided by the South Pole Telescope in the region of spherical harmonics of degree
$\,\ell\approx 4000$ lying at the level of $37.3\,\mu\hbox{K}^2$. In order that
Eq (\ref{MinCMB}) predicts just this value, the scale factor $\alpha(\tau_B)
\equiv \widetilde{\alpha}(\widetilde{\tau}_B)$ must be $\cong 2.05\times 10^{-13}$,
which is about 0.55 times smaller than the minimum entropy--ratio graphically
determined in Fig.\,\ref{crit_time_det} of \S\,\ref{crossroad}, but is $1.08$
times greater than the doubled entropy--ratio determined in Fig.\,\ref{scfactonentr}.
This is consistent with the existence of sterile neutrinos contributing to the
entropy density of the CMB background.

Despite the femto--granularity of the collapsed matter, the spectral power of the hot spots
is observable because it is magnified by a factor of $1/\alpha(\tau_B)^2= 2.38\times 10^{25}$
by the cosmic evolution of the gravitational coupling constant after big bang.

\centerline{---------------------------------}

\centerline{---------------}

\newpage

\appendix
\markright{R.Nobili, Conformal General Relativity - {\bf \ref{VacDynApp}}  Dynamical vacuum  of CGR}
\renewcommand\thefigure{\thesection.\arabic{figure}}
\setcounter{figure}{0}
\section{THE DYNAMICAL VACUUM OF CGR}
\label{VacDynApp}
In this Appendix we analyze and solve numerically Eqs (\ref{varphieqB}) and (\ref{sigmaeqB}) for the VEVs,
$\varphi(\tau)$ and $\sigma(\tau)$ of Higgs field $\varphi(x)$ and ghost field $\sigma(x)$, respectively,
in the kinematic--time representation. Here, for notational convenience, we cast such equations in the form:
\vspace{-1mm}
\begin{eqnarray}
\label{varphieq}
&&\hspace{-14mm}\ddot\varphi(\tau) + \frac{3}{\tau}\,
\dot\varphi(\tau)=\lambda\bigg[\frac{\mu^2}{\lambda}\,\alpha(\tau)^2-
\varphi(\tau)^2\bigg]\varphi(\tau),\quad 0< \varphi(\tau) \le \frac{\mu}{\sqrt{\lambda}}\,,\\
\label{seq}
&&\hspace{-14mm}\ddot\alpha(\tau) +  \frac{3}{\tau}\,\dot\alpha(\tau)=
\frac{\mu^2}{\sigma_0^2}\bigg[\frac{\mu^2}{\lambda}\,\alpha(\tau)^2-
\varphi(\tau)^2\bigg]\alpha(\tau),\quad 0< \alpha(\tau) \le 1\,.
\vspace{-1mm}
\end{eqnarray}

As in \S\,\ref{Evolvingvac}, we assume that $\sigma_0 =\lim_{\tau \rightarrow \infty}\sigma(\tau)$ is finite,
so that $\alpha(\tau)=\sigma(\tau)/\sigma_0$ can be regarded as the scale factor of the spacetime inflation
mentioned in \S\,\ref{introduction}; $\mu$ is related to Higgs--boson mass $\mu_H \cong 125.1$\,GeV
by equation $\mu=\mu_H/\sqrt{2}\cong 88.47$\,GeV; $\lambda\cong 0.1291$ is the self--coupling constant of the
Higgs--boson field.

In the semi--classical approximation of the SMEP, $\lambda$ is related to the Fermi coupling
constant $G_F \cong 1.16637\times 10^{-5}$ GeV$^{-2}$ by equation $\lambda=\mu_H^2 G_F/\sqrt{2}$.

This important relation is discussed in \S\,\ref{Fermicoupling} of Appendix
{\bf \ref{PathIntApp}} and made diagrammatically evident in Fig.\,\ref{FermiConst}.

The role of constant $\sigma_0$ is commented in \S\,\ref{CGR&SMEP}: in order that CGR approaches GR at large
$\tau$, we must take $\sigma_0 =\sqrt{6/\kappa}$, where $\kappa$ is the gravitational
coupling constant of GR. Therefore, from $\kappa = 6/\sigma_0^2 \cong 2.435\!\times \!10^{18}$GeV
we obtain $\mu^2/\sigma_0^2 \cong 2.26\!\times\!10^{-34}$.

To solve Eqs (\ref{varphieq}) and (\ref{seq}) we need appropriate initial conditions:
\begin{itemize}
\vspace{-1mm}
\item[{\em i})] We must exclude $\varphi(0)=0$ and/or $\alpha(0)=0$, since otherwise one or two solutions
would be trivial.
\vspace{-1mm}
\item[{\em ii})] We must assume $\dot\varphi(0)=\dot\alpha(0)=0$, since otherwise the frictional terms
proportional to $1/\tau$ would diverge at $\tau=0$.
\vspace{-1mm}
\item[{\em iii})] To make sure that the initial state is very close to that in which $\varphi(0) =0$,
we shall assume $0 < \varphi(0)\!\ll\!\mu\,\alpha(0)/\sqrt{\lambda}\,$. As will be hereafter shown,
this condition also ensures that $a(\tau)$ increases monotonically for increasing $\tau$.
\vspace{-1mm}
\item[{\em iv})] In order that the asymptotic limit of $\varphi(\tau)$ is $\varphi(\infty)=
\mu/\sqrt{\lambda}$, we must take $a(0)< 1$.
\end{itemize}

Note that the left--hand sides of Eqs (\ref{varphieq}) (\ref{seq}) include respectively
the frictional terms $3\,\dot\varphi(\tau)/\tau$ and $3\,\dot\alpha(\tau)/\tau$, which
have the effect of making the potential--energy density of $\varphi$--$\alpha$ interaction,
$U\big[\varphi(\tau), \alpha(\tau)\big] = (\lambda/4)\bigl[\varphi(\tau)^2-
\mu^2\alpha(\tau)^2/\lambda\bigr]^2$, approach zero for $\tau\rightarrow\infty$.

\newpage

As is evident from the structure of Eq  (\ref{varphieq}), if $0<\varphi(0)\ll \mu\,
\alpha(0)/\sqrt{\lambda}$ and $\partial_\tau\varphi(\tau)=0$ at $\tau=0$, $\varphi(\tau)$
will take a long time to reach appreciable values, because $\varphi=0$ is a stagnation
point of $U(\,\varphi, \alpha)$ for any value of $\alpha >0$. However, even if $\varphi(0)$
is not so small, but in any case less than $\mu \,\alpha(0)/\sqrt{\lambda}$, $\varphi(\tau)$
tends initially to decrease with increasing $\tau$ because for small $\tau$ the frictional
term $3\,\dot \varphi(\tau)/\tau$ acts as a strong damping agent.

Numerical simulations showed that, even for moderately small values of $\varphi(0)$,
$\varphi(\tau)$ first becomes smaller than $\varphi(0)$, and then, at a certain time
$\tau_B$, called the {\em big--bang time}, jumps suddenly to a certain value
$\varphi(\tau_B)$,  close to $\varphi_{\hbox{\tiny max}}(\tau_B)=\sqrt{2}\,\mu\,
\alpha(\tau_B)/\sqrt{\lambda}$. After $\tau_B$, $\varphi_{\hbox{\tiny max}}(\tau)$
oscillates with decreasing amplitude getting closer and closer to $\varphi(\infty)
=\mu/\sqrt{\lambda}$.

Note that as long as $\lambda\,\varphi^2(\tau)\ll \mu^2 \alpha^2(\tau)$, Eq (\ref{seq})
is well--approximated by equation
\begin{equation}
\label{seq0}
\ddot\alpha(\tau) + \frac{3}{\tau}\,\dot\alpha(\tau)=
\frac{\mu^4}{\lambda\,\sigma^2_0}\,\alpha^3(\tau)\,,\quad 0< \alpha(\tau) \le 1\,,
\end{equation}
the general solution of which is exactly %5.0929e-10
\begin{equation}
\label{s0oftau}
\alpha(\tau)=\frac{\alpha(0)}{1-\tau^2/\tau^2_c},\,\,\mbox{with $\alpha(0)$ arbitrary
and }\,\tau_c = \frac{\sqrt{8\,\lambda}\,\sigma_0}{\alpha(0)\,\mu^2}\cong
\frac{5.0929\times 10^{-10}}{\alpha(0)}\,\hbox{sec}.
\end{equation}
Here, the dimensional equivalence 1 GeV $\cong 1.5192\times 10^{24}$sec, listed at
page iv is used. This equation shows that the magnitude of $\alpha(0)$ has an
important role in determining the {\em critical time} $\tau_c> \tau_B$ at which
$\alpha(\tau)$ becomes explosive.

Also note that, if $\varphi(0)=\dot\varphi(0)=0$, we have identically $\varphi(\tau)=0$.
In this case, Eq  (\ref{seq0}) is exact and, as Eq (\ref{s0oftau}) shows, the explosion of
$\alpha(\tau)$ at $\tau =\tau_c$ is unavoidable.

However, provided that the second member of Eq  (\ref{seq}) remains sufficiently small,
which is indeed the case because $\mu^2/\sigma_0^2$ is very small, Eq (\ref{varphieq})
ensures that $\alpha(\tau)$  may take a very long time to reach the critical point.

If $\dot\varphi(0)=0$ and $\varphi(0)$ is positive, although negligible with respect to
$\mu\,\alpha(0)/\sqrt{\lambda}$, $\varphi(\tau)$ starts increasing more and more, as is
evident from Eq (\ref{varphieq}). Therefore, in the long run the behavior of $\alpha(\tau)$
departs considerably from that described by Eq (\ref{s0oftau}). This is due to the fact that
the kinetic energy of ghost fields VEV $\sigma(\tau)\equiv \sigma_0\alpha(\tau)$ grows
uncontrollably.

More in general, no matter how small is $\varphi(0)>0$, at some kinematic time $\tau_B <\tau_c$,
$\varphi(\tau)$ jumps abruptly to a relative maximum $\varphi_B$. Starting from a very small
value of $\varphi(0)$ is crucial for obtaining a value of $\tau_B$ very close to $\tau_c$.
In doing so, in fact, the abruptness of the jump, together with the slope of the
$\varphi$--amplitude profile, can be increased at will.

As can easily be inferred from Eq (\ref{seq}), as long as $c(\tau)\equiv\mu^2\alpha^2(\tau)
-\lambda\,\varphi^2(\tau)$ is positive, the curvature of the $\alpha(\tau)$ profile remains
positive, but when $c(\tau)$ changes sign, the curvature tends to become negative.

However, during the time interval $[0, \tau_i]$ in which $\mu^2\alpha^2(\tau)\gg\lambda\,
\varphi^2(\tau)$, $\alpha(\tau)$ is well--approximated by the solution to Eq (\ref{seq0}),
here renamed for notational convenience as $\alpha_i(\tau)$. But when, damped by the
frictional forces, $c(\tau)$ becomes sufficiently small, the remaining portion of
$\alpha(\tau)$, $\alpha_f(\tau)$, tends to satisfy equation $\ddot\alpha_f(\tau)
+ 3\,\dot\alpha_f(\tau)/\tau=0$, the general solution of which, from a certain
kinematic time $\tau_f > \tau_B$ on, can be written as
\begin{equation}
\label{squeue}
\alpha_f(\tau)= 1+\bigl(\alpha_f -1\bigr)\frac{\tau^2_f}{\tau^2}\,,
\end{equation}
where $\alpha_f(\tau_f)=a_f$ and $\alpha_f(\infty) = 1$.
\vspace{3mm}
\begin{figure}[!h]
\centering
\vspace{-4mm}
\mbox{%
\begin{minipage}{0.55\textwidth}
\includegraphics[scale=0.73]{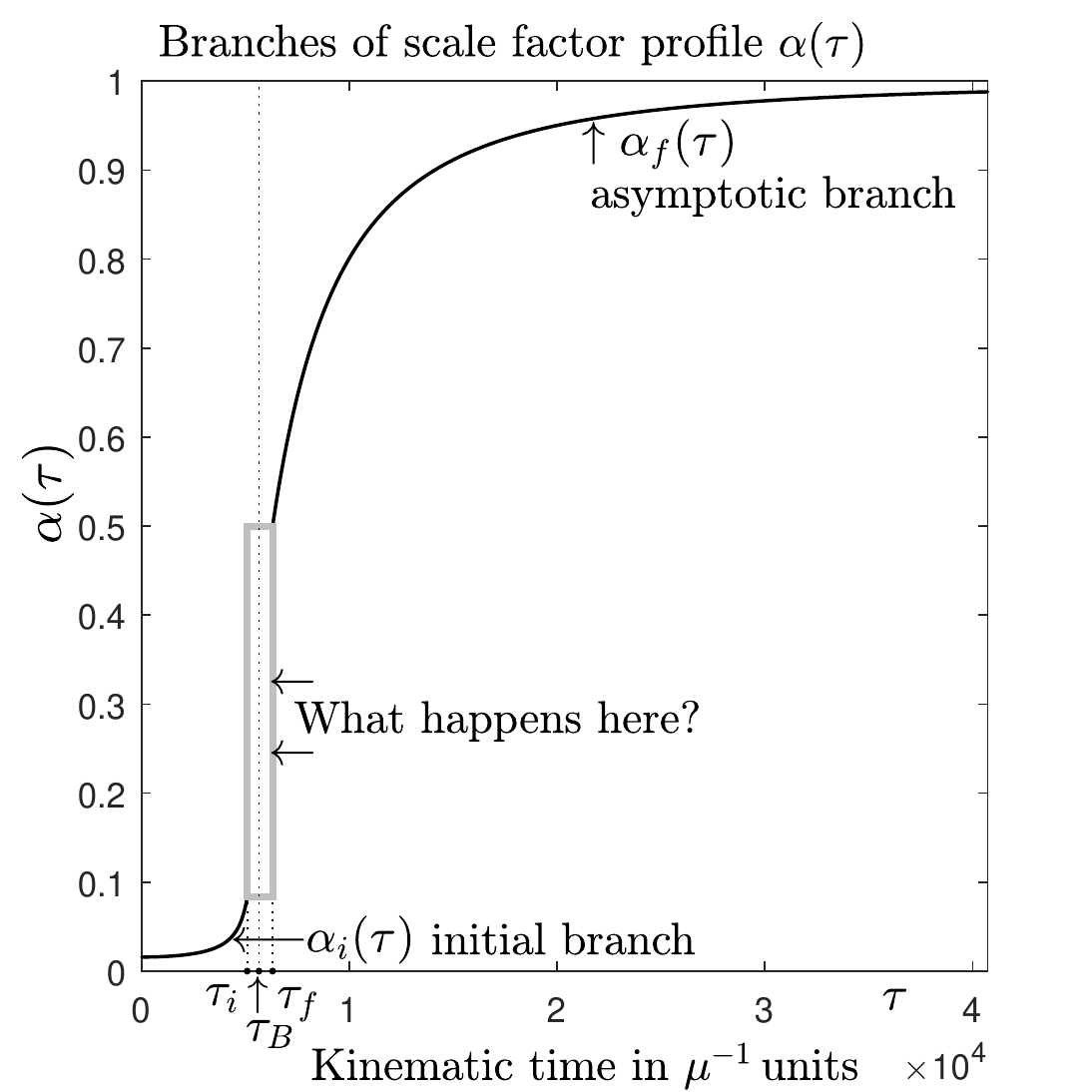}
\end{minipage}%
\begin{minipage}[c]{0.42\textwidth}
\caption{\small The initial and final branches $\alpha_i(\tau)$ and $\alpha_f(\tau)$
of the scale--factor profile are respectively determined by Eqs (\ref{seq0}) and
(\ref{squeue}), but the intermediate portion, within the gray box delimited by points
$\tau_i$ and $\tau_f$, is determined by Eq (\ref{seq}) combined with Eq (\ref{varphieq}).
The exact behaviors of $\alpha(\tau)$ and $\varphi(\tau)$ in this region is crucial for
understanding what happens in the neighborhood of big bang time $\tau_B$.
}
\label{VacStFig1}
\end{minipage}
}
\end{figure}
\vspace{-4mm}

As shown in Fig.\ref{VacStFig1}, the missing portion of scale--factor profile $\alpha(\tau)$
lie in between the curvilinear branches $\alpha_i(\tau)$ and $\alpha_f(\tau)$: the first of
these extends from $\tau = 0$ to $\tau = \tau_i$ and is characterized by a positive curvature,
i.e., $\ddot\alpha_i(\tau)>0$; the second extends from $\tau = \tau_f$  to $\tau = \infty$ and is
characterized by a negative curvature, i.e., $\ddot\alpha_f(\tau)<0$.

The behavior of $\alpha(\tau)$ in the joining region $\tau_i <\tau < \tau_f$, as well as
the precise values of $\tau_B$ and $\alpha_B$, and consequently of $\alpha(0)$, cannot be
determined in this way because they depend on the details of the interaction between
$\varphi(\tau)$ and $\alpha(\tau)$ \cite{BROUT}.

\newpage

\subsection{Dynamical vacuum equations in kinematic--time representation}
\label{kinemtimesolutions}
\begin{figure}[!ht]
\centering\includegraphics[scale=0.98]{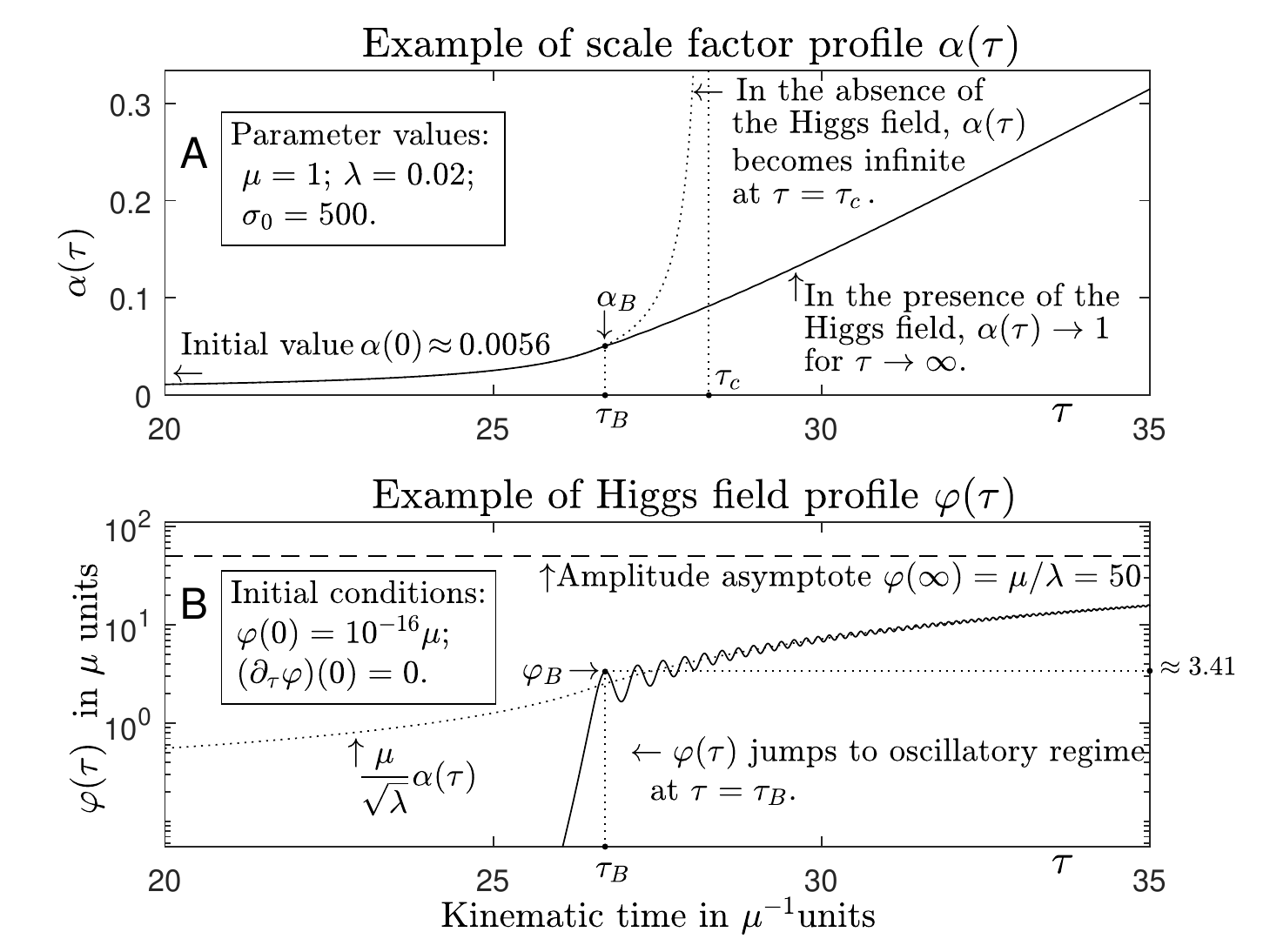}
\vspace{-4mm}
\caption{\small Example of scale factor profile $\alpha(\tau)$ and Higgs field amplitude
$\varphi(\tau)$ in the kinematic--time representation as functions of $\tau$. {\bf A}.
Solid line: scale factor $\alpha(\tau)$. Dotted line: scale factor in the absence of
the Higgs field; it becomes infinite at critical time $\tau_c$. At big bang time
$\tau_B$, where $\sigma(\tau_B)/\sigma_0=\alpha_B$, accelerated expansion transits
smoothly to decelerated expansion. {\bf B}. Solid line: profile of Higgs field amplitude
$\varphi(\tau)$, asymptotically adhering to profile of $\mu\,\alpha(\tau)/\sqrt{\lambda}$
(dotted line). Dashed line: scale factor asymptote times $\mu/\sqrt{\lambda}$. All profiles
are computed for the parameters reported in the inset of panel {\bf A}: they are rendered
very poorly because a plausible value of $\alpha(0)$ is about $10^{-27}$, not value $0.0056$
indicated on the left--bottom corner of panel {\bf A}.}
\label{VacStFig2}
\vspace{-4mm}
\end{figure}

Fig.\,\ref{VacStFig2} shows a numerical solution to Eqs (\ref{varphieq}) (\ref{seq})
with non--realistic parameters, for the only purpose of exhibiting the qualitative features of
$a(\tau)$ and $\varphi(\tau)$ profiles.
\vspace{-4mm}

Starting from about $\varphi(0) =10^{-16}\mu$, $\varphi(\tau)$ jumps almost instantly to its
maximum at big--bang time $\tau\cong \tau_B$, where it reaches its maximum $\varphi_B =\sqrt{2}\,\mu \,
\sigma(\tau_B)/\sqrt{\lambda}\,\sigma_0\approx 3.41\,\mu$ with excess potential--energy density $\Delta U_B =
(\lambda/4)\big(\varphi_B^2-\mu^2\alpha_B^2/\lambda\big)^2\approx 0.0331\mu^4$. After $\tau_B$, it
oscillates about curve $\mu/\sqrt{\lambda}\,\alpha(\tau)$ with a progressively decreasing amplitude.
Meanwhile, $\Delta U_B$ converts to kinetic--energy density during a sort of rarefaction--condensation
process.

Unfortunately, solving numerically Eqs  (\ref{seq}) and (\ref{varphieq}) for more significant
values of the parameters is prohibitive. This is due to the fact that the slope of $\alpha(\tau)$
at $\tau=\tau_B$ is so large and the oscillation of $\varphi(\tau)$ is so furious that they cannot
be graphically rendered. A better visualization of their behaviors is obtained by solving the vacuum
stability equations in the proper--time representation described in the following three subsections.

\subsection{From kinematic--time representation to proper--time representation}
\label{proptimesolutions}
The proper--time representation of Eqs (\ref{varphieq}) and (\ref{seq}) we can be obtained
as follows:
\noindent 1) Put $\varphi(\tau)= y(\tau)\,\alpha(\tau)$ in these equations and combine the
results so as to obtain
\begin{equation}
\label{yforvarphi}
\partial^2_\tau\, y(\tau) +\bigg[\frac{3}{\tau} +
2\,\frac{\partial_\tau\alpha(\tau)}{\alpha(\tau)}\bigg]\partial_\tau y(\tau) =
\alpha(\tau)^2\bigg(\lambda -\frac{\mu^2}{\sigma_0^2}\bigg)\bigg[\frac{\mu^2}{\lambda} -
y(\tau)^2\bigg]y(\tau)\,.
%\vspace{-2mm}
\end{equation}

\noindent 2) Define the proper time of vacuum--stability equation and its differential
$d{\widetilde{\tau}}$ as
\vspace{-1mm}
\begin{equation}
\label{tildetauoftau}
\widetilde{\tau}(\tau) = \int_0^\tau \alpha(\tau')\,d\tau'\,;\quad d{\widetilde{\tau}} =
\alpha(\tau)\,d\tau\,;
\vspace{-1mm}
\end{equation}
since $\widetilde{\tau}$ is one--to--one with $\tau$, the inverse $\tau=\tau(\widetilde{\tau})$
of Eq (\ref{tildetauoftau}) does exist. Therefore, for any function $\widetilde{f}(\widetilde{\tau})$
belonging to the proper--time representation, we have
\vspace{-1mm}
\begin{equation}
\label{tau2tiltauder}
\partial_\tau \widetilde{f}(\widetilde{\tau}) = \partial_{\widetilde{\tau}} \widetilde{f}(\widetilde{\tau})\,
\frac{d\widetilde{\tau}(\tau)}{d\tau} = \alpha(\tau)\,\partial_{\widetilde{\tau}}
\widetilde{f}(\widetilde{\tau})\equiv
\alpha[\tau(\widetilde{\tau})]\,\partial_{\widetilde{\tau}}\widetilde{f}(\widetilde{\tau})\equiv
\widetilde{\alpha}(\widetilde{\tau})\,\partial_{\widetilde{\tau}}\widetilde{f}(\widetilde{\tau})\,.
\vspace{-1mm}
\end{equation}

\noindent 3) Define the inflation factor and the Higgs--field amplitude in the proper--time
representation respectively as $\tilde\alpha(\tilde\tau)\equiv \alpha[\tau(\tilde\tau)]$ and
$\widetilde{\varphi}(\widetilde{\tau}) = y[\tau(\widetilde{\tau})]\equiv
\widetilde{\alpha}(\widetilde{\tau})^{-1}\varphi[\tau(\widetilde{\tau\,})]$.

\noindent 4) Using Eq (\ref{tau2tiltauder}) in Eq  (\ref{yforvarphi}) and re--organizing algebraically
the result, we obtain the proper--time representations of Eq (\ref{varphieq}) in the form:
\begin{equation}
\label{propvarphieq}
\partial^2_{\widetilde{\tau}}\,\widetilde{\varphi}(\widetilde{\tau})+
3\bigg[\frac{1}{\tau(\widetilde{\tau})}+
\frac{\partial_{\,\widetilde{\tau}}\,\widetilde{\alpha}(\widetilde{\tau})}
{\widetilde{\alpha}(\widetilde{\tau})}\bigg]\partial_{\,\widetilde{\tau}}\,
\widetilde{\varphi}(\widetilde{\tau})=\bigg(\lambda-\frac{\mu^2}{\sigma_0^2}\bigg)
\bigg[\frac{\mu^2}{\lambda}- \tilde\varphi^2(\tilde\tau)\bigg]\widetilde{\varphi}
(\widetilde{\tau}\,)\,.
\end{equation}

Differently from Eq (\ref{varphieq}), the frictional force term on the left side of this equation
includes the additional term $\partial_{\widetilde{\tau}}\log \widetilde{\alpha}(\widetilde{\tau})$,
which is clearly proportional to the rate of scale--factor variation for increasing proper time.
The right--hand side of the same equation represents the driving force generated by potential--energy
density $\widetilde{U}(\widetilde{\varphi}) = \frac{1}{4}\lambda'\big(\widetilde{\varphi}^2 -
\mu^2/\lambda\big)^2$, where $\lambda' =\lambda- \mu^2/\sigma_0^2$. However, since $\mu^2/\sigma_0^2
\approx \lambda\times 10^{-14}$, $\lambda'$ can be safely replaced by $\lambda$. The motion equation
for $\widetilde{\alpha}(\widetilde{\tau})$ seems to have disappeared. But actually is not, because
it is obtained from Eq (\ref{seq}) by carrying out the substitutions described in point 3).

The kinematic--time representation can easily be recovered by performing the inverse transformations
$\tau(\widetilde{\tau})\rightarrow \tau$, $\widetilde{\alpha}(\widetilde{\tau})\rightarrow \alpha(\tau)$,
$\partial_{\widetilde{\tau}} \rightarrow \alpha^{-1}\,\partial_{\tau}$, $\widetilde{\varphi}(\widetilde{\tau})
\rightarrow\alpha(\tau)\varphi(\tau)$.

\subsection{Dynamical vacuum equations in proper--time representation}
\label{VacPotenDens}
\begin{figure}[ht]
\centering\includegraphics[scale=0.65]{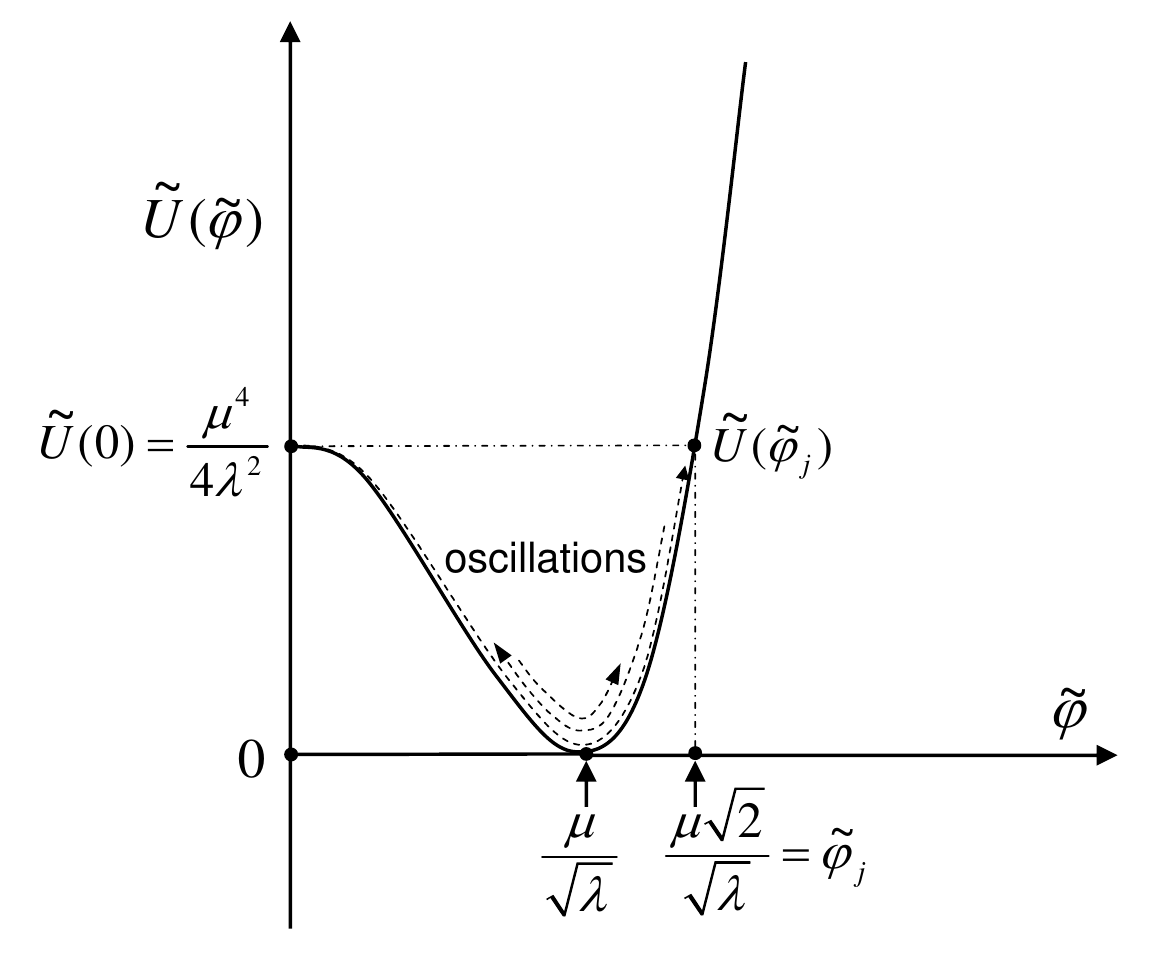} %F14
\vspace{-4mm}
\caption{\small  {\em Higgs--field amplitude oscillations in proper--time representation}.
Assume that $\widetilde{\varphi}(0)$ is very small and positive and $\dot{\widetilde{\varphi}}(0)
=0$, so $\widetilde{U}(0)$ is very close to $\mu^4/4\lambda$. Then $\widetilde{\varphi}$ rolls
down to the bottom of the well and climbs up to a value $\widetilde{\varphi}_j$ close from below to
$\sqrt{2} \mu/\sqrt{\lambda}$, where $\widetilde{U}(\widetilde{\tau})$ has a value very close
to $\widetilde{U}(\widetilde{\varphi}_j)=\widetilde{U}(0)$. From this moment on, at big--bang time
$\widetilde{\tau}_B$, $\widetilde{\varphi}(\widetilde{\tau})$ moves back toward the bottom of the
well and then, damped by the frictional force appearing in Eq (\ref{propvarphieq}), oscillates with
decreasing amplitude about the bottom of the well, where $\widetilde{\varphi}=\mu/\sqrt{\lambda}$
and $\widetilde{U}=0$.}
\label{VacStFig3}
\end{figure}

In the kinematic--time representation, the potential--energy density of the
dynamic vacuum has the form $U\big[\varphi(\tau), \alpha(\tau)\big]=
(\lambda/4)\big[\varphi(\tau)^2-\mu^2\sigma(\tau)^2/\lambda\,\sigma_0^2\big]^2$.
Since in the proper--time representation $\varphi(\tau)$ is replaced by
$\widetilde{\varphi} (\widetilde{\tau})= \varphi(\widetilde{\tau})/\alpha
(\widetilde{\tau})$ and $\sigma(\tau)$ by $\widetilde{\sigma}(\widetilde{\tau})
= \sigma(\widetilde{\tau})/\alpha(\widetilde{\tau})\equiv\sigma_0$, the
potential--energy density takes the form $\widetilde{U}\big[\widetilde{\varphi}
(\widetilde{\tau})\big]=(\lambda/4)\big[\widetilde{\varphi}(\widetilde{\tau}))^2
-\mu^2/\lambda\big]^2$, which depends only on $\widetilde{\varphi}(\widetilde{\tau})$.
We can therefore visualize the behavior of the Higgs field as a damped oscillation
of a ball in a well, as described in Fig.\,\ref{VacStFig3}.
\vspace{-3mm}

Actually, the oscillation of Higgs--field amplitude represented in the figure is misleading.
Despite the magnitude of the frictional forces and the proximity to the stagnation point,
the initial rolling down of $\widetilde{\varphi}(\widetilde{\tau})$ is not so slow as one might believe.
This happens because the time course of the proper time, compared to kinematic time, is initially
highly compressed, because of the enormous initial smallness of $\alpha(\tau)$ in Eq (\ref{tildetauoftau}).
Rather, the evolution of the vacuum state after $\tau_B$ is so fast that the initial potential energy--density
of the vacuum, $\widetilde{U}(0)= \mu^4/4\,\lambda$, converts almost instantly into thermal--energy
density through an irreversible thermodynamic process of the type described in Appendix {\bf \ref{PathIntApp}}.

\newpage

In Fig.\,\ref{VacStFig4}A, the scale factor profile shown in Fig.\,\ref{VacStFig2}A is plotted for
comparison as a function of proper time $\widetilde{\tau}$. Fig.\,\ref{VacStFig4}B shows the profile
of Higgs--field--amplitude $\widetilde{\varphi}(\widetilde{\tau})$ as defined in \S\,\ref{proptimesolutions}.
At big-bang time $\widetilde{\tau}_B$, $\widetilde{\varphi}$ jumps to a maximum close to
$\sqrt{2}\mu/\sqrt{\lambda}$, then oscillates up and down about the straight line with ordinate at
$\widetilde{\varphi}(\infty)=\mu/{\sqrt{\lambda}}$, as described in Fig.\ref{VacStFig3}. Note that,
since $\widetilde{\sigma}(\widetilde{\tau}) = \sigma_0$, the effective scale factor after $\widetilde{\tau}_B$
is equal to one.
\begin{figure}[ht]
\vspace{-4mm}
\centering\includegraphics[scale=0.98]{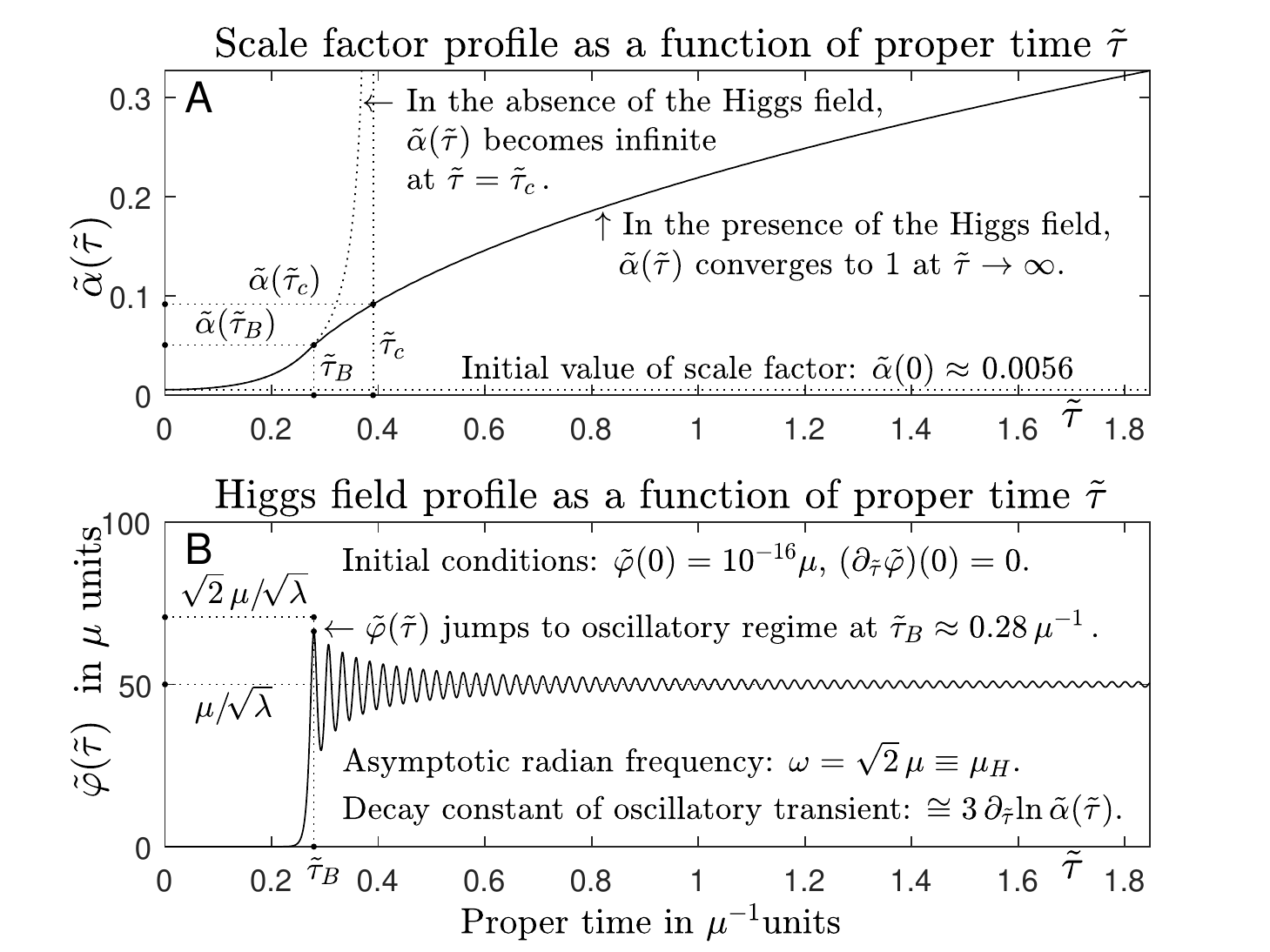}
\caption{\small Example of scale factor profile $\tilde\alpha(\tilde\tau)$ and
Higgs field amplitude $\tilde\varphi(\tilde\tau)$ for comparison with $\alpha(\tau)$
and $\varphi(\tau)$ profiles shown in Fig.\,\ref{VacStFig2}. -- {\bf A} Solid line:
scale factor profile during inflation. Dotted line: scale factor in the absence
of Higgs field; it coincides with that of an empty spacetime, which becomes infinite
at critical time $\tilde \tau_c$. At big--bang time $\tilde \tau_B$, spacetime
expansion stops abruptly and transits to a decelerated regime, which lasts until
$\tilde\alpha(\tilde\tau)$ approaches 1. -- {\bf B} Solid line: Higgs field
amplitude $\tilde\varphi(\tilde\tau)$ converges to its VEV $\mu/\sqrt{\lambda}$
at $\tilde\tau =\infty$. Compression of proper--time scale, relative to
conformal--time scale, makes residual oscillation of $\tilde\varphi$
amplitude approach a sinusoid.}
\label{VacStFig4}
\end{figure}

Of note, the reason why $\tilde\varphi(\tilde\tau)$ exhibits pronounced oscillations,
whereas $\tilde a(\tilde\tau)$ does not, is due to the fact that factor
$\mu^2/\sigma_0^2$ on the right side of Eq (\ref{seq}) is enormously smaller
than factor $\lambda$ on the same side of Eq (\ref{varphieq}).

\subsection{The time course of the scalar factor for $\tau_B/\tau_c$ very close to 1}
\label{predictions}
\begin{figure}[!ht]
\vspace{-2mm}
\hspace{-4mm}
\vspace{-4mm}
\includegraphics[scale=0.53]{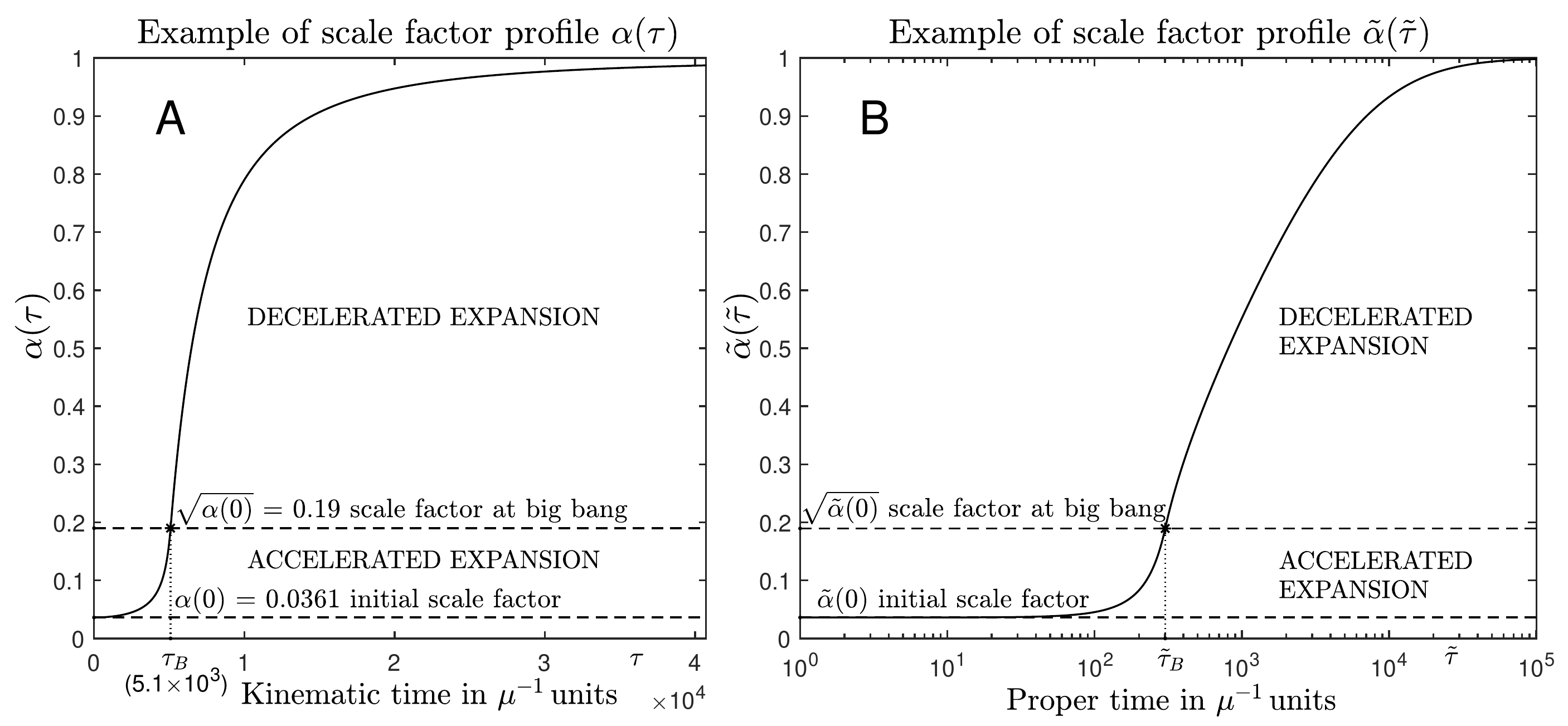} %F16
\caption{\small {\bf A}: Example of a smoothly joined scale--factor profile as a function of kinematic
time~$\tau$. {\bf B}: The same profile as a function of proper time $\tilde\tau$. Note the strong
scale compression of $\tilde\tau$ relative to $\tau$. Since slope of the profile at $\tau_B$
is expected to be enormously greater than that here shown, time taken by Higgs field amplitude
jump is negligible. Remarkably, smooth join of accelerated--expansion branch and
decelerated--expansion branch leads to equation $\alpha(\tau_B)= \sqrt{\alpha(0)}$.}
\label{VacStFig5}
\end{figure}

Numerical simulations of $\varphi (\tau)$  and $\tilde\varphi(\tilde\tau)$ profiles showed that
the time interval of appreciable oscillation amplitude shrinks more and more as $\tau_B$ and
$\tilde\tau_B$ get closer and closer to $\tau_c$ and $\tilde\tau_c$, respectively. Thus, if
$\tau_c-\tau_B$, and $\tilde\tau_c-\tilde\tau_B$, are very small, we can presume that the
smooth join of the initial branch $\alpha_i(\tau)$ and the asymptotic branch $\alpha_f(\tau)$
represented in Fig.\,\ref{VacStFig1}, provides a good  approximation of the true scale--factor
profiles $\alpha(\tau)$ and $\tilde\alpha(\tilde \tau)$ (Brout {\em et al}, 1978). An example
of sigmoidal profiles obtained by this method is shown in Figs.\,\ref{VacStFig5}\,A and B.

Since the order of magnitude of scale expansion across inflation estimated by the cosmologists
is in the order of magnitude of $10^{28}$, we conclude that the jump from a state of very small
Higgs field amplitude $\varphi(0)$ to one of large amplitude $\varphi(\tau_B)$ is capable of
producing an almost instantaneous huge amount of Higgs quanta per unit volume at conformal
time $\tau_B$, which may therefore be interpreted as the big bang.

The smooth junction of branches $\alpha_i(\tau)$ and $\alpha_f(\tau)$ at $\tau = \tau_B$ can easily be obtained
by imposing the joining conditions $\alpha_i(\tau_B)=\alpha_f(\tau_B)$ and $\dot\alpha_i(\tau_B)=\dot\alpha_f(\tau_B)$.
From these conditions, and first using Eq (\ref{squeue}) and then Eq (\ref{s0oftau}), we obtain
\begin{eqnarray}
\label{s0tosj}
\hspace{-14mm}&& \alpha_i(\tau)=\frac{(1-\tau_B^2/\tau_c^2)^2}{1-\tau^2/\tau_c^2}
\quad\mbox{for } 0 \leq \tau \leq\tau_B;\quad\alpha_f(\tau)\!=\! 1\!-\!\frac{\tau^4_B}
{\tau_c^2\tau^2}\quad\mbox{for }\tau_B\leq\tau<\infty; \\
\label{s0stauj}
\hspace{-14mm}&& \alpha_i(\tau_B)=\alpha_f(\tau_B)=\sqrt{\alpha(0)}=
1-\frac{\tau_B^2}{\tau_c^2};\quad \dot\alpha_i(\tau_B) =\dot\alpha_f(\tau_B)\!=\!
\frac{2\tau_B}{\tau_c^2}\cong\frac{2}{\tau_B};\\
\label{sdot}
\hspace{-14mm}&&\dot\alpha_i(\tau)=\frac{2\tau (1-\tau_B^2/\tau_c^2)^2}
{\tau_c^2(1-\tau^2/\tau_c^2)^2}\quad\mbox{for } 0 \leq \tau \leq\tau_B;\quad
\dot\alpha_f(\tau)=\frac{2\tau^4_B}{\tau_c^2\tau^3}\quad\mbox{for }\tau_B\leq\tau<\infty\,.
\end{eqnarray}

However, although $\tau_B$ is very close to $\tau_c$, the scale expansion during interval
$[\tau_B, \tau_c]$ is not negligible, as is evident for instance from the levels reached by
$\alpha(\tau)$ at $\tau=\tau_B$ and $\tau=\tau_c$ in Fig.\,10A. In fact, as $\tau_B$ approaches
$\tau_c$ from above we have
\begin{equation}
\label{sc2sj}
\alpha_f(\tau_c)-\alpha_f(\tau_B)\!=\!\frac{\tau^2_B}{\tau_c^2}\Big(1\!-\!\frac{\tau^2_B}{\tau_c^2}\Big)
\!=\!  \frac{\tau^2_B}{\tau_c^2}\,\alpha_f(\tau_B),\quad\alpha(\tau_c)\!\cong\!2\,\alpha_f(\tau_B)\,\,
\mbox{and}\,\,\tau_c -\tau_B\!\cong\! \frac{\tau_B}{2} \sqrt{\alpha(0)}.\nonumber
\end{equation}
Since $\alpha(0)$ is estimated to be in the order of magnitude of $10^{-24}$,
we find $(\tau_c -\tau_B)/\tau_c \approx 10^{-13}$, which also is very small.
Similar relations also hold for $\tilde\alpha(\tilde\tau_c)$ and $\tilde\alpha(\tilde\tau_B)$.

Unfortunately, due to the discontinuity of $\ddot\alpha(\tau)$ at $\tau = \tau_B$,
the scale factor constructed in this way is not perfectly smooth. In fact, from
\begin{equation}
\label{ddotalpha}
\ddot\alpha_i(\tau) = \frac{2\,\alpha(0)}{\tau_c^2(1-\tau^2/\tau_c^2)^2} +
\frac{8\,\tau^2\alpha(0)}{\tau_c^4 (1-\tau^2/\tau_c^2)^3}\quad
\mbox{and}\quad\ddot\alpha_f(\tau) = -\frac{6\tau_B^4}{\tau_c^2 \tau^4}\,,
\end{equation}
we derive
$$
\ddot\alpha_i(\tau_B) =
\frac{2}{\tau_c^2}\bigg[1+ \frac{4\,\tau_B^2}{\tau_c^2 \sqrt{\alpha(0)}}\bigg]
\cong \frac{8}{\tau_c^2\sqrt{\alpha(0)}}\,,\quad \ddot\alpha_f(\tau_B)= -\frac{6}{\tau_c^2}\,,
$$
as $\tau_B/\tau_c \cong 1$ and $\sqrt{\alpha(0)} \ll 1$, where we would like to find
$\ddot\alpha_0(\tau_B)= \ddot\alpha_f(\tau_B)=0$ instead.

This contrasts with the expected flatness of true scale factor $\alpha(\tau)$ at the moment of the
accelerated--to--decelerated transition. Of the two second derivatives, the more deceptive is clearly
$\ddot\alpha_i(\tau_B)$, as it is greater than $\ddot\alpha_f(\tau_B)$ by about $8/\tau_c^2
\sqrt{\alpha(0)}$. This implies that the true $\alpha(\tau_B)$ and $\dot\alpha(\tau_B)$ are
actually a little smaller than $\alpha_i(\tau_B)$ and $\dot\alpha_i(\tau_B)$, respectively.

However, the discrepancy is negligible. In fact, as can be evinced from the coefficients of
$\big(\alpha^2 - \lambda\,\varphi^2/\mu^2\big)$ -- in Eqs (\ref{varphieq}) and (\ref{seq})
-- the ratio between the rising times of $\varphi(\tau)$ and $\alpha(\tau)$ at $\tau=\tau_B$
is $\sqrt{\lambda}\sigma_0/\mu \cong 2.42\times 10^{16}$. This means that the time taken by
$\sqrt{\lambda}\,\varphi(\tau)/\mu$ to pass from a very small value $a(0)$ to $a(\tau)$, as $\tau$
approaches $\tau_B$, is in the order of magnitude of $\tau_B\times 10^{-16}$. Correspondingly,
the time taken by $\ddot\alpha(\tau)$ to deviate from $\ddot\alpha(\tau) \cong \ddot\alpha_i(\tau) > 0$
to $\ddot\alpha(\tau)\cong \ddot\alpha_f(\tau)<0$, across $\tau=\tau_B$, is negligible.
We can therefore regard $\alpha(\tau_B)$ and $\dot\alpha(\tau_B)$ as virtually equal to
$\alpha_i(\tau_B)=\alpha_f(\tau_B)$ and $\dot\alpha_i(\tau_B)=\dot\alpha_f(\tau_B)$, respectively

\subsection{Energy density of the universe at big bang}
\label{Unievol}
\vspace{-1mm}
Performing measurement--unit conversions

\smallskip
\centerline{
\begin{tabular}{l r @{.} l}
1 eV  as mass ($\times \,c^{-2}$)  & $\rightarrow \quad $  1&78$\times 10^{-36}$ Kg\,,\\
1 eV$^{-1}$ as length ($\times \,\hslash\,c$)  & $\rightarrow \quad $ 1&97$\times 10^{-7}$ m\,,\\
1 eV$^{-1}$ as time ($\times \,\hslash$)  & $\rightarrow\quad$ 6&58$\times 10^{-16}$ s\,,
\end{tabular}}
\smallskip

\noindent where $c$ is the speed of light and $\hslash$ the Planck constant divided by $2\pi$,
we derive
\vspace{-2mm}
\begin{eqnarray}
%& & \mu_H\cong 125.6\,\, \mbox{GeV} \cong 2.252 \times 10^{-25} \,\mbox{Kg}\,; \quad \tau_B
%\cong \frac{4.91\times 10^{-10}}{s(0)}\, \mbox{sec}\,;\nonumber\\
& & 1 \,\mbox{Kg} \cong 5.62 \times 10^{26}\, \mbox{GeV} \,;\quad 1\, \mbox{GeV}\cong
1.78\times 10^{-27}\,\mbox{Kg}\cong 1.52\times 10^{24}\mbox{s}^{-1}\,; \nonumber\\
& & 1 \,\mbox{GeV} \cong  5.076\times 10^{15}\,\mbox{m}^{-1} \,;\quad 1 \,\mbox{GeV}^{-1}
\cong 1.97 \times 10^{-16}\,\mbox{m} \cong 6.58\times 10^{-25}\,\mbox{s} \,; \nonumber\\
& & 1 \, \mbox{m}^{-1} \cong 1.97\times 10^{-16}\,\mbox{GeV}\,;\quad 1\,\mbox{s}^{-1}
\cong 6.58\times 10^{-25} \,\mbox{GeV}\,; \nonumber \\
& & 1 \, \mbox{m} \cong 5.076\times 10^{15}\,\mbox{GeV}^{-1}\,;\quad 1\,\mbox{s}\cong
1.52\times 10^{24}\,\mbox{GeV}^{-1} \,; \nonumber \\
& & 1 \,\mbox{Kg/m}^{3}\cong 4.297\times 10^{-21}\, \mbox{GeV}^4 \,;\quad 1\, \mbox{GeV}^4\cong
 2.327\times 10^{20}\,\mbox{Kg/m}^{3}\,. \nonumber
\vspace{-6mm}
\end{eqnarray}

Now, using Eq (\ref{s0oftau}) and recalling that for very small ratios $a(0)/a(\infty)$ it is
$\tau_B\cong \tau_c$, we obtain
\vspace{-1mm}
\begin{equation}
\label{Sotauj}
\tau_B \cong  \tau_c = \frac{\sqrt{8\,\lambda}\,\sigma_0}{\alpha(0)\,\mu^2} \equiv
\frac{\sqrt{32\lambda}\,M_{rP}}{\alpha(0)\,\mu^2_H}\cong \frac{9.44\times 10^{16}}{\alpha(0)}
\mbox{GeV}^{-1}\cong \frac{5.097\times 10^{-10}}{\alpha(0)}\, \mbox{s}\,.
\end{equation}

Here, $M_{rP}= \sigma_0/\sqrt{6} = 2.435\times 10^{18}$ GeV is the reduced Planck mass,
$\mu_H = \sqrt{2}\,\mu\cong  125.1$GeV $\cong 2.243 \times 10^{-25}$Kg  is the mass of the
Higgs and $\lambda \cong 0.1291$ is the self--coupling constant determined as
explained below Eqs (\ref{varphieq}) and (\ref{seq}).

As exemplified in Fig.\,8B for the approximate proper--time representation, $\tilde\varphi(\tilde\tau)$
remains virtually zero for $\tilde\tau <\tilde\tau_B$, then jumps almost abruptly to $\tilde
\varphi_{\hbox{\tiny max}}= \mu\,\sqrt{2/\lambda}\equiv \mu_H/\sqrt{\lambda}$ at $\tilde\tau=\tilde\tau_B$
and, for $\tilde\tau >\tilde\tau_B$, oscillates with decreasing amplitude and approximate radian frequency
$\mu/\sqrt{\lambda} \equiv\mu_H /{\sqrt{2\lambda}}$. A number of cycles after $\tilde\tau_B$, as
$\tilde \varphi(\tilde\tau)$ approaches $\mu/\sqrt{\lambda}$, the damped oscillation tends to become
harmonic with proper--time period
\begin{equation}
\label{ptimeperiod}
\Delta\tilde\tau_H = \frac{2^{3/2}\pi}{\mu_H} \cong 4.67\times 10^{-26}\,\hbox{s}\,.
\end{equation}

As evidenced in Figs.\,\ref{VacStFig3} and \ref{VacStFig4}, the maximum of energy density
$\tilde U_{\hbox{\tiny max}} (\tilde\tau_B)$ is attained immediately after $\tilde\tau_B$ and,
since $\tilde\tau_B$ is very close to $\tilde\tau_c$, it is very close to
\vspace{-1mm}
\begin{equation}
\label{HiggsEdens}
\tilde U_{\hbox{\tiny max}} (\tilde\tau_B) = \frac{\mu^4}{4\lambda}\equiv \frac{\mu_H^4}{16\lambda} =
\frac{(125.1)^4}{16\times 0.1291} \cong 1.186\times 10^8 \,\mbox{GeV$^4$}\,.
%\vspace{-1mm}
\end{equation}

Because in the proper time representation the total energy of the system is conserved, the sum of the kinetic and
potential energy densities of the Higgs field fades away as $1/\tilde\tau^3$.

\newpage

\markright{R.Nobili, Conformal General Relativity - {\bf \ref{PathIntApp}} Path integrals and effective action}
\renewcommand\thefigure{\Alph{section}\arabic{figure}}
\setcounter{figure}{0}
\section{PATH INTEGRALS AND EFFECTIVE ACTIONS}
\label{PathIntApp}
The functional method of {\em effective action} was originally introduced by Jona--Lasinio in 1964 to
clarify the mechanism of the spontaneous breakdown of a symmetry in relativistic field theories.
A few years later it was used by Coleman \& E.Weinberg (1973), Jackiw (1974) and other authors
\cite{COLEMAN1} to achieve important results in the path--integral representation of quantum field
theories (QFT). Concise introductions to this subject are available in Coleman's book
{\em Aspects of Symmetry} (1985) and in S.Weinberg's treatise on {\em Quantum Field Theory}, Vol II,
Ch.16, (1996).  Normally the implementation of the method is based on the representation of the
path integral in the Minkowskian spacetime. We call this the {\em standard approach} to the
effective action. Unfortunately it is not a well--fitting approach to CGR because in this case the
spacetime is conical (see \S\,1.1 of P2), i.e., the action integral extends from an initial
time $\tau=0$ to $\tau=\infty$ (see \S\,\ref{futconegeom}). In this Appendix, we describe only the
standard method, presuming that it fits the behavior of CGR over time.

\subsection{Path integral and Green's functions}
\label{PathInTAndGRFunx}
For simplicity, let us we exemplify the path--integral approach to the effective action method for a
theory of a self--interacting scalar field $\phi(x)$ whose classical dynamics is described by a
Lagrangian density ${\cal L}(x)\equiv{\cal L}[\phi(x), \partial_\mu\phi(x)]$. The method can easily be
generalized to the case of many boson fields, and also of many fermion fields provided that the
formalism of Grassmann variables is used, as described in Appendix {\bf \ref{GrassAlgApp}}.

Let ${\cal A}[\phi]$ be the classical action of ${\cal L}(x)$ and assume that an external classical current
$J(x)$ coupled to $\phi(x)$ is turned on.  According to the path--integral method, the complete quantum--field
amplitude from the vacuum state in the far past to the vacuum state in the far future, is given by a
functional integral of the form
\begin{equation}
\label{020amplA}
Z[J\,] \equiv\langle 0^+|0^-\rangle_J =\int{\cal D} \phi\, e^{\textstyle\, \frac{i}{\hbar}\,\{ {\cal A}[\phi]
+\int \phi(x)\,J(x)\,d^4x\}},
\end{equation}
where ${\cal D}\phi\equiv \prod_x d\phi(x)$ is the path measure over the space of function $\phi(x)$.
Actually, $Z[J\,]$ describes only the sum of all quantum--bubbles of the vacuum state elicited by $J(x)$.

The functional derivatives of $Z[J\,]$ with respect to $J(x)$ provide the vacuum expectation values
(VEVs) of the time--ordered operators
\begin{equation}
\label{disconnGreensFucts}
G^{(n)}(x_1,x_2, \dots, x_n) =  \langle 0^+|T[\phi(x_1)\,\phi(x_2)\dots\phi(x_n)]|0^-\rangle_J
\end{equation}
in the presence of external current $J(x)$. By functional derivations we obtain in fact
\begin{eqnarray}
\label{functJders}
&& \langle 0^+|T[\phi(x_1)\,\phi(x_2)\dots\phi(x_n)]|0^-\rangle_J =\bigg(\frac{\hbar}{i}
\bigg)^n\!\!\frac{\delta^n Z[J\,]}{\delta J(x_1)\,\delta J(x_2)\cdots\delta J(x_n)} =\nonumber \\
& & \int\!{\cal D}\phi\,\phi(x_1)\,\phi(x_2)\cdots\phi(x_n)\,e^{\textstyle \frac{i}{\hbar}\,
\{{\cal A}[\phi]\! +\!\int\!\phi(x)J(x)d^4x\}}\,.\nonumber
\end{eqnarray}

Eqs (\ref{disconnGreensFucts}) represent the sum of connected and disconnected Feynman diagrams,
with external lines (including propagators) corresponding to the fields $\phi(x_1), \phi(x_2),
\dots, \phi(x_n)$.

Let us call both vertices and ending points of a Feynman diagram as {\em dots}. A Feynman diagram
or sub--diagram is called {\em connected} if it is possible to go from a dot to another dot via
internal lines.  A diagram or sub--diagram which is not connected is called {\em disconnected}.

We can therefore expand $Z[J\,]$ as a functional Taylor series of $J(x)$,
\begin{equation}
\label{Gnseries}
Z[J\,]= \sum_n\!\frac{1}{n!}\Big(\frac{i}{\hbar}\Big)^{\!n}\!\!\!
\int^{(n)}\!\!\!G^{(n)}(x_1,x_2, \dots, x_n\!)\,J(x_1) J(x_2) \dots J(x_n)\, d^4x_1 d^4x_2 \cdots d^4x_n.\!
\vspace{-2mm}
\end{equation}

However, for most purposes, it is more convenient to put $Z[J]= e^{\textstyle \frac{i}{\hbar}W[J]}$ and
consider, instead of $G^{(n)}(x_1,x_2, \dots, x_n)$, the functional derivatives
of $W[J]$ with respect to $J(x)$,
\begin{equation}
\label{GRNSFUNCTS}
G^{(n)}_c(x_1,x_2, \dots, x_n)=\frac{\delta^n W[J\,]}{\delta J(x_1)\,\delta J(x_2)\cdots\delta J(x_n)}\,,
\end{equation}
which represent all possible {\em connected} Feynman diagrams with $n$ external lines, not counting
as different those diagrams that differ only by a permutation of vertices. These are called the
{\em connected Green's functions} in the presence of an external current $J(x)$. With other
conventions, the right--hand side of Eq (\ref{GRNSFUNCTS}) may have a factor of $(\hbar/i)^{n-1}$.

We can therefore expand $W[J\,]$ as a functional series of $J(x)$,
\begin{equation}
W[J\,] = \sum_n\frac{1}{n!}\!\int^{(n)}\!\!\!G^{(n)}_c(x_1,x_2, \dots, x_n)
\,J(x_1)\,J(x_2)\cdots J(x_n)\, d^4x_1 d^4x_2 \cdots d^4x_n\,.\nonumber
\end{equation}

To explain why functions (\ref{GRNSFUNCTS}) are connected, there is nothing better than quoting a
passage from S.Weinberg's treatise {\em Quantum Field Theory}, Vol II, pp.64--65, (1996):
\begin{itemize}
\item[]
``$Z[J]$ is given by the sum of {\em all} vacuum--vacuum amplitudes in the presence of current $J(x)$,
including disconnected as well as connected diagrams, but not counting as different those diagrams that
differ only by a permutation of vertices in the same or different connected subdiagrams.
A general diagram that consists of $N$ connected components will contribute to $Z[J]$ a term
equal to the product of the contribution of these components, divided by the number $N!$ of
permutations of vertices that merely permute all the vertices in one connected component
with all the vertices in another.$^*$ Hence, the sum of all diagrams is
$$
Z[J] = \sum_{N=0}^\infty \frac{1}{N!}\Big(\frac{i}{\hbar} W[J]\Big)^N = e^{\textstyle\,\frac{i}{\hbar}\,W[J]},
$$
where $(i/\hbar) W[J]$ is the sum of all connected vacuum--vacuum amplitudes, again not counting
as different those diagrams that differ only by a permutation of vertices.''
\item[]
{\small Footnote$^*$: ``The contribution of a Feynman diagram with $N$ connected components containing $n_1, n_2,
\cdots, n_N$ vertices is proportional to a factor $1/(n_1+n_2+\cdots n_N)!$ from the Dyson  expansion, and
a factor $(n_1+n_2+\cdots n_N)!/N!$ equal to the number of permutations of those vertices, counting as
identical those permutations that merely permute all the vertices in one component with all the vertices
in another}.''
\end{itemize}

\subsection{Quantum excitations of a scalar field}
\label{QFExcitations}
Denote for simplicity the classical action of $\phi(x)$, in the presence of external current $J(x)$,
as ${\mathbb A}[\phi]\equiv{\cal A}[\phi]+ \int\phi(x)J(x)\,d^4x$; the classical solution $\phi_c(x)$
to the Euler--Lagrange equation is satisfies equation
$$
\frac{\delta {\mathbb A}[\phi]}{\delta\phi_c(x)}= 0\,,\quad \hbox{i.e.,}\,
-\frac{\delta {\cal A}[\phi]}{\delta\phi_c(x)} = J(x)\,.
$$
Putting $\phi(x)= \phi_c(x) + \hat\phi(x)$, where $\hat\phi(x)$ is an arbitrary smooth variation of
$\phi_c(x)$, we can interpret the totality of $\hat\phi(x)$ as the set of quantum excitations of
$\phi_c(x)$. This allows us to rewrite the right--hand side of Eq (\ref{020amplA}) in the equivalent form
\begin{equation}
\label{020amplB}
e^{\textstyle\,\frac{i}{\hbar}\,W[J\,]}=\int {\cal D}\hat\phi\, e^{\textstyle \,\frac{i}{\hbar}
\{{\cal A}[\phi_c+\hat\phi]+\int[\phi_c(x)+\hat\phi(x)]J(x)\, d^4x\}-
\frac{1}{2\hbar}\!\int\!\epsilon\,\hat\phi^2(x)\,d^4x }\,.
\end{equation}
The replacement of ${\cal D}\phi\equiv \prod_x d\phi(x)$ with ${\cal D}\hat\phi\equiv \prod_x
d\hat\phi(x)$ is indeed possible because $\phi_c(x)$ is fixed. The term proportional to $\epsilon$
has been added to damp the amplitude of the path integral when the quantum excitation tends to infinity.

To facilitate the integration of Eq (\ref{020amplB}), it is useful to denote the functional
derivatives of ${\mathbb A}[\phi_c]$ with respect to $\phi_c$ as
$$
\frac{\delta^n {\mathbb A}[\phi]} {\delta \phi(x_1)\,\delta\phi(x_2) \cdots
\delta \phi(x_n)}\bigg|_{\phi= \phi_c} \equiv {\mathbb A}^{(n)}[\phi_c;x_1, x_2,\dots ,x_n]\,,
$$
and expand ${\mathbb A}[\phi_c+\hat\phi]$ into the functional Taylor series
\begin{eqnarray}
\label{TaylorAct}
&&\hspace{-12mm}{\mathbb A}[\phi_c+\hat\phi]\!=\!{\mathbb A}[\phi_c]+\!\int\!\!{\mathbb A}'[\phi_c;x]
\,\hat\phi(x)\,d^4x!+\!\frac{1}{2}\!\!\iint\!\!{\mathbb A}''[\phi_c; x_1, x_2]\hat\phi(x_1)\,
\hat\phi(x_2)\,d^4x_1 d^4x_2\!+\\
&&\hspace{7mm}\sum_{n>2} \frac{1}{n!}\iint\!\!\cdots\!\! \int\!\! {\mathbb A}^{(n)}
[\phi_c; x_1, x_2,\dots, x_n]\,\hat\phi(x_1)\,\hat\phi(x_2)\cdots \hat\phi(x_n)\,
d^4x_1 d^4x_2\cdots d^4 x_n. \nonumber
\end{eqnarray}
Note that ${\mathbb A}'[\phi_c;x]=0$ because $\phi_c(x)$ is the solution to the classical motion equation.

Assume, for instance, that the Lagrangian density of ${\mathbb A}[\phi]$ is that of a Higgs field, i.e.,
\begin{equation}
\label{mulambdaLagDens}
{\cal L}(x) = -\frac{1}{2}\,\phi(x)\big(\Box_x -i\epsilon\big)\phi(x)-
\frac{\lambda}{4}\,\big[\,\phi^2(x)- v^2\big]^2+J(x)\,\phi(x)\,,
\end{equation}
where $\Box_x$ is the d'Alembert operator, $\lambda$ is the self--coupling constant and $v$
the resting value of $\phi$ in the absence of $J$, so we have ${\mathbb A}''[\phi_c; y, x] =
\delta^4(y-x)\big[\,\Box_x +3\lambda\,\phi_c^2(x) - \lambda\,v^2  -i\epsilon \big]$,
and Eq (\ref{TaylorAct}) can be written as
\begin{eqnarray}
\label{mathbbA-exp}
{\mathbb A}[\phi_c+\hat\phi\,] & = & {\mathbb A}[\phi_c]-\frac{1}{2}\!\iint\!\hat\phi(y)\,
\delta^4(y-x)\big[\,\Box_x+3\lambda\,\phi_c^2(x) - \lambda\,v^2
-i\epsilon \big]\hat\phi(x)\,d^4y\,d^4x +\nonumber\\
&& \int\!J(x)\,\hat\phi(x)\,d^4x -\lambda\!\int\!\big[\,3\,\phi_c(x)\,\hat\phi^3(x) +
\frac{1}{4}\,\hat\phi^4(x)\big]\,d^4x\,.
\end{eqnarray}

The `continuous matrix' ${\cal K}[\phi_c; x,y] = -\delta^4(y-x)\big[\,\Box_x +3\lambda\,
\phi_c^2(x) - \lambda\,v^2-i\epsilon \big]$ will be called the {\em kinetic kernel} of
field excitation $\hat\phi(x)$. We can define its inverse ${\cal K}^{-1}[\phi_c; x,y]$
as the continuous matrix that satisfies equation $\int\!{\cal K}[\phi_c; x,z]\,
{\cal K}^{-1}[\phi_c; z,y]\,d^4z = \delta^4(x-y)$.

Since the kinetic kernel satisfies equation $\int{\cal K}[\phi_c;x,y]\,\hat\phi(y)\,d^4y = - J(x)$,
the Feynman propagator of $\hat\phi(y)$, $\Delta[\phi_c; x, y]$, satisfies equation
\begin{equation}
\label{calKDelta}
\int {\cal K}[\phi_c; x, z]\,\Delta[\phi_c; z, y]\,d^4z= -\hbar\,\delta^4(x-y)\,.
\end{equation}

Therefore, we can write
\begin{equation}
\label{FeynPropagA}
\Delta[\phi_c; y, x] =-\hbar\,{\cal K}[\phi_c;x,y]^{-1}\,,\quad {\cal K}[\phi_c;x,y] =-\hbar\,\Delta^{-1}
[\phi_c; x, y]\,,
\end{equation}
and the second term on the left--hand side of Eq (\ref{mathbbA-exp}) as
\begin{equation}
\label{KineticOperat}
-\frac{1}{2}\!\int\!\hat\phi(x)\big[\,\Box_x+3\lambda\,\phi_c^2(x) -
\lambda\,v^2  -i\epsilon \big]\hat\phi(x)\,d^4x= \frac{1}{2}\!\iint\! \hat\phi(y)\,{\cal K}[\phi_c; y, x]\,
\hat\phi(x)\,d^4y \,d^4x.
\end{equation}

Now, let us apply the expansion (\ref{mathbbA-exp}) to path integral (\ref{020amplB}).
Since the classical action ${\mathbb A}[\phi_c]$ is independent of $\hat\phi$, we
can factor $e^{\textstyle \frac{i}{\hbar}{\mathbb A}[\phi_c]}$ out of the path integral
and put for brevity ${\mathbb A}_I[\hat\phi] =  -\lambda\!\int\!\big[\,3\,\phi_c(x)
\,\hat\phi^3(x) +\frac{1}{4}\,\hat\phi^4(x)\big]\,d^4x$. So the path integral takes the form
\begin{equation}
\label{020ampl}
e^{\textstyle\, \frac{i}{\hbar}W[J\,]}\!= e^{\textstyle \frac{i}{\hbar}{\mathbb A}[\phi_c]}
\!\!\int\!{\cal D}\hat\phi\, e^{\textstyle \frac{i}{\hbar}{\mathbb A}_I[\hat\phi]}\,
e^{\textstyle \frac{i}{2\hbar}\!\iint\!\hat\phi(y)\,{\cal K}[\phi_c; y, x]\,\hat\phi(x)
\,d^4y\,d^4x+\frac{i}{\hbar}\!\int\!J(x)\,\hat\phi(x)\,d^4x }.\\
\end{equation}

${\mathbb A}_I[\hat\phi]$ also can be taken out of the path integral, provided we
replace it with the functional operator % ${\mathbb A}_I[\phi_c;\frac{\delta}{\delta J}]$.
\begin{equation}
\label{interactionAction}
{\mathbb A}_I\Big[\phi_c;\frac{\delta}{\delta J}\Big] \displaystyle{
=-\lambda\!\int\!\bigg[3\,\phi_c(x)\frac{\delta^3}{\delta J(x)^3}  +
\frac{1}{4}\,\frac{\delta^4}{\delta J(x)^4}\bigg]d^4x}\,.
\end{equation}
Thus, in summary, we obtain the useful factorization
\begin{equation}
\label{020amplC}
e^{\textstyle \, \frac{i}{\hbar}W[J\,]} =e^{\textstyle \frac{i}{\hbar}\,{\mathbb A}[\phi_c]}
\,e^{\textstyle \frac{i}{\hbar}\,{\mathbb A}_I[\phi_c;\frac{\delta}{\delta J}]}
\,e^{\textstyle \frac{i}{\hbar}\,S[\phi_c]}, \nonumber
\end{equation}
where, in consideration of Eq (\ref{KineticOperat}), we can express the last factor as
\begin{equation}
\label{S0ofPhiC}
e^{\textstyle \,\frac{i}{\hbar}S[\phi_c]} = \int\! {\cal D}\hat\phi\,e^{\textstyle
\,\frac{i}{2\hbar}\!\iint\!\hat\phi(y)\,{\cal K}[\phi_c;y, x]\,\hat\phi(x)\,d^4y\, d^4x
+\frac{i}{\hbar}\!\int\!J(x)\,\hat\phi(x)\,d^4x}.
\end{equation}

This path integral can be integrated by generalizing the complex Gaussian integral
\begin{equation}
\label{xAxboson}
I_{\bf B}\!=\!\int\!e^{\textstyle i\big[\frac{1}{2}\,\tilde{\bf x}({\bf B}\!+\!i\epsilon){\bf x}\!
+\!\tilde{\bf x}\!\cdot\!{\bf y}\!+\!\tilde{\bf y}\!\cdot\!{\bf x}\big]}d^n {\bf x}=\!
\frac{e^{\textstyle -\frac{i}{2}\,\tilde{\bf y}\, {\bf B}^{\!-1}{\bf y}}}
{\sqrt{\hbox{Det}(i{\bf B}/2\pi)}}\!=\! e^{\textstyle  -\frac{1}{2}\,\hbox{Tr}\ln(i{\bf B}/2\pi)
\!-\! \frac{i}{2}\,\tilde{\bf y}\,{\bf B}^{-1}{\bf y}}\!,
\end{equation}
where $\bf B$ is an $n\times n$ symmetric matrix, $\bf y$ is an $n$--dimensional vector,
$\tilde{\bf y}$ its transposed and the identities, where the following identities are used.
\begin{equation}
\label{LnDet2Trln}
\big[\,\hbox{Det}(i{\bf B}/2\pi)\big]^{-1/2}=
e^{\textstyle -\frac{1}{2}\ln\hbox{Det}(i{\bf B}/2\pi)}
= e^{\textstyle -\frac{1}{2}\,\hbox{Tr}\ln(i{\bf B}/2\pi)}
\end{equation}

Transferring Eq (\ref{xAxboson}) and the second of Eqs (\ref{LnDet2Trln}) to the functional
calculus domain, and exploiting Eq (\ref{FeynPropagA}), we obtain for Eq (\ref{S0ofPhiC})
the generalized Gaussian integral
\begin{equation}
\label{GaussianIntWithJ}
e^{\textstyle \,\frac{i}{\hbar} S[\phi_c]} =  e^{\textstyle -\frac{1}{2}\,\hbox{Tr}
\ln\frac{i\,{\cal K}[\phi_c]}{2\pi}}\,
e^{\textstyle -\frac{i}{2\hbar}\iint\! J(y)\,\Delta[\phi_c;\,y, x]\,J(x)\,d^4y\,d^4x},
\end{equation}
where ${\cal K}[\phi_c]$ is a shorthand for ${\cal K}[\phi_c;x,y]$.
We can therefore rewrite Eq (\ref{020ampl}) in the compact form,
\begin{equation}
\label{020amplD}
e^{\textstyle \, \frac{i}{\hbar}W[J\,]}\!=e^{\textstyle \frac{i}{\hbar}\big\{{\mathbb A}[\phi_c]\! +
\!\frac{i\hbar}{2}\hbox{Tr}\ln\frac{i\,{\cal K}[\phi_c]}{2\pi} \!\big\}}
e^{\textstyle \frac{i}{\hbar}{\mathbb A}_I[\phi_c;\frac{\delta}{\delta J}]}\!\!\int\!\!{\cal D}\hat\phi\,
e^{\textstyle \frac{-i}{2\hbar}\!\iint\!\!J(y)\,\Delta[\phi_c;y, x]\,J(x)\,d^4y\,d^4x}.
\end{equation}

Note that, if ${\mathbb A}_I[\hat\phi]=0$, i.e., if in Eq (\ref{mulambdaLagDens}) it is $\lambda =0$,
Eq (\ref{020amplD}) simplifies to
\begin{equation}
\label{020amplE}
e^{\textstyle \, \frac{i}{\hbar}W[J\,]}=e^{\textstyle \frac{i}{\hbar}\big\{{\mathbb A}[\phi_c]\! +
\!\frac{i\,\hbar}{2}\,\hbox{Tr}\ln\frac{i\,{\cal K}[0]}{2\pi} \!\big\}}\!\int\!\!{\cal D}\hat\phi\,
e^{\textstyle -\frac{i}{2\hbar}\!\iint\!\!J(y)\,\Delta[\phi_c;y, x]J(x)\,d^4y\,d^4x}.
\end{equation}
Here ${\cal K}[0]$ is a shorthand for ${\cal K}[0;x,y] = -\delta^4(y-x)[\Box_x+i\epsilon]\equiv
-\hbar\,\Delta^{-1}[0; x, y]$, where $\Delta[0; x, y]$ is the Feynman propagator of a massless
scalar field. In this case, $Z[J\,]$ becomes a pure phase--factor times the infinite series of
massless--field propagators elicited by $J(x)$.

\subsection{The Gaussian term of the path--integral Lagrangian density}
\label{digressOnlnDet}
Note that, both in the interacting--field case (\ref{020amplD}) and in the free--field case
(\ref{020amplE}), the classical action ${\mathbb A}[\phi_c]$ is modified by the addition of
the Gaussian term
\begin{equation}
\label{AGterm}
{\mathbb A}_G[\phi_c] = \frac{i}{2}\,\ln\hbox{Det}\frac{\Delta^{-1}[\phi_c]}{2\pi i} =
\frac{i\hbar}{2}\,\hbox{Tr}\ln\frac{i\,{\cal K}[\phi_c]}{2\pi} \equiv
\frac{i\hbar}{2}\hbox{Tr}\!\iint\ln i\,{\cal K}[\phi_c; x,y]\,d^4y\,d^4y\,.
\end{equation}

In general, since $\phi_c$ depends on $x$,  we cannot expect ${\mathbb A}_G[\phi_c]$ to have a simple
analytical form. But if $\phi_c$ is a constant, we can put $m^2 = 3\lambda\,\phi_c^2 - \lambda\,v^2$ and
diagonalize the continuous matrix ${\cal K}[\phi_c; x,y]$ by passing to the momentum space. So we obtain
$$
\tilde{\cal K}[\phi_c; p,q] = \frac{1}{(2\pi\hbar)^4}\iint e^{\textstyle \frac{i}{\hbar}(p\!\cdot\!x -
q\!\cdot\!y)}{\cal K}[\phi_c; x,y]\,d^4x\,d^4y =\big[p^2 -m^2 -i\epsilon \big]\delta^4(p-q).
$$
The logarithm of this diagonal matrix is just the diagonal matrix with the logarithm in
the main diagonal. We can therefore cast Eq (\ref{AGterm}) in the form
$$
{\mathbb A}_G[\phi_c]= -\frac{i\hbar}{2 (2\pi\hbar)^4}\hbox{Tr}\!\iint \ln \frac{i\,[p^2 -
m^2 - i\epsilon] }{2\pi}\,\delta^4(p-q)\,d^4p\,d^4q.
$$

Since $\delta^4(p-q)$ is related to spacetime volume ${\cal V}_4$ as indicated by the following steps,
$$
(2\pi\hbar)^4 \delta^4(p-q) =\int e^{\textstyle\, \frac{i}{\hbar}(p-q)\!\cdot\! x}d^4x \quad\hbox{and }\,
\lim_{p\rightarrow q} \int e^{\textstyle\, \frac{i}{\hbar}(p-q)\!\cdot\! x}d^4x  = \int d^4x \equiv {\cal V}_4,
$$
we can express the Gaussian contribution to the path--integral Lagrangian density
\begin{equation}
\label{calGm2}
{\cal G}(m^2)=\frac{ {\mathbb A}_G[\phi_c]}{{\cal V}_4}= -
\frac{i\hbar}{2}\hbox{Tr}\!\int \ln \frac{i\,[p^2 - m^2 - i\epsilon] }{2\pi}\,d^4p\,.
\end{equation}
Term $i\epsilon$ in the squared brackets tell us that we must rotate the $p_0$ contour
counterclockwise, carry out the Wick rotation $p_0\rightarrow ip_0$ and integrate $p_0$ from
$-\infty$ to $\infty$, which yields
\begin{equation}
\label{calGm2Eucl}
{\cal G}(m^2) = \frac{\hbar}{2(2\pi)^4}\int\!\ln\frac{p^2+m^2}{2\pi}\,d^4 p\,.
\end{equation}
Putting $\int\!f(p^2)\,d^4p  =\pi^2\int\!f(p^2)p^2 dp^2= \pi^2\int\! f(x)x\, dx$, where $x=p^2$
and $2\pi^2$ is the unit--4D--sphere surface, we obtain the definite integral of Eq (\ref{calGm2}),
from infrared cut--off $p^2=\epsilon^2$ to ultraviolet cut--off $p^2=\Lambda^2$, and
its first three derivatives with respect to $m^2$,
\begin{eqnarray}
&&{\cal G}(m^2) =
\frac{\hbar}{32\,\pi^2}\!\int_0^{\Lambda^2}\!\!x \ln\frac{x+m^2}{2\pi}
dx;\quad {\cal G}'(m^2) = \frac{\hbar}{32\,\pi^2}\!\int_0^{\Lambda^2}\!\!\frac{x\,dx}
{x+m^2}\,;\nonumber\\
&&{\cal G}''(m^2) = -\frac{\hbar}{32\,\pi^2}\!\int_0^{\Lambda^2}\!\!\frac{x\,dx}
{\big(x+m^2\big)^2}\,;\quad  {\cal G}'''(m^2) =
\frac{\hbar}{16\,\pi^2}\!\int_0^{\Lambda^2} \frac{x\,dx}{(x+m^2)^3}\,.\nonumber
\end{eqnarray}
To simplify this integration we can subtract from ${\cal G}(m^2)$ the part independent of $\Lambda^2$,
$\overline{\cal G}(m^2) = {\cal G}(m^2)- {\cal G}(0) - m^2 {\cal G}'(0) - (m^4/2){\cal G}''(0)$.
The subtracted terms are
\begin{eqnarray}
&& {\cal G}(0)=\frac{\hbar}{32\,\pi^2}\bigg[\!\int_{\epsilon^2}^{\Lambda^2}\!\!
x\,(\ln x)\,dx-(\ln 2\pi)\,\Lambda^4 \bigg]=\frac{\hbar\Lambda^4}{64\,\pi^2}
\ln\frac{\Lambda^2}{\epsilon^2(2\pi)^2\!\sqrt{e}}\,; \nonumber\\
&& {\cal G}'(0) = \frac{\hbar}{32\,\pi^2}\!\int_{\epsilon^2}^{\Lambda^2}\!\!dx
=\frac{\Lambda^2-\epsilon^2}{32\,\pi^2}\,;\quad {\cal G}''(0) =-\frac{\hbar}{32\,\pi^2}
\!\int_{\epsilon^2}^{\Lambda^2}\!\!\frac{dx}{x}= -\frac{\hbar}{32\,\pi^2}
\ln\frac{\Lambda^2}{\epsilon^2} \,;\nonumber
\end{eqnarray}
where $e$ the Neper constant. $\overline{\cal G}(m^2)$  can easily be calculated
by triple integration of ${\cal G}'''(m^2)$ from $\epsilon^2$ to $m^2$, which gives
$\overline{\cal G}(m^2) =\hbar\, m^4 \ln (m^2/\epsilon^2)/64\,\pi^2$.

In summary, for $\epsilon^2\rightarrow 0$ we have
\begin{equation}
\label{lnDetTerm}
{\cal G}(m^2) = \hbar\bigg(\frac{\Lambda^4}{64\,\pi^2}\ln\frac{\Lambda^2}{(2\pi)^2\!\sqrt{e}}
+ \frac{m^2\Lambda^2}{32\,\pi^2} +\frac{m^4}{64\,\pi^2}\ln\frac{m^2}{\Lambda^2}\bigg)\,.
\end{equation}
The first term in the round brackets on the left hand side gives the path integral a mere phase factor.
Its removal is equivalent to normalizing Eq~(\ref{AGterm}) by replacing $\hbox{Det}(\Delta^{-1}/2\pi i)$
with $\hbox{Det}\,\Delta_0\Delta^{-1}$, where $\Delta_0(x,y)$ is the propagator of a free massless field.
In this case we simply obtain
\begin{equation}
\label{lnDetTerm2}
{\cal G}(m^2) = \hbar\bigg(\frac{m^2}{64\,\pi^2}\,\Lambda^2-\frac{m^4}{64\,\pi^2}\ln \Lambda^2
+\frac{m^4\ln m^2}{64\,\pi^2}\bigg).
\end{equation}

If $\phi_c$ depends on $x$, we shall have $m^2(x)=3\lambda\,\phi_c^2(x) - \lambda\,v^2$, and
therefore, in place of Eq (\ref{calGm2Eucl}), we shall have an integral with the integrand
still in diagonal form,
\begin{equation}
\label{calGm2xEucl}
{\cal G}[\widetilde m^2(p)] = \frac{\hbar}{2(2\pi)^4}\int\!\ln\frac{p^2+\widetilde m^2(p)}{2\pi}\,d^4 p\,,
\end{equation}
where $\widetilde m^2(p)$ is the Fourier transform of $m^2(x)$ in the Euclidean momentum space. Of course,
we cannot expect that integrals of this sort can always be expressed in a compact analytical form.

\subsection{The tree diagrams of the semiclassical approximation}
\label{Treediags}
Consider the classical action of a self--interacting scalar field $\phi(x)$
\begin{equation}
\label{GenClassAction}
{\cal A}[\phi] = -\frac{1}{2}\int\! \phi(x)\big(\Box + m^2\big)\phi(x)\,d^4x
-\! \int\! V(x)\, d^4x \,,
\end{equation}
in which the potential energy density $V(x)$ is a polynomial in $\phi(x)$ of degree greater than two,
and call vertex of order $n>2$ the function
\begin{equation}
\label{V(n)x}
V^{(n)}(x)\!=\!\int\!\!\cdots\!\!\int\!\! \frac{\delta^n {\cal A}[\phi]}{\delta \phi(x_1)\cdots
\delta \phi(x_n)}\,\delta^4(x\!-\!x_1)\cdots \delta^4(x\!-\!x_n)\,d^4x_1\!\cdots d^4x_n
\! =\! \frac{\delta^n V(x)}{\delta \phi(x)^n}\,.
\end{equation}

Denote as ${\mathbb A}[\phi] =  {\cal A}[\phi]+\int\!J(x)\,\phi(x)\,d^4x$ the action of the
same system in the presence of an external current $J(x)$. The variational equation
\begin{equation}
\label{classSolut}
\frac{\delta {\mathbb A}[\phi]}{\delta \phi_c(x)} =0;\quad\hbox{i.e., }\,\big(\Box +
m^2\big)\phi_c(x) + V^{(1)}(x)=J(x)\,,
\end{equation}
provides the classical solution to the motion equation, $\phi_c(x)$. The semiclassical approximation
to the path integral over action ${\mathbb  A}[\phi_c]$ is defined by equation
\begin{equation}
\label{semiclass}
e^{\textstyle\,\frac{i}{\hbar}\,\bar{W}[J\,]}= e^{\textstyle \,\frac{i}{\hbar}\,\{{\cal A}[\phi_c]+
\int\!J(x)\,\phi_c(x)\, d^4x\}}\,.
\end{equation}
In accordance with Eqs (\ref{GRNSFUNCTS}), the functional derivatives of $\bar W[J\,]$ with
respect to $J(x)$,
\begin{equation}
\label{Treediagrams}
\frac{\delta^n \bar W[J\,]}{\delta J(x_1)\,\delta J(x_2)\cdots\delta J(x_n)}=\bar
G^{(n)}_c(x_1,x_2, \dots, x_n)\,,
\end{equation}
provide the semiclassical connected Green's functions $\bar G^{(n)}_c(x_1,x_2, \dots, x_n)$.
In particular,
\begin{equation}
\label{barG2xy}
\frac{\delta^2 \bar W[J\,]}{\delta J(x)\,\delta J(y)} =
\frac{\delta \phi_c(x)}{\delta J(y)}=  \bar G^{(2)}_c(x, y)=\hbar\,\bar\Delta(x, y)\,,
\end{equation}
provides the propagator in the form of a continuous matrix $\bar\Delta(x, y)$ with
$x,y$ as indices. This matrix is manifestly related, via $\delta \phi_c(x)/\delta J(y)$, to
the classical kinetic kernel of ${\cal A}[\phi_c]$,
\begin{equation}
\label{barKxy}
\bar{\cal K}(x,y) =  \frac{\delta^2  {\cal A}[\phi_c]}{\delta \phi_c(x)\,\delta \phi_c(y)}
=-\frac{\delta J(x)}{\delta \phi_c(y)}= - \bar G^{(2)}_c(x, y)^{-1} \,,
\end{equation}
by equations $\bar\Delta(x, y) = -\hbar\,\bar{\cal K}^{-1}(x,y)$  and $\bar{\cal K}(x,y) =
-\hbar\,\bar\Delta^{-1}(x, y)$ [{\em cf} Eqs (\ref{FeynPropagA})]. Basing on these two equations,
we can easily verify that the functional variation of $\bar{\cal K}^{-1}(x,y)$ with respect to
$J(z)$ can be cast in the form
\begin{equation}
\label{deltabarKxy}
\frac{\delta \bar{\cal  K}^{-1}(x,y)}{\delta J(z)} = \frac{1}{\hbar^2}\!\int\! \bar\Delta(x, x')\,
\frac{\delta \bar {\cal K}(x', y')}{\delta J(z)}\,\bar \Delta(y', y)\, d^4x'\,d^4y\,'.
\end{equation}

On the other hand, using Eqs (\ref{Treediagrams}) and (\ref{barKxy}), we obtain the connected
3--point Green's function
$$
\frac{\delta\bar {\cal K}(x', y')}{\delta J(z)} = \int\!\frac{\delta\bar {\cal K}(x', y')}
{\delta \phi_c(z')}\,\frac{\delta \phi_c(z')}{\delta J(z)}\, d^4z' =\int\frac{\delta^3
{\cal A}[\phi_c]}{\delta \phi_c(x')\,\delta \phi_c(y')\,\delta \phi_c(z')}\,\bar\Delta(z',z)\,d^4z'\,.
$$
Inserting this function into Eq (\ref{deltabarKxy}) and using Eq (\ref{V(n)x}), we arrive at
\begin{equation}
\label{delta3WJ}
\frac{\delta^3 \bar W[J\,]}{\delta J(x_1)\delta J(x_2)\delta J(x_3)}= \bar G^{(3)}_c(x, y, z)=
\frac{1}{\hbar^3}\,V^{(3)}(x)\,\bar\Delta(x, x_1)\,\bar\Delta(x, x_2)\,\bar\Delta(x, x_3),
\end{equation}
where we have put $V^3(x)= \delta^3 {\cal A}[\phi_c]/\delta \phi_c(x)^3.$
By further computing the functional variation of Eq (\ref{delta3WJ}) with respect to $J(x_4)$,
we obtain the tree expansion of the connected 4--point Green's function
\begin{eqnarray}
\label{delta4WJ}
&&\hspace{-6mm} \frac{\delta^4 \bar W[J\,]}{\delta J(x_1)\,\delta J(x_2)\,\delta J(x_3)\,
\delta J(x_4)}= \frac{1}{\hbar^4}V^{(4)}(x)\,\bar\Delta(x, x_1)\,\bar\Delta(x, x_2)\bar\Delta(x, x_3)
\,\bar\Delta(x, x_4)+ \nonumber \\
&&\!\!\frac{1}{\hbar^5}\bigg[\!\iint\!\!V^{(3)}(y_1)\bar\Delta(y_1, y_2) V^{(3)}(y_2)
\bar\Delta(y_1,x_1)\bar\Delta(y_1,x_2)
\bar\Delta(y_2, x_3)\bar\Delta(y_2, x_4) d^4y_1d^4y_2 +\nonumber\\
&&\!\!\iint\!\! V^{(3)}(y_1)\bar\Delta(y_1, y_2) V^{(3)}(y_2)
\bar\Delta(y_1,x_1)\bar\Delta(y_1,x_3) \bar\Delta(y_2, x_2)\bar\Delta(y_2, x_4)
d^4y_1d^4y_2 +\nonumber\\
&&\!\!\iint\!\! V^{(3)}(y_1)\bar\Delta(y_1, y_2) V^{(3)}(y_2) \bar\Delta(y_1,x_4)\bar\Delta(y_1,x_3)
\bar\Delta(y_2, x_1)\bar\Delta(y_2, x_2) d^4y_1d^4y_2\!\bigg]\!.
\end{eqnarray}

\begin{figure}[!h]
\vspace{-6mm}
\centering
\includegraphics[scale=1]{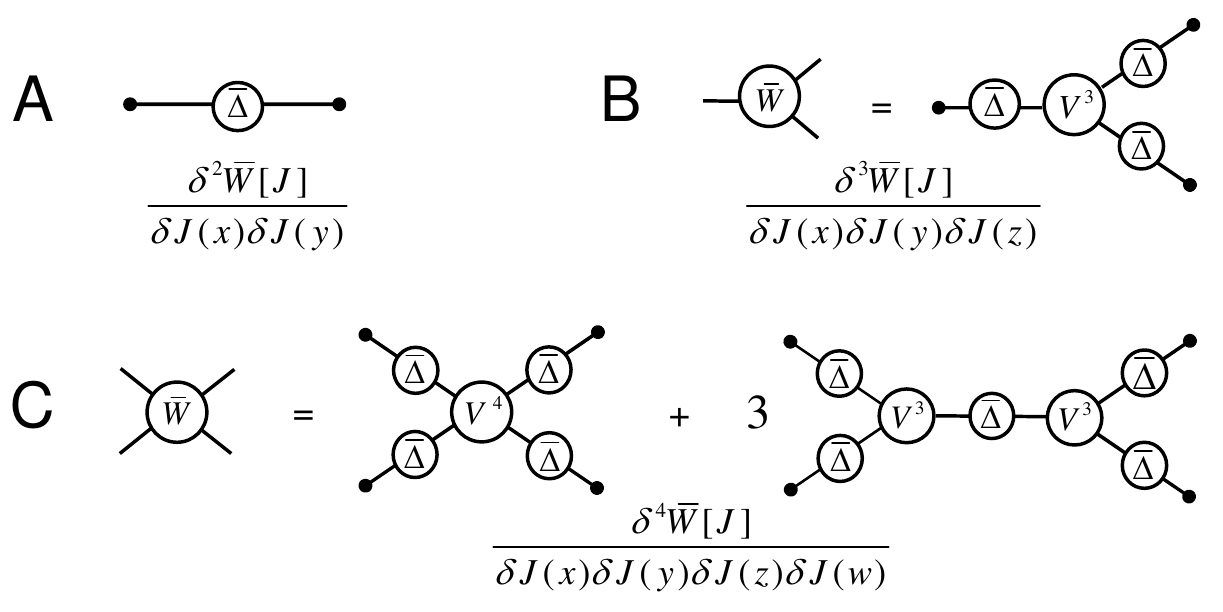} %F17
\vspace{-4mm}
\caption{\small
{\em The tree diagrams of the semiclassical approximation.} {\bf A}: The propagator as a vertex of order one.
{\bf B}: The 3--point connected Green's function as three propagators stemming from a vertex of order three.
{\bf C}: Connected 4--point Green's functions: the first is a vertex of order four, the
other three are formed by two vertices of order three connected by a propagator.  The mechanism of
tree generation is clear: an additional variation with respect to $J(x)$ increases the order of
each vertex by one and adds a new propagator.}
\label{PathintFig1}
\vspace{-2mm}
\end{figure}

All these results are graphically described in Fig.\,\ref{PathintFig1}.

Connected diagrams of a general path integral which cannot disconnected by cutting through any one
internal line  are called {\em one--particle irreducible} (1PI). In particular, all the external
propagators stemming from the diagrams are amputated at their insertion points. These diagrams play
a role similar to that of vertices $V^{(n)}(x)$ in the tree expansion of a semiclassical path integral.
With the difference that the local polynomials of the classical field $\phi_c(x)$ are replaced by
suitable $n$--point functionals $\Gamma^{(n)}(x_1, x_2, \cdots, x_n)$ representing the sum of all the
$n$--point 1PI diagrams with the legs stemming from the same set of quantum fields.

In fact, all the connected Green's functions generated by an exact path--integral functional $W[J\,]$
have the topological structure of a tree which can be obtained by connecting 1PI diagrams  with a
quantum field propagators $\Delta(x, y)$ that include higher corrections.

\begin{figure}[!h]
\vspace{-4mm}
\centering
\includegraphics[scale=1]{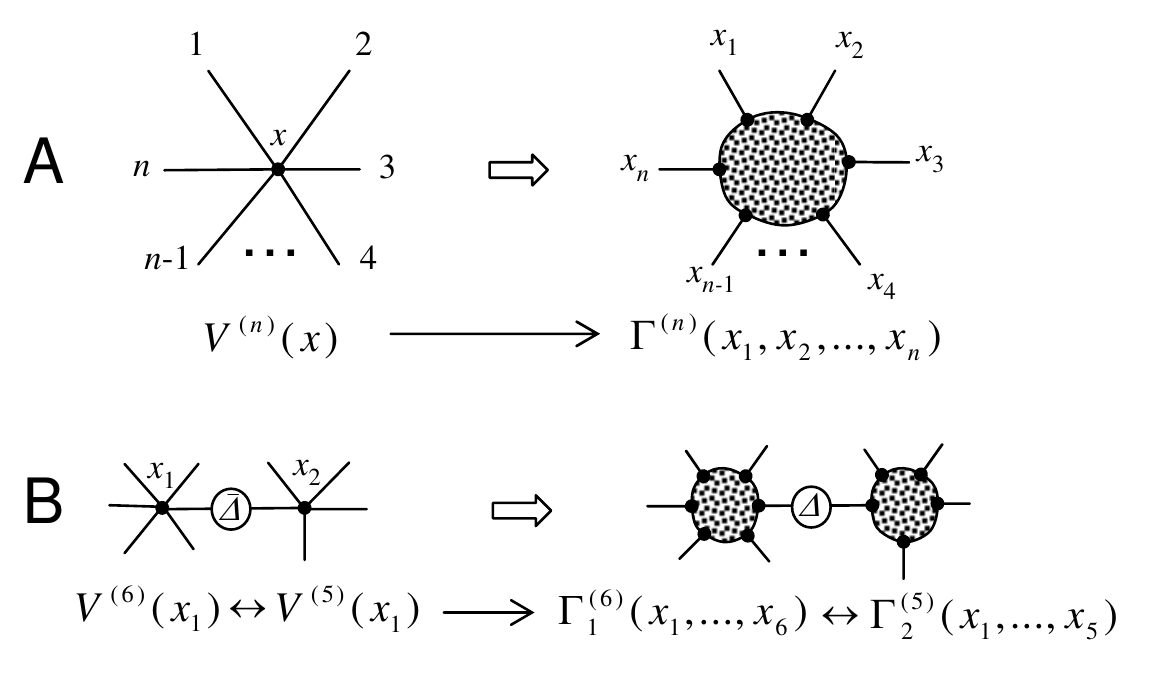} %F18
\vspace{-4mm}
\caption{\small {\bf A}:  One--particle irreducible  (1PI) diagrams of order $n$; on the left, a pointlike
vertex $V^{(n)}(x)$ of order $n$ in the tree expansion of a semiclassical functional $\bar W[J\,]$
[see Eq (\ref{semiclass})]; on the right, corresponding to a 1PI diagram $\Gamma^{(n)}(x_1,x_2,\dots,x_n)$
in the tree expansion of a path--integral functional $W[J\,]$ [see Eq (\ref{020amplD})]. {\bf B}:
Two one-particle reducible (1PR) diagrams connected by a single propagator;  on the left, two pointlike
vertices of the semiclassical tree expansion of $\bar W[J\,]$ connected by one classical propagator
$\bar\Delta$; on the right, two 1PI diagrams of the tree expansion of $W[J\,]$, connected by one
quantum--field propagator $\Delta$.}
\label{PathintFig2}
\vspace{-2mm}
\end{figure}

The relationship between the pointlike vertices and propagators of a semiclassical trees and
those of the connected Green's function of a true quantum path integral is sketched and
exemplified in Fig.\,\ref{PathintFig2}.

To fully appreciate the richness of the decomposition of connected diagrams into 1PI vertices and propagators, we must
carry out the Legendre transformation of the generating functional $W[J\,]$. This will be done in the following section.

Let us premise that, in carrying out this program, we will avoid discussing the renormalization problems arising in
the path integral method, which may be found in other papers (Coleman and E. Weinberg, 1973). In this regard,
the only serious problem arises from the irreducible divergences of the Gaussian term described
in \S\,\ref{digressOnlnDet}.

\subsection{Effective action and loop expansion}
\label{EffLang-loopExp}
The simplest of connected Green's functions is the VEV of the quantum field $\phi(x)$, i.e.,
\begin{equation}
\label{meanphi}
G^{(1)}_c(x) = \frac{\delta W[J\,]}{\delta J(x)} =
\frac{\langle 0^+|\phi(x)|0^-\rangle_J}{\langle 0^+|0^-\rangle_J} \equiv \bar\phi(x)\,.
\end{equation}
In Coleman's book (1985), $\bar\phi(x)$ is denoted as $\phi_c(x)$ and called the `classical field'; but here
it will be call the {\em effective field} because it is generally different from the Euler--Lagrange solution to
classical field equation (\ref{classSolut}). As shown by Eqs (\ref{GRNSFUNCTS}), the connected two--point
Green's function $G^{(2)}_c(x, y)$ coincides with the functional derivative of $\bar\phi(x)$
with respect to $J(y)$, which is related to the complete propagator of $\phi(x)$,
$\Delta[\bar\phi; x, y]$, by equation
\begin{equation}
\label{Gofxy}
G^{(2)}_c(x, y) = \frac{\delta^2 W[J\,]}{\delta J(x)\,\delta J(y)} =
\frac{\delta \bar\phi(x)}{\delta J(y)} = \hbar\, \Delta[\bar\phi; x, y]\,.
\end{equation}

Eq (\ref{meanphi}) can be reciprocated by defining $J[\bar\phi; x]$ as the current that yields a prescribed
value of $\bar\phi(x)$; so $J$ will depend functionally on $\bar\phi$, not $\bar\phi$ on $J$. This reciprocation
allows us to interchange the role of $W[J\,]$ with that of a functional $\Gamma[\bar\phi]$  of $\bar\phi(x)$,
called the {\em effective action} of the system, by introducing the Legendre transformation
\begin{equation}
\label{GAMMAfunct}
\Gamma[\bar\phi\,] = W[J\,] - \int\! J[\bar\phi; x]\,\bar\phi(x)\, d^4x \,.
\end{equation}

A necessary condition for the existence of this transformation is that there is a one--to--one mapping between
$\bar\phi(x)$ and $J(x)$ in suitable domains of these functions. This condition is equivalent to requiring
that $W[J\,]$ is positively or negatively convex in suitable domains of the functional space of $J(x)$
\cite{ZIA}. We shall assume this condition to be satisfied for all theories of interest. Since $\phi_c(x)$
also is one--to--one with $J(x)$, also the relation between $\phi_c(x)$ and $\bar\phi(x)$ is one--to--one in
suitable domains of these functions.

The relevant property of $\Gamma[\bar\phi\,]$ is that it can be expanded in series of
$\bar\phi(x)$,
\begin{equation}
\label{Gammaexpansion}
\Gamma[\bar\phi\,]\! =\! \sum_{n=0}^{\infty}\frac{1}{n!}\!\iint\!\!\dots\!\!\int\Gamma^{(n)}
(x_1,x_2,\dots, x_n)\,\bar\phi(x_1)\,\bar\phi(x_2)\cdots \bar\phi(x_n)\,d^4x_1\,d^4x_2\cdots d^4x_n,
\end{equation}
where $\Gamma^{(n)}(x_1,x_2,\dots, x_n)$ are the complete 1PI diagrams of order $n$ described in the previous section.
From Eqs (\ref{meanphi}) and  (\ref{GAMMAfunct}), we obtain
\begin{equation}
\label{deltaGAMMAdeltaphi}
\frac{\delta\Gamma[\bar\phi\,]}{\delta \bar\phi(x)} = - J[\bar\phi;x]\,.
\end{equation}
A general proof that $\Gamma^{(n)}(x_1,x_2,\dots, x_n)$ represent 1PI diagrams is found in
S.Weinberg's book, Vol II, Ch.16, pp.66--67 (1996), already quoted in the beginning of
this Appendix.

$\Gamma[\bar\phi\,]$ is an `effective action' not only in the sense that the value for $\bar\phi(x)$,
in the absence of the external current $J(x)$, is given by the stationarity condition
$\delta \Gamma[\bar\phi]/\delta \bar\phi(x) = 0$ for $J(x)=0$, but also in the sense that $W[J]$
may be represented as a sum of connected {\em tree} diagrams, with vertices calculated as if the
action were $\Gamma[\bar\phi\,]$ instead of ${\cal A}[\phi_c]$.

Now note that the functional derivative of $-J[\bar\phi;x]$ with respect to $\bar\phi(y)$ is
the complete kinetic kernel of the effective Lagrangian ${\cal K}[\hat\phi; x, y]$.
In fact, from Eq (\ref{Gofxy}), we obtain
\begin{equation}
\label{Deltaxy}
\frac{\delta^2 \Gamma[\bar\phi\,]}{\delta \bar\phi(x)\,\delta\bar\phi(y)} =
- \frac{\delta J[\bar\phi;x\,]}{\delta\bar\phi(y)}=
-\bigg\{\frac{\delta \bar\phi[J;y\,]}{\delta J(x)}\bigg\}^{-1} = -\hbar\,\Delta[\bar\phi;x, y]^{-1}=
{\cal K}[\bar\phi; x, y]\,.
\end{equation}

We can proceed further by taking the functional derivative of (\ref{Deltaxy}) with respect
to $\delta\bar\phi(z)$; in this way, after suitable manipulations, we find
\begin{eqnarray}
\label{1PI3VERTEX}
&&\hspace{-6mm}\frac{\delta^2 \Gamma[\bar\phi\,]}{\delta \bar\phi(x)\,\delta \bar\phi(y)\,
\delta \bar\phi(z)}\!=\!-\!\iiint\!\!\frac{\delta^3 W[J\,]}{\delta J(\bar x)\delta J(\bar y)
\delta J(\bar z)}\,\frac{\delta J(\bar x)}{\delta\bar\phi(x)}\,\frac{\delta J(\bar y)}
{\delta\bar\phi(y)}\,\frac{\delta J(\bar z)}{\delta\bar\phi(z)}\,d^4\bar x\, d^4\bar y\,
d^4\bar z \equiv \nonumber\\
&&\iiint\!{\cal K}[\bar\phi; \bar x, x]\,{\cal K}[\bar\phi; \bar y, y]\,
{\cal K}[\bar\phi; \bar z, z]\,G^{(3)}_c(\bar x,\bar y,\bar z)
\,d^4\bar x\, d^4\bar y\, d^4\bar z =\Gamma^{(3)}(x, y, z)\,,
\end{eqnarray}
in accordance with Eq (\ref{Gammaexpansion}). This is the connected 3--point Green's function, with
the external lines amputated by the complete kinetic kernels ${\cal K}[\bar\phi;\bar x, x]$,
${\cal K}[\bar\phi;\bar y, y]$, ${\cal K}[\bar\phi;\bar z, z]$.

Further deriving this equation with respect $\delta\bar\phi(w)$ yields
\begin{eqnarray}
\label{1PI4VERTEX}
&&\hspace{-8mm}\Gamma^{(4)}(x, y, z, w)\! =\!\!\int^{(4)}\!\!\!\!\frac{\delta^4 W[J\,]}
{\delta J(\bar x)\delta J(\bar y)\delta J(\bar z)\delta J(\bar w)}\,
\frac{\delta J(\bar x)}{\delta\bar\phi(x)}\,\frac{\delta J(\bar y)}{\delta\bar\phi(y)}\,
\frac{\delta J(\bar z)}{\delta\bar\phi(z)}\frac{\delta J(\bar w)}{\delta\bar\phi(w)}\,d^4\bar x\,
d^4\bar y\, d^4\bar z\,d^4\bar w +
\nonumber\\
&&\int^{(4)}\!\!\!\frac{\delta^3 W[J\,]}{\delta J(\bar x)\delta J(\bar y)\delta J(\bar z)}
\bigg\{\frac{\delta J(\bar y)}{\delta\bar\phi(y)}\,\frac{\delta J(\bar z)}{\delta\bar\phi(z)}\,
\frac{\delta^2 J(\bar x)}{\delta\bar\phi(x)\delta\bar\phi(w)}\,+
\frac{\delta J(\bar x)}{\delta\bar\phi(x)} \frac{\delta J(\bar z)}{\delta\bar\phi(z)}
\frac{\delta^2 J(\bar y)}{\delta\bar\phi(y)\delta\bar\phi(w)}+\nonumber \\
&& \hspace{6mm}\frac{\delta J(\bar x)}{\delta\bar\phi(x)}\frac{\delta J(\bar y)}{\delta\bar\phi(y)}
\frac{\delta^2 J(\bar z)}{\delta\bar\phi(z)\delta\bar\phi(w)}\bigg\}\,d^4\bar x\,
d^4\bar y\, d^4\bar z\,d^4\bar w\,.
\end{eqnarray}
The second integral on the right--hand side cancels the three 1PR connected diagrams generated by the
first integral; these are the analogs of the semiclassical 1PR terms analytically described in Eq
(\ref{delta4WJ}) and graphically sketched in Fig.B1--C.

This procedure can be continued to represent a $\Gamma$--function of any order as a functional combination
of connected Greens's functions of the same order or less, kinetic kernels and functional derivatives
of these, arranged in such a way that all 1PR diagrams cancel out.

This is exactly the reciprocal of the expansion of a connected Greens's function into a tree of 1PI
diagrams connected by complete propagators.

Another important property of $\Gamma[\bar\phi]$ is that it can be Taylor--expanded in series of $\hbar$,
\begin{equation}
\label{barhexpans}
\Gamma[\bar\phi]= \sum^{\infty}_{L=0}\hbar^L\,\Gamma_L[\bar\phi]\,,
\end{equation}
where $\Gamma_L[\bar\phi\,]$ represent 1PI diagrams with $L$ closed loops; $\Gamma_0[\bar\phi]$ is the
sum of all diagrams with no closed loop, $\Gamma_1[\bar\phi]$ is the sum of all diagrams
with one closed loop, etc.

Comparing Eq (\ref{barhexpans}) with Taylor expansion (\ref{Gammaexpansion}), we see that each loop term
$\Gamma_L[\bar\phi]$ is in turn an infinite summations of terms of the form
$\Gamma_L[\bar\phi]=\sum_{n=1}^{\infty} \frac{1}{n!}\,\Gamma_L^{(n)}[\bar\phi]$, where
\begin{equation}
\label{GammaLoopExp}
\Gamma_L^{(n)}[\bar\phi]  =\!\iint\!\!\dots\!\!\int\Gamma^{(n)}_L
(x_1,x_2,\dots, x_n)\,\bar\phi(x_1)\,\bar\phi(x_2)\cdots \bar\phi(x_n)\,d^4x_1\,d^4x_2\cdots d^4x_n\,.
\end{equation}

To prove Eq (\ref{barhexpans}), let us denote as $I$ the number of internal lines, as $V$ the number
of vertices and as $P$ the power of $\hbar$ associated with any given 1PI diagram; then, we have $P=I-V$.
This is because every propagator carries a factor of $\hbar$ and every term of the interaction Lagrangian
density, including that produced by the current $J$, carries a factor of $\hbar^{-1}$. It is important that
there are no propagators attached to the external lines.

On the other hand, the number of loops in a diagram is equal to the number of independent integration
momenta; every internal line contributes one integration momentum, while every vertex contributes a $\delta$
function that reduces the number of independent momenta, except for one $\delta$ function that is left
over for over--all energy--momentum conservation; thus, the number of loops is $L=I-V+1$. Combining this
with the previous result, we obtain the desired result $L=P+1$ (Coleman \& E.Weinberg, 1973).

\subsection{Evaluation of the effective action from the path integral}
\label{PathIntEval}
We have so far represented the QFT of a scalar field $\phi(x)$ in the presence of an external
current $J(x)$ in two different but equivalent ways: (1) as a path integral
\begin{equation}
\label{WJ2calA}
e^{\textstyle\,\frac{i}{\hbar}\,W[J\,]}=
\int {\cal D}\hat\phi\, e^{\textstyle \,\frac{i}{\hbar}
\big\{{\cal A}[\phi_c+\hat\phi]+\int[\phi_c(x)+\hat\phi(x)]J(x)\, d^4x\big\}}
\end{equation}
over the variations $\hat\phi(x)$ of $\phi(x)$ from the classical field $\phi_c(x)$;
(2) as an exponential of the effective action,
\begin{equation}
\label{Gamma2WJ}
e^{\textstyle\,\frac{i}{\hbar}\,\Gamma[\bar\phi\,]} = e^{\textstyle\,\frac{i}{\hbar}\,\big\{W[J\,]
-\int \bar\phi(x)J(x)\,d^4x \big\}}\,,
\end{equation}
functionally dependent on the quantum--field VEV $\bar\phi(x)$.

In the first case, $W[J\,]$ can be expanded in a functional series of connected Green's functions
representing Feynman diagrams with both 1PI and 1PR subdiagrams and $\phi_c(x)$ is regarded as a
functional of $J(x)$. In the second case, $\Gamma[\bar\phi\,]$ can be expanded into a functional
series of terms representing the sum of all 1PI diagrams and $J(x)$ is regarded as a functional
of $\bar\phi(x)$. It is therefore clear that, at least in restricted domains of their respective
functional spaces, the relation between $\phi_c(x)$ and $\bar\phi(x)$ also is one--to--one.

We have also seen in the previous section that, in passing from the expansion of $W[J\,]$ in series of
connected function $G^{(n)}_c$, which dependent functionally on $\phi_c(x)$ through $J(x)$, to the loop
expansion of $\Gamma[\bar\phi\,]$ in series of 1PI $L$--loop diagrams $\hbar^L\Gamma_L[\bar\phi\,]$, which
depend functionally on $\bar\phi(x)$, all 1PR diagrams and subdiagrams of $G^{(n)}_c$ disappear. The problem
then arise of whether is it possible to further expand the connected Green's functions of $W[J\,]$ in series
of the difference $\hat\phi_c(x) = \bar\phi(x)- \phi_c(x)$.

This problem has been first solved by Jackiw in 1974, who showed -- explicitly up to the second order
in $\hbar$ and logically for all subsequent orders -- that in carrying out this expansion all the 1PR
diagrams of $W[J\,]$ cancel out. We introduce here his method for the sole purpose of deriving the exact
expressions of $\Gamma_0[\bar\phi\,]$ and $\Gamma_1[\bar\phi\,]$.

Since the terms of the loop expansion, $\Gamma_L[\bar\phi]$, acquires a factor of $\hbar^L$,
$\Gamma[\bar\phi\,]$ expands as indicated by Eq (\ref{barhexpans}).  We can therefore establish
the equivalence of Eqs (\ref{WJ2calA}) and (\ref{Gamma2WJ}) in the form:
\begin{equation}
\label{GammaWrel}
e^{\textstyle \frac{i}{\hbar}\,\Gamma[\bar\phi] +\!
\frac{i}{\hbar}\int\!\bar\phi(x)J(x)d^4x}\!\!= e^{\textstyle \frac{i}{\hbar}\big\{{\cal A}[\phi_c]+
\int\phi_c(x)J(x)\, d^4x\!+\!W_1[\phi_c]\big\}}e^{\textstyle \frac{i}{\hbar} W_2[\phi_c]},
\end{equation}
where $W_1[\phi_c] =(i\hbar/2)\ln\hbox{Det}\{\Delta_0\Delta^{-1}[\phi_c]\}$ is the
normalized Gaussian term described in \S\,\ref{digressOnlnDet}, near Eq (\ref{lnDetTerm2}),
and
\begin{equation}
\label{W2exp}
e^{\textstyle  \frac{i}{\hbar}  W_2[\phi_c]}= e^{\textstyle \frac{i}{\hbar}{\cal A}_I[\phi_c;
\frac{\delta}{\delta J}]}\!\!\int\!{\cal D}\hat\phi\,e^{\textstyle -\frac{i}{2\hbar}\!\iint\!\!J(y)\,
\Delta[\phi_c;y, x]J(x)\,d^4y\,d^4x}\,.
\end{equation}
Here, ${\cal A}_I[\phi_c;\frac{\delta}{\delta J}]$ is given by Eq (\ref{interactionAction}),
and the double integral in the exponent of the integrand on  the right--hand
is retrieved from Eq (\ref{020amplD}).

Taking the logarithms of both sides of Eqs (\ref{GammaWrel}), we obtain the equality
\begin{equation}
\label{GammaAcomparison}
\Gamma[\bar\phi]+\!\int\!\bar\phi(x)J(x)\,d^4x = W[J\,]\equiv {\cal A}[\phi_c]+\!\int\!\phi_c(x)J(x)\,d^4x
+ W_1[\phi_c]+  W_2[\phi_c]\,.
\end{equation}
The first two terms on the right--hand side of this equation are of order $0$ in $\hbar$, while
$W_1[\phi_c]$ and $W_2[\phi_c]$ are respectively of order $\hbar$ and $\hbar^2$ (hence the subscripts).

If $\hbar$ were zero, we would have $\delta W[J\,]/\delta J(x)= \bar\phi(x) = \phi_c(x)$.
But since $W_1[J\,]$ is of order $\hbar$, the difference $\hat\phi_c(x)= \bar\phi(x) - \phi_c(x)$
also is of this order.  In fact, to the first order in $\hbar$ we have
\begin{equation}
\label{hatphi_c}
\hat\phi_c(x)\!=\!\int\!\!\frac{\delta W_1[\phi_c]}{\delta \phi_c(y)}\,\Delta[\phi_c;y,x]\,d^4y
\equiv\frac{i\hbar}{2}\!\int\!\frac{\delta\ln\hbox{Det}\{\Delta_0\Delta^{-1}[\phi_c]\}}
{\delta \phi_c(y)}\,\Delta[\phi_c;y,x]\,d^4y.
\end{equation}
Generalizing the Jacobi formula $d\ln\hbox{Det}{\bf A}= \hbox{Tr}{\bf A}^{\!-1} d{\bf A}$
to the functional case, we obtain
\begin{equation}
\label{fundlndetc}
\frac{\delta W_1[\phi_c]}{\delta \phi_c(x)}= \frac{i\hbar}{2}\frac{\delta\ln\hbox{Det}
\{\Delta_0\Delta^{-1}[\phi_c]\}}{\delta \phi_c(x)}=\frac{i\hbar}{2}\hbox{Tr}
\bigg\{\Delta[\phi_c]\,\frac{\delta\Delta^{-1}[\phi_c]}{\delta \phi_c(x)}\bigg\}\,,
\end{equation}
which is proportional to $\hbar$; so, Eq (\ref{hatphi_c}) becomes
\begin{equation}
\label{hatphic2DDD}
\hat\phi_c(x) =\frac{i\hbar}{2}\!\int \hbox{Tr}\bigg\{\Delta[\phi_c]\,
\frac{\delta\Delta^{-1}[\phi_c]}{\delta \phi_c(y)}\bigg\}\,\Delta[\phi_c;z,x]\,d^4y\,.
\end{equation}

It is therefore clear that, to prove explicitly equality (\ref{GammaAcomparison}), we
must expand
\begin{equation}
\label{WJexpansion}
W[J\,]={\cal A}[\bar\phi-\hat\phi_c] +\int\![\bar\phi(x)-\hat\phi_c(x)]J[\bar\phi-
\hat\phi_c;\,x]\,d^4x+W_1[\bar\phi-\hat\phi_c]+W_2[\bar\phi-\hat\phi_c]
\end{equation}
in series of $\hat\phi_c(x)$ and compare the coefficients of order $\hbar^n$ with those of the
same orders in the loop expansion of $\Gamma[\bar\phi]+\!\int\!\bar\phi(x)J(x)\,d^4x$, in accordance
with Eq (\ref{barhexpans}).

In particular,  $\Gamma_0[\bar\phi\,]$ must coincide with the term of order zero in $\hat\phi_c(x)$
in the right--hand side of Eq (\ref{WJexpansion}); i.e., the classical action of the
Higgs--boson Lagrangian density (\ref{mulambdaLagDens}) with $\phi_c(x)$ replaced by $\bar\phi(x)$.
We obtain therefore the exact expression
\begin{equation}
\label{Gamma0barphi}
\Gamma_0[\bar\phi] \equiv  {\cal A}[\bar\phi] =\int\Big\{\frac{1}{2}\,\bar\phi(x)
\big(-\Box_x +i\epsilon\big)\bar\phi(x)-\frac{\lambda}{4}\,\big[\,\bar\phi^2(x)- v^2\big]^2+
\bar\phi(x)J(x)\Big\}\,d^4x\,.
\end{equation}

Considering that $W_1[\bar\phi] = W_1[\phi_c+\hat\phi_c; x, y] + O(\hbar^2)$ is of order $\hbar$,
and that $\Gamma_1[\bar\phi\,]$ is the unique term of this order in the effective action, we also
infer the equality
\begin{equation}
\label{barfundlndet}
\Gamma_1[\bar\phi\,] = W_1[\bar\phi] = \frac{i\hbar}{2} \ln\hbox{Det}\{\Delta_0\Delta^{-1}[\bar\phi]\}\,.
\end{equation}
Here, in accordance with Eq (\ref{FeynPropagA}), with the $\hbar$ factor for dimensional consistency,
it is
\begin{equation}
\label{FeynPropag-barphi}
\Delta[\bar\phi; y, x] =\frac{i\,\hbar}{-\Box_y -3\,\lambda\,\bar\phi^2(y)+\lambda\,v^2+i\epsilon}
\,\delta^4(x-y)\,.
\end{equation}

Now, in place of Eqs (\ref{fundlndetc}) and (\ref{hatphic2DDD}) we have the exact expressions
\begin{eqnarray}
\label{deltaGamma1deltabarphi}
&&\hspace{-10mm} \frac{\delta \Gamma_1[\bar\phi\,]}{\delta \bar\phi(x)}=\frac{i\hbar}{2}
\frac{\delta\ln\hbox{Det}\{\Delta_0\Delta^{-1}[\bar\phi]\}}{\delta \bar\phi(x)}=
\frac{i\hbar}{2}\hbox{Tr}\bigg\{\Delta[\bar\phi]\,\frac{\delta\Delta^{-1}[\bar\phi]}
{\delta \bar\phi(x)}\bigg\}\,,\\
\label{barphi2DDD}
&&\hspace{-10mm}\hat\phi_c(x)\!=\!\int\!\frac{\delta \Gamma_1[\bar\phi]}{\delta \bar\phi(y)}\,
\Delta[\bar\phi_c;y,x]\,d^4y=\frac{i\hbar}{2}\!\int \hbox{Tr}\bigg\{\Delta[\bar\phi]\,
\frac{\delta\Delta^{-1}[\bar\phi]}{\delta \bar\phi(y)}\bigg\}\Delta[\bar\phi;y,x]\,d^4y\,,
\end{eqnarray}
which differ from Eqs  (\ref{fundlndetc}) and (\ref{hatphic2DDD}) by terms of order $\hbar^2$. It is
therefore evident that $\Gamma_1[\bar\phi]=0$ and/or $\delta\Gamma_1[\bar\phi]/\delta\bar\phi(x)=0$
entails $\phi_c(x)=\bar\phi(x)$.

\subsection{1--loop terms in multi--field effective Lagrangian densities}
\label{1looptermsgen}
As described in \S\,\ref{digressOnlnDet} near Eqs (\ref{calGm2}) and (\ref{lnDetTerm2}), and on
account of Eqs  (\ref{barfundlndet}), the normalized Gaussian contribution to the path--integral
Lagrangian density, as a function of field VEV $\bar\phi$, coming from a scalar field $\phi(x)$,
has the form
\begin{eqnarray}
\label{FullGamma1barphi}
G\big[m^2(\bar\phi)\big] &=&  \frac{i}{2\,{\cal V}_4}\,\ln\hbox{Det}\frac{\Delta_0\Delta^{-1}[\bar\phi]}
{2\pi i}\equiv\frac{\Gamma_1[\bar\phi]}{{\cal V}_4} = \nonumber\\
&& \hbar\,\bigg[\frac{m^2(\bar\phi)\,\Lambda^2}{32\,\pi^2} -\frac{m^4(\bar\phi)}{64\,\pi^2}\ln\Lambda^2
+\frac{m^4(\bar\phi)\ln m^2(\bar\phi)}{64\,\pi^2}\bigg]\,,
\end{eqnarray}
where ${\cal V}_4$ is the spacetime volume, $\Gamma_1[\bar\phi]$ is the 1--loop term of the effective
action. One might think that the cut--off dependent parts of this expression could be calmly removed by
standard renormalization procedures; but actually this removal would be seriously questionable because the
cut--off dependent terms are present also in the free--field case.

Independently of this incongruence, the real problem with $G\big[m^2(\bar\phi)\big]$ is that this expression
provides an additional contribution to the classical action which distorts rather strongly the potential profile
of the scalar field. For example, for a Higgs field, the minimum of the effective potential may migrate so far
away from that of the classical potential that it becomes impossible to implement the Standard Model of
elementary particles.

The only way to avoid this problem is to take advantage of the fact that the Gaussian terms of fermion
fields bear a negative sign so that, if there are several boson and fermion fields, it may happen that
in certain conditions the sum of all the one--loop terms of all these fields be free from cut-off dependent
terms, if not vanishing.

Since all Gaussian integrals are similarly obtained from the kinetic kernels of bosonic or fermionic
excitations, as shown by Eq (\ref{AGterm}), the normalized 1--loop term for the Lagrangian density of
any quantum field of mass $m$ will have the same general form,
\begin{equation}
\label{GenFullGamma0}
G(m^2)  =  \hbar\,D\bigg(\frac{m^2\Lambda^2}{32\,\pi^2} - \frac{m^4}{64\,\pi^2}\ln\Lambda^2 +
\frac{m^4}{64\,\pi^2}\ln m^2\bigg),
\end{equation}
where $D$ is the dimension of the Gaussian integral: $D=1$ for a massive scalar field;
$D=3$ for a massive vector field.

Extracting from Eqs (\ref{GaussianIntWithJ}), (\ref{020amplD}) the Gaussian integral
for a boson field $\phi(x)$ of mass $m(\bar\phi)$, and denoting the kinetic operator
$\Delta^{-1}[\bar\phi]$ as $-\big[\,\square+m^2(\bar\phi)\big]$, we get
$$
I_B = \bigg[\hbox{Det}\frac{\square+m^2(\bar\phi)+i\epsilon}{\square+i\epsilon}\bigg]^{-1/2}
=e^{\textstyle -\frac{1}{2}\,\hbox{Tr}\ln\frac{\square+m^2(\bar\phi)+i\epsilon}{\square+i\epsilon}}\,,
$$
and for a gauge vector field $V^\mu(x)$,
$$
I_V = \bigg[\hbox{Det}\frac{\square+m^2(\bar V)+i\epsilon}{\square+i\epsilon}\bigg]^{-3/2} =
e^{\textstyle -\frac{3}{2}\,\hbox{Tr}\ln\frac{\square+m^2(\bar V)}{\square+i\epsilon}}\,.
$$

From Appendix {\bf \ref{GrassAlgApp}}, we retrieve the Gaussian integrals for a Dirac field $\nu_D(x)$ of
mass $\mu(\bar\nu_D)$, for a left--or right--handed Majorana neutrino field $\nu_M(x)$ of mass $\mu(\bar\nu_M)$,
and for a hybrid neutrino field $\nu_{MD}(x)$ composed by a Dirac neutrino of mass $\mu(\bar\nu_D)$, a
right--handed Majorana neutrino field $\nu_R(x)$ of mass $\mu(\bar\nu_R)$ and a left--handed neutrino
$\nu_L(x)$ of $\mu(\bar\nu_L)$,
\begin{eqnarray}
&&\hspace{-2mm} I_D =\hbox{Det}\, \bigg[\frac{\square+\mu^2(\bar\nu_D)+i\epsilon}{\square+i\epsilon^2}\bigg]^2 =
e^{\textstyle -\frac{1}{2}\hbox{Tr}\ln\Big[\frac{\square+\mu^2(\bar\nu_D)+i\epsilon}{\square+i\epsilon}
\Big]^{-4}}\!\!\!\!\!;\nonumber\\
&&\hspace{-2mm} I^M =\hbox{Det}\,\frac{\square+ \mu^2(\bar\nu_M)+i\epsilon}{\square+i\epsilon}=
e^{\textstyle -\frac{1}{2}\hbox{Tr}\ln\Big[\frac{\square+ \mu^2(\bar\nu_M)+i\epsilon}{\square+i\epsilon}
\Big]^{-2}}\!\!\!\!\!;\nonumber\\
&&\hspace{-2mm} I^{DM} = \hbox{Det}\bigg[\frac{\big(\square+ m^2_++i\epsilon\big)(\square+ m^2_-+i\epsilon)}
{\big(\square +i\epsilon)^2}\bigg]^2
= e^{\textstyle -\frac{1}{2}\hbox{Tr}\ln\Big[\frac{(\square+ m^2_+ +i\epsilon)(\square+ m^2_- +i\epsilon)}
{(\square+i\epsilon)^2}\Big]^{-4}}\!\!\!; \nonumber
\end{eqnarray}
where
$$
m^2_\pm=\frac{2\,\mu^2(\bar\nu_D)+\mu^2(\bar\nu_L) + \mu^2(\bar\nu_R) \pm\big[\,\mu(\bar\nu_L)
+\mu(\bar\nu_R)\big]\sqrt{4\,\mu^2(\bar\nu_D)+\big[\,\mu(\bar\nu_L)-\mu(\bar\nu_R)\big]^2}}{2}
$$
are the masses of the Dirac--Majorana neutrinos. Thus we obtain $D=-4$ for a Dirac field, $D= -2$
for a Majorana field and $D= -8$ for a Dirac--Majorana field.

From a general methodological standpoint, provided that the number of particle types and
mass parameters is sufficiently large and well--balanced, there is no reason why the condition
for the vanishing of the sum of all 1--loop terms,
\begin{equation}
\label{mathbbG}
{\mathbb G}(\bar\phi) =\hbar \Big[\sum_S G(m^2_S)  + 3\sum_V G(m^2_V) -
 4\sum_F G(m^2_F) - 2\sum_M G(m^2_M)\Big] =0,
\end{equation}
could not be satisfied. In this regard, it is worth noting the paper of Alberghi {\em et al}. (2008), who proved that,
in the framework of the SMEP, Eq (\ref{mathbbG}) can be satisfied provided that at least one fermion term, no matter
whether Dirac or Majorana, is added to the sum. It is therefore evident from Eqs (\ref{deltaGamma1deltabarphi})
and (\ref{barphi2DDD}) that ${\mathbb G}(\bar\phi)=0$ implies $\phi_c(x)=\bar\phi(x)$.

Here is the most important result of this Subsection. In general, the classical potential of a multi--field
theory is destroyed by the addition of the 1--loop terms; just what suffices to invalidate entirely the SMEP,
where the Higgs' field VEV, $v$, is naively determined by minimizing the classical potential
$U(x)= \frac{\lambda}{4} \big[\phi^2_c(x) - v^2\big]^2$ with respect to $\phi_c(x)$. But,
provided that ${\mathbb G}(\bar\phi)=0$, this determination is correct also for the quantized theory.

\subsection{1--loop terms of the effective potential in the general case}
\label{1loopgencase}
Here, we briefly report on a more general and accurate computation of the effective potential, carried out
by Coleman and E.Weinberg in 1973, for a renormalizable field theory which involves a set of real scalar--fields
$\varphi^a(x)$, Yukawa couplings of these fields with a set of fermions $\psi^a(x)$ (not necessarily parity--conserving),
and minimal gauge--invariant interactions of $\vec\varphi(x)$ with a set of vector fields $A_{\mu}^a(x)$. All these
fields are massless and the index $a$ runs over the appropriate range in each case. Sometime we will find it convenient
to assemble the scalar fields into a vector $\vec\varphi(x)$. In the 1--loop approximation, these interactions
contribute additively to the potential $V$ of the effective Lagrangian density. Therefore, we have
$V = V_0+ V_s+ V_f +V_g $, where $ V_0$ is the 0--loop effective potential and the next three terms are
the 1--loop contributions of the mentioned interactions.

\begin{figure}[!h]
\vspace{-2mm}
\centering
\includegraphics[scale=0.8]{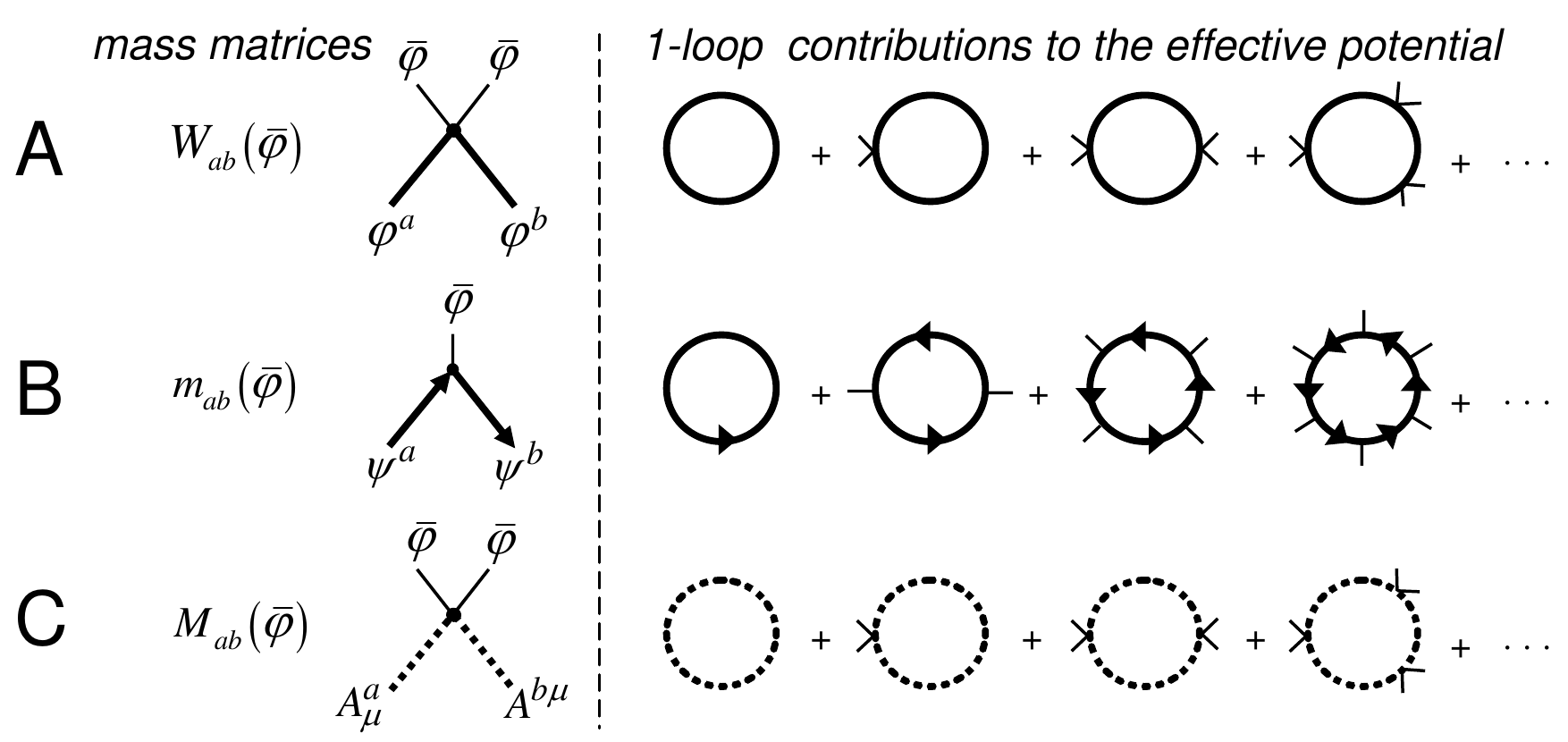} %F19
\vspace{-4mm}
\caption{\small
{\em 1--loop contributions to the effective potential generated by the spontaneous breakdown of a symmetry.}
The Higgs field is defined as $\varphi=\sqrt{\vec\varphi \!\cdot\!\vec\varphi}$ and its VEV is denoted as
$\bar\varphi$. The little bars in the loops represent the vacuum interactions of various
orders in $\bar\varphi$  arising from the mass--matrix graphs depicted on the left. {\bf A}: (left) mass--matrix
of scalar fields; (right) contributions to the effective potential from scalar--field loops. {\bf B}: (left)
mass--matrix of fermions; (right) contributions to the effective potential from fermion loops. {\bf C}:
(left) mass matrix of gauge fields; (right) contributions to the interaction potential from gauge--field loops.}
\label{PathintFig3}
\vspace{-2mm}
\end{figure}

\begin{figure}[!h]
\centering
\includegraphics[scale=0.8]{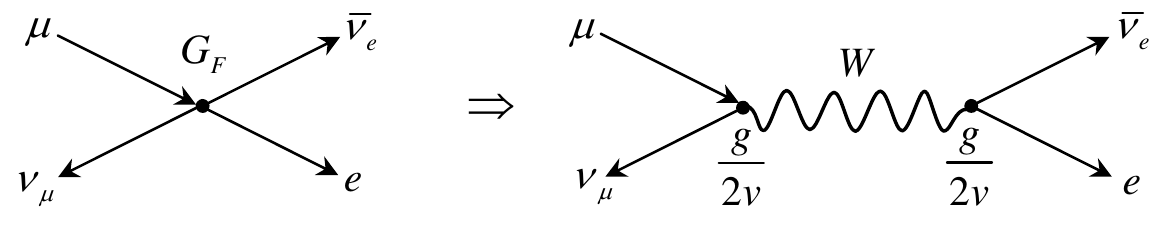}
\caption{\small Replacing Fermi coupling--constant with W gauge--field.}
\vspace{-2mm}
\label{FermiConst}
\end{figure}

If we quantize the theory with the gauge fields in the Landau gauge, the propagators of the gauge fields
have the form $D_{\mu\nu} = -i\,\big[g_{\mu\nu} - k_\nu k_\nu/k^2\big]/(k^2+i\epsilon)$, whence
$D^\mu_\mu = -3/(k^2+i\epsilon)$, and the only graphs we need to consider are those represented in
Fig.\,\ref{PathintFig3}

{1.\em Scalar--field contributions}. The Lagrangian density for a set of scalar fields
$\vec\varphi(x)$, interacting with a set of gauge fields $A^a_\mu(x)$, has the general form
\begin{equation}
\label{calLofs}
{\cal L}_s  = \frac{1}{2}\, g^{\mu\nu}\, \vec\varphi\Big(\overleftarrow{\partial}_\mu
+ig_a\,{\mathbb T}_a A^a_\mu \Big)\Big(\partial_\nu -ig_b\, A^b_\nu {\mathbb T}_b \Big)\vec\varphi-
U(\vec\varphi)\,,
\end{equation}
where ${\mathbb T}_a$ is the (Hermitian) representation of the $a$th infinitesimal transformation
of the gauge group which acts on $\vec \varphi$, $g_a$ is the coupling constant associated to
$A^a_\mu$; if the group is simple all $g$'s are equal; otherwise this is not the case;
\begin{equation}
\label{Uvarphi&mab}
U(\vec\varphi)=\frac{\lambda}{4}\Big(\vec\varphi\!\cdot\!\vec\varphi-v^2\Big)^2\,;
\quad m_{ab}(\vec\varphi) = \frac{\partial^2 U (\vec\varphi)}{\partial{\varphi}^a
\partial{\varphi}^b}\,,
\end{equation}
where $v$ is a constant, are respectively the potential and the mass matrix of the scalar
fields. We can identify $\varphi=\sqrt{\vec\varphi\!\cdot\!\vec\varphi\,}$ with the Higgs
field and $\bar\varphi= \langle\varphi\rangle$ with its VEV.

{2.\em Fermion contributions}. The Yukawa couplings are ruled by the Lagrangian density
\begin{equation}
\label{calLoff}
{\cal L}_f = i\,\bar\psi^a{\slashed\partial}\,\psi^a +\bar\psi^a m_{ab}(\vec\varphi)\,\psi^b\,,
\quad m_{ab}(\vec\varphi)= A_{ab}(\vec\varphi)+ i B_{ab}(\vec\varphi)\,\gamma_5\,,
\end{equation}
where, for $A_{ab}$, $B_{ab}$ and $\gamma_5$, we use Hermitian matrices, and
$m=\big[m_{ab}(\vec\varphi)\big]$ is the fermion--mass matrix. Exploiting the fact that only loops
with an even number of internal fermions contribute to the sum (Fig.3B,
Appendix {\bf \ref{PathIntApp}}), we can group pairwise and condense the terms in matrix product
as follows
$$
\cdots\, m\, \frac{1}{\slashed p}\, m\, \frac{1}{\slashed p}\,\cdots  \equiv
\cdots\, m\, m^{\dag} \frac{1}{p^2}\,\cdots
$$

Thus, for instance, from a loop with $2n$ internal fermions, we get Tr$\big(m m^{\dag}/p^2\big)^n$.
These  can be not the only internal line in a non--Abelian gauge theory, but also ghost fields
of Faddeev and Popov, which in the Landau gauge have not direct coupling with possible scalar
fields of the theory (Coleman \& E.Weinberg, Appendix {\bf \ref{VacDynApp}}; 1973).

{3.\em Gauge--field contributions}. The contributions of the gauge--field loops to the effective potential
may be computed in a similar way. The mass matrix of the gauge fields, ${\mathbb M}^2(\vec\varphi)$,
is provided by the nonderivative couplings of Lagrangian density (\ref{calLofs}),
\begin{equation}
\label{gfieldmassmatr}
{\cal L}_s = \cdots \frac{1}{2} \sum_{ab} \big[{\mathbb M}^2(\vec\varphi)\big]_{ab} A_{\mu}^a  A^{\mu\, b} + \cdots ,
\quad \hbox{where }\, \big[{\mathbb M}^2(\vec\varphi)\big]_{ab}= g_a g_b \big(T_a\,\vec\varphi)
\!\cdot\! (T_b\, \vec\varphi\big).
\end{equation}
% g_a g_b \big(\,\widetilde{\!\vec\varphi}\,T_a T_b\, \vec\varphi\big).

Like $W$, ${\mathbb M}^2$ is a real symmetric matrix and a quadratic function of $\vec\varphi$. We call
this matrix ${\mathbb M}^2$ because the vector fields are minimally coupled to the scalar fields and
${\mathbb M}^2(\vec{\bar\varphi})$ is the squared mass matrix of the gauge field, with the propagators in
Landau gauge.

In summary, the contributions to the effective potential coming from the 1--loop terms, including
the cut--off--dependent terms, depend only on the mass--matrices of the three following types of
interaction
\vspace{-1mm}
\begin{eqnarray}
\label{Vofs-terms}
&&\hspace{-15mm}\hbox{\bf A}\!:\quad V_s = \hbar\,\hbox{Tr}\bigg\{\frac{W(\bar\varphi)\,\Lambda^2}{32\,\pi^2}-
\frac{W^2(\bar\varphi)}{64\,\pi^2}\ln\Lambda^2
+\frac{W^2(\bar\varphi)\ln W(\bar\varphi)}{64\pi^2}\bigg\};\\
\label{Voff-terms}
&&\hspace{-15mm}\hbox{\bf B}\!:\quad V_f = -\hbar\,\hbox{Tr}\bigg\{\frac{ m m^\dag (\bar\varphi)\,\Lambda^2}{32\,\pi^2} -
\frac{\big[m m^\dag (\bar\varphi)\big]^2}{64\,\pi^2}\ln\Lambda^2 +
\frac{\big[m m^\dag (\bar\varphi)\big]^2
\ln\, m m^\dag (\bar\varphi)}{64\pi^2}\bigg\};\\
\label{Vofg-terms}
&&\hspace{-15mm}\hbox{\bf C}\!:\quad V_g =  3\,\hbar\,\hbox{Tr}
\bigg\{\frac{{\mathbb M}^2(\bar\varphi)\,\Lambda^2}{32\,\pi^2}-
\frac{{\mathbb M}^4(\bar\varphi)}{64\,\pi^2}\ln\Lambda^2+
\frac{{\mathbb M}^4(\bar\varphi)\ln {\mathbb M}^2
(\bar\varphi))}{64\pi^2} \bigg\}.
\vspace{-1mm}
\end{eqnarray}

Note that the fermion contribution $V_f$ has a sign which is opposite to that of
all other terms, and that factor 3 in the gauge--field contribution $V_g$
comes from the trace of the numerator of gauge--field propagators.

For the reasons explained in \S\S\,\ref{AsympConfInvinCGR}, \ref{mainproblems},
\ref{ShiftInvCGR} of the main text and in \S\S\,\ref{1looptermsgen} and
\ref{1loopgencase} of this Appendix, in order for the SMEP to survive
quantization, the total 1--loop term $\Gamma_1(\bar\varphi)$ of the effective
action must vanish, i.e., it must be $V_s+V_f +V_g=0$. This clearly requires
that the bosonic and fermionic mass--terms of different orders of magnitude,
appearing in Eqs (\ref{Vofs-terms})--(\ref{Vofg-terms}), be perfectly balanced
for any value of momentum cut--off $\Lambda$.

\subsection{The self--coupling constant of the Higgs boson field}
\label{Fermicoupling}
The vanishing of $\Gamma_1(\bar\varphi)$ entails two important facts:
the mass spectrum of the SMEP comes from the spontaneous breakdown of
conformal symmetry (see \S\ref{AsympConfInvinCGR}) and the classical limit
of the path integral is preserved (see \S\,\ref{ShiftInvCGR}). In
particular, the effective potential of the Higgs field $\varphi(x)$
equals the classical potential, as naively assumed in some approaches
to the SMEP. This circumstance allows us to establish an important
relation between the Fermi coupling--constant of the weak currents
$G_F \cong 1.16637\times 10^{-5}$ GeV$^{-2}$ and the self--coupling
constant $\lambda$ of the Higgs boson field,
namely  $\lambda=\mu_H^2 G_F/\sqrt{2}$ via the VEV of the Higgs
field \cite{WEINBERG67}
$v= 2^{-1/4}G_F^{-1/2} \cong 246$ GeV.

\newpage

%%%%%%%%%%%%%%%%%%%%%%%%%%%%%%%%%%%%%%%%%%%%%%%%%%%%%%
\markright{R.Nobili, Conformal General Relativity - {\bf \ref{ThermVacApp}} Thermal vacua}
\renewcommand\thefigure{\Alph{section}\arabic{figure}}
\setcounter{figure}{0}
\section{BRIEF INTRODUCTION TO THERMAL VACUA}
\label{ThermVacApp}
\label{onthermalvacua}
The connection between thermodynamics and quantum field theory has been investigated by several authors
since the early 1960s  \cite{ARAKI} \cite{KUBO} \cite{HAAG}. The entire subject is rooted in the theory of
infinite direct--products of quantum--field representations over a continuum of vacuum states
\cite{VNEUMANN} \cite{BRATTELI}. In this view, quantum fields are regarded as unitarily inequivalent
representations of purely algebraic entities called fundamental fields (Umezawa {\em et al.}, 1982), which
differ from each other in the VEVs of one or more scalar fields. In this view, physical particles are
regarded as quantum excitations of a particular vacuum state. The most familiar vacuum state is that of
the Fock representation, which is characterized by zero densities of particles and unphysical zero temperature
of the system.

But to achieve the full control of this complex matter we must elevate our view over the realm of non--separable
Hilbert spaces, which, for example, allows us to include, in the input--output states of the $S$ matrix, coherent
swarms of infrared photons (Kibble, 1968). Since the unitarily inequivalent representations of a non--separable
Hilbert space form a continuum of mutually orthogonal spaces, each of which has its own fundamental state,
this higher level of mathematical complexity is suitable for describing the classical limit of the macroscopic
world and even the irreversible process of its continuous  evolution.

Although unitarily inequivalent, two representations may be mutually related by an {\em algebraic map} ${\cal U}_B$,
called Bogoliubov transformations \cite{BOGOLIUBOV} -- generally depending on one, several or even infinite
parameters $\theta$ -- which preserves the canonical commutation relations of the fundamental fields and
can be formally manipulated as a unitary operator.

Any ${\cal U}_B$ can be viewed in two ways: (1) {\em \`a la} Heisenberg, as an invertible map
between of a set of bounded operators $X$, algebraically constructed  from the fundamental fields
represented in a Hilbert space $\cal H$, and a set of bounded operators $X'$, represented, in a
non--unitarily equivalent way, in the same Hilbert space. This is formally represented as $X\rightarrow
X'= {\cal U}_B X\,{\cal U}_B^{-1}$; or (2) {\em \`a la} Schr\"odinger, by replacing the vacuum
state $|\Omega\rangle$ of $ {\cal H}$ with the vacuum state $|\Omega' \rangle$ of a second Hilbert
space ${\cal H}'$. In this case, we shall write $|\Omega' \rangle = {\cal U}_B^{-1}|\Omega\rangle$.
The two modes are clearly equivalent because $\langle \Omega| X'|\Omega\rangle = \langle \Omega| {\cal U}_B\, X\,
{\cal U}_B^{-1} |\Omega\rangle =\langle \Omega'| X |\Omega'\rangle$.

The simplest example of Bogoliubov transformations is formally defined by
\begin{equation}
\label{lingen}{\cal U}(\theta) = e^{iG(\theta)}\,,\quad \mbox{where }\, G(\theta)=
- i\sum_k\theta[a(k) - a^\dag(k)]\,,
\end{equation}
which maps the annihilation--creation operators $a(k), a^\dag(k)$ of a fundamental scalar field,
represented in a Fock space with vacuum state $|\Omega\rangle$, into the representation
\begin{equation}
a'(k) = {\cal U}(\theta)\,a(k)\,{\cal U}^\dag(\theta) = a(k) + \theta\,, \quad
a'^{\,\dag}(k) = {\cal U}(\theta)\,a^\dag(k)\,{\cal U}^\dag(\theta) = a^\dag(k) + \theta\,,\nonumber
\end{equation}
of the same fundamental field in a second Fock space with vacuum state $|\Omega'\rangle =
{\cal U}(\theta)|\Omega\rangle$. Hence ${\cal U}(\theta)$ performs a simple translation
of the boson field amplitude.

Denoting by $N(k) = a(k)\,a^\dag(k)$ and $N'(k) = a'(k)\,a'^{\,\dag}(k)$ the particle--number
operators, respectively in the first and second representations, we can easily verify equations
$\langle \Omega| N(k) |\Omega\rangle = \langle \Omega'| N'(k)|\Omega'\rangle =0$ and $\langle \Omega| N'(k)
|\Omega\rangle = |\theta|^2$.  Since ${\cal U}(\theta)$ changes the particle--number 0 into $|\theta|^2$
without modifying the spectrum of the Hamiltonian, it may be interpreted as an adiabatic transformation
at zero temperature. Therefore, the thermal properties of $|\Omega\rangle$ and $|\Omega'\rangle$ are trivial.

\begin{figure}[!h]
\centering
\includegraphics[scale=0.8]{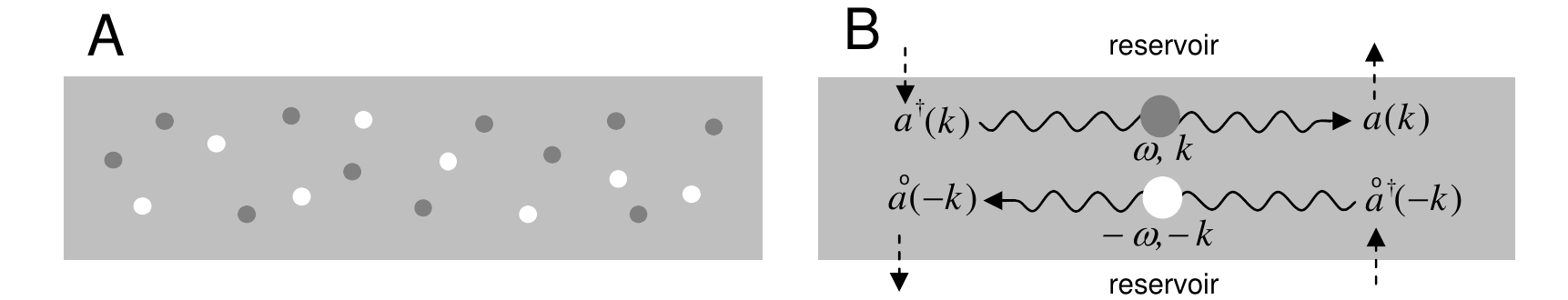} %F20
\caption{\small {\bf A}: Thermal vacuum as an incoherent superposition of
particles (dark spots) and holes (white spots). {\bf B}: Exchange of thermal
quanta with reservoir occurs in two modes: 1) by creation and annihilation
of particles of energy--momentum $\{\omega, k\}$, respectively represented
by operators $a^\dag(k)$ and $a(k)$; 2) by annihilation and creation of holes
of energy--momentum $\{-\omega, -k\}$, respectively represented by operators
$\mathring{a}(-k)$ and $\mathring{a}^\dag(-k)$. Both modes result in same
amount of energy--momentum exchanged with reservoir. Since particles and
holes are independent degrees of freedom, all $\mathring{a}(-k)$ and
$\mathring{a}^\dag(-k)$ commute with all $a(k)$ and $a^\dag(k)$.
Simultaneous creations or annihilations of particles and holes of
opposite energy--momentum represent thermal fluctuations.}
\label{ThermVacFig1}
\end{figure}

Vacuum states with non--trivial thermal properties are called
thermal vacua. These are characterized by the unboundedness from
below of the number of possible quantum annihilations. Thus, in
order for a thermal vacuum to be a cyclic state, the fundamental--field
representation needs a twofold number of degrees of freedom
(Araki \& Woods, 1963): one representing "positive" thermal
excitations -- say {\em particles} -- the other representing
"negative" thermal excitations -- say {\em particle holes}. For
instance, the state of an empty box immersed in a thermal reservoir
of temperature $T$ is of this sort (Fig.\,\ref{ThermVacFig1}).

If a number of particles of energy--momentum  $(\omega, k)$ and an equal number of holes of
energy--momentum $(-\omega, -k)$ are simultaneously created or annihilated, the energy--momentum
exchanged between system and reservoir is zero. We can regard these zero--sum processes as
internal fluctuations of the thermal vacuum \cite{MANN}. Since these are unobservable,
the Heisenberg indetermination relations of the matter fields appear to be affected by an
additional entropic indetermination representing {\em thermal noise} with Gaussian standard
deviation of both field amplitudes and their time--derivatives (Umezawa, 1993).

On this basis, the thermal vacuum of an infinite system can be ideally obtained by expanding the volume
of the box to infinity. Since, at this limit, the reservoir disappears, the vacuum itself must be
regarded as its own reservoir. In this case, the thermal fluctuations are more appropriately described
as quantum fluctuations of a mixture of virtual particles and holes. In the following, we only refer to
infinite systems.

It is intuitive that the ratio between hole density and particle density  varies with temperature
and approach zero as $T\rightarrow 0$. If this limit could be reached, all holes would disappear,
which is impossible, in accordance with the third principle of thermodynamics.

Let $a^\dag(k)$, $a(k)$ respectively be the creation and annihilation operators of a boson of energy--momentum
$\omega, k$, and $\mathring{a}^\dag(-k)$, $\mathring{a}(-k)$ respectively be the creation and annihilation
operators of a boson--hole of energy--momentum $-\omega, -k$. Since particles and holes are independent
degrees of freedom, all $\mathring{a}(-k)$, $\mathring{a}^\dag(-k)$ commute with all $a(k)$, $a^\dag(k)$.
Therefore, as far as energy--momentum balance is concerned, the actions of $a(k)$ and $\mathring{a}(-k)$
produce the same effects. It is therefore natural to introduce, as creation and annihilation operators of
thermal fluctuations, linear combinations
\begin{equation}
\label{parthole}
a(k, T)={\cal C}(k, T)\,a(k) +{\cal S}(k,T)\,\mathring{a}^\dag(-k);\,\,\, a^\dag(k, T) =
{\cal C}(k, T)\, a^\dag(k) +{\cal S}(k,T)\,\mathring{a}(-k),
\end{equation}
where ${\cal C}(k, T), {\cal S}(k, T)$ are real and positive coefficients. This is  because possible phase
factors can be canceled by a redefinition of $a(k)$ and  $\mathring{a}^\dag(-k)$.

The requirement that $a(k, T), a^\dag(k, T)$ should satisfy the canonical commutation relations (c.c.r.)
$[a(k, T), a^\dag(k', T)] = \delta^3(k-k')$ leads to equations ${\cal C}(k, T)^2 - {\cal S}(k, T)^2 = 1$.
Eqs  (\ref{parthole}) can be written as $a(k, T) = {\cal U}_{\,T}\,a(k)\,{\cal U}_{\,T}^{-1}$, $a^\dag(k, T) =
{\cal U}_{\,T}\,a^\dag(k)\,{\cal U}_{\,T}^{-1}$, as if the Bogoliubov operator were unitary, by defining
\begin{equation}
\label{quadgen} {\cal U}_{\,T}= e^{i\,G_T},\,\, \mbox{with }\, G_T=
- i\sum_k {\cal S}(k, T) \big[\mathring{a}(-k)\,a(k)  - \mathring{a}^\dag(-k)\,a^\dag(k)\big]\,.
\end{equation}

Denoting by $N(k)=a(k)\,a^\dag(k)$ and $\mathring{N}(-k)=\mathring{a}(-k)\,\mathring{a}^\dag(k)$,
respectively, the number operators of particles and holes of momentum $k$ in the
Fock space representation, we find $[G_T, N(k)] =[G_T, \mathring{N}(-k)]$, showing that
$N(k) - \mathring{N}(-k)$ are the invariants of ${\cal U}_{\,T}$.

Let us denote by $|\Omega_F \rangle$ the Fock vacuum state of $a(k), a^\dag(k), \mathring{a}(-k),
\mathring{a}^\dag(-k)$, the thermal vacuum as $|\Omega_T \rangle\! =\!{\cal U}_{\,T}^{-1}|\Omega_F\rangle$
and the number of thermal excitations of momentum $k$ in the Fock representation as
$N(k,T) = a(k,T)\,a^\dag(k, T)$. We thus have $a(k)|\Omega_F \rangle\! =\!\mathring{a}(-k)|\Omega_F \rangle
\!=\!a(k, T)|\Omega_T \rangle\!=\!\mathring{a}(-k,T)|\Omega_T \rangle\!=0$; hence $N(k)|\Omega_F \rangle\!=\!
\mathring{N}(-k)|\Omega_F \rangle\!=\!N(k,T)|\Omega_T \rangle=0$.

Developing ${\cal U}_{\,T}$ in series of powers of $G_T$ and rearranging the terms by repeated commutations
\cite{UMEZAWA2}, we can prove the equation
$$
|\Omega_T \rangle = {\cal U}_{\,T}^{-1}|\Omega_F\rangle =
\sum_{n,k}\frac{{\cal S}(k,T)^n}{n!\,\exp[\,\ln\cosh {\cal S}(k,T)]}
\big[\mathring a^\dag(-k)\,a^\dag(k)\big]^n |\Omega_F \rangle\,,
$$
showing that, in the Fock--space representation, the thermal vacuum is a quantum--entan\-gled superposition
of particle--hole pairs of zero energy and zero momentum fluctuations.

For each operator $X$ in the algebra of $\{a(k), a^\dag(k)$, $\mathring{a}(-k), \mathring{a}^\dag(-k)\}$,
there is an operator $X(T)= {\cal U}[\theta]\,X\,{\cal U}[\theta]^{-1}$ in the algebra of $\{a(k,T),
a^\dag(k,T)$, $\mathring{a}(-k,T), \mathring{a}^\dag(-k,T)\}$, which satisfies equation $\langle\Omega_T|
X |\Omega_T \rangle = \langle \Omega_F|X(T)|\Omega_F \rangle$. In particular, we have $\langle \Omega_T|N(k)|
\Omega_T \rangle =\langle \Omega_F|N(k,T)|\Omega_F \rangle = {\cal S}(k,T)^2  V$, where $V =
(2\pi)^3\delta^3(0)= \int\! e^{ikx}|_{k=0}d^3x$ is the space volume.

Since, in accordance with Bose--Einstein's statistics, particle density $n(k)=N(k)/V$ at thermal equilibrium is
$\langle \Omega_T |n(k)|\Omega_T \rangle =[e^{\omega(k)/T} -1]^{-1}$,
we find for Eqs (\ref{parthole})
$$
{\cal S}(k, T) = \frac{1}{\sqrt{e^{\omega(k)/T} -1}}\,,\quad  {\cal C}(k, T) =\sqrt{1+{\cal S}(k,T)^2}=
\frac{e^{\omega(k)/2\,T}}{\sqrt{e^{\omega(k)/T} -1}}\,.
$$
We thereby realize that ${\cal U}_{\,T}$ makes a boson field in the Fock--space at temperature $T=0$,
jump to a boson gas at temperature $T>0$ in the space of particle--hole representation.
Similar results are obtained for fermionic particles and holes, in which case the coefficients
are ${\cal S}(k, T)= 1/\sqrt{e^{\omega(k)/T} +1},\,\, {\cal C}(k, T)= e^{\omega(k)/2T}/\sqrt{e^{\omega(k)/T} +1}$.

In other terms, ${\cal U}_{\,T}^{-1}$ makes the zero--temperature vacuum of the Fock representation jump to a
thermal vacuum at temperature $T$, leaving formally unvaried the algebra of the fundamental fields. It is also
possible to build a thermal Bogoliubov map, ${\cal U}_{\,T}(t)$, that depends on time $t$. In this way, it would
then be possible to represent a continuous thermal evolution of the vacuum state. When applied to the
fundamental state of an initially empty system, ${\cal U}_{\,T}(t)$, would be able to generate a gas that
remains in thermodynamic equilibrium at a continuously varying temperature (Umezawa, 1993).

\newpage

\markright{R.Nobili, Conformal General Relativity - {\bf \ref{DirMajorApp}} Majorana neutrinos}
\renewcommand\thefigure{\Alph{section}\arabic{figure}}
\setcounter{figure}{0}
\section{DIRAC AND MAJORANA NEUTRINOS}
\label{DirMajorApp}
For several decades, for lack of experimental evidence of right--handed neutrinos,
it was often believed that these important partners of charged leptons were
massless and perhaps of the Weyl type. The discovery of neutrino oscillations
\cite{SUPERKAMIOKANDE} showed that the left--handed neutrinos differ in mass
by less than 1 eV, thus proving they are fermions of the Dirac type. So, each
of them can be partitioned in four states: 1) a left--handed neutrino $\nu_L(x)$,
only interacting with the left--handed component of its charged partner; 2) its
antiparticle $\bar \nu_R(x)$; 3) a right--handed neutrino $\nu_R(x)$ not
interacting with a charged lepton, therefore called {\em sterile}; 4) its
antiparticle $\bar \nu_R(x)$. However, other sterile neutrinos may exist, in
particular those of the Majorana type, which are the subject of this Appendix.

To fix notations and conventions, let us consider the free--field Lagrangian density ${\cal L} =
\bar\psi\big(i\slashed{\partial}-m\big)\psi$ of a Dirac field $\psi(x)$, and its adjoint $\bar\psi(x)$,
in a Minkowski spacetime of metric signature $\{1, -1,-1,-1\}$. Null variation with respect to
$\bar\psi(x)$ yields motion equation $\big(i\slashed{\partial} -m\big)\psi\equiv
\big(i\gamma^\mu\partial_\mu-m\big)\psi=0$ and null variation with respect to $\psi(x)$ yields the
adjoint equation $\psi^\dag\big(i\overset{\leftarrow}{\partial}_\mu \gamma^\mu+m\big)=0$, where
$\gamma^\mu$ are the gamma matrices in the standard Pauli--Dirac (PD) representation. The solutions
to these equations can be written
as \cite{BJORKEN64}
\begin{eqnarray}
\label{Diracpsi}
&& \hspace{-10mm}\psi(x) =  \int\frac{d^3\vec p}{(2\pi)^{3/2}}\sqrt{\frac{m}{E_p}}\sum_{s=\pm1}\!\!
\big[u_s(\vec p\,)\,a_s(\vec p\,)\,e^{-i p\cdot x} +v_s(\vec p\,)\,b^\dag_s(\vec p\,)
\,e^{i p\cdot x}\big],\\
&&
\label{Diracpsidag}
\hspace{-10mm}\bar\psi(x) = \int\frac{d^3\vec p}{(2\pi)^{3/2}}\sqrt{\frac{m}{E_p}} \sum_{s=\pm1}\!\!
\big[\bar u_s(\vec p\,)\,a^\dag_s(\vec p\,)\,e^{-i p\cdot x} +\bar v_s(\vec p\,)\,b_s(\vec p\,)
\,e^{i p\cdot x}\big]\,,
\end{eqnarray}
where $E_p\equiv \sqrt{m^2+|\vec p\,|^2}$, while $a^{\dag}_s(\vec p\,)$, $a_s(\vec p\,)$ are respectively
the creation and annihilation operators of particles with momentum $\vec p\,$, and $z$--axis
spin projections $s =\pm 1/2$, while $b^{\dag}_s(\vec p\,)$, $b_s(\vec p\,)$ are those of
antiparticles. From the canonical anticommutation relations
\begin{eqnarray}
& & \hspace{-10mm}\big\{a_s(\vec p), a^\dag_{s'}({\vec p\,}')\big\} = \big\{b_s(\vec p),
b^\dag_{s'}({\vec p\,}')\big\} =\delta^3(\vec p- {\vec p\,}')\,\delta_{s s'}  \,;\nonumber\\
& & \hspace{-10mm}\big\{a_s(\vec p), a_{s'}({\vec p\,}')\big\} = \big\{b_s(\vec p),
b_{s'}({\vec p\,}')\big\} =\big\{a_s(\vec p), a_{s'}({\vec p\,}')\big\}^\dag =
\big\{b_s(\vec p), b_{s'}({\vec p\,}')\big\}^\dag=0\,;\nonumber\\
& & \hspace{-10mm}\big\{\psi(x), \psi^\dag(x')\big\}\,\delta(x^0-{x^0}')=\delta^4(x - x')\,,
\quad \big\{\psi(x),\psi(x')\big\}\,\delta(x^0 - {x^0}')= 0\,, \quad\hbox{etc};\nonumber
\end{eqnarray}
and Lorentz--group representations for spinors, we derive the normalization conditions and momentum--space
equations for matrices $u_s(p)$, $v_s(p)$, $\bar u_s(p)\equiv u^\dag_s(p)\,\gamma^0$,
$\bar v_s(p)\equiv v^\dag_s(p)\,\gamma^0$,
\begin{eqnarray}
\label{BARnormalconds}
& & \hspace{-16mm}\bar u_s(\vec p\,)\,u_{s'}(\vec p\,) = - \bar v_s(\vec p\,)\,v_{s'}(\vec p\,) = \delta_{s s'}\,;
\quad u ^\dag_s(\vec p\,)\,u_{s'}(\vec p\,) = v^\dag_s(\vec p\,)\,v_{s'}(\vec p\,) = \frac{E_p}{m}\,\delta_{s s'},\\
\label{momentumequats}
& & \hspace{-16mm}(\slashed p -m)\, u_s(\vec p\,) =\bar u_s(\vec p\,)\,(\slashed p -m) =0,\quad
(\slashed p+m)\,v_{s'}(\vec p\,) = \bar v_s(\vec p\,)\,(\slashed p+m) = 0.
\end{eqnarray}

At variance with Dirac neutrinos, Majorana neutrinos exist in two distinct {\em elicities}, or {\em chiralities},
and coincide with their own antiparticles, or {\em conjugate} particles. Let us recall how the
chiral form, $\Psi$, and the conjugate form, $\psi^c$, of a Dirac
field $\psi$ can be determined.

The chiral form can  easily be obtained by decomposing first $\psi$ into its left--handed and right--handed
components, respectively $\psi_L= P_L\psi$ and $\psi_R= P_R\psi$, by the chiral projectors $P_L= \frac{1}{2}
\,(1- \gamma^5)$ and $P_R= \frac{1}{2}\,(1+ \gamma^5)$, where $\gamma^5 =i\gamma^0\gamma^1\gamma^2\gamma^3$.
Since $\gamma^5\gamma^\mu = -\gamma^\mu\gamma^5$, we obtain the relationship $P_R\gamma^\mu = \gamma^\mu P_L$
follows. We can then convert the PD representation of gamma matrices $\gamma^\mu$ to their chiral form
by means of the involutory transformation
\vspace{-1mm}
\begin{equation}
\label{chiralgammas}
T\!=\!\frac{1}{\sqrt{2}}\!
\begin{bmatrix}
I & I \\
I &\!\! - I
\end{bmatrix}\!\!,\,\,\,
\hbox{yielding }\,
T\gamma^\mu T \!=\!
\begin{bmatrix}
0 & \sigma^\mu \\
\bar \sigma^\mu & 0
\end{bmatrix}\,\,\hbox{with }
\sigma^\mu\!\!= \{I, -\vec{\boldsymbol\sigma}\},\,
\bar\sigma^\mu\!\!=\{I,\vec{\boldsymbol\sigma}\}.
\vspace{-1mm}
\end{equation}
$I$, $0$ and $\vec {\boldsymbol\sigma}\equiv\big\{\sigma^1, \sigma^2, \sigma^3 \big\}$ stand respectively
for the $2\times2$ unit, zero and Pauli  matrices. We can then derive the following chiral representation
for $\gamma^5$, $P_L$, $P_R$ and $\gamma^0$
\begin{equation}
\label{chiralprojs}
T\gamma^5 T =
\begin{bmatrix}
-I & 0 \\
0 & I
\end{bmatrix}\!,\quad
T P_L T \!=\!
\begin{bmatrix}
I &  0 \\
0 & 0
\end{bmatrix}\!,\quad
T P_R T \!=\!
\begin{bmatrix}
0 & 0 \\
0 & I
\end{bmatrix}\!,\quad
T\gamma^0 T =
\begin{bmatrix}
0 & I \\
I & 0
\end{bmatrix}\!,
\end{equation}
which allow us to represent $\psi$ and its Dirac adjoint $\bar\psi$ in the chiral form
\begin{eqnarray}
\label{chiralcompsA}
& &\hspace{-16mm}
\Psi = T\psi=
\begin{bmatrix}
\Psi_L \\
\Psi_R
\end{bmatrix}\!,\quad \overline\Psi =
\bar\psi\,T= \psi^{\dag}T\,(T\gamma^0 T) = \Big[\Psi^\dag_L,\Psi_R^\dag\Big]\!
\begin{bmatrix}
0 & I \\
I & 0
\end{bmatrix}\! =\Big[\Psi^\dag_R,\Psi_L^\dag\Big],\\
\label{chiralcompsB}
&&\hspace{-16mm}
T\psi_L =
\begin{bmatrix}
\Psi_L \\
0
\end{bmatrix}\!,\quad
T\psi_R=
\begin{bmatrix}
0 \\
\Psi_R
\end{bmatrix}\!,
\quad
\bar\psi_L T= \Big[0,\,\Psi^\dag_L\Big]\!,\quad  \bar\psi_R T = \Big[\Psi^\dag_R,\,0\Big]\!.
\end{eqnarray}

If $\psi$ is represents a massless neutrino, $\Psi_L$ and $\Psi_R$ can be regarded as two
independent Weyl spinors of opposite elicities and respective Lagrangian densities
\begin{equation}
\label{WeylLagrs}
\mathcal{L}_{L}^{\hbox{\tiny Weyl}}  =
\Psi_L^\dag i\bar\sigma^\mu\partial_\mu\Psi_L\,;\quad \mathcal{L}_{R}^{\hbox{\tiny Weyl}}  =
\Psi_R^\dag i\sigma^\mu\partial_\mu\Psi_R\,.
\end{equation}

In spacetime coordinates, these satisfy respectively motion equations $\partial_0 \Psi_L = \vec\sigma\cdot
\vec\partial\, \Psi_L$ and $\partial_0 \Psi_R = - \vec\sigma\cdot \vec\partial\, \Psi_R$; or, in the
momentum space, $\vec\sigma\!\cdot\!\vec p\,/|\vec p\,| = 1$ and $\vec\sigma\!\cdot\!\vec p\,/|\vec p\,|
= -1$.

If the neutrino is massive, ${\cal L} = \bar\psi\big(i\slashed{\partial}-m\big)\psi$ can be expressed
in the chiral form
\begin{equation}
\label{chiralDirac}
{\mathcal L} = \mathcal{L}_{L}^{\hbox{\tiny Weyl}} + \mathcal{L}_{R}^{\hbox{\tiny Weyl}} -
m\,\bar\psi\,\psi =\Psi_L^\dag i\bar\sigma^\mu\partial_\mu\Psi_L +
\Psi_R^\dag i\sigma^\mu\partial_\mu\Psi_R - m\big(\Psi_R^\dag\Psi_L+ \Psi_L^\dag\Psi_R\big),
\end{equation}
i.e., as the Lagrangian density of two Weyl neutrinos of opposite chiralities coupled by the Dirac
mass term. In fact, we have $\bar\psi\,\psi = (\psi^\dag T) (T\gamma^0 T) (T\psi) =
\Psi_R^\dag\Psi_L+ \Psi_L^\dag\Psi_R$.

From elementary quantum mechanics, we know that the antiparticle--conjugate $\psi^{\,c}$ of a spinor
field $\psi$ is related to its Dirac adjoint $\bar\psi$ by equations
\begin{equation}
\label{psictopsi*}
\psi^{\,c} =C\,\widetilde{\!\bar\psi}=
C \gamma^0\psi^*,\quad\bar\psi^c = -\tilde\psi\, C^{-1}\,,
\end{equation}
where the tilde superscription denotes transposition. The $4\times 4$ matrix $C$
is determined, up to an arbitrary phase factor $e^{i\theta}$, by requiring that $\psi^{\,c}$
behaves like $\psi$ under Lorentz transformations. The usual choice is $C=i\,\gamma^2\gamma^0 =
-\,C^{-1} = - C^\dag = -\,\widetilde{C}$.

By expressing $\bar\psi$ and $\psi$ as functions of $\psi^c$ and $\bar\psi^c$, Lagrangian density
${\cal L} =\bar\psi\big(i\slashed{\partial}-m\big)\psi$ turns into ${\cal L}^c = \widetilde{\cal L}
\equiv {\cal L}$ up to a surface term and an anti--commutation. Thus we have
\begin{equation}
\label{PDCConj}
{\cal L}  = \bar\psi\,i\slashed{\partial}\,\psi -m\,\bar\psi\,\psi =
\bar\psi^{\,c}\,i\slashed{\partial}\,\psi^{\,c}- m\,\bar\psi^{\,c} \psi^{\,c}.
\end{equation}
%$\binom{\,\,0 \,\,\,\, 1\,}{-1\,\,\, 0\,}$
To obtain the second line of this equation, it is suitable to represent $C$ in its chiral form
\begin{equation}
\label{chiralC}
T C T = \,T\!\!
\begin{bmatrix}
 0 & i\sigma^2 \\
-i\sigma^2 &0
\end{bmatrix}\!
\begin{bmatrix}
 I & 0 \\
0 & -I
\end{bmatrix}\! T =
\begin{bmatrix}
0 & -i\sigma^2 \\
-i\sigma^2 & 0
\end{bmatrix}\!,
\,\,\hbox{where}
-i\sigma^2\! =
\begin{bmatrix}
0 & 1 \\
-1 & 0
\end{bmatrix}\!,
\end{equation}
by mean of which we can express Eqs (\ref{chiralDirac}) and (\ref{PDCConj}) in the equivalent ways
\begin{eqnarray}
\label{ChiralLagdens}
\hspace{-15mm}{\cal L}\!\!&=&\!\!\bar\psi\, i\slashed{\partial}\,\psi -m\,\bar\psi\,\psi =
\Psi_L^\dag i\bar\sigma^\mu\partial_\mu\Psi_L +
\Psi_R^\dag i\sigma^\mu\partial_\mu\Psi_R - m \,\big(\Psi_R^\dag\Psi_L+ \Psi_L^\dag\Psi_R\big) =
\nonumber \\
\hspace{-15mm}
\!\!& &\!\! \bar\psi^c i\slashed{\partial}\,\psi^c -m\,\bar\psi^c\psi^c = \Psi_L^{c\dag}
i\bar\sigma^\mu\partial_\mu\Psi^c_L +\Psi_R^{c\dag} i\sigma^\mu\partial_\mu\Psi^c_R
- m \,\big(\Psi_R^{c\dag}\Psi^c_L+ \Psi_L^{c\dag}
\Psi^c_{R}\big).
\end{eqnarray}
Using the first of (\ref{psictopsi*}), we can express the chiral components of $\psi^{\,c}$
in the chiral form
\begin{equation}
\label{RLflip}
T\psi^c=\!
\begin{bmatrix}
\Psi^{\,c}_L\\
\Psi^{\,c}_R
\end{bmatrix}\! =
T C\gamma^0 T^2 \psi =
\!
\begin{bmatrix}
0 &\!\! -i\sigma^2 \\
-i\sigma^2 &\!\! 0
\end{bmatrix}\!\!
\begin{bmatrix}
\Psi^*_L \\
\Psi^*_R
\end{bmatrix}\!=\!
\begin{bmatrix}
-i\sigma^2 \Psi^*_R \\
-i\sigma^2\Psi^*_L
\end{bmatrix}\!,
\end{equation}
i.e., $\Psi^{\,c}_L = -i\sigma^2 \Psi^*_R$ and $\Psi^{\,c}_R = -i\sigma^2 \Psi^*_L$.
The flip of chirality is due to the fact that the relation between chirality and
helicity is reversed for antiparticles.

Separating the chiral components of Eq (\ref{ChiralLagdens}), we can easily realize that
the kinetic terms of Lagrangian densities (\ref{WeylLagrs}) are
respectively equivalent to
\begin{equation}
\label{LKinTerms}
{\cal K}_L=\frac{1}{2}\,\big(\Psi_L^\dag i\bar\sigma^\mu\partial_\mu\Psi_L +
\Psi^{c\,\dag}_R i\bar\sigma^\mu\partial_\mu\Psi^c_R \big);
\quad {\cal K}_R=\frac{1}{2}\,\big(\Psi_R^\dag i\sigma^\mu\partial_\mu\Psi_R +
\Psi^{c\,\dag}_L i \sigma^\mu\partial_\mu\Psi^c_L \big).
\end{equation}

By null variations of Lagrangian density (\ref{ChiralLagdens}) with respect to $\Psi^\dag_L$,
$\Psi^\dag_R$, $\Psi^{c\dag}_L$ and $\Psi^{c\dag}_R$, we obtain the motion
equations $i\bar\sigma^\mu\partial_\mu\Psi_L - m\,\Psi_R= 0$, $i\sigma^\mu
\partial_\mu\Psi_R - m\,\Psi_L=0$, $i\bar\sigma^\mu\partial_\mu\Psi^c_L - m\,\Psi^c_R= 0$,
$i\sigma^\mu\partial_\mu\Psi^c_R - m\,\Psi^c_L=0$, the respective solutions of which are
\begin{eqnarray}
\label{chiralphiL}
&&\hspace{-8mm} \Psi_L(x) =\int\!\frac{d^3\vec p}{(2\pi)^{3/2}}
\sqrt{\!\frac{m}{E_p}}\,\bigg[\mathcal{U}_L(\vec p\,)\,a_L(\vec p\,)\,e^{-ip\cdot x}
+\mathcal{V}_R(\vec p\,)\,b^{\dag}_R(\vec p\,)\,\,e^{ip\cdot x}\bigg];\\
\label{chiralphiR}
&&\hspace{-8mm} \Psi_R(x) =\int\!\frac{d^3\vec p\,}{(2\pi)^{3/2}}
\sqrt{\!\frac{m}{E_p}}\,\bigg[\mathcal{U}_R(\vec p\,)\,a_R(\vec p\,)\,
e^{-ip\cdot x} +\mathcal{V}_L(\vec p\,)\,b^{\dag}_L(\vec p\,)\,\,e^{ip\cdot x}\bigg];\\
\label{antichiralphiL}
&&\hspace{-8mm} \Psi^c_L(x) =\int\!\frac{d^3\vec p}{(2\pi)^{3/2}}
\sqrt{\!\frac{m}{E_p}}\,\bigg[\mathcal{U}^{\,c}_L(\vec p\,)\,a^{c}_L(\vec p\,)\, e^{-ip\cdot x}
+\mathcal{V}^{\,c}_R(\vec p\,)\,b^{c \dag}_R(\vec p\,)\,\,e^{ip\cdot x}\bigg];\\
\label{antichiralphiR}
&&\hspace{-8mm} \Psi^c_R(x) =\int\!\frac{d^3\vec p}{(2\pi)^{3/2}}
\sqrt{\!\frac{m}{E_p}}\,\bigg[\mathcal{U}^{\,c}_R(\vec p\,)\,a^c_R(\vec p\,)\,e^{-ip\cdot x}
+\mathcal{V}^{\,c}_L(\vec p\,)\,b^{c\dag}_L(\vec p\,)\,\,e^{ip\cdot x}\bigg].
\end{eqnarray}
From canonical anticommutation relations  among $\Psi_{L, R}(x)$ and $\Psi^\dag_{L, R}(x')$ at $x^0 = {x'}^0$,
and their respective Lorentz group representations, we can derive the following normalization conditions and
momentum--space equations for the spinor components
\begin{eqnarray}
\label{NormCond}
& &\hspace{-10mm} \mathcal{U}^\dag_L(\vec p\,)\,\mathcal{U}_L(\vec p\,) =\mathcal{V}^\dag_L(\vec p\,)
\,\mathcal{V}_L(\vec p\,)= \mathcal{U}^\dag_R(\vec p\,)\,\mathcal{U}_R(\vec p\,) =
\mathcal{V}^\dag_R(\vec p\,)\,\mathcal{V}_R(\vec p\,)=\frac{E_p}{m}\,, \nonumber\\
& &\hspace{-10mm} \mathcal{U}^\dag_L(\vec p\,)\,\mathcal{U}_R(\vec p\,) =
\mathcal{V}^\dag_L(\vec p\,)\,\mathcal{V}_R(\vec p\,) =\mathcal{U}^\dag_R(\vec p\,)
\,\mathcal{U}_L(\vec p\,)= \mathcal{V}^\dag_R(\vec p\,)\,\mathcal{V}_L(\vec p\,) =0\,,\\
\label{TheUs}
& & \hspace{-10mm} p_\mu \bar\sigma^\mu \mathcal{U}_L(\vec p\,) = m\,\mathcal{U}_R(\vec p\,) \,,
\quad p_\mu\sigma^\mu \mathcal{U}_R(\vec p\,) = m\,\mathcal{U}_L(\vec p\,)\,.
\end{eqnarray}

Identifying in Eqs (\ref{chiralphiL})--(\ref{antichiralphiR}) $a^{\,c}_L(\vec p)$ as $b_R(\vec p)$,
$a^{\,c}_R(\vec p)$ as $b_L(\vec p)$, $b^{\,c}_L(\vec p)$ as $a_R(\vec p)$ and $b^{\,c}_R(\vec p)$
as $a_L(\vec p)$, we also derive the flipping relationships among spinor components:
\begin{equation}
\vspace{-1mm}
\label{FlippingRels}
\mathcal{U}^{\,c}_{L,\, R}(\vec p\,) = -i\,\sigma^2 \mathcal{V}^{\,*}_{R,\,L}(\vec p\,);\quad
\mathcal{V}^{\,c}_{L,\,R}(\vec p\,) = -i\,\sigma^2 \mathcal{U}^{\,*}_{R,\, L}(\vec p\,)\,.
\vspace{-1mm}
\end{equation}

Now assume that $\Psi(x)$ is a Dirac--neutrino field and denote it as $\nu^D(x)$. The basic difference
between this field and a Majorana neutrino field $\nu^M(x)$ with the same mass is that the latter
coincides with its own antiparticle \cite{KAYSER} \cite{LANGACKER}. So, in accordance with
Eq (\ref{FlippingRels}), the chiral components of $\nu^{Mc}$ obey equations $\nu^{Mc}_L = -i\,\sigma^2
\nu^{M*}_R$ and $\nu^{Mc}_R = -i\,\sigma^2 \nu^{M*}_L$. Therefore, in place of
(\ref{chiralphiL})--(\ref{antichiralphiR}) we have only two chiral components,
\begin{eqnarray}
\label{nuMajorL}
&&\hspace{-20mm} \nu^M_L(x)\! =\! -i\sigma^2\nu^{Mc\,*}_R(x)\! =\!\int\!
\frac{d^3\vec p}{(2\pi)^{3/2}}\sqrt{\frac{m}{E_p}}\bigg[\mathcal{U}_L(\vec p\,)
\,a_L(\vec p\,)\, e^{-ip\cdot x}\!+\! \mathcal{V}_R(\vec p\,)\,
a^{\dag}_R(\vec p\,)\, e^{ip\cdot x}\bigg]\!, \\
\label{nuMajorR}
&& \hspace{-20mm}\nu^M_R(x)\! =\! -i\sigma^2\nu^{Mc\,*}_L(x)\!
=\!\int\!\frac{d^3\vec p}{(2\pi)^{3/2}}\sqrt{\frac{m}{E_p}}
\bigg[\mathcal{U}_R(\vec p\,)\,a_R(\vec p\,)\, e^{-ip\cdot x}
\!+\! \mathcal{V}_L(\vec p\,)\, a^{\dag}_L(\vec p\,)\,
e^{ip\cdot x}\bigg]\!,
\end{eqnarray}
and there is no reason why the masses of fields $\nu^M_L(x)$ and $\nu^M_R(x)$ should be the same.

A left--handed Weyl field $\nu_L$ is Lorentz--transformed by the $2\times 2$ complex matrices
$\Lambda_L=\begin{psmallmatrix}a & b\\c & d\end{psmallmatrix}$,  with Det$(\Lambda_L)=1$;
the same Lorentz transformations for the chiral companion $\nu_R$ require instead the conjugate
matrices $\Lambda_R = -\sigma^2\,{\Lambda}^*_L \sigma^2$ \cite{BECCHI}.

Since $\Lambda_L$ and $\Lambda_R$ are not unitary, the products $\nu^\dag_L\nu_L$
and $\nu^\dag_R\nu_R$ are not Lorentz invariant; but $\nu^\dag_L\sigma^2\nu^*_L=
i\,\nu^\dag_L \nu^c_R$ and $\nu^\dag_R\sigma^2\nu^*_R =i\,\nu^\dag_R\nu^c_L$
instead are, as we can easily verify. Therefore the requirement that the
mass terms of chiral fields be Lorentz invariant and hermitian restricts
to two groups: ({\em i}) $\nu^\dag_L\nu_R + \hbox{h.c.}= {\nu^c_L}^\dag \nu^c_R
+ \hbox{h.c.}$; ({\em ii}) $\nu^\dag_L\nu^c_R + \hbox{h.c.}$, $\nu^\dag_L\nu^c_R
+ \hbox{h.c.}$. Those of the first group {\em couple states of same fermion
number and opposite chiralities}; those of the second group {\em couple states
of opposite fermion number and same chirality} \cite{YANAGIDA}. The first is
the case of Dirac neutrinos, the second is that of Majorana neutrinos.

Expressing the kinetic--energy terms of left-- and right--handed neutrinos as
in Eqs (\ref{LKinTerms}) -- but with $\nu$ in place of $\Psi$ -- we can
write the Lagrangian densities of a Dirac neutrino of mass $m_D$, $\nu^D$,
that of a left--handed (active) Majorana neutrino of mass $m_L$, $\nu^M_L$,
and that of a right--handed (sterile) Majorana neutrino of mass $m_R$,
$\nu^M_L$, as follows:
\begin{eqnarray}
\label{DLangr}
\hspace{-6mm}{\cal L}^D &=& \frac{1}{2}\,\Big[\nu_L^{D\,\dag}i\,
\bar\sigma^\mu\partial_\mu\nu^D_L+\nu_R^{D\,\dag}i\,
\sigma^\mu\partial_\mu\nu^D_R + \nu_L^{Dc\,\dag}i\,
\bar\sigma^\mu\partial_\mu\nu^{Dc}_L+ \nu_R^{Dc\,\dag}i\,
\sigma^\mu\partial_\mu\nu^{Dc}_R-\nonumber \\
\hspace{-6mm}&& m_D\Big(\nu^{D\,\dag}_L\nu^D_R+
\nu^{D\,\dag}_R\nu^D_L +\nu^{Dc\,\dag}_L\,\nu^{D\,c}_R+
\nu^{Dc\,\dag}_R\nu^{Dc}_L\Big)\Big];\\
\label{LLangr}
\hspace{-6mm}{\cal L}^M_L&=&\frac{1}{2}\,\Big[\nu_L^{M\,\dag}i\,\bar\sigma^\mu\partial_\mu\nu^M_L
+ \nu_R^{Mc\,\dag}i\,\bar\sigma^\mu\partial_\mu\nu^{Mc}_R -m_L\Big(\nu^{Mc\,\dag}_R\nu^M_L
+\nu^{M\,\dag}_R\nu_L^{Mc}\Big)\Big];\\
\label{RLangr}
\hspace{-6mm}{\cal L}^M_R&=&\frac{1}{2}\,\Big[\nu_R^{M\,\dag}i\,\sigma^\mu\partial_\mu\nu^M_R
+ \nu_L^{Mc\,\dag}i\,\sigma^\mu\partial_\mu\nu^{Mc}_L -
m_R\,\Big(\nu^{Mc\,\dag}_L\nu^M_R + \nu^{M\dag}_R\nu^{Mc}_L\Big)\Big].
\end{eqnarray}

\begin{figure}[!h]
\vspace{-2mm}
\centering
\includegraphics[scale=0.7]{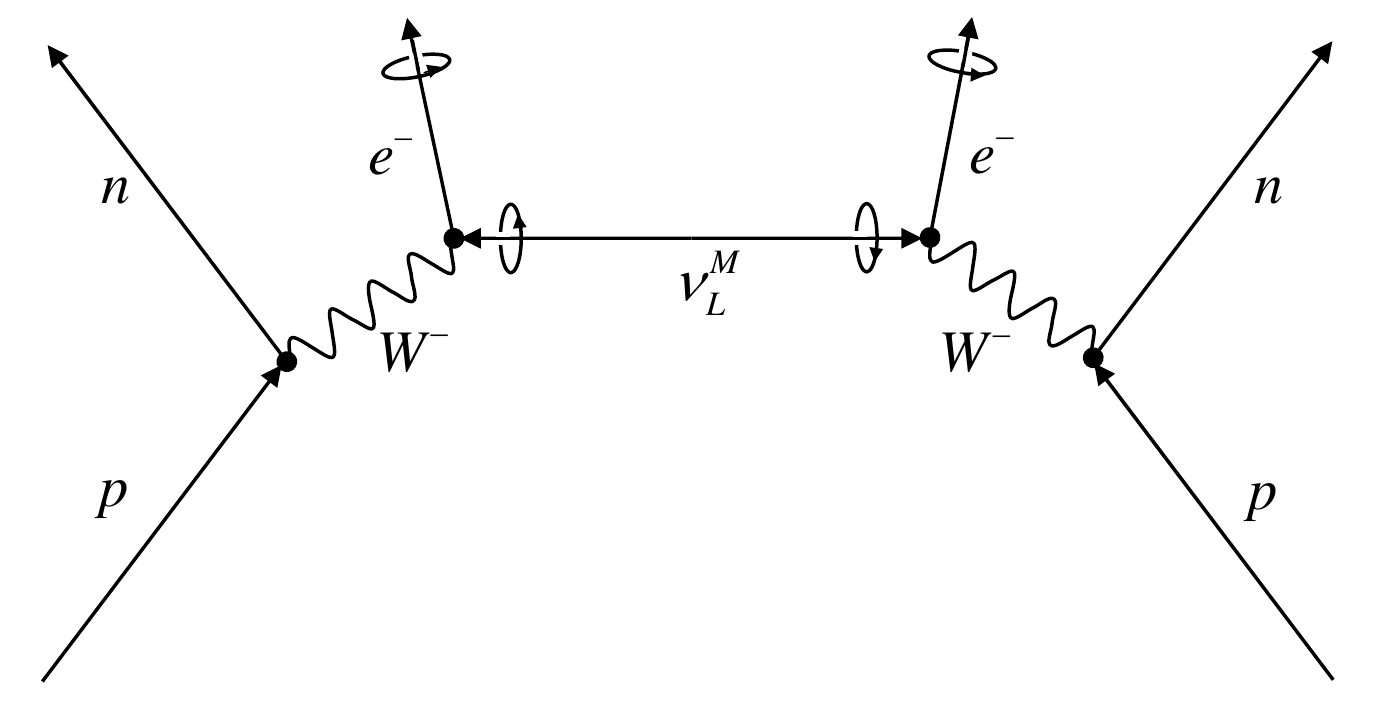} %F21
\vspace{-5mm}
\caption{\small
{\em Feynman diagram of neutrinoless double--beta decay}. Two protons $p$ turn into two neutrons $n$
and two electrons $e^-$, thus violating lepton number $L$ by two units, $\Delta L=2$, and isospin $t^3_L$
by one unit, $\Delta t^3_L=1$ \cite{KLAPDOR}. The process is mediated by a left--handed Majorana
neutrino $\nu^M_L$ of mass $m_L$. Since $\nu^M_L$ coincides with its conjugate $\nu^{\,cM}_L$, its
propagator is bidirectional. Therefore, it delivers left--handed helicity at both electroweak vertices
$\nu^M_L W^-e^-$.}
\label{DirMajFig1}
\vspace{-3mm}
\end{figure}

In Fig.\,\ref{DirMajFig1}, the Feynman diagram of a process involving a left--handed Majorana
neutrino is represented.

\subsection{Majorana neutrinos and Dirac--Majorana hybrids}
\label{D-M-HYBRIDS}
The mass terms of Dirac and Majorana neutrinos can be mixed to form composed states of particles.
Hybrids of this sort occur in most extensions of SMEP as $SU(2)$ singlets. Right--handed
Majorana neutrinos cannot interact except by mixing. If these sterile particles were
sufficiently massive they would be good candidates for dark matter.
In this subsection we describe the basic properties of the neutrinos and their
possible mixings.

The Lagrangian densities of a left--handed Majorana neutrino field $\nu_L(x)$ and
of a right--handed Majorana neutrino field $\nu_R(x)$ are respectively
\begin{eqnarray}
\label{LMajL}
&& {\cal L}^M_L = \nu^\dag_L\,i\,\bar\sigma^\mu\partial_\mu\nu_L
 -\frac{1}{2}\,m_L\big(\nu^{c\,\dag}_R\,\nu_L+\nu^\dag_L\,\nu_R^{\,c}\big)\,;\\
\label{LMajR}
&&{\cal L}^M_R = \nu^\dag_R\,i\,\sigma^\mu\partial_\mu\nu_R -
\frac{1}{2}\, m_R\,\big(\nu^{c\,\dag}_L\,\nu_R+\nu^\dag_R\,\nu^{c}_L\big)\,.
\end{eqnarray}
We know from (\ref{nuMajorL}) and (\ref{nuMajorR}) that the two fields are mutually
related by equations $\nu_L =i\,\sigma^2\nu^{\,c*\,}_R$ and $\nu_R=i\,\sigma^2\nu^{\,c\,*}_L$.
Indicating the two spin components of $\nu_L(x)$ as $z_1(x)$ and $z_2(x)$ and the two
spin components of $\nu_R(x)$ as  $z_3(x)$ and $z_4(x)$, we can rewrite the above equations
in the form
\begin{eqnarray}
\label{UL&UCR}
&& \hspace{-10mm}\nu_L(x)=
\begin{bmatrix}
z_1(x)\\
z_2(x)
\end{bmatrix}\!;
\quad
\nu^{\,c}_R(x) = -i\,\sigma^2 \nu^{\,*}_L(x) =
\begin{bmatrix}
0 & 1\\
-1 & 0
\end{bmatrix}\!\!
\begin{bmatrix}
z_1^*(x)\\
z_2^*(x)
\end{bmatrix}\!=
\begin{bmatrix}
z_2^*(x)\\
-z_1^*(x)
\end{bmatrix}\!;
\\
\label{UR&UCL}
&&\hspace{-10mm}\nu_R(x)=
\begin{bmatrix}
z_3(x)\\
z_4(x)
\end{bmatrix}\!,
\quad
\nu^{\,c}_L(x) = -i\,\sigma^2 \nu^{\,*}_R(x) =
\begin{bmatrix}
0 & 1\\
-1 & 0
\end{bmatrix}\!\!
\begin{bmatrix}
z_3^*(x)\\
z_4^*(x)
\end{bmatrix}\!
=\begin{bmatrix}
z_4^*(x)\\
-z_3^*(x)
\end{bmatrix}\!;
\end{eqnarray}
and their hermitian conjugates, as follows
\begin{eqnarray}
\label{UL&UCRDAG}
&& \nu^\dag_L(x)=
\Big[z^*_1(x),\, z^*_2(x)\Big];
\quad
\nu^{\,c^\dag}_R(x) = -i\,\widetilde{\nu}_L(x)\,\sigma^2 =
\Big[z_2(x),\,-z_1(x)\Big];
\\
\label{UR&UCLDAG}
&&\nu^\dag_R(x)=\Big[z^*_3(x),\, z^*_4(x)\Big];
\quad
\nu^{\,c\dag}_L(x) = -i\,\widetilde{\nu}_R(x)\,\sigma^2 =
\Big[z_4(x),\,-z_3(x)\Big].
\end{eqnarray}

We can therefore express Eqs (\ref{LMajL}) and (\ref{LMajR}) in the
$2\times 2$ matrix form:
\begin{eqnarray}
&&{\cal L}^M_L=
\widetilde{
\begin{bmatrix}
z^*_1\\
z^*_2
\end{bmatrix}}\!\!
\begin{bmatrix}
i\,\partial_0-i\,\partial_3 & -i\,\partial_1- \partial_2\\
-i\,\partial_1+ \partial_2 & i\,\partial_0+i\,\partial_3
\end{bmatrix}\!\!
\begin{bmatrix}
z_1\\
z_2
\end{bmatrix}\!-
\frac{m_L}{2} \left\{
\widetilde{
\begin{bmatrix}
z_1\\
z_2
\end{bmatrix}}\!\!
\begin{bmatrix}
z_2\\
-z_1
\end{bmatrix} +
\widetilde{
\begin{bmatrix}
z^*_1\\
z^*_2
\end{bmatrix}}\!\!
\begin{bmatrix}
z^*_2\\
-z^*_1
\end{bmatrix}\right\};
\nonumber \\
&&{\cal L}^M_R=
\widetilde{
\begin{bmatrix}
z^*_3\\
z^*_4
\end{bmatrix}}\!\!
\begin{bmatrix}
i\,\partial_0+i\,\partial_3 & i\,\partial_1+ \partial_2\\
i\,\partial_1-\partial_2 & i\,\partial_0-i\,\partial_3
\end{bmatrix}\!\!
\begin{bmatrix}
z_3\\
z_4
\end{bmatrix}\!-
\frac{m_R}{2} \left\{
\widetilde{
\begin{bmatrix}
z_3\\
z_4
\end{bmatrix}}\!\!
\begin{bmatrix}
z_4\\
-z_3
\end{bmatrix} +
\widetilde{
\begin{bmatrix}
z^*_3\\
z^*_4
\end{bmatrix}}\!\!
\begin{bmatrix}
z^*_4\\
-z^*_3
\end{bmatrix}\right\}.
\nonumber
\end{eqnarray}
Unfortunately, this expressions cannot be cast in the form ${\cal Z}^\dag_L {\mathbb L}_L{\cal Z}_L$
and ${\cal Z}^\dag_R {\mathbb L}_R{\cal Z}_R$.

But this difficulty can be circumvented if we express the above equations in a mix of two
independent Majorana neutrinos. Consider, in fact, the mix of a left--handed and a
right--handed Majorana neutrino described by the total
Lagrangian density
\begin{equation}
\label{MLMajoranaLagr}
{\cal L}^M = {\cal K}_L + {\cal K}_R  -\frac{1}{2} \Big[m_L\Big(\nu^\dag_L\,\nu_L^{\,c}+
\nu^{c\,\dag}_L\,\nu_L\Big)+m_R\,\Big(\nu^{c\,\dag}_R\,\nu_R + \nu^\dag_R\,\nu^{c}_R\Big)\Big]\,,
\end{equation}
where, in virtue of Eqs (\ref{LKinTerms}), we may put
$$
{\cal K}_L= \frac{1}{2}\,\Big(\nu^\dag_L\,i\,\bar\sigma^\mu\partial_\mu\nu_L+
\nu^{c\,\dag}_R\,i\,\bar\sigma^\mu\partial_\mu\nu^{c}_R\Big)\,; \quad
{\cal K}_R = \frac{1}{2}\,\Big(\nu^\dag_R\,i\,\sigma^\mu\partial_\mu\nu_R+
\nu^{c\,\dag}_L\,i\,\sigma^\mu\partial_\mu\nu^{c}_L\Big)\,.
$$
Then, Eq (\ref{MLMajoranaLagr}) condense into the $8\times 8$ matrix
\begin{equation}
\label{DiracMatrix}
{\cal L}^{M}= \frac{1}{2}\,\Big[\nu^\dag_L,\,\nu^{c\,\dag}_L,\,\nu^\dag_R,\,\nu^{c\,\dag}_R\Big]\!\!
\begin{bmatrix}
i\bar\sigma^\mu\partial_\mu & 0 & -{\mathbb I}_2\,m_L &0 \\
0 & i\sigma^\mu\partial_\mu &0 &-{\mathbb I}_2\,m_R \\
-{\mathbb I}_2\,m_L & 0 & i\sigma^\mu\partial_\mu &0\\
0 & -{\mathbb I}_2\,m_R & 0 & i\bar\sigma^\mu\partial_\mu
\end{bmatrix}\!\!
\begin{bmatrix}
\nu_L\\
\nu^{\,c}_L\\
\nu_R\\
\nu^{\,c}_R
\end{bmatrix}\!.
\end{equation}

If the mix comprises a Dirac neutrino of mass $m_D$, in place of ${\cal L}^{M}$ we have instead
$$
{\cal L}^{DM}\!\!\!\!=\!{\cal K}\!-\!\frac{1}{2}\!\Big[m_D\Big(\nu^\dag_L\nu_R
\!+\!\nu^{c\,\dag}_R \nu^c_L\!+\!\nu^\dag_R\nu_L\!+\!\nu^{c\,\dag}_L\nu^c_R\Big)
\!\!+\!m_L\Big(\nu^\dag_L\nu_L^{\,c}\!+\!\nu^{c\,\dag}_L\!\nu_L\Big)\!\!+\!
m_R\Big(\nu^{c\,\dag}_R\nu_R\!+\!\nu^\dag_R\nu^{\,c}_R\Big)\!\Big],
$$
which we condense into
\begin{equation}
\label{DiracMajoranaMatrix}
{\cal L}^{DM}= \frac{1}{2}\,\Big[\nu^\dag_L,\,\nu^{c\,\dag}_L,\,\nu^\dag_R,\,\nu^{c\,\dag}_R\Big]\!\!
\begin{bmatrix}
i\bar\sigma^\mu \partial_\mu & -{\mathbb I}_2\,m_D & -{\mathbb I}_2\,m_L &0 \\
-{\mathbb I}_2\,m_D & i \sigma^\mu\partial_\mu &0 &-{\mathbb I}_2\,m_R \\
-{\mathbb I}_2\,m_L & 0 & i\sigma^\mu \partial_\mu &-{\mathbb I}_2\,m_D\\
0 & -{\mathbb I}_2\,m_R & -{\mathbb I}_2\,m_D & i \bar\sigma^\mu\partial_\mu
\end{bmatrix}\!\!
\begin{bmatrix}
\nu_L\\
\nu^{\,c}_L\\
\nu_R\\
\nu^{\,c}_R
\end{bmatrix}\!.
\end{equation}

Eq (\ref{DiracMajoranaMatrix}) can be written as ${\cal L}^{DM}(x) = \frac{1}{2} \,{\cal Z}^\dag(x)
\big[\,\mathbb{I}_8\,i\partial_0-\mathbb{E}(\,i\vec{\partial}\,)\big]{\cal Z}(x)$, where
$\mathbb{I}_8$ is the $8\times 8$ unit matrix,  ${\cal Z}(x)$ is the eightfold multiplet of
chiral components, ${\cal Z}^\dag(x)$ that of its hermitian conjugates, and the hermitian operator
$\mathbb{E}(\,i\vec{\partial}\,)$ represents the energy density operator of the hybrid neutrino
field as a functional of $i\vec{\partial}$. Therefore, the null variations of action integral
${\cal A}^{DM}=\int{\cal L}^{DM}(x)\,d^4x$ with respect to $\nu^\dag_L(x),\,\nu^{c\,\dag}_L(x),
\,\nu^\dag_R(x),\,\nu^{c\,\dag}_R(x)$ provide the motion equations of the neutrino components
$\nu_L(x),\,\nu^{c}_L(x),\,\nu_R(x),\, \nu^{c}_R(x)$.

In the 4--momentum space we have ${\cal A}^{DM}= \frac{1}{2}\!\int{\cal Z}^\dag(-p) \big[\,\mathbb{I}_8
p_0-\mathbb{E}(\vec{p}\,)\big]{\cal Z}(p)d^4p$, where
\begin{equation}
\label{EnergyDensMatrix}
\mathbb{E}(\vec{p}\,)=
\begin{bmatrix}
-\sigma^i p_i & {\mathbb I}_2\,m_D & {\mathbb I}_2\,m_L &0 \\
{\mathbb I}_2\,m_D & \sigma^i p_i &0 &{\mathbb I}_2\,m_R \\
{\mathbb I}_2\,m_L & 0 & \sigma^i p_i &{\mathbb I}_2\,m_D\\
0 & {\mathbb I}_2\,m_R & {\mathbb I}_2\,m_D & -\sigma^i p_i
\end{bmatrix}\!\!
\end{equation}
is a hermitian matrix depending on $\vec p$. To further simplify the algebraic computation, we can diagonalize
the $2\times 2$ matrices $\vec\sigma\cdot\vec p$, thus making $\mathbb{E}(\vec p)$ depend on $|\vec p\,|$,
by means of a unitary diagonal operator $\mathbb{U}(\theta, \vec n) = \mathbb{I}_8\, e^{\textstyle
-i[\vec \sigma\cdot \vec n(\vec p)]\,\theta/2}$, where $\vec n(\vec p)$ is a 3D vector depending on $\vec p$
and satisfying the condition $|\vec n(\vec p)| =1$. The computation yields 4 possible pairs of
degenerate eigenvalues,
\begin{equation}
E_0 =  \pm \sqrt{\frac{2\,|\vec p\,|^2+ 2\,m_D^2+m_L^2 + m_R^2 \pm\big(m_L+m_R\big)
\sqrt{4\,m_D^2+\big(m_L-m_R\big)^2}}{2}}\,,\nonumber
\end{equation}
with the two $\pm$ regarded as independent. By squaring these eigenvalues we obtain
\begin{equation}
\label{HybridDet}
\hspace{-0.5mm}p^2 \equiv p_0^2 -|\vec p\,|^2 = \frac{2\,m_D^2+m_L^2 + m_R^2
\pm\big(m_L+m_R\big)\sqrt{4\,m_D^2+\big(m_L-m_R\big)^2}}{2} =
m^2_\pm\,,
\end{equation}
showing that the Dirac--Majorana hybrid splits in two fields with squared masses $m^2_\pm$. So,
the determinant of $\mathbb{I} p_0-\mathbb{E}(\vec{p}\,)$ is $(m_- m_+)^4$.

In particular, for $m_L=0$  we obtain two hybrids of squared masses
\begin{equation}
\label{rightseesaw}
m^2_{\pm} =  \frac{1}{2}\bigg(2\,m_D^2 + m_R^2 \pm m_R\sqrt{4\,m_D^2+m_R^2}\,\bigg)\,.
\end{equation}
For $m_R\gg m_D$, the hybrid splits into a nearly left--handed Majorana neutrino of mass $m_+ \cong m_R$
and a nearly Dirac neutrino of mass $m_- \cong m_D^2/m_R \ll m_D$. Since the larger  $m_+$ the smaller
$m_-$, this decomposition is called the {\em left--handed seesaw}. It has been speculated that this
mechanism may explain the smallness of leptonic neutrinos and that the nearly
Majorana neutrino may be the sterile constituent of dark matter.

If $m_D=0$, Eq (\ref{HybridDet}) becomes
\begin{equation}
\label{MajoranaDets}
p^2 = \frac{m_L^2 + m_R^2 \pm\big(m_L^2-m_R^2\big)}{2}=
\begin{cases}
m^2_+ = m^2_L\\
m^2_- = m^2_R
\end{cases},
\end{equation}
showing that the determinant of $\mathbb{I}_8\,i\partial_0-\mathbb{E}(\,i\vec{\partial})\,$
factorizes in the product of $p^2-  m^2_L$ and $p^2-  m^2_R$, which can be respectively identified
as the determinants of  ${\cal L}^M_L$ and $ {\cal L}^M_R$.

\newpage

\subsection{Yukawa potentials of Dirac and Majorana neutrinos}
\begin{figure}[!h]
\vspace{-2mm}
\centering
\includegraphics[scale=0.75]{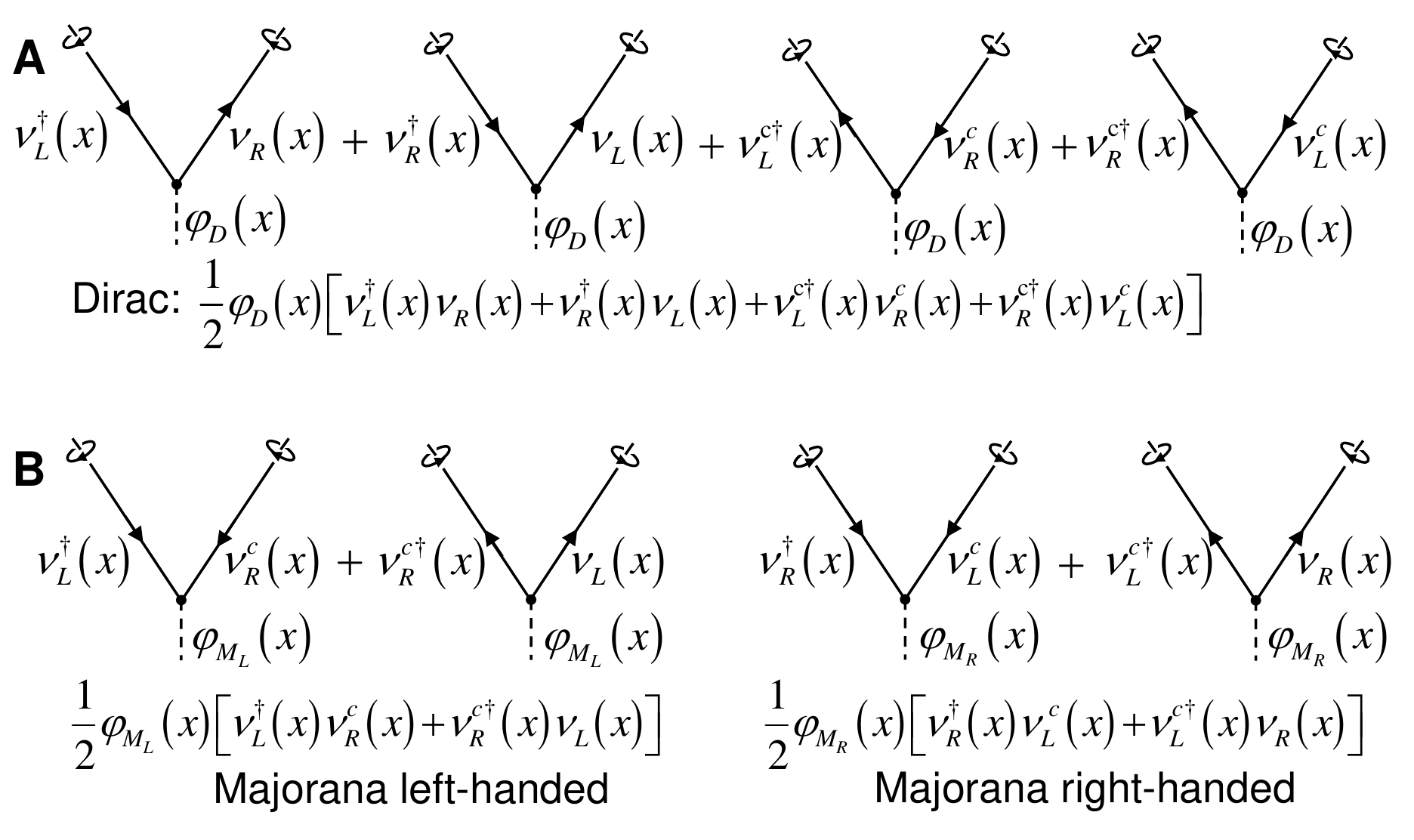}  %F22
\vspace{-6mm}
\vspace{1mm}
\caption{\small
{\em Scalar--field interactions with the chiral components of Dirac and Majorana neutrinos}.
{\bf A}: Scalar field $\phi_D(x)$ connects the left--handed and right-handed chiral components
$\nu_L(x)$, $\nu_R(x)$  (arrows running from left to right) of a Dirac neutrino as well as those
of the anti--neutrino, $\nu^c_L(x)$ and $\nu^c_R(x)$ (arrows running from right to left). The interaction
preserves the lepton number $L$ but flips the opposite chiralities. {\bf B}: Scalar fields $\phi_{M_L}(x)$,
$\phi_{M_R}(x)$ connect respectively the left--handed and right--handed components of a Majorana neutrino,
$\nu_L(x)$ and $\nu_R(x)$, and their charge conjugate counterparts, $\nu^c_L(x)$ and $\nu^c_R(x)$.
The interactions preserves chirality but violate the lepton number by two units, $\Delta L =2$.}
\label{DirMajFig2}
\vspace{-1mm}
\end{figure}

Replacing in Eq (\ref{DiracMajoranaMatrix}) $m_D$, $m_L$ and $m_R$ with three scalar fields
$\varphi_D(x)$, $\varphi_L(x)$ and $\varphi_R(x)$, respectively, we obtain the three Yukawa
potentials which connect in different ways the chiral components of the Dirac and the Majorana neutrinos.
The diagrams of the Yukawa couplings are represented in Fig.\,\ref{DirMajFig2}.

If the VEVs of the three scalar fields are not zero, we obtain a Dirac--Majorana neutrino hybrid with the mass
terms of the type described by Lagrangian density (\ref{DiracMajoranaMatrix}). More sophisticated types of mixings
may be obtained by the interaction of a scalar field multiplet with a multiplet of a Dirac and/or Majorana
neutrinos giving rise to a wide collection of interesting effects such as neutrino oscillations and strongly
unbalancing between light active neutrinos and heavy sterile Majorana neutrinos  \cite{YANAGIDA}
\cite{KLAPDOR} \cite{MOHAPATRA} \cite{DREWES}.

\newpage

\markright{R.Nobili, Conformal General Relativity - {\bf \ref{GrassAlgApp}}  Grassmann algebra}
\renewcommand\thefigure{\Alph{section}\arabic{figure}}
\setcounter{figure}{0}
\section{\hspace{-2mm} GRASSMANN ALGEBRA AND BEREZIN INTEGRAL}
\label{GrassAlgApp}
The original idea of path integral is to express the wave amplitude of a quantum field as an integral
over the histories of a classical field. This method works well with boson fields because the states of
these are linearly superposable, but not with fermions fields, because in this case the superposition
is precluded by Pauli's exclusion principle. The classical analog for fermionic fields can nevertheless
be implemented provided that the ordinary commutative algebra of $c$--numbers be replaced by an algebra
of anti--commuting units, known as Grassmann units -- in short $G$--units -- since the far 1832.
These units may form a discrete set $z_1, z_2, \dots$, a set of continuum functions $z(x), z(y), \dots$,
or a mixed set $z_i(x), z_j(y), \dots$ etc. From their anti--commutative properties $\big\{z_i, z_j\big\}
\equiv z_i z_j+ z_j z_i=0$, $\big\{z_j(x),z_k(y)\big\}=0$, we derive the nihilpotency properties
$z_i^2=0$, $z_i(x)^2=0$ etc.

Linear combinations of $G$--units in the complex domain are called Grassmann variables, or G--numbers.
The nihilpotency property of the $G$--units makes the Grassmann algebra very simple: any function
of $n$ G--units, $f(z_1, z_2, \dots , z_n)$, can be expanded in the form
$$
f(z_1, z_2, \dots , z_n)  = c_0 + \sum_i\!c_i\, z_i + \sum_{i<j}\!c_{ij}\, z_i z_j +
\sum_{i<j<k}\!c_{ijk}\, z_i z_j z_k + \dots c_{1\,2\dots \,n}\,  z_1 z_2\cdots z_n\,.
$$
For continuous indices, summations must be replaced by integrations.

If we choose $c_{ij\dots k}$ to be totally antisymmetric, we can write
\begin{eqnarray}
\label{G-expans}
f(z_1, z_2, \dots , z_n) &=& c_0 + \sum_i c_i\,z_i + \frac{1}{2!}\!\sum_{ij}c_{ij}\, z_i z_j +
\frac{1}{3!}\!\sum_{ijk}\!c_{ijk}\, z_i z_j z_k + \nonumber \\
& & \dots +\frac{1}{n!}\!\sum_{i_1 i_2\dots i_n} c_{i_1\,i_2\dots \,i_n}\,  z_1\, z_2\cdots z_n\,,
\end{eqnarray}
since symmetric coefficients do not contribute anyway.

Let us perform a change of $G$--units by a linear transformation $z_i \rightarrow z'_i = \sum_j T_i^j z_j$,
where $\{z_1, z_2, \dots z_n\}$ is an ordered set of $G$--units and $T=[T^i_j]$ a squared matrix. Then,
using Eq (\ref{G-expans}), we can easily prove the following equalities:
\begin{eqnarray}
\label{GDeterminant}
\prod_i z'_i &=& \frac{1}{n!}\,\epsilon^{i_1 i_2\dots i_n}z'_{i_1} z'_{i_2}...z'_{i_n} =
\frac{1}{n!}\,\epsilon^{i_1 i_2\dots i_n} T^{j_1}_{i_1} T^{j_2}_{i_2}\dots T^{j_n}_{i_n}
z_{j_1}z_{j_2}\dots z_{j_n}  = \nonumber \\
& & \frac{1}{n!}\,\epsilon^{i_1 i_2\dots i_{\,n}}\,T^{j_1}_{i_1} T^{j_2}_{i_2}\dots
T^{j_{\,n}}_{i_{\,n}}\,\epsilon_{j_1 j_2\dots j_{\,n}} \prod_i z_i = \hbox{Det}(T)\prod_j z_i\,.
\end{eqnarray}
where Det($T$) is the determinant of matrix $T$.

Basing on these results, we will be able to implement an integro--differential in the $G$--number domain,
in analogy to the bosonic case. The appropriate technique for doing this has been described by Berezin
in 1969 \cite{BEREZIN} \cite{PESKIN} \cite{RAMOND}.

The differentials of a set of $G$--numbers $z_1, z_2, \dots , z_n$ can be formally introduced as a set
of $G$--numbers, $dz_1, dz_2, \dots , dz_n$, whose basic purpose is  allow us to describe the
$m$--dimensional integral of a $G$--number function  $f(z_1, z_2, \dots , z_n)$ in the form
\begin{equation}
\label{BerezinInt}
I_m= \int dz_1 dz_2 \dots dz_m f(z_1, z_2, \dots , z_n)\,.
\end{equation}
Since the range of a $G$--unit does not exist, the integration symbol $\int$ is here introduced
only as the the formal analog of an indefinite integral in the $c$--number domain.

Since in the $c$--number domain the integral $I_1 = \int dx\,(A + B\,x)=  A\!\int dx + B\!\int\!dx\,x$,
where $A$ and $B$ are independent of $x$, is invariant under translations $x\rightarrow x+c$, we
will transfer the same property to the $G$--number domain by imposing the condition
$$
I_1\!=\!\int\! dz\,(A + B\,z)\!=\!A\!\int\! dz\! +\! B\!\int dz\,z \!=\! \int dz\,[A + B\,(z+c)]\!
 =\!(A \!+\! B\,c)\!\int dz + B\!\int\! dz\,z \,.
$$
This clearly implies $\int dz =0$ and $\int dz\,z = C\neq 0$, where $C$ is independent of $z$.
In the lack of any criterion of choice, we assume $C=1$. In summary, we have the simple rule
\begin{equation}
\label{Berezinrule}
I_1\! =\! \int dz\,(A + B\,z) = B\,, \quad\hbox{i.e.,}\quad  \int dz = 0\,, \quad \int dz\, z = 1\,.
\end{equation}

To extend this result to the general case,  it is sufficient to recall Eq (\ref{G-expans})
and consider the particular case
\begin{equation}
\label{MNrule}
I_m = \int dz_1\, dz_2 \cdots dz_m\,z_1\, z_2 \cdots z_n.
\end{equation}
Clearly, if $m>n$, it is $I_m=0$. If $m\le n$, we meet a problem of sign depending on the
ordering of $dz_i$ and $z_j$. For example, for $m=1$ and $n=2$, we have $\int dz_1\, z_1 z_2 = z_2$
and $\int dz_1 z_2 z_1 = - \int dz_1 z_1 z_2 = -z_2$. To disambiguate the integration for $n\ge m$,
we can arrange the product $z_1 z_2 \dots z_n$ in the form $(-1)^P z_m\, z_{m-1} \cdots z_1\,
z_{m+1}\,z_{m+2}\cdots z_n$, where $P$ is the number of permutations needed to produce the desired
arrangement; so we obtain
$$
I_m = \int  dz_1 dz_2 \dots dz_m\,z_m\, z_{m-1} \cdots z_1\, z_{m+1}\,z_{m+2}\cdots z_n =
z_{m+1}\,z_{m+2}\cdots z_n\,.
$$
Therefore, in particular, for $m=n$, we simply obtain $I_m=1$.

To represent the classical equivalent of chiral spinor fields, it is convenient to introduce
pairs of complex conjugated $G$--units formed by two standard (real) units, $z$ and $z'$:
$$
z=\frac{z+i z'}{\sqrt{2}}\,, \quad z^*=\frac{z-i z'}{\sqrt{2}}\,,
\quad\hbox{with the ordering convention }\,\int(dz^*dz)\,zz^*= 1\,.
$$
So, for two complex units $z_1$ and $z_2$, we have $(z_1 z_2)^* = z_2^* z_1^* = -z_1^* z_2^*$
and, for $n$ complex units, $(z_1z_2\dots z_n)^* =(-1)^n z^*_n z^*_{n-1}\dots z^*_1$.
The following commutation relations among complex conjugate bilinears can be easily proven
\begin{equation}
\label{commutbilin}
\big[z^*_iz_i,\, z^*_jz_j\big]=0\,.
\end{equation}

The simplest Gaussian integral in a complex $G$--number domain is
\begin{equation}
\label{simplestgaussian}
\iint dz^* dz\, e^{i\, z^* a\, z} = \iint dz^* dz\,(1+i\,z^*z\,a) =
 \iint dz^* dz\,(1- i\, z\,z^* a) =-i\, a\,.
\end{equation}
If $z, z^*$ were ordinary $c$--numbers, the integration would give $2\pi/a$. Ignoring the
factor $2\pi$, which is unimportant in path integral computations, we see that the
substantial difference with $G$--numbers is that $a$ appears in the numerator rather
than in denominator.

To perform general Gaussian integrals in higher dimensions, it is suitable to regard $z_i$ as
the components of a covariant $G$--vector ${\bf z}$, and ${z^*}^i$ as those of a the hermitian
conjugate vector ${\bf z}^\dag$. We can represent the ordered products of differentials $dz_i$
and $dz^\dag_i$ in the form
$$
dz_1 dz^*_1\, dz_2 dz^*_2 \cdots dz_n dz^*_n \equiv \prod_{j=1}^n dz^*_j\, dz_j =
d{\bf z}^\dag\!\cdot d{\bf z}\,.
$$
These bilinears are invariant under unitary transformations. Let us perform the linear transformations
$z_i\rightarrow z'_i = U_i^j z_j$ and $z_i\rightarrow {{z'}^*}^i={U^\dag}^i_{\!j}\,{z^*}^j$, where
$U=\big[U_i^j\big]$ is a unitary matrix;  in short, ${\bf z}\rightarrow {\bf z}' = {\bf U} {\bf z}$;
${\bf z}^\dag\rightarrow {{\bf z}^\dag}' ={\bf z}^\dag {\bf U}$. Then,  applying Eq (\ref{GDeterminant}),
we obtain
\begin{equation}
\label{unitinvariance}
{d{\bf z}'}^\dag\!\!\cdot d{\bf z}' = \hbox{Det}(U)\,\hbox{Det}(U^\dag)\prod_{j=1}^n dz^*_j\, dz_j =
\hbox{Det}(UU^\dag)\prod_{j=1}^n dz^*_j\, dz_j =d{\bf z}^\dag\!\cdot d{\bf z}\,.
\end{equation}

Now consider a fermionic Gaussian integral of the form
$$
I_{\bf F} =\iint\!\prod_j dz^*_j dz_j\, e^{\textstyle i\,{z^*}^r\! F_r^s\, z_s} = \iint\!\prod_j dz^*_j dz_j\,
e^{\textstyle i\,{\bf z}^\dag\! {\bf F}\,{\bf z}} \,,
$$
where ${\bf F}= \big[F_r^s\big]$ is a hermitian matrix of dimension $n$. Hence,
a unitary matrix ${\bf U}$ exists, which satisfies equation (\ref{unitinvariance}) and ${\bf UFU}^\dag =
{\bf F}'= \hbox{Diag}(f_1, f_2, \dots, f_n)$, where $\{f_1, f_2, \dots, f_n\}$ are the eigenvalues
of $\bf F$.

Then, expanding in series of Taylor the exponential and using Eq (\ref{simplestgaussian}), we find
\begin{equation}
\label{biggaussian}
I_{\bf F}=\iint\!\prod_i dz^*_i dz_i e^{\textstyle i\sum_j\! z^*_j z_j\, f_j}\! =\!
\iint \prod_i dz^*_i dz_i\prod_j\big(1\!-\!i\, z_j\, z^*_j\, f_j\big) =\hbox{Det}(-i{\bf F})\,.
\end{equation}

Now consider a Gaussian integral of the form
\begin{equation}
\label{IGwithetas}
I_{\bf F} =\iint \prod_j dz^*_j\, dz_j\,e^{\textstyle i\,\big({\bf z}^\dag {\bf F} {\bf z} +
{\bf z}^\dag\!\cdot\!{\boldsymbol \eta} + {\boldsymbol \eta}^\dag\!\cdot\!{\bf z}\big)}\,,
\end{equation}
where $G$--vectors ${\boldsymbol \eta}^\dag$ and ${\boldsymbol \eta}$ do not contain any $z_i$
or $z^*_i$. Using the identity
$$
{\bf z}^\dag {\bf F}\, {\bf z} + {\bf z}^\dag\!\cdot{\boldsymbol \eta}
+ {\boldsymbol \eta}^\dag\!\cdot{\bf z} \equiv
\big({\bf z}^\dag+ {\boldsymbol \eta}^\dag{\bf F}^{\!-1}\big) {\bf F}\big( {\bf z} +
{\bf F}^{\!-1}{\boldsymbol \eta}\big)-{\boldsymbol \eta}^\dag{\bf F}^{\!-1}{\boldsymbol \eta}\,,
$$
and exploiting the commutative property of bilinears, as exemplified in Eq (\ref{commutbilin}),
we can rewrite Eq (\ref{IGwithetas}) in the form
$$
I_{\bf F}  =e^{\textstyle -i\,{\boldsymbol \eta}^\dag{\bf F}^{\!-1}{\boldsymbol \eta}}\iint
\prod_j dz^*_j\, dz_j\, e^{\textstyle i\,\big({\bf z}^\dag +{\boldsymbol\eta}^\dag{\bf F}^{\!-1}\big)
{\bf F}\big( {\bf z} +{\bf F}^{\!-1}{\boldsymbol \eta}\big)}\,.
$$
Expanding the exponential in the double integral, and then carrying out the integration,
we easily realize that the terms proportional to ${\boldsymbol \eta}$ and
${\boldsymbol \eta}^\dag$ vanish. So we obtain
\begin{equation}
\label{biggaussian2}
I_{\bf F}  =\hbox{Det}(-i{\bf F})\,
e^{\textstyle -i\,{\boldsymbol \eta}^\dag{\bf F}^{\!-1}{\boldsymbol \eta}}=e^{\textstyle \hbox{Tr}\ln(-i{\bf F})
-i\,{\boldsymbol \eta}^\dag{\bf F}^{\!-1}{\boldsymbol \eta}}\,,
\end{equation}
which is the fermionic analog of the bosonic Gaussian integral (\ref{xAxboson}), i.e.,
\begin{equation}
I_{\bf B}=\!\int e^{\textstyle i\big[\frac{1}{2}\,\tilde{\bf x}\,({\bf B}\!+\!i\epsilon){\bf x}\!
+\!\tilde{\bf x}\!\cdot\!{\bf y}\!+\!\tilde{\bf y}\!\cdot\!{\bf x}\big]}d^n{\bf x}\!=\!
\frac{e^{\textstyle -\frac{i}{2}\,\tilde{\bf y}\,{\bf B}^{-1}{\bf y}}}{\sqrt{\hbox{Det}(i{\bf B}/2\pi)}}
= e^{\textstyle  -\frac{1}{2}\,\hbox{Tr}\ln(i{\bf B}/2\pi)\!-\! \frac{i}{2}\,\tilde{\bf y}\,
{\bf B}^{-1}{\bf y}}.\nonumber
\end{equation}

\subsection{Gaussian integrals of Dirac fields and Majorana neutrinos}
Path integrals over fermion fields require that the Lagrangian densities be implemented in the
$G$--number domain and the fermion fields be represented as linear combinations of $G$--units.
The relevant issue with this state of affairs is that, while in the $c$--number domain the Dirac
Lagrangian density ${\cal L}_D(x) =\psi^\dag(x)\,\gamma^0 \big(i\,\slashed\partial - m\big)\,
\psi(x)$ is functionally equivalent to ${\cal L}^\dag_D(x)$, in the $G$--number domain it is
instead functionally equivalent to $-{\cal L}^\dag_D(x)$. Therefore, the appropriate expression
for the Lagrangian density in the second domain is ${\cal L}_D(x)=\psi^\dag(x)\, \gamma^0
\big(\slashed\partial + i\,m\big)\,\psi(x)$. Since it is often preferable to decompose
the fermion field in its chiral components, $\{\psi_L(x)$, $\psi_R(x)\}$, or in its
conjugate components $\{\psi^c_L(x)$, $\psi^c_R(x)\}$, the $G$--number Lagrangian density
shall be written as  ${\cal L}_D(x) =\psi^\dag_L(x)\,\bar\sigma^\mu\partial_\mu\,\psi_L(x) +
\psi^\dag_R(x)\,\sigma^\mu\partial_\mu\,\psi_R(x) +i\,m\big[\,\psi^\dag_R(x)\,\psi_L(x)+
\psi^\dag_L(x)\,\psi_R(x)\big]$, or as ${\cal L}_D(x) = \psi^{c\,\dag}_R(x)\,
\bar\sigma^\mu\partial_\mu\,\psi^c_R(x)+\psi^{c\,\dag}_L(x)\,\sigma^\mu\partial_\mu\,
\psi^c_L(x) +i\,m \big[\,\psi^{c\,\dag}_L(x)\,\psi^c_R(x)+\psi^{c\,\dag}_R(x)\,
\psi^c_L(x)\big]$.
\newpage
Here are a few examples.
\begin{itemize}
\item[1.] {\em Lagrangian density of a Dirac field in G--units}:
Let ${\cal Z}(x) = [z_1(x); z_2(x); z_3(x); z_4(x)]$ be the column of $G$--units.
representing the the chiral components of a Dirac field of mass $m_D$, and ${\cal Z}^\dag =
[z^*_1(x), z^*_2(x), z^*_3(x), z^*_4(x)]$ the row of complex conjugate units
that represent their hermitian conjugate components. The Grassmann action integral can be
written as
$$
{\cal A}^D=i\!\int{\cal Z}^\dag(x)\big[\,\mathbb{I}\,\partial_0-
\mathbb{E}^D(\vec\nabla)\big]{\cal Z}(x)\,d^4x\,,
$$
where $\mathbb{I}$ is the $4\times 4$ unit matrix and
$$
{\mathbb E}^D(\vec\nabla)=
\begin{bmatrix}
\vec\sigma\cdot\vec\nabla & -i\,m_D\,{\mathbb I}_2\\
-i\, m_D\,{\mathbb I}_2 & -\vec\sigma\cdot\vec\nabla
\end{bmatrix}\equiv
\begin{bmatrix}
\partial_3 & \partial_1-i \partial_2 & -i\,m_D &0 \\
\partial_1+i \partial_2 &-\partial_3 &0 & -i\,m_D \\
-i\,m_D & 0 & -\partial_3 & -\partial_1+i\,\partial_2 \\
0 & -i\,m_D &  -\partial_1-i\,\partial_2 & \partial_3
\end{bmatrix}\!,
$$
where ${\mathbb I}_2$ is the $2\times 2$ unit matrix, is the hermitian operator which represents
the energy density of the free fermion field. Using a suitable unitary diagonal operator of the
form $\mathbb{U}(\theta, \vec n)= \mathbb{I}_4\, e^{-i(\vec \sigma\,\cdot\, \vec n)\,\theta/2}$, with
$|\vec n| =1$, we may bring the Lagrangian density matrix $\mathbb{L}^D(x)=\mathbb{I}\,
\partial_0-\mathbb{E}^D(\vec\nabla)$ to the diagonal form
$$
\mathbb{D}(x)=\begin{bmatrix}
{\mathbb I}_2\, [\partial_0-\sqrt{\nabla^2} ] & -i\, m_D\,{\mathbb I}_2\\
-i\,m_D\,{\mathbb I}_2 &  {\mathbb I}_2\,[\partial_0 +\sqrt{\nabla^2}]
\end{bmatrix}\,.
$$
The Gaussian integral is
\begin{eqnarray}
\label{IDIntegral}
I_D &\!=\!& \!\prod_x d {\cal Z}^\dag(x)\,d{\cal Z}(x)\,e^{\textstyle i{\cal A}^D}\!\! =\!
\prod_{i,x} d {\cal Z}^\dag(x)\,d{\cal Z}(x)\,{\cal Z}^\dag(x)\Big[i\,\mathbb{D}(x)
-\frac{1}{2}\,\mathbb{D}^2(x)\Big]{\cal Z}(x) =
\nonumber \\
&&\hbox{Det}\big[\,i\,(\square+m^2_D)^2\big] = e^{\textstyle \hbox{Tr}\ln\big[\,i\,(\square+m^2_D)^2\big]}\,.
\end{eqnarray}

In this case the Grassmann action integral has the form
\begin{equation}
\label{CalLML}
{\cal A}^M_L =\int\!\bigg\{\nu^\dag_L(x)\, \bar\sigma^\mu \partial_\mu\,\nu_L(x)
+\frac{i}{2}\, m_L\,\Big[\nu^\dag_L(x)\,\nu_R^{\,c}(x)
+\,\nu^{c\,\dag}_R(x)\, \nu_L(x)\Big]\bigg\}\,d^4x.
\end{equation}

Using Eqs (\ref{UL&UCRDAG}) (\ref{UR&UCLDAG}), we can rewrite Eq (\ref{CalLML}) in the form
\begin{eqnarray}
\label{CalLMLMatrix}
{\cal L}^M_L(x)&=&
\widetilde{
\begin{bmatrix}
z^*_1(x)\\
z^*_2(x)
\end{bmatrix}}\!\!
\begin{bmatrix}
\partial_0-\partial_3 & -\partial_1+ i\,\partial_2\\
 -\partial_1-i\, \partial_2 &\partial_0+\partial_3
\end{bmatrix}\!\!
\begin{bmatrix}
z_1(x)\\
z_2(x)
\end{bmatrix}\!+
\nonumber \\
&& \frac{i}{2}\,m_L\left\{
\widetilde{
\begin{bmatrix}
z^*_1(x)\\
z^*_2(x)
\end{bmatrix}}\!\!
\begin{bmatrix}
z_2^*(x)\\
-z_1^*(x)
\end{bmatrix} +
\widetilde{
\begin{bmatrix}
z_2(x)\\
z_1(x)
\end{bmatrix}}\!\!
\begin{bmatrix}
z_1(x)\\
-z_2(x)
\end{bmatrix}\right\}\,.
\nonumber
\end{eqnarray}

Since the mass term is a scalar, we can perform a Lorentz transformation of spin components $z_i(x)$ and $z^*_i(x)$
to diagonalize the matrix operator of the first term in the left--hand side of Eq (\ref{CalLMLMatrix}), so as to obtain
\begin{eqnarray}
{\cal L}^M_L(x)&=& z_1^*(x)\,\Big(\partial_0- \sqrt{\nabla^2}\,\Big) \, z_1(x) +
z_2^*(x)\, \Big(\partial_0+ \sqrt{\nabla^2}\,\Big)\, z_2(x)+\nonumber\\
& & i\, m_L\Big[z^*_1(x)\, z^*_2(x)+z_2(x)\, z_1(x)\Big]\,.\nonumber
\end{eqnarray}
Since only the terms of second and  fourth order in $z_i(x)$ and $z^*_i(x)$ contribute
to the Berezin integration of  $e^{\textstyle i\int{\cal L}^ML(x)\,d^4x} $,
we obtain the Gaussian integral
\begin{eqnarray}
I^M_L &=& \int \prod_{i,\,x} dz^*_i(x)\,dz_i(x)\, e^{\textstyle\, i\int\!{\cal L}^M_L(y)\,d^4y }=\nonumber\\
&&\int\prod_{i,\,x}  dz^*_i(x)\,dz_i(x)\,\bigg\{i\,{\cal L}^M_L(x)-\frac{1}{2}\,\Big[{\cal L}^M_L(x)\Big]^2\bigg\}
= \nonumber\\
&&\hbox{Det}[i\,(\square+ m_L^2)]= e^{\textstyle \hbox{Tr}\ln i\,(\square+ m_L^2)}\,.\nonumber
\end{eqnarray}

In a similar way we obtain for the Lagrangian density of a right--handed Majorana--neutrino
the Berezin path integral $I^M_R =\hbox{Det}[i\,(\square+ m_R^2)]=
e^{\textstyle \hbox{Tr}\ln i\,(\square+ m_R^2)}$.

\item[3.]{\em Lagrangian density of a hybrid Dirac--Majorana neutrino in G--units}.

Basing on the results of \S\,\,\ref{D-M-HYBRIDS}, prove that the Gaussian integral
of a Dirac field of mass $m_D$ mixed with a left--handed Majorana field mass $m_L$ and
a right--handed Majorana field of $m_R$ is
$$
I^{DM} = \hbox{Det}[i\,(\square+ m^2_+)^2\,(\square+ m^2_-)^2]
= e^{\textstyle \hbox{Tr}\ln \big[i\,(\square+ m^2_+)^2\,(\square+ m^2_-)^2\big]}.
$$
where
$$
m^2_\pm=\frac{2\,m_D^2+m_L^2 + m_R^2 \pm\big(m_L+m_R\big)\sqrt{4\,m_D^2+\big(m_L-m_R\big)^2}}{2}\,.
$$

\end{itemize}
\newpage

\markright{R.Nobili, Conformal General Relativity - {\bf \ref{BasFormApp}} Basic formulas of tensor calculus}
\section{BASIC FORMULAS OF TENSOR CALCULUS}
\label{BasFormApp}
For clarity, and also for the purpose of indicating sign conventions, we report from the treatise of Eisenhart (1949)
\cite{EISENHART} the basic formulae of standard tensor calculus.
It is presumed that the reader is already acquainted with the notions of metric tensor $g_{\mu\nu}$,
Christoffel symbols and covariant derivatives, here respectively denoted as $\Gamma_{\mu\nu}^\lambda$
and $D_\mu$.

\subsection{Formulary}
\label{Formulary}
\paragraph*{\centerline{\em Covariant and contravariant derivatives of mixed tensors}}
Covariant derivatives $D_\mu$ and contravariant derivatives $D_\mu$ act on $T^{\sigma\dots}_{\quad\,\lambda\dots}$
as follows:
\begin{eqnarray}
\label{Dvmu}
D_\mu T^{\sigma\dots}_{\quad\,\lambda\dots} & = & \partial_\mu T^{\sigma\dots}_{\quad\,\lambda\dots}
+ \Gamma_{\mu\rho}^\sigma T^{\rho\dots}_{\quad\,\lambda\dots} + \dots - \Gamma_{\mu\lambda}^\rho
T^{\sigma\dots}_{\quad\,\rho\dots} - \dots\\
\label{contraDvmu}
D^\mu T^{\sigma\dots}_{\quad\,\lambda\dots} & = & \partial^{\mu}T^{\sigma\dots}_{\quad\,\lambda\dots}
+ \Gamma_{\rho}^{\mu\sigma} T^{\rho\dots}_{\quad\,\lambda\dots} + \dots - \Gamma_{\lambda}^{\mu\rho}
T^{\sigma\dots}_{\quad\,\rho\dots} - \dots
\end{eqnarray}
where $ \Gamma_{\rho}^{\mu\sigma} \equiv g^{\mu\nu}\Gamma_{\nu\rho}^\sigma $. Since by definition $
D_\mu$ satisfy equations
\begin{eqnarray}
\label{Dmugdownmunu} & & D_\rho g_{\mu\nu} \equiv \partial_\rho g_{\mu\nu}-\Gamma^\lambda_{\rho\nu}
g_{\lambda\mu} -\Gamma_{\rho\mu}^\lambda g_{\lambda\nu}= 0\,,\\
\vspace{-1mm}
\label{Dmugupmunu} & & D_\mu g^{\sigma\lambda} \equiv \partial_\mu g^{\sigma\lambda}+
\Gamma_{\mu\rho}^\sigma g^{\rho\lambda} + \Gamma_{\mu\rho}^\lambda g^{\sigma\rho}= 0\,,
\vspace{-1mm}
\end{eqnarray}
we have $D_\mu \bigl(g_{\nu\lambda}\,T^{\sigma\dots}_{\quad\,\lambda\dots}\bigr) =
g_{\nu\lambda} D_\mu T^{\sigma\dots}_{\quad\,\lambda\dots}$ and $D_\mu \bigl(g^{\nu\lambda}\,
T^{\sigma\dots}_{\quad\,\lambda\dots}\bigr) = g^{\nu\lambda} D_\mu T^{\sigma\dots}_{\quad\,\lambda\dots}$
for any tensor $T^{\sigma\dots}_{\quad\,\lambda\dots}$. Since $D^\mu\dots  =
g^{\mu\nu}D_\nu \dots =D_\nu g^{\mu\nu}\dots$, the same property holds also for contravariant derivatives.
In short,  covariant derivatives carry through $g_{\mu\nu}$, $g^{\mu\nu}$ and any function of
these tensors. Thus, in particular, we have $D_\mu \big(\sqrt{-g}\,T^{\sigma\dots}_{\quad\,\lambda\dots}\big)=
\sqrt{-g}\, D_\mu T^{\sigma\dots}_{\quad\,\lambda\dots}$, where $g$ is the determinant of matrix
$\bigl[g_{\mu\nu}\bigr]$.

\paragraph*{\centerline{\em Covariant and contravariant divergences}}
Let $T^{\mu\nu\dots\rho}$ be a contravariant tensor and
$T^\mu_{\,\cdot\,\nu\dots\rho} \equiv g^{\mu\sigma}T_{\sigma\nu\dots\rho}$.
The contravariant divergence acts as follows:
\begin{eqnarray}
\label{covdiv} D_\mu T^{\mu\nu\dots\rho} &\!\! =\!\! &
\frac{1}{\sqrt{-g}}\,\partial_\mu \big(\sqrt{-g}\, T^{\mu\nu\dots\rho}\big)
+ \Gamma_{\mu\lambda}^\nu T^{\mu\lambda\dots\rho} + \dots + \Gamma_{\mu\lambda}^\rho
T^{\mu\nu\dots\lambda};\nonumber\\
\vspace{-1mm}
D^\mu T_{\mu\nu\dots\rho} &\!\! = \!\! & \frac{1}{\sqrt{-g}}
\partial^\mu \big(\sqrt{-g}\, T_{\mu\nu\dots\rho}\big)
-\Gamma_{\nu}^{\lambda\mu} T_{\mu\lambda\dots\rho} - \dots -
\Gamma_{\rho}^{\lambda\nu} T_{\mu\nu\dots\lambda}\,;\nonumber\\
\vspace{-1mm}
D_\mu T^{\mu}_{\,\,\cdot\,\nu\dots\lambda} & = &  \frac{1}{\sqrt{-g}}\,
\partial_\mu \big(\sqrt{-g}\,T^{\mu}_{\,\,\cdot\,\nu\dots\lambda}\big)  -
\Gamma_{\mu\nu}^\rho T^{\mu}_{\,\,\cdot\,\rho\dots\lambda}
- \dots-\Gamma_{\mu\lambda}^\rho T^{\mu}_{\,\,\cdot\,\nu\dots\rho}\,.
\vspace{-1mm}
\end{eqnarray}

Using Eq (\ref{Dmugdownmunu}) or (\ref{Dmugupmunu}), and Eq (\ref{Dvmu}) or (\ref{contraDvmu})
we immediately obtain
$$
\frac{1}{\sqrt{-g}}\,\partial_\mu \sqrt{-g}= \Gamma^\lambda_{\mu\lambda}\,;
\quad \frac{1}{\sqrt{-g}}\,\partial_\mu\big(\sqrt{-g}\,g^{\mu\nu}\big) =
\Gamma^\nu_{\rho\sigma}g^{\rho\sigma};\quad \Gamma^{\nu}_{\nu\mu}=\partial_\mu
\ln \sqrt{-g}\,,
$$
in the last of which also identity $\partial_\mu \ln g =g^{\rho\sigma}\,\partial_\mu g_{\rho\sigma}=
- g_{\rho\sigma}\,\partial_\mu g^{\rho\sigma}$ is exploited.

\paragraph*{\centerline{\em Shorthand notations for partial and covariant derivatives}}
Here are a few self--explanatory examples of abbreviated index notations:
\begin{eqnarray}
&&\hspace{-6mm}\partial_\mu T_\nu \equiv T_{\nu,\,\mu};\quad \partial_\mu \partial_\nu T_\lambda
\equiv T_{\lambda,\,\mu\nu};\quad D_\mu D_\nu T_\lambda \equiv T_{\lambda\,;\mu\nu};
\quad D_\mu T^\mu \equiv T^\mu_{\,\,\,;\mu};\quad D^\mu T_\mu \equiv T_\mu^{\,\,;\mu};\nonumber\\
&&\hspace{-6mm} D_\mu D^\nu T^\mu \equiv T^{\mu\, ;\nu}_{\hspace{4mm};\mu};\quad D_\mu g_{\nu\lambda}
\equiv  g_{\nu\lambda\,;\mu} =0; \quad  D_\mu g^{\nu\lambda}
\equiv  g^{\nu\lambda}_{\hspace{3mm};\mu} =0;\quad D_\mu  T^{\mu\nu\dots\rho}\equiv
T^{\mu\nu\dots\rho}_{\hspace{8mm};\mu};\nonumber \\
&&\hspace{-6mm} D_\mu T^{\mu\nu\dots\rho}\equiv  T^{\mu\nu\dots\rho}_{\hspace{8mm};\mu};
\quad D_\mu \partial_\lambda T^{\mu\nu}\equiv  T^{\mu\nu}_{\hspace{4mm},\lambda;\mu};\quad
D_\mu \partial^\lambda T^{\mu\nu}\equiv  T^{\mu\nu,\lambda}_{\hspace{6mm};\mu}\,.\nonumber
\end{eqnarray}

\paragraph*{\centerline{\em Shorthand notations for tensors with symmetric indices}}
Let us denote cyclic permutations of indices $a b c \dot\lambda$ as $[a b c \dot\lambda]$,
then the summations over a set of terms with permuted indices will be written as:
$$
T_{\mu\nu\rho} + T_{\mu\rho\nu} \equiv T_{\mu[\nu\rho]};\quad T_{\mu\nu\rho\lambda}
+ T_{\mu\lambda\rho\nu}+  T_{\mu\rho\lambda\nu} =  T_{\mu[\nu\rho\lambda]},\,\,\hbox{etc.}
$$
\paragraph*{\centerline{\em Christoffel--symbol variations}}
Let us study the relationship between the following formulae:
\begin{eqnarray}
\label{Christoff}
& & \hspace{-8mm}\Gamma_{\mu\nu}^\lambda =  \frac{1}{2}\, g^{\rho\lambda}
\bigl(\partial_\mu g_{\rho\nu} + \partial_\nu g_{\rho\mu}-\partial_\rho g_{\mu\nu}\bigr);\\
\label{deltaChristoff}
& & \hspace{-8mm}\delta\Gamma_{\mu\nu}^\lambda = \frac{1}{2}\,g^{\rho\lambda}\bigl( D_\mu
\delta g_{\rho\nu} + D_\nu \delta g_{\rho\mu}-D_\rho \delta g_{\mu\nu}\bigr);
\vspace{-2mm}
\end{eqnarray}
where $\delta g_{\mu\nu}$ are small arbitrary variations of $g_{\mu\nu}$. Since
$g_{\mu\nu}+\delta g_{\mu\nu}$ also is a metric tensor, we must have
$(g_{\mu\rho}+\delta g_{\rho\mu})(g^{\rho\nu}+\delta g^{\rho\nu})= \delta_\mu^\nu$,
which implies $g_{\rho\nu}\delta g^{\rho\mu} = -g^{\rho\mu}\delta g_{\rho\nu}$. Because
$\Gamma^\rho_{\mu\nu}+ \delta \Gamma^\rho_{\mu\nu}$ also are Christoffel symbols,
to the first order in $\delta g_{\rho\nu}$ we shall have
\vspace{-2mm}
\begin{eqnarray}
\label{ChristSymbVarApp}
&&\hspace{-6mm} \delta\Gamma^\rho_{\mu\nu}= \frac{1}{2}\, \big(g^{\rho\sigma} + \delta g^{\rho\sigma} )
\big[\partial_\mu(g_{\sigma\nu}+\delta g_{\sigma\nu})+ \partial_\nu(g_{\sigma\mu}+\delta g_{\sigma\mu})
- \partial_\sigma(g_{\mu\nu}+\delta g_{\mu\nu})\big]-\Gamma^\rho_{\mu\nu} = \nonumber\\
&&\hspace{-6mm} \frac{1}{2}\, g^{\rho\sigma}\big(\partial_\mu\delta g_{\sigma\nu}+ \partial_\nu
\delta g_{\sigma\mu}- \partial_\sigma\delta g_{\mu\nu}\big) -g^{\rho\sigma}\Gamma^\lambda_{\mu\nu}
\delta g_{\sigma\lambda},
\end{eqnarray}
where equation $g_{\rho\nu}\delta g^{\rho\mu} = -g^{\rho\mu}\delta g_{\rho\nu}$ has been
exploited in the last step.

\quad Proof: Since $\delta g_{\mu\nu}$ is a tensor (not a pseudo--tensor), we can write
$D_\mu \delta g_{\nu\lambda}= \partial_\mu \delta g_{\nu\lambda}-\Gamma_{\mu\nu}^\rho
\delta g_{\rho\lambda} - \Gamma_{\mu\lambda}^\rho \delta g_{\nu\rho}$. Eq (\ref{deltaChristoff})
can then be obtained by replacing $\partial_\mu \delta g_{\nu\lambda}$ with $D_\mu \delta g_{\nu\lambda}+
\Gamma_{\mu\nu}^\rho\delta g_{\rho\lambda} +\Gamma_{\mu\lambda}^\rho \delta g_{\nu\rho}$
in equation $\delta \Gamma_{\mu\nu}^\rho = \frac{1}{2}\,g^{\rho\lambda} \big(\partial_\mu
\delta g_{\nu\lambda} +\partial_\nu \delta g_{\mu\lambda} -  \partial_\lambda
\delta g_{\mu\nu}\big)-g^{\rho\sigma}\Gamma_{\mu\nu}^\lambda
\delta g_{\sigma\lambda}$.

\paragraph*{\centerline{\em The Christoffel symbols of a diagonal metric}}
\quad For diagonal metrics $[g_{\mu\nu}] = \mbox{diag}[h_0, h_1, \dots, h_{n-1}]$,
Eqs (\ref{Christoff}) simplify to
\begin{equation}
\label{Nonzerogammas}
\Gamma^\rho_{\mu\nu} = 0\,\,(\rho,\mu,\nu\neq);\,\,\,\Gamma^{\rho}_{\mu\rho} =
\frac{\partial_{\mu} h_{\rho}}{2\,h_{\rho}}\,\, (\rho \neq\mu);\,\,\, \Gamma^\rho_{\mu\mu} =
-\frac{\partial_\rho h_\mu}{2h_\rho}\,\,(\rho\neq \mu);
\,\,\, \Gamma^{\rho}_{\rho \rho} = \frac{\partial_{\rho} h_{\rho}}{2\,h_{\rho}};
\end{equation}
where repeated indices are not summed.

\paragraph*{\centerline{\em The Riemann tensor and its variations}}
From Eqs (\ref{deltaChristoff}) and (\ref{RiemannTensor}) we derive the
Riemann tensor and its variations
\begin{eqnarray}
\label{RiemannTensor}
& & \hspace{-10mm}R^\lambda_{\, .\,\mu\sigma\nu}  = \partial_\sigma \Gamma^\lambda_{\nu\mu}
-\partial_\nu \Gamma^\lambda_{\mu\sigma}+ \Gamma^\rho_{\mu\nu}\Gamma^\lambda_{\rho\sigma}-
\Gamma^\rho_{\mu\sigma}\Gamma^\lambda_{\rho\nu}\,;\\
\label{RiemannTensVar}
& & \hspace{-10mm}\delta R^\lambda_{\, .\,\mu\sigma\nu}  =
\frac{1}{2}\,g^{\lambda\rho}\big( D_\sigma D_\mu \delta g_{\nu\rho} + D_\sigma
D_\nu\delta g_{\rho\mu}-D_\sigma D_\rho \delta g_{\nu\mu} +\nonumber \\
& & \qquad\quad  D_\nu D_\rho\delta g_{\mu\sigma}-D_\nu D_\mu\delta g_{\rho\sigma}-
D_\nu D_\sigma\delta g_{\rho\mu} \big)\,;
\end{eqnarray}
where $\delta R^\lambda_{\, .\,\mu\sigma\nu}$ are the variations of $R^\lambda_{\, .\,\mu\sigma\nu}$
caused by metric--tensor variations $\delta g_{\nu\nu}$. Note that the Riemann tensor is skew symmetric
in the last two lower indices, $R^\lambda_{\, .\,\mu\sigma\nu}=- R^\lambda_{\, .\,\mu\nu\sigma}$.

\paragraph*{\centerline{\em The Riemann tensor of a diagonal metric}}
Riemann tensors of diagonal metrics of the form $[g_{\mu\nu}] = \mbox{diag}[h_0, h_1,
\dots, h_{n-1}]$ the simplifies as follows (Eisenhart, p.44; 1949),
%\vspace{-2mm}
\begin{eqnarray}
& & \hspace{-6mm}R_{\rho\mu\sigma\nu}=0 \quad (\rho,\mu,\sigma, \nu \neq)\,;\nonumber\\
& & \hspace{-6mm}R_{\rho\mu\mu\nu}= |h_\mu|^\frac{1}{2}\Big[\partial_\rho\partial_\nu |h_\mu|^\frac{1}{2}-
\bigl(\partial_\rho |h_\mu|^\frac{1}{2}\bigr)\partial_\nu \ln |h_\rho|^\frac{1}{2}-
\bigl(\partial_\nu |h_\mu|^\frac{1}{2}\bigr)\partial_\rho \ln |h_\nu|^\frac{1}{2}\Big]\,\,
(\rho, \mu, \nu \neq);\nonumber\\
& & \hspace{-6mm}R_{\rho\mu\mu\rho} = |h_\rho h_\mu|^\frac{1}{2}\bigg[\partial_\rho
\bigg(\frac{\partial_\rho |h_\mu|^\frac{1}{2}}{|h_\rho|^{\frac{1}{2}}}\bigg) +
\partial_\mu\bigg(\frac{\partial_\mu |h_\rho|^\frac{1}{2}}{|h_\mu|^{\frac{1}{2}}}\bigg)+
{\sum}'_\lambda \bigg(\frac{\partial_\lambda |h_\rho|^\frac{1}{2}}{|h_\lambda|^{\frac{1}{2}}}\bigg)
\partial_\lambda |h_\mu|^\frac{1}{2}\bigg]\,\, (\rho\neq\mu);\nonumber
%\vspace{-2mm}
\end{eqnarray}
where ${\sum}'_\mu$ indicates the sum for $\lambda = 0, 1,\dots , n-1$ excluding $\lambda = \mu$ and $\lambda = \rho$.

\paragraph*{\centerline{\em  Ricci tensor, Ricci scalar and their variations}}
The following equation can easily be obtained from Eqs (\ref{RiemannTensVar}):
\begin{eqnarray}
\label{RicciTensor}
&&\hspace{-12mm} R_{\mu\nu}\equiv  R^\rho_{\, .\,\mu\rho\nu}  = \partial_\rho
\Gamma^\rho_{\mu\nu} -\partial_\nu \Gamma^\rho_{\mu\rho}+
\Gamma^\sigma_{\mu\nu}\Gamma^\rho_{\sigma\rho}-
\Gamma^\sigma_{\mu\rho}\Gamma^\rho_{\sigma\nu}\,;\\
\label{RicciScalar}
&&\hspace{-12mm} R= g^{\mu\nu}\big(\partial_\rho
\Gamma^\rho_{\mu\nu} -\partial_\nu \Gamma^\rho_{\mu\rho}+
\Gamma^\sigma_{\mu\nu}\Gamma^\rho_{\sigma\rho}-
\Gamma^\sigma_{\mu\rho}\Gamma^\rho_{\sigma\nu}\big);\\
\label{Rmunuvariations}
&&\hspace{-12mm} \delta R_{\mu\nu}  =  \frac{1}{2} \bigl(D_\mu D^\rho\delta
g_{\rho\nu}+D_\nu D^\rho \delta g_{\rho\mu} - D^2\, \delta g_{\mu\nu}-D_\mu D_\nu\,
g^{\rho\sigma}\delta g_{\rho\sigma}\bigr)\,;\\
&&\hspace{-12mm}
\label{Rvariation}
\delta R \equiv \delta\big(R_{\mu\nu}g^{\mu\nu}\big) =
R_{\mu\nu}\, \delta g^{\mu\nu}+\frac{1}{2}\big(g_{\mu\nu} D^2-D_\mu D_\nu\big)\delta g^{\mu\nu}\,.
\vspace{-4mm}
\end{eqnarray}
The sign convention of the Ricci tensor adopted here matches that of
Landau--Lifchitz (1970), but is opposite to that of Eisenhart,
$R_{\mu\nu}=R^\rho_{\, .\,\mu\nu\rho} = -R^\rho_{\, .\,\mu\rho\nu}$.

From Eq (\ref{Rvariation}) we obtain the important formula:
\vspace{-2mm}
\begin{equation}
\label{useful formula}
\frac{1}{\sqrt{-g}} \frac{\delta}{\delta g^{\mu\nu}}\int \sqrt{-g}\,f\,R\,d^nx
= f\bigl(R_{\mu\nu} - \frac{1}{2}g_{\mu\nu}R\bigr)+\bigl(g_{\mu\nu}D^2 - D_\mu
D_\nu \bigr)f\,.
\vspace{-2mm}
\end{equation}

\paragraph*{\centerline{\em The identities of Bianchi}}
Here are introduced the celebrated identities discovered by L. Bianchi in his investigations of
group--theoretical properties of the Riemann tensor and its contractions \cite{BIANCHI}:

{\em The first, or algebraic, identities} (proven in \S\,\ref{LocalFlatness}):
\begin{equation}
\label{AlgBianIds}
R^\lambda_{.\,[\mu\nu\rho]} \equiv R^\lambda_{.\,\mu\nu\rho} + R^\lambda_{.\,\nu\rho\mu} +
R^\lambda_{.\,\rho\mu\nu} =0\,.
\end{equation}

{\em The second, or differential, identities} (proven in \S\,\ref{LocalFlatness}):
\begin{equation}
\label{SecBianIds}
R^\lambda_{\mu [\rho \nu;\sigma]} \equiv R^\lambda_{\mu \rho \nu;\sigma}+
R^{\lambda}_{\mu\nu\sigma;\rho}+R^{\lambda}_{\mu\sigma\rho;\nu}=0
\end{equation}
Contracting over indices $\mu$  and $\nu$, then over $\lambda$ and $\rho$, and
using the antisymmetry of the Riemann tensor in the last two indices, we arrive at
$R_{;\sigma}-R^{\rho}_{\sigma ;\rho }-R^{\nu }_{\sigma ;\nu}\equiv R_{;\sigma}-2
R^{\rho}_{\sigma ;\rho }=0$, or $R^{\rho}_{\sigma;\rho}-\frac{1}{2} R_{;\sigma}$,
i.e., the conservation equation for Einstein's gravitational tensor $G^\rho_\sigma$,
\begin{equation}
\label{SecContrBianIds}
D_\rho G^\rho_\sigma \equiv \big(R^{\rho}_{\sigma}-\frac{1}{2}\,
\delta_\sigma^\rho R\big)_{;\rho} =0\,.
\end{equation}

\paragraph*{\centerline{\em Beltrami--d'Alembert operators}}
It is the generalization of the d'Alembert operator in curved spacetimes.
\vspace{-1mm}
\begin{eqnarray}
\label{D2f}
D^2 f  & \equiv & D^\mu D_\mu f = \partial_\mu\partial^\mu f - \Gamma^{\mu}_{\mu\rho} \partial^\rho f=
\frac{1}{\sqrt{-g}}\,\partial_\mu \bigl(\sqrt{-g}\,g^{\mu\nu}\partial_\nu f\bigr)\,;\\
\vspace{-2mm}
\label{D2vrho}
D^2  v_\rho & \equiv & D_\mu D^\mu v_\rho = \frac{1}{\sqrt{-g}}\, \partial_\mu \bigl(\sqrt{-g}\,
\partial^\mu v_\rho\bigr)-\Gamma_{\rho\mu}^{\lambda} \partial^\mu v_\lambda\,;\\
\label{D2vrhoup}
D^2  v^\rho & \equiv & D_\mu D^\mu v_\rho = \frac{1}{\sqrt{-g}}\, \partial_\mu \bigl(\sqrt{-g}\,
\partial^\mu v^\rho\bigr)+\Gamma^{\rho}_{\mu\lambda} \partial^\mu v^\lambda\,.
\vspace{-2mm}
\end{eqnarray}

\paragraph*{\centerline{\em Commutators of covariant derivatives}}
Commutators of covariant derivatives act on vectors vectors $v_\rho$
and scalars $f$ as follows:
\vspace{-2mm}
\begin{equation}
\label{covdevcomms}
(D_\mu D_\nu - D_\nu D_\mu)\, v_\rho = R^\sigma_{\,\cdot\,\rho\mu\nu}\,v_\sigma\,;\quad
(D_\mu D_\nu - D_\nu D_\mu)\,f=0\,.
\vspace{-2mm}
\end{equation}
the second of which implies $D^2 D_\nu f =D_\nu D^2f$.

\newpage

\subsection{The geometric significance of the Ricci tensor}
\label{RicciTensSign}
To clarify the geometric meaning of $R_{\mu\nu}$ and $R(x)$, let us consider matrix equation
$$
\vspace{-1mm}
\big[R_{\mu\nu}(x) - c(x)\, g_{\mu\nu}(x)\big]\,\lambda^\mu(x) \equiv R_{\mu\nu}(x)\lambda^\mu(x)
- c(x)\,\lambda_\nu(x) = 0\,,
\vspace{-1mm}
$$
This has $n$ solutions, $\lambda^\mu_k(x)$, respectively associated to eigenvalues $c_k(x)$
$(k=1,2,\dots, n)$, which satisfy the orthonormalization conditions $\lambda^\mu_k(x)\lambda_{\mu h}(x)
= \delta_{kh}$. We can therefore write $R_{\mu\nu}(x) = \sum_k c_k(x) \lambda_{\mu k}(x)\lambda_{\nu k}(x)$
and interpret $c_k(x)$ as the spacetime curvatures at $x$ along {\em principal direction} $\lambda^\mu_k(x)$.
The interesting formula $R(x) = \sum_k c_k(x)$ thence follows. Since it may happen that the curvatures at
$x$ conspire to make $\sum_k c_k(x)=0$, we see that $R(x)=0$ does not necessarily imply $R_{\mu\nu}(x)=0$.
A Ricci tensor with one or more $c_k(x)=0$ will be be called degenerate. If $R_{00}(x)= R_{i0}(x)=0$, but
$R_{ij}(x)\neq 0$ ($i,j=1,2,3$), the Ricci tensor will be called {\em temporally flat}, in which case the
curvature of the spacetime is purely spatial. If $c_k(x)= \rho(x)$ for all $k$, we have $R_{\mu\nu}(x)= c(x)\,
g_{\mu\nu}(x)$, in which case the Ricci tensor is called {\em isotropic}. If $c_k$ do not depend on $x$,
we have $R_{\mu\nu}(x)= \sum_k c_k \lambda_{\mu k}(x) \lambda_{\nu k}(x)$, in which case the Ricci tensor
is called {\em homogeneous}. In $n$D, a homogeneous and isotropic Ricci tensor has the form
\begin{equation}
\label{homisoriccitensors}
R_{\mu\nu}(x)= \frac{R}{n}\, g_{\mu\nu}(x)\,,
\end{equation}
where $R$ is a constant, and the Riemann tensor has the form (Eisenhart, pp. 83, 203):
\begin{equation}
\label{constcurv}
R_{\mu\nu\rho\sigma}(x) = \frac{R}{n(n-1)}\big[ g_{\mu\rho}(x)\,g_{\nu\sigma}(x)
- g_{\mu\sigma}(x)\,g_{\nu\rho}(x)\big]\,.
\end{equation}

\subsection{Conformal--tensor calculus}
\label{ConfTensCalc}
The tensor calculus of CGR is enriched by new features, which are due to Weyl
transformation. Carrying out a Weyl transformation with scale factor $e^{\beta(x)}$,
the standard tensors of GR are transformed as follows:
\begin{eqnarray}
\label{scalefactorgmunu}
&& \hspace{-17mm}g_{\mu\nu}(x)\rightarrow \widehat{g}_{\mu\nu}(x)=
e^{2\beta(x)}g_{\mu\nu}(x);\\
\label{Gammavariations}
&& \hspace{-17mm}\Gamma^\lambda_{\mu\nu} \rightarrow \widehat{\Gamma}^\lambda_{\mu\nu}=
\Gamma^\lambda_{\mu\nu} + \delta^\lambda_\nu \partial_\mu \beta +
\delta^\lambda_\mu \partial_\nu \beta -g_{\mu\nu}\partial^\lambda\beta \,; \\
\label{tildeRiemann}
&& \hspace{-17mm}R_{\mu\rho\sigma\nu} \rightarrow \widehat{R}_{\mu\rho\sigma\nu} =
e^{2\beta}\big[R_{\mu\rho\sigma\nu}+g_{\mu\nu} A_{\rho\sigma}+g_{\rho\sigma} A_{\mu\nu}
-g_{\mu\sigma}A_{\rho\nu}-g_{\rho\nu}A_{\mu\sigma} + \nonumber\\
&&\qquad(g_{\mu\nu} g_{\rho\sigma} -g_{\mu\sigma} g_{\rho\nu})(\partial^\lambda\beta)\,
\partial_\lambda \beta\big],\,\hbox{where } A_{\mu\nu}\!=\!D_\mu\partial_\nu\beta -
(\partial_\mu\beta)\,\partial_\nu\beta;\\
\label{RmunutotildeRmu}
&& \hspace{-17mm}R_{\mu\nu}\rightarrow \widehat{R}_{\mu\nu}= R_{\mu\nu}\!
-\!(n\!-\!2)\big[D_\mu\partial_\nu\beta\!-\!(\partial_\mu\beta)\,\partial_\nu\beta\big]
\! -\!g_{\mu\nu}\big[D^2\beta+(n\!-\!2)(\partial^\rho \beta)\,\partial_\rho\beta\big]\!;\\
\label{RtotildeR}
&& \hspace{-17mm}R \rightarrow \widehat{R} = e^{-2\beta}\bigl[R-2(n\!-\!1) D^2\beta
-(n\!-\!1)(n\!-\!2)(\partial^\rho \beta)\,\partial_\rho \beta\bigr];
\end{eqnarray}
where $\delta_\mu^\nu$ is the Kronecker delta (from Eisenhart's treatise, 1949, pp.89--90,
but with opposite sign convention for $R_{\mu\nu}$ and $R$). In this subsection, all
symbols superscripted by a bar indicate the tensors changed by the Weyl transformation.

Thus, for example, the conformal counterpart of Einstein's gravitational tensor in $n$D,
$G_{\mu\nu}\equiv R_{\mu\nu} - \frac{1}{2}g_{\mu\nu}R$,  is
\begin{equation}
\label{GmunutotildeGmunu}
\widehat{G}_{\mu\nu} =G_{\mu\nu} - (n-2)\big[D_\mu\partial_\nu\beta -
(\partial_\mu\beta)\,\partial_\nu\beta\big]+g_{\mu\nu}(n-2)\Big[D^2\beta +
\frac{n-3}{2}\,(\partial^\rho\beta)\,\partial_\rho\beta\Big].
\end{equation}

Using identities
\begin{eqnarray}
& & \!\!D_\mu\partial_\nu \beta = e^{-\beta} D_\mu\partial_\nu e^{\beta} -
e^{-2\beta} (\partial_\mu e^\beta)\,\partial_\nu e^\beta= e^{-2\beta}
\big[D_\mu(e^{\beta}\partial_\nu e^{\beta}) -
2\,(\partial_\mu e^\beta)\,\partial_\nu e^\beta\big],\nonumber\\
& & \!\!D^2\beta = e^{-\beta} D^2 e^{\beta}-e^{-2\beta}(\partial^\rho
e^{\beta})\,\partial_\rho e^{\beta}= e^{-2\beta}\big[D^\rho(e^{\beta}
\partial_\rho e^{\beta})-2\,(\partial^\rho e^{\beta})\,
\partial_\rho e^{\beta}\big],\nonumber
\end{eqnarray}
Eqs (\ref{RmunutotildeRmu}) can be cast respectively in the form
\begin{eqnarray}
\label{RmunutotildeRmu3}
&&\hspace{-8mm}\widehat{R}_{\mu\nu}= R_{\mu\nu}
-(n-2)\,e^{-2\beta} \big[e^{\beta} D_\mu\partial_\nu e^{\beta} -
2\,(\partial_\mu e^\beta)\,\partial_\nu e^\beta\big]- \nonumber \\
&&\quad g_{\mu\nu}e^{-2\beta}\big[e^{\beta} D^2 e^{\beta} +(n-3)(\partial^\rho e^{\beta})
\,\partial_\rho e^{\beta}\big] \equiv \nonumber\\
&& \quad R_{\mu\nu}-(n-2)\,e^{-2\beta}\big[D_\mu(e^{\beta}\partial_\nu e^{\beta}) -
3\,(\partial_\mu e^\beta)\,\partial_\nu e^\beta\big] -\nonumber \\
&& \hspace{4mm} g_{\mu\nu}\, e^{-2\beta}\big[D^\rho(e^{\beta}\partial_\rho e^{\beta})-
(n-4)(\partial^\rho e^\beta)\,\partial_\rho e^\beta\big],
\end{eqnarray}
Eq (\ref{RtotildeR}) in the form
\begin{eqnarray}
\label{RtotildeR2}
&&\hspace{-10mm}\widehat{R} = e^{-2\beta}R -(n-1)\,e^{-4\beta}\big[(n-4)(\partial^\rho
e^{\beta})\,\partial_\rho e^{\beta}+2\,e^{\beta} D^2 e^{\beta}\big]\equiv\nonumber\\
&&\hspace{-2mm} e^{-2\beta}R -(n-1)\,e^{-4\beta}\big[(n-6) (\partial^\rho
e^{\beta})\,\partial_\rho e^{\beta}+2\,D_\mu (e^{\beta}\partial^\mu e^{\beta})\bigr],
\end{eqnarray}
and Eq (\ref{GmunutotildeGmunu}) in the form
\begin{eqnarray}
\label{RmunutotildeRmu2}
&&\hspace{-10mm}\widehat{G}_{\mu\nu} = G_{\mu\nu} -(n-2)\,e^{-2\beta}\big[e^{\beta}D_\mu\partial_\nu
e^{\beta} -2\,(\partial_\mu e^\beta)\,\partial_\nu e^\beta\big] +  \nonumber \\
&& \hspace{2mm} g_{\mu\nu} (n-2)\,e^{-2\beta}\bigg[e^{\beta}D^2 e^{\beta}+\frac{(n-5)}{2}
(\partial^\rho e^{\beta})\partial_\rho e^{\beta}\bigg]\equiv \nonumber\\
&& \hspace{2mm} G_{\mu\nu}- (n-2)\, e^{-2\beta}\big[D_\mu(e^{\beta}\partial_\nu e^{\beta})-
3\,(\partial_\mu e^{\beta})(\partial_\nu e^{\beta})\big]+ \nonumber \\
& & \hspace{2mm} g_{\mu\nu} (n-2)\, e^{-2\beta}\bigg[D_\rho(g^{\rho\tau}e^{\beta}\partial_\tau e^{\beta})+
\frac{n-7}{2}\, (\partial^\rho e^{\beta})\,\partial_\rho e^{\beta}\bigg] \,.
\end{eqnarray}

In particular, for $n=4$, we obtain
\begin{eqnarray}
\label{TildeRmunu4}
&&\hspace{-14mm}\widehat{R}_{\mu\nu}= R_{\mu\nu}+ e^{-2\beta}\!\big[4\,(\partial_\mu  e^{\beta})
\,\partial_\nu  e^{\beta}- g_{\mu\nu} (\partial^\rho  e^{\beta})
\,\partial_\rho  e^{\beta}\big]-\nonumber \\
&&\hspace{2mm}e^{-\beta}\big(2\,D_\mu\partial_\nu\ e^{\beta}+g_{\mu\nu}D^2 e^{\beta}\big);\\
\label{TildeR4}
&&\hspace{-14mm}\widehat{R} = e^{-2\beta}\bigl(R - 6\, e^{-\beta}D^2 e^{\beta}\bigr)
\equiv e^{-2\beta} R + 6\, e^{-4\beta}\bigl[(\partial^\rho e^{\beta})\,
\partial_\rho  e^{\beta} -D_\mu( e^{\beta}\,\partial^\mu e^{\beta})\bigr];\\
\label{GmunutotildeGmunu4}
&&\hspace{-14mm}\widehat{G}_{\mu\nu} =
R_{\mu\nu} - \frac{1}{2}\,g_{\mu\nu}R + e^{-2\beta}\big[4\,(\partial_\mu e^{\beta})\,
\partial_\nu e^{\beta} -\nonumber\\
&&\hspace{12mm} g_{\mu\nu}(\partial^\rho e^{\beta})\,\partial_\rho e^{\beta}\big]+2 e^{-\beta}
(g_{\mu\nu}D^2-D_\mu\partial_\nu)\, e^{\beta}.
\vspace{-2mm}
\end{eqnarray}

The covariant and contravariant derivatives of the conformal tensors mimic the standard ones:
\vspace{-2mm}
\begin{eqnarray}
\label{tildeDmutildeTheta}
\widehat{D}_\mu \widehat{T}^{\,\sigma\dots}_{\quad\,\lambda\dots} & = & \widehat{\partial}_\mu
\widehat{T}^{\,\sigma\dots}_{\quad\,\lambda\dots} + \widehat{\Gamma}_{\mu\rho}^\sigma
\widehat{T}^{\,\rho\dots}_{\quad\,\lambda\dots} + \dots - \widehat{\Gamma}_{\mu\lambda}^\rho
\widehat{T}^{\,\sigma\dots}_{\quad\,\rho\dots} - \dots \\
\label{contratildeDmutildeTheta}
\widehat{D}^{\,\mu} \widehat{T}^{\,\sigma\dots}_{\quad\,\lambda\dots} & = & \widehat{\partial}^{\,\mu}
\widehat{T}^{\,\sigma\dots}_{\quad\,\lambda\dots}
+ \widehat{\Gamma}_{\rho}^{\mu\sigma} \widehat{T}^{\,\rho\dots}_{\quad\,\lambda\dots}
+ \dots - \widehat{\Gamma}_{\lambda}^{\mu\rho}
\widehat{T}^{\,\sigma\dots}_{\quad\,\rho\dots} - \dots
\end{eqnarray}
with $\widehat{\partial}_\mu = e^{-\alpha} \partial_\mu$, $\widehat{\partial}^{\,\mu} =
e^{\alpha} \partial^\mu$, $\widehat{\Gamma}_{\rho}^{\mu\sigma} =
\widehat{g}^{\,\mu\nu}\widehat{\Gamma}_{\mu\rho}^\sigma$. Conformal--covariant derivatives carry
through $\widehat{g}_{\mu\nu}$, $\widehat{g}^{\,\mu\nu}$ and any function of these tensors

The vanishing of conformal--covariant derivatives
$\widehat{D}_\mu\widehat{g}_{\nu\lambda} = 0$, as well as the
carrying--through  properties $\widehat{D}_\mu (\widehat{g}_{\nu\lambda}
\widehat{T}^{\cdots}_{\cdots})=\widehat{g}_{\nu\lambda} \widehat{D}_\mu
\widehat{T}^{\cdots}_{\cdots}$, $\widehat{g}_{\nu\lambda} \widehat{D}_\mu
\widehat{T}^{\cdots}_{\cdots}$, $\widehat{D}_\mu (\!\sqrt{-\widehat{g}\,}\,
\widehat{T}^{\cdots}_{\cdots})=\sqrt{-\widehat{g}\,}\,\widehat{D}_\mu
\widehat{T}^{\cdots}_{\cdots}$ still hold. In particular, the conformal
covariant divergence of a conformal covariant tensor with two indices can
be written as
\begin{equation}
\label{tildeDvT2}
\widehat{D}^{\,\mu} \widehat{T}_{\mu\nu} = \frac{1}{\sqrt{-\widehat{g}}}\,
\widehat{\partial}_\mu \big(\sqrt{-\widehat{g}\,}\, \widehat{g}^{\,\mu\sigma}
\widehat{T}_{\sigma\nu}\big)-\widehat{\Gamma}_{\nu}^{\sigma\lambda}\,
\widehat{T}_{\sigma\lambda} =  \frac{1}{\sqrt{-\widehat{g}}}\,
\widehat{\partial}_\mu \big(\sqrt{-\widehat{g}\,}\,\widehat{T}^{\,\mu}_\nu\big)
-\widehat{\Gamma}_{\nu\lambda}^{\sigma} \widehat{T}^{\,\lambda}_{\sigma}\,.
\end{equation}

\paragraph*{\centerline{\em The conformal--curvature tensor of Weyl}}
The existence of three tensors $R_{\mu\nu\rho\sigma}$, $R_{\mu\nu}$ and $R$
accounting for the metric structure of spacetime poses the problem of their
characterization as components of the local curvature and of their possible
relationships. This can be evidenced by decomposing the first two tensors into
traceless components. Since $R$ is the contraction of $R_{\mu\nu}$, it is
natural to perform the decomposition $R_{\mu\nu}= E_{\mu\nu} + g_{\mu\nu} R/n$,
where $E_{\mu\nu}$ is the traceless part of $R_{\mu\nu}$.

In the same way,  we can decompose the Riemann tensor as
\vspace{-1mm}
\begin{equation}
\label{Rdecomposition} R_{\mu\nu\rho\sigma} = C_{\mu\nu\rho\sigma} +
E_{\mu\nu\rho\sigma} + F_{\mu\nu\rho\sigma}\,,
\vspace{-1mm}
\end{equation}
where $E_{\mu\nu\rho\sigma}$ depend linearly on $E_{\mu\nu}$ and
$F_{\mu\nu\rho\sigma}$ on $R$, so as to satisfy the traceless conditions
$C_{\mu\nu\rho\sigma}g^{\mu\rho}= C_{\mu\nu\rho\sigma}g^{\nu\rho} =
C_{\mu\nu\rho\sigma}g^{\mu\sigma} = C_{\mu\nu\rho\sigma}g^{\nu\sigma} =0$.

$R_{\mu\nu\rho\sigma}$ represents the local spacetime curvature caused both by
matter and gravitational waves. The Ricci tensors $R_{\mu\nu}$ and $R$ are
tied to local distribution of matter through the Einstein gravitational
equation. By contrast, in spacetime regions empty of matter and
non--gravitational fields,  $C_{\mu\nu\rho\sigma}$ represents the curvature
due to distant matter distributions and gravitational waves. As accounted for
by the traceless properties of this tensor, these waves behave as
volume--preserving tidal oscillations of spacelike volumes. Spacetime regions
where the Weyl tensor vanishes are devoid of gravitational radiation. These
are called {\em conformally flat} as the metric tensor takes the general form
$g_{\mu\nu}(x) = e^{2\alpha(x)} \bar g_{\mu\nu}$, where $\bar g_{\mu\nu}$ is a
metric tensor of Special Relativity. In these regions, all gravitational
effects are due to the immediate presence of matter or non--gravitational
fields and the variation of gravitational fields in distant regions have no
effects.

The decomposition described by Eq.(\ref{Rdecomposition}) leads to the formulae
\begin{eqnarray}
& & E_{\mu\nu\rho\sigma} = \frac{1}{n-2}(g_{\mu\rho} E_{\nu\sigma} - g_{\nu\rho}
E_{\mu\sigma} - g_{\mu\sigma} E_{\nu\rho} + g_{\nu\sigma} E_{\mu\rho})\,;\nonumber \\
& & F_{\mu\nu\rho\sigma}
=\frac{g_{\mu\rho}g_{\nu\sigma}-g_{\mu\sigma}g_{\nu\rho}}{n(n-1)}R\,.\nonumber
\end{eqnarray}
These tensors being pairwise orthogonal, we have
$$
R_{\mu\nu\rho\sigma}R^{\mu\nu\rho\sigma} = C_{\mu\nu\rho\sigma}C^{\mu\nu\rho\sigma} +
E_{\mu\nu\rho\sigma} E^{\mu\nu\rho\sigma} + F_{\mu\nu\rho\sigma}F^{\mu\nu\rho\sigma} \,,
$$
and, after simplification,
$$
R_{\mu\nu\rho\sigma}R^{\mu\nu\rho\sigma} =
C_{\mu\nu\rho\sigma}C^{\mu\nu\rho\sigma} + \frac{4}{n-2} E_{\mu\nu} E^{\mu\nu}
+ \frac{2}{n(n-1)}\,R^2\,.
$$
In terms of Ricci tensors, we have
\begin{equation}
\label{riccidecomp} R_{\mu\nu\rho\sigma}R^{\mu\nu\rho\sigma} =
C_{\mu\nu\rho\sigma}C^{\mu\nu\rho\sigma} + \frac{4}{n-2} R_{\mu\nu} R^{\mu\nu} -
\frac{2}{(n-1)(n-2)}\,R^2\,.
\end{equation}

For the purpose of evidencing a singular property of the Weyl tensor, it is preferable
to consider the {\em canonical decomposition }
\begin{equation}
\label{einsteindecomp} R^\mu_{\,\cdot\,\nu\rho\sigma} =
C^\mu_{\cdot\,\nu\rho\sigma} + G^\mu_{\cdot\,\nu\rho\sigma} +
H^\mu_{\cdot\,\nu\rho\sigma}\,,
\end{equation} with
\begin{eqnarray}
& & G^\mu_{\,\cdot\,\nu\rho\sigma} = \frac{1}{n-2}\,(\delta^\mu_\rho
R_{\nu\sigma} - g_{\nu\rho} R^\mu_{\,\cdot\,\sigma}- \delta^\mu_\sigma
R_{\nu\rho} + g_{\nu\sigma} R^\mu_{\,\cdot\,\rho})\,;\nonumber \\
& & H^\mu_{\,\cdot\,\nu\rho\sigma} = -\frac{1}{(n-1)(n-2)}\,(\delta^\mu_\rho
g_{\nu\sigma}-\delta^\mu_\sigma g_{\nu\rho})\,R\,.\nonumber
\end{eqnarray}
\newpage

\markright{R.Nobili, Conformal General Relativity - {\bf \ref{GravGaugInvApp}} Perturbations of cosmic background}
\section{PERTURBATIONS OF COSMIC BACKGROUND}
\label{GravGaugInvApp}
The need to modify the standard approach to the theory of gravitational perturbations
and gauge transformations comes from the fact that the cosmic background of CGR is flat
but not Minkowskian. In practice, this means that we must generalize the linearized
gravitational equations  excellently described for instance in Ref.\cite{MISNER}, by
replacing the partial derivatives after spacetime parameters with covariant
derivatives depending on the Christoffel symbols of a conical spacetime. We shall start
from a reformulation of a few little theorems on trace reversal and Lorentz gauge, the
revision of the important theorems on local flatness, to arrive to an improved treatment
of the theory of Newtonian potentials.

\subsection{Trace reversal of symmetric tensors in 4D--spacetime}
\label{TraceRev}
Let us introduce here a simple operator that may be of help in dealing with gravitational equations and
infinitesimal gauge transformations. For any symmetric tensor $A_{\mu\nu}$ of a 4--D spacetime we can
define its {\em trace reverse} $\overline{A}_{\mu\nu}$ characterized by the following properties:
\begin{equation}
\label{tracerevpropos}
\overline{A}_{\mu\nu} \equiv A_{\mu\nu} -\frac{1}{2}\,g_{\mu\nu}A^\lambda_\lambda\,,
\quad \overline{A}^\lambda_\lambda= -A^\lambda_\lambda\,,\quad
\overline{\overline{A}}_{\mu\nu} \equiv A_{\mu\nu}\,,\quad \overline{D^2 A_{\mu\nu}}\equiv D^2  \overline{A}_{\mu\nu}\,,
\end{equation}
where $D^2$ is the operator of Beltrami--d'Alembert. This operation can be profitably applied to
equations between symmetric tensors of GR. So, for instance, the trace reverse of equation
\begin{equation}
\label{NOTREVER}
G_{\mu\nu} = \overline{R}_{\mu\nu}\equiv R_{\mu\nu}-\frac{1}{2}\,g_{\mu\nu} R^\lambda_\lambda =
\kappa\, \mathbb{T}_{\mu\nu},\,\,\,\hbox{or }\,
G^\mu_\nu = \overline{R\,}^\mu_\nu \equiv R^\mu_\nu -\frac{1}{2}\,\delta^\mu_\nu R^\lambda_\lambda =
\kappa\, \mathbb{T}^\mu_\nu,
\end{equation}
where $\mathbb{T}_{\mu\nu}$ is an EM--tensor and $\delta^\mu_\nu$ the Kronecker delta, provides
another important gravitational equation of GR,
\begin{equation}
\label{TREVER}
R_{\mu\nu} = \kappa\,\bigg(\mathbb{T}_{\mu\nu}-\frac{1}{2}\,g_{\mu\nu}\,
\mathbb{T}^\lambda_\lambda\bigg),\,\,\,\hbox{or }\,R^\mu_\nu = \kappa\,\bigg(\mathbb{T}^\mu_\nu-
\frac{1}{2}\,\delta^\mu_\nu\,\mathbb{T}^\lambda_\lambda\bigg).
\end{equation}

\subsection{Gravitational gauge invariance and Lorentz gauge}
\label{GaugeInvGmunu}
In \S\,\ref{introduction}, the fundamental principle of GR has been introduced as
action invariance under the coordinate diffeomorphisms $x^\mu\rightarrow \bar x^\mu
= \bar x^\mu(x)$; i.e., the invertible smooth mappings of spacetime
parameters $x \equiv \{x^0, x^1, x^2, x^3\}$ that change a metric tensor
$g_{\mu\nu}(x)$ into the gravitationally equivalent metric tensor
$\bar g_{\mu\nu}[\bar x(x)]$ so as to satisfy equations
\begin{equation}
\label{GaugeTransfs}
\bar g_{\mu\nu}[\bar x(x)]\,d\bar x^\nu(x)\,d\bar x^\mu(x)
= g_{\rho\sigma}(x)\,d x^\rho\,d x^\sigma.
\end{equation}

Since these transformations are arbitrary, we can construct them in such a way that the metric tensor, or
the Christoffel symbols, will satisfy particular conditions. The most interesting and simple of these
is perhaps the {\em harmonic--gauge condition} that spacetime parameters $x^\mu$ be harmonic, i.e., satisfy
the covariant d'Alembert equation $D^2 x^\rho = 0$. By contracting the indices $\mu$ and $\nu$
of equation $D_\mu D_\nu x^\rho=\partial_\mu \partial_\nu x^\rho-\Gamma^\sigma_{\mu\nu}
\partial_\sigma x^\rho\equiv-\Gamma^\rho_{\mu\nu}$  with $g^{\mu\nu}$, we obtain the equivalent
condition
\begin{equation}
\label{covharmcondition} g^{\mu\nu}\Gamma^\sigma_{\mu\nu} =0\,.
\end{equation}
This is also known as the {\em Lorentz gauge condition}, because it is the gravitational analog of the
familiar Lorentz gauge of electrodynamics (Peacock, 1999, p.41).

An alternative formulation of this condition is provided by equation
\vspace{-1mm}
\begin{equation}
\label{Dmuginv}
\partial_\mu\big(\sqrt{-g}\,g^{\mu\nu}\big) =0\,,
\vspace{-1mm}
\end{equation}
which can be derived from equation
$$
D_\mu \big(g^{\mu\lambda}\sqrt{-g}\,\big) \equiv \partial_\mu \big(g^{\mu\lambda}\sqrt{-g}\,\big)+
\big(g^{\rho\lambda}\Gamma^\mu_{\rho\mu} + g^{\mu\rho}\Gamma^{\lambda}_{\rho\mu}-
g^{\mu\lambda}\Gamma^{\rho}_{\rho\mu}\big)\,\sqrt{-g} =0,
$$
by using Eq (\ref{covharmcondition}) in Eqs (\ref{Dmugupmunu}) of Appendix {\bf \ref{BasFormApp}}.

Now assume that metric $g_{\mu\nu}(x)$ undergoes the perturbation $g_{\mu\nu}(x) \rightarrow g_{\mu\nu}(x)
+h_{\mu\nu}(x)$, where $h_{\mu\nu}(x)$ is regarded as a deviation from the gravitational field included
in $g_{\mu\nu}(x)$. By carrying out a metric diffeomorphism of the form $x^\mu\rightarrow\bar x^\mu =
x^\mu - \xi^\mu(x)$, where $\xi^\mu(x)$ are suitable functions of $x$, we obtain the gauge transformation
$h_{\mu\nu}(x)\rightarrow h'_{\mu\nu}(x)$, where
\begin{eqnarray}
\label{SmallDiffeos}
&&\hspace{-7mm}h'_{\mu\nu}(x) = h_{\mu\nu}(x)- g_{\rho\nu}(x)\,
\partial_\mu\xi^{\rho}(x)- g_{\rho\mu}(x)\,\partial_\nu\xi^{\rho}(x)- \xi^{\rho}(x)\,
\partial_\rho g_{\mu\nu}(x)=\nonumber\\
&&\hspace{8.5mm}h_{\mu\nu}(x)- g_{\rho\nu}(x)\big[\partial_\mu-\Gamma^\lambda_{\rho\mu}(x)\,
g_{\lambda\nu}(x)\big]\xi^{\rho}(x)- g_{\rho\mu}(x)\big[\partial_\nu-
\Gamma_{\rho\nu}^\lambda(x)\,g_{\lambda\mu}(x)\big]\,\xi^{\rho}(x)=\nonumber\\
&&\hspace{8.5mm}h_{\mu\nu}(x)- D_\mu \xi_\nu(x)-D_\nu\xi_\mu(x)\,.
\vspace{-2mm}
\end{eqnarray}
Here,  we have used identities $\partial_\mu g_{\nu\lambda}-\Gamma_{\mu\nu}^\rho g_{\rho\lambda} -
\Gamma_{\mu\lambda}^\rho g_{\nu\rho}= 0$ and $\partial_\rho g_{\mu\nu}-\Gamma_{\rho\nu}^\lambda
g_{\lambda\mu} -\Gamma_{\rho\mu}^\lambda g_{\lambda\nu}= 0$ followed by a suitable rearrangement of terms.

Since four--vectors  $\xi^\mu(x)$ are arbitrary, we can cast $h'_{\mu\nu}(x)$ in some form of
particular interest. For example, by solving for $\xi_\nu(x)$ the Beltrami--d'Alembert equations,
\begin{equation}
\label{CovLorentzCond}
D^2 \xi_\nu(x) = \frac{1}{2}\, D^\mu h_{\mu\nu}(x)\,,
\end{equation}
we obtain from Eq (\ref{SmallDiffeos}) the covariant Lorentz--gauge condition $D^\mu h'_{\mu\nu}(x)=0$.
Note that the solution to Eq (\ref{CovLorentzCond}) exists because, for any function $a(x)$ and
any Beltrami--d'Alembert operator $D^2$, there is always a function $b(x)$ satisfying equation
$D^2\,b(x) = a(x)$.

A gauge transformation of this sort will be used in \S\,\ref{NwtnnApprox} to simplify the expressions of
the gravitational perturbations of a metric tensor in standard GR.

\subsection{The kinematic--time structure of the flat cosmic background}
\label{ConGraveq1}
Let us represent the cosmic background of CGR as a flat conical spacetime parameterized by
{\em polar--hyperbolic coordinates} $x=\{\tau,\varrho, \theta, \phi\}$; where $\tau$ is the kinetic
time and $\varrho, \theta, \phi$ the components of the {\em hyperbolic--Euler angle} described in
Fig.1 of \S\,~\ref{futconegeom}.  Therefore its metric and its contravariant form can be written as
\begin{eqnarray}
\label{FRWmet1}
&&\hspace{-10mm}g_{\mu\nu}(x) = \hbox{diag}\big[1, -\tau^2, -\tau^2\sinh\!\varrho^2,
- \tau^2(\sinh\!\varrho\,\sin\theta)^2\big];\\
\label{invFRWmet1}
&&\hspace{-10mm}g^{\mu\nu}(x) = \hbox{diag}\Big[1, -\frac{1}{\tau^2},
-\frac{1}{\tau^2\sinh\!\varrho^2},- \frac{1}{\tau^2(\sinh\!\varrho\,\sin\theta)^2}\Big]\,;
\end{eqnarray}
which gives $\sqrt{-g(x)}= \tau^3(\sinh \varrho)^2\sin\theta$. Therefore, we can write the 4D--volume
element as $dV = \sqrt{-g(x)}\,d\tau\,d\Omega$, where $d\Omega = d\varrho\, d\theta\, d\phi$.

This metric is not foliated into a set of parallel 3D--hyperplanes, as is the case for
the Minkowskian spacetime, but into a set of 3D--hyperboloids whose shape evolves in time.

The Beltrami--d'Alembert operator constructed from this metric, already described by Eqs
(\ref{hyperbdalambert}) and (\ref{unitlaplop}), has the form
\begin{equation}
\label{hyperbdalambertc}
D^2 f\equiv\frac{1}{\sqrt{-g}}\,\partial_\mu\Bigl[\sqrt{-g}\,g^{\mu\nu}\partial_\nu f\Bigr]=
\partial_\tau^2 f+ \frac{3}{\tau}\,\partial_\tau f- \Delta_\Omega f,
\end{equation}
where term $3\,\partial_\tau f/\tau$ works as a frictional term and
\begin{equation}
\label{unitlaplopc1}
\Delta_\Omega f \equiv \frac{1}{\tau^2\, (\sinh\varrho)^2} \bigg\{\partial_\varrho
\big[(\sinh\varrho)^2\partial_\rho f\big]+\frac{1}{\sin\theta}\,\partial_\theta
(\sin\theta\, \partial_\theta f) +\frac{1}{(\sin\theta)^2}\,\partial^2_\phi f\bigg\}
\end{equation}
is the Laplace operator already introduced in \S\,\ref{futconegeom}.

Since the metric is diagonal, we can use Eqs (\ref{Nonzerogammas}) to obtain the only
nonzero Christoffel symbols:
\begin{eqnarray}
\label{curvchristlist}
&&\hspace{-12mm} \Gamma_{01}^1=\Gamma_{10}^1=\Gamma_{02}^2 =\Gamma_{20}^2 =\Gamma_{03}^3 =
\Gamma_{30}^3 =\frac{1}{\tau};\quad\Gamma^0_{1 1} = \tau; \quad \Gamma^0_{2 2} =
\tau\,(\sinh\varrho)^2;\nonumber\\
&& \hspace{-12mm} \Gamma^0_{3 3} = \tau\,(\sinh\!\varrho\,\sin\theta)^2\,;\quad \Gamma^2_{21} =
\Gamma^2_{12} = \Gamma^3_{31} = \Gamma^3_{13} = \frac{\cosh\varrho}{\sinh\varrho}\,;
\quad \Gamma^2_{11} = -\frac{\cosh\!\varrho}{\sinh\!\varrho}\,;
\nonumber\\
&&\hspace{-12mm} \Gamma^1_{22} =  -\sinh\varrho\,\cosh\varrho;\quad \Gamma^1_{33} =
-\sinh\!\varrho\, \cosh\!\varrho\sin^2\!\theta\,;\quad \Gamma^2_{33} =
-\frac{\cosh\varrho}{\sinh\varrho}\sin\theta^2\,.
\end{eqnarray}

\subsection{The local flatness theorem and the proof of Bianchi identities}
\label{LocalFlatness}
One of the most important theorems of GR is that we can cast any given
metric tensor $g_{\mu\nu}(x)$ in the form
\begin{equation}
\label{LocFlatness}
g_{\mu\nu}(x) = \eta_{\mu\nu} + c_{\mu\nu\rho\sigma}\,x^\rho x^\sigma + \cdots,
\end{equation}
where $\eta_{\mu\nu}= \hbox{diag}[1, -1, -1, -1]$, so that $\partial_\lambda g_{\mu\nu}(x)=0$ and
$\Gamma^\lambda_{\mu\nu}(x) =0$ at $x=0$ \cite{SCHUTZ} \cite{CHENG}.

The physical meaning of this theorem is clear: {\em  at any given point $x_0$
of a spacetime, any small body can be put into an inertial state of free fall}.

The importance of this theorem lie in that, due to the vanishing  of the Christoffel symbols
at $x_0$, precisely at this point, the covariant derivatives $D_\mu$ can be replaced by $\partial_\mu$,
so that the Beltrami--d'Alembert operator $D^2$ takes the form $\square = \partial_0^2 -
\partial_1^2 -\partial_2^2 - \partial_3^2$.

%R^\lambda_{.\,[\mu\nu\rho]} \equiv R^\lambda_{.\,\mu\nu\rho} + R^\lambda_{.\,\nu\rho\mu} +
%R^\lambda_{.\,\rho\mu\nu} =0\,.

Let us apply this theorem to prove both the first and the second Bianchi identity, $R^\lambda_{.\,[\mu\nu\rho]}=0$
and $R^\lambda_{\mu [\rho \nu;\sigma]}=0$, introduced in Eqs (\ref{AlgBianIds}) and (\ref{SecBianIds}).
If we flatten the metric at a point $x_0$, Eq (\ref{RiemannTensor}) becomes $R^\lambda_{\, .\,\mu\sigma\nu}(x_0)
= \big[\partial_\sigma\Gamma^\lambda_{\nu\mu}(x)-\partial_\nu\Gamma^\lambda_{\mu\sigma}(x)\big]_{x=x_0}$, which
makes it evident that the sum of the three similar expressions, obtained by cyclical permutation of
indices $[\mu\sigma\nu]$, vanishes; which proves the validity of Eq (\ref{AlgBianIds}) at $x_0$.

In a similar way, by flattening the metric at $x_0$, the covariant derivative contracted with index
$\rho$ of Eq (\ref{RiemannTensor}), gives $\big[R^\lambda_{\, .\,\mu\sigma\nu;\rho}(x)\big]_{x=x_0}=
\big[\partial_\rho\partial_\sigma \Gamma^\lambda_{\nu\mu}(x)-\partial_\rho\partial_\nu
\Gamma^\lambda_{\mu\sigma}(x)\big]_{x=x_0}$. Here again, we find that the sum of the three similar
expressions obtained by cyclical permutations of indices $[\sigma\nu\rho]$ is just zero; which proves
the validity of Eq (\ref{SecBianIds}) at $x_0$.

It is then evident that by carrying out the inverse gauge transformation, we can restore
both the original metric tensor and the original Christoffel symbols at $x_0$. Which proves the
validity of Bianchi identities in all reference frames.

\subsection{The method of linearized gravity}
\label{NwtnnApprox}
The local flatness theorem allows us to replace the tensor calculus of GR with its weak field
approximation; i.e., replacing $g_{\mu\nu}(x)$  with $\eta_{\mu\nu}$ and the covariant derivatives
$D_\mu$ with the partial derivatives $\partial_\mu$. After carrying out all the computations on
the formalism so linearized, the covariant formalism can be fully restored by applying the
inverse procedure.

By linearizing  Eq (\ref{ChristSymbVarApp}), after replacing $\delta g_{\mu\nu}(x)$ with $h_{\mu\nu}(x)$,
we obtain the linearized Christoffel--symbol $\delta\Gamma^\rho_{\mu\nu}= \frac{1}{2}\,
\big(\partial_\mu h^\rho_\nu+\partial_\nu h^\rho_\mu- \partial^\rho h_{\mu\nu}\big)$.  Here and next,
the $\delta$ before any quantity represented by a capital letter indicates that the quantity is small,
not a variation.

By applying the same procedure to Eq (\ref{Rmunuvariations}), we obtain
\vspace{-1mm}
\begin{equation}
\label{deltaRmunu}
\delta R_{\mu\nu} = \delta\Gamma^\rho_{\mu\nu,\rho}-\delta\Gamma^\rho_{\mu\rho,\nu}
=\frac{1}{2}\big(h^\rho_{\mu,\rho\nu}+h^\rho_{\nu,\rho\mu}-
h_{\mu\nu,\rho}^\rho-h_{,\mu\nu}\big),
\vspace{-1mm}
\end{equation}
where the indices after comma denote partial derivatives and  $h\equiv h^\rho_\rho = \eta^{\rho\sigma}h_{\rho\sigma}$.

Since the linearized form of the Ricci--scalar is $\delta R =\eta^{\rho\sigma}\delta R_{\rho\sigma}$,
we can write extensively the gravitational tensor as
\vspace{-1mm}
\begin{equation}
\label{dGmunuAtx0}
\delta G_{\mu\nu}(x) =\delta R_{\mu\nu}(x)-\frac{1}{2}\,\eta_{\mu\nu}\delta R(x)
\equiv \delta\bar{R}_{\,\mu\nu}(x)\,,
\vspace{-1mm}
\end{equation}
where $\delta \overline{R}_{\,\mu\nu}$ is the trace reverse of Ricci tensor $\delta R_{\mu\nu}$,
as already defined in \S\,\ref{TraceRev}.

Thereby, we find that the linearized expression of the perturbative gravitational equation has the general form
\vspace{-1mm}
\begin{equation}
\label{flattGmunu}
\delta G_{\mu\nu}(x)=  \frac{1}{2}\,\partial_\rho\big[\bar h^\rho_{\mu,\nu}(x) +\bar h^\rho_{\nu,\mu}(x)-
\bar h_{\mu\nu}^\rho(x)\big]-\bar h_{,\mu\nu}(x)= \kappa\,{\mathbb T}_{\mu\nu}(x)\,.
\vspace{-1mm}
\end{equation}
where $\bar h_{\mu\nu}(x)= h_{\mu\nu}(x)- \frac{1}{2}\,\eta_{\mu\nu} h(x)$ is the trace reverse of
$h_{\mu\nu}(x)$, $T_{\mu\nu}(x)$ is the linearized EM tensor at $x$ and $\kappa$ is the gravitational
coupling constant.

By recovering the covariant form of Eq (\ref{flattGmunu}), we obtain the gravitational equation
\vspace{-1mm}
\begin{equation}
\label{Rmunu&Tmunu}
\delta G_{\mu\nu}(x)= \frac{1}{2}\,\big[\bar h^\rho_{\mu;\rho\nu}(x) +\bar h^\rho_{\nu;\mu\rho}(x)-
\bar h_{\mu\nu;\rho}^\rho(x)\big]-\bar h_{;\mu\nu}(x)= \kappa\,{\mathbb T}_{\mu\nu}(x)\,,
\vspace{-1mm}
\end{equation}
where the indices after semicolons denote covariant derivatives. Since all terms in squared brackets
that appear in the second step, are obtained by carrying out covariant derivatives of
$\bar h^\rho_{\mu;\rho}(x)$, we can impose the Lorentz--gauge condition $\bar h^\rho_{\mu;\rho}(x) =0$.

Therefore, on account of Eqs (\ref{tracerevpropos}) and (\ref{TREVER}), Eq (\ref{Rmunu&Tmunu})
and its trace reverse can be rewritten as
\begin{equation}
\label{DeltaCvsBarT00}
- \bar h_{;\mu\nu}(x) \equiv - D^2 \bar h_{\mu\nu}(x)
=\kappa\,{\mathbb T}_{\mu\nu}(x)\,,\quad - D^2 h_{\mu\nu}(x)
=\kappa\bigg[{\mathbb T}_{\mu\nu}(x) - \frac{g_{\mu\nu}(x)}{2}\,{\mathbb T}^\lambda_\lambda(x)\bigg].
\end{equation}
where $D^2$ is the Beltrami--d'Alembert operator (Misner {\em et al.}, p.436, 1973).
\newpage

\markright{R.Nobili, Conformal General Relativity - {\bf \ref{HomIsotSptApp}} Gravitation in expanding spacetimes}
\section{GRAVITATION IN EXPANDING SPACETIMES}
\label{HomIsotSptApp}
The {\em cosmological principle} asserts that, disregarding the {\em peculiar motions} of stars
and galaxies, which differ from each other by not more than a few hundreds Km/sec, the universe
on the large scale expands isotropically with respect to all comoving observers of today, and --
by extrapolation -- to all ideal comoving observers of the previous epochs.

This suffices to state that the average matter density $\rho(t)$ and pressure $p(t)$ on the large
scale is homogeneous in each spacelike surface $\Sigma(t)$ at any  time $t$. To prove
this, consider a pair of intersecting spheres centered at two comoving observers; since the density
in each sphere is the same by isotropy, it is same also in the intersection. By using spheres
with different radii around each point, we can cover with intersections the entire universe.

This principle, however, does not determine the shape of $\Sigma(t)$, neither does it predict how
these spacelike surfaces should evolve in time. These depend instead on the topological structure of
the spacetime, which is presumed to be {\em cylindrical} in the Standard Model of Modern Cosmology
(SMMC), but {\em conical} in CGR. In this Appendix, we investigate the the structural difference
between these two sorts of spacetimes with regard to the dependence of gravitational field depend on
$\rho(t)$ and $p(t)$.

As we shall show, the relevant difference is that in CGR the Hubble parameter depends on the
expansion factor of the universe, $a(t)$, very differently from the SMMC, to the point
that the gravitational potentials themselves come to depend explicitly on $a(t)$.

\subsection{Gravitation in cylindrical spacetimes}
\label{CylGraveq}
If we assume that the universe is homogeneous and isotropic on large scales, and governed by the
gravitational equation of GR, we are lead to Friedmann--Robertson--Walker (FRW) metrics of the form
\begin{equation}
\label{standardRWmetric}
ds^2 = dt^2 - a(t)^2\bigg[\frac{dr^2}{1- k r^2} + r^2 d\theta^2 +r^2\big(\sin\theta\big)^2 d\phi^2 \bigg]\,.
\end{equation}
Here, $t$ is the proper time of comoving observers, $a(t)$ is a non--negative expansion factor, which in the SMMC
is called the scale factor, with dimension of length, $r$ is not the radius of a polar coordinate system,
but an adimensional parameter; so, we can regard $R(t)=a(t)\,r$ as the evolving radius of curvature of the
3D--space;  $\{\theta, \phi\}$ are Euler's angles; $k = 1, 0, -1$, according as the 3D--space curvature is
positive, zero or negative. % (Peacock, 1999).

These metrics can be called {\em cylindrical} because they represent the spacetime as foliated into
a sets of hyperplanes orthogonal to the time axis. Here we focus only on FRW models with $k=0$ and
zero cosmological constant because different values of these constants are incompatible with CGR.
Therefore, the metric matrix of this model, its inverse and the squared line element can be written as
\begin{eqnarray}
\label{RWmet}
&&\hspace{-14mm}g_{\mu\nu}(x) = \hbox{diag}\big[1, -a(t)^2, -a(t)^2, -a(t)^2\big];\\
\label{invRWmet}
&&\hspace{-14mm}g^{\mu\nu}(x) = \hbox{diag}\Big[1, -\frac{1}{a(t)^2}, -\frac{1}{a(t)^2},
-\frac{1}{a(t)^2}\Big];\\
\label{ds2RW}
&&\hspace{-14mm} ds^2(t) = dt^2 - a(t)^2\big(dx^2 + dy^2 + dz^2\big) \equiv
dt^2 - a(t)^2 dr^2;
\end{eqnarray}
whence, $\sqrt{-g(x)}= a(t)^3$. So, if $a(\tau)=1$ the metric is Minkowskian.

The Beltrami--d'Alembert operator constructed out of this metric is
\vspace{-1mm}
\begin{equation}
\label{RWdalambert}
D^2 f\equiv\frac{1}{\sqrt{-g}}\,\partial_\mu\Bigl[\sqrt{-g}\,g^{\mu\nu}\partial_\nu f\Bigr]=
\ddot f + 3\,\frac{\dot a}{a}\,\dot f- \frac{\Delta f}{a^2},
\vspace{-1mm}
\end{equation}
where $f$ is any function of spacetime parameters $x$, $\dot f \equiv \partial_t f$,
$\ddot f \equiv \partial_t^2 f$, and $\Delta$ denotes the  Laplace operator $\nabla^2=
\partial^2_x+\partial^2_y +\partial^2_z$. Note that $3\,(\dot a/a)\,\dot f$ has the form of a
frictional term.

To obtain the gravitational equation associated with this metric, we must first derive the
Christoffel symbols. Since metric (\ref{RWmet}) is diagonal, we can use the formulas described
by Eqs (\ref{Nonzerogammas}) of Appendix {\bf \ref{BasFormApp}}, which yield the only nonzero symbols
\vspace{-1mm}
\begin{equation}
\label{RWchristlist}
\Gamma_{01}^1=\Gamma_{02}^2 =\Gamma_{03}^3  =\Gamma_{10}^1=\Gamma_{20}^2 = \Gamma_{30}^3
=\frac{\dot a}{a};\quad\Gamma^0_{1 1} = \Gamma^0_{2 2} = \Gamma^0_{3 3} =a \dot a\,,
\vspace{-1mm}
\end{equation}
then Eqs (\ref{RicciTensor}), to derive the components of the Ricci tensor and Ricci scalar
\vspace{-1mm}
\begin{equation}
\label{RWRicciTens}
R_{00}=- 3\,\frac{\ddot a}{a}\,;\quad R_{ij}= g_{ij}\bigg(2\,\frac{\dot a^2}{a^2}+
\frac{\ddot a}{a}\bigg)\,\,\,(i, j =1,2,3);\quad R\equiv R^\mu_\mu =
-6\bigg[\bigg(\frac{\dot a}{a}+\frac{\ddot a}{a}\bigg)^{\!\!2}\bigg].
\vspace{-1mm}
\end{equation}

Therefore, the components of the gravitational tensor, $G_{\mu\nu} = R_{\mu\nu} -\frac{1}{2}\,g_{\mu\nu}R$,
are
\vspace{-1mm}
\begin{equation}
\label{RWRGravTens}
G_{00} = 3\,\bigg(\frac{\dot a}{a}\bigg)^2\,;\quad G_{ij} = g_{ij}\bigg(2\,\frac{\ddot a}{a}+
\frac{\dot a^2}{a^2}\bigg)\,\,\,(i, j =1,2,3);\quad G \equiv G^\mu_\mu = 6\,\frac{\ddot a}{a}\,.
\vspace{-1mm}
\end{equation}

If we regard the matter field as a homogeneous and isotropic fluid with energy density $\rho(t)$,
pressure $p(t)$ and 4--velocity $u_\mu(t)$, we can denote the EM--tensor ${\mathbb T}_{\mu\nu}$,
as a function depending only on $t$, and write the gravitational equation
\vspace{-2mm}
\begin{equation}
\label{CSTTmunu}
G_{00}(x)  = \kappa\,{\mathbb T}_{\mu\nu}(t)\equiv\kappa \big[\rho(t)+p(t)\big]\, u_\mu(t)\,
u_\nu(t) - g_{\mu\nu}(t)\, p(t)\,,
\vspace{-2mm}
\end{equation}
where $\kappa$ is the gravitational coupling constant.

\newpage

\noindent Since for non--relativistic fluid velocities we have $u_\mu\cong \{1,0,0,0\}$,
Eq (\ref{CSTTmunu}) simplifies to
\begin{equation}
\label{GReqsOfFluid}
G_{00}(t) \equiv 3\bigg[\frac{\dot a(t)}{a(t)}\bigg]^2\!\!\!=\!\kappa\,\rho(t);
\,\,\,G_{ij}(t) \equiv 2 g_{ij}\!\bigg[\frac{\ddot a(t)}{a(t)}+\frac{\dot a(t)^2}{a(t)^2}\bigg] =
\kappa\,g_{ij}p(t)\,\,\,(i, j =1,2,3).
\end{equation}
The first of these provides the the expansion rate of the universe, i.e., the {\em Hubble parameter}
\begin{equation}
\label{RWHubblePar}
H(t) = \frac{\dot a(t)}{a(t)} =  \sqrt{\frac{\kappa}{3}\,\rho(t)}\,.
\end{equation}

The trace reverse of Eqs (\ref{GReqsOfFluid}) provides the Friedmann--Lema\^itre equation
\cite{FRIEDMANN} \cite{LEMAITRE},
\begin{equation}
\label{FLRWeqs}
\hspace{18mm}R_{00}(t)= -3\,\frac{\ddot a(t)}{a(t)} = \frac{\kappa}{2}\,\big[\rho(t)  + 3\,p(t)\big];
\quad \hbox{(cf. \S\,\ref{TraceRev} of Appendix {\bf \ref{GravGaugInvApp}})}.
\end{equation}

Note that the expansion factor $a(t)$ may vary considerably in the course of time and that both
$H(t)$ and $R_{00}(t)$ remain unvaried if $a(t)$ is multiplied by a constant. This means that the
peculiar values of this factor have no objective meaning. It is therefore customary and convenient
to assume $a(t_U)=1$ at the present age of the universe $t_U$.

Equations (\ref{GReqsOfFluid}) and (\ref{FLRWeqs}) describe the expansion of the universe as a
side effect of the average energy density and pressure of the matter field. Since in this
representation the celestial bodies and their peculiar motions are neglected, Eqs (\ref{GReqsOfFluid})
and (\ref{FLRWeqs}) provide only a description of the {\em cosmic background} of the RW universe
(cf. \S\S\,\,\ref{CGRafterBB} and \ref{MachEinstPrinc}).

To correct the representation of this desolate landscape, we should add to $T_{\mu\nu}(t)$ one or more
terms representing the contributions from the celestial bodies. The simplest of which is a point--like
particle of mass $m$; for instance, a star or a black hole resting at $\vec r =0$.
In this case, the energy density of the cosmic background changes as follows:
\begin{equation}
\label{rhoPlusDeltaRho}
\rho(t) \rightarrow \rho'(x) = \rho(t) + m\, \delta^3(\vec r);\quad p(t) \rightarrow p(t)\,.
\end{equation}

More sophisticate corrections can be introduced by adding to $T_{\mu\nu}(t)$ a Thirring EM tensor
$t_{\mu\nu}(x)$, describing a set of pointlike particles of mass $m_i$ moving along trajectories
of equation $x^\mu= z^\mu_i(s_i)$, where $s_i$ is the proper times of particle $i$ as measured
by an ideal observer whose reference system is solid with it \cite{THIRRING} \cite{LANDAU1}.
Hence, in summary, we have
\begin{equation}
\label{ThirrCorr}
t_{\mu\nu}(x)=\sum_i m_i \!\int_0^\infty\!\delta^3\big[\vec x - \vec z_i(s_i)\big]
u_{i\mu}(s_i)\, u_{i\nu}(s_i)\, ds_i \,,
\end{equation}
where $\delta^3$ is the 3D Dirac delta and $u_{i\mu}(s_i)$ is the covariant 4--velocity
of particle $i$.

\subsection{Gravitational perturbations of cylindrical spacetimes}
\label{CylGravPerts}
Consider a metric tensor of the form
\vspace{-1mm}
\begin{equation}
\label{PertRWmet}
\bar g_{\mu\nu}= \hbox{diag}\big[1, -a(t)^2, -a(t)^2, -a(t)^2\big] + h_{\mu\nu}
\vspace{-1mm}
\end{equation}
where $h_{\mu\nu}$ represents the gravitational perturbation
caused by the celestial bodies. Since $\bar g_{\mu\nu}$ must satisfy equation  $\bar g^{\mu\lambda}\,
\bar g_{\lambda\nu} = \delta^\mu_\nu$, we find that, to the first order in $h_{\mu\nu}$,
the contravariant form of $\bar g_{\mu\nu}$ can be written as
$$
\bar g^{\mu\nu}=
\hbox{diag}\big[1, -\frac{1}{a(t)^2}, -\frac{1}{a(t)^2}, -\frac{1}{a(t)^2}\big] -h^{\mu\nu}.
$$
Then, using  Eq~(\ref{ChristSymbVarApp}) of Appendix~{\bf \ref{BasFormApp}}
with $\delta g_{\mu\nu}$ replaced by  $h_{\mu\nu}$, we obtain the Christoffel--symbol
variations $\delta \Gamma_{\mu\nu}^\rho =\frac{1}{2}\,g^{\rho\lambda}
\big(\partial_\mu h_{\nu\lambda} + \partial_\nu h_{\mu\lambda}-
\partial_\lambda h_{\mu\nu}\big)- g^{\rho\sigma}\Gamma_{\mu\nu}^\lambda h_{\sigma\lambda}$,
and from these, using Eq~(\ref{Rmunuvariations}), we obtain the perturbation of the Ricci
tensor and its trace reversed form, defined in \S\,\ref{tracerevpropos} of Appendix
{\bf \ref{GravGaugInvApp}}, i.e., respectively
\vspace{-1mm}
\begin{eqnarray}
\label{CylRicciTensPert}
&&\hspace{-8mm} \delta R_{\mu\nu}  =  \frac{1}{2} \bigl(D_\mu D^\rho\,h_{\rho\nu}+
D_\nu D^\rho h_{\rho\mu} - D^2 h_{\mu\nu}-D_\mu D_\nu\,
g^{\rho\sigma}h_{\rho\sigma}\bigr)\,,\\
\label{CylRicciTensPertRev}
&&\hspace{-8mm} \delta G_{\mu\nu}\equiv\delta \overline{R}_{\mu\nu}  =  \frac{1}{2}
\bigl(D_\mu D^\rho\,\bar h_{\rho\nu}+D_\nu D^\rho \bar h_{\rho\mu} -
D^2 \bar h_{\mu\nu}- D_\mu D_\nu\,g^{\rho\sigma}\bar h_{\rho\sigma}\bigr)\,,
\vspace{-1mm}
\end{eqnarray}
where $D_\rho$ denote covariant derivatives, $D^2 = D^\rho D_\rho$ denotes the Beltrami--d'Alembert
operator described by Eq (\ref{RWdalambert}) and $\delta G_{\mu\nu}$ denotes the perturbation of
the gravitational tensor. Therefore, the first component of the total perturbed Ricci tensor
and that of its trace reversed counterpart are
\vspace{-1mm}
\begin{eqnarray}
\label{CylR00Appwithc}
&&\hspace{-8mm}  R^0_0+ \delta R^0_0=  -3\,\frac{\ddot a}{a}+ \frac{D^2 h^0_0}{2\,a^2}-
3\,\frac{\dot a}{a}\,\dot h^0_0+\partial_i \dot h^i_0 -\frac{1}{2}\,\ddot h^i_i\,,\\
\label{CylG00Appwithc}
&&\hspace{-8mm} G^0_0+ \delta  G^0_0= 3\bigg[\frac{\dot a(t)}{a(t)}\bigg]^2+
\frac{D^2 \bar h^0_0}{2\,a^2}-3\,\frac{\dot a}{a}\,\dot {\bar h}^0_0+
\partial_i \dot {\bar h}^i_0 -\frac{1}{2}\,\ddot {\bar h}^i_i\,,
\vspace{-1mm}
\end{eqnarray}
where $D^2$ is the operator of Beltrami--d'Alembert and
\vspace{-1mm}
\begin{equation}
R^0_0(t)=  -3\,\frac{\ddot a(t)}{a(t)};\quad  G^0_0(t) =
3\bigg[\frac{\dot a(t)}{a(t)}\bigg]^2,
\nonumber \\
\vspace{-1mm}
\end{equation}
in accordance with Eqs (\ref{CSTTmunu}) and (\ref{FLRWeqs}),

In addition, in virtue of Eqs (\ref{DeltaCvsBarT00}) of Appendix {\bf \ref{GravGaugInvApp}}, we have
\vspace{-1mm}
\begin{eqnarray}
\label{deltaR00ofx}
&&\hspace{-12mm}\delta R^0_0(x)\equiv\frac{D^2 h^0_0(x)}{2\,a(t)^2}-3\,\frac{\dot a(t)}{a(t)}\,
\dot h^0_0(x)+\partial_i \dot h^i_0(x) -\frac{1}{2}\,\ddot h^i_i(x)=\kappa\,\frac{\delta \rho(x)+
3\,\delta p(x)}{2}; \\
\label{deltaG00ofx}
&&\hspace{-12mm} \delta  G^0_0(x) \equiv \frac{D^2 \bar h^0_0(x)}{2\,a(t)^2}-3\,\frac{\dot a(t)}{a(t)}\,
\dot {\bar h}^0_0(x)+\partial_i \dot {\bar h}^i_0(x) -\frac{1}{2}\,\ddot {\bar h}^i_i(x)
=\kappa\,\delta\rho(x);
\vspace{-1mm}
\end{eqnarray}
where $\delta\rho(x)$ and $\delta p(x)$ are respectively the corrections to energy
density and pressure caused by the matter field in the weak field approximation.

\newpage

If the matter field is formed by celestial bodies slowly moving with respect to the speed
of light, and their gravitational effects are weak and independent of time -- in other
terms, if the celestial bodies can be regarded as static spheres -- Eqs~(\ref{deltaR00ofx})
and (\ref{deltaG00ofx}) can be further simplified by expressing the gravitational field as
a Newtonian potential $\Phi(x)$, regarded as a perturbation of the cosmic background.
In these circumstances, $\delta p(x)=0$, and $\dot{\bar h}^0_0(x)$, $\bar h^i_0(x)$,
$\bar h^i_j(x)$ are negligible, and $\bar h^\mu_\nu(x)=\hbox{diag}\big[\bar h^0_0(x),
0,0,0\big]$. Consequently, tensor $h^\mu_\nu$ is related to its trace reverse
$\bar h^\mu_\nu$ as follows:
$$
h^0_0 =  \bar h^0_0 - \frac{\bar h^0_0}{2}=\frac{\bar h^0_0}{2};
\quad h^1_1 =h^2_2 =h^3_3 =  - \frac{\bar h^0_0}{2};
\quad h^i_j=0,\,\,\hbox{for } i\neq j\,;
$$
i.e., extensively, $h^\mu_\nu(x)=h^0_0(x)\,\hbox{diag}\big[1,1,1,1\big]$. Since in the Newtonian
approximation $\delta G^0_0(x)$ is related to  $\bar h^0_0(x)$ and $h^0_0(x)$ and $\delta\rho(x)$
by equations
\begin{equation}
\label{CylGravPotEq}
\delta G^0_0(x) = -\frac{\nabla^2 \bar h^0_0(x)}{2\,a(t)^2} =  -\frac{\nabla^2 h^0_0(x)}{a(t)^2}=
\kappa\,\delta\rho(x)\,,
\end{equation}
where $\nabla^2$ is the operator of Laplace, putting $x=\{t, \vec r\}$ and  $\delta\rho(x) =
\delta\rho(t, \vec r)$, where $\vec r = \{x,y,z\}$, we obtain by integration
\begin{equation}
\label{NewtPotInExpST}
\frac{\bar h^0_0(\tau, \vec r)}{2} =  h^0_0(\tau, \vec r) =\kappa\!\!\int \frac{\delta\rho(t, \vec{r\,}')}
{a(t)^2|\vec r - \vec{r\,}'|}\, d^3 r'=2 \Phi(t, \vec r)\,,
\end{equation}
where $\Phi(\tau,\vec r)\equiv \Phi(x)$ is the Newtonian potential in the
expanding spacetime and $|\vec r - \vec{r\,}'|$ is the distance of the
potential amplitude from its source--element $\kappa\,\delta\rho(t,\vec{r\,}')\, d^3 r'$.

If $h_{\mu\nu}(x)$ is diagonal, the squared line element of perturbed metric (\ref{PertRWmet}) has
the form
$$
d\bar s^2(x) = dt^2\big[1+  h_{00}(x)\big] - a(t)^2\big[1- h_{00}(x)\big]\big(dr^2 +r^2d\theta^2+
r^2\sin\theta^2 d\phi^2\big),
$$
we can write
\begin{equation}
\label{ds2hmunu}
d\bar s^2(x) = dt^2\big[1+2\,\Phi(x)\big]-a^2(t)\big[1-2\,\Phi(x)\big]\big(dr^2 +r^2d\theta^2+
r^2\sin\theta^2 d\phi^2\big)\,.
\end{equation}

However, since the scale factor is defined up to a constant factor, we can assume
this factor to be such that $a(t_U)=1$ at the age of the universe today, $t_U$.
\begin{equation}
\label{ds2hmunutoday}
d\bar s^2(t_U, \vec r)  = dt^2\big[1+2\,\delta \Phi(t_U, \vec r)\big]-
\big[1-2\,\delta \Phi(t_U, \vec r)\big]\big(dr^2 +r^2 d\theta^2+r^2\sin\theta^2 d\phi^2\big)\,,
\end{equation}
where
\begin{equation}
\label{NewtPotInExpST2}
\Phi(t_U, \vec r) = \kappa\!\int\!\frac{\delta\rho(t_U, \vec {r\,}')}
{2\,|\vec r - \vec{r\,}'|}\, d^3 r'
\end{equation}
is the Newtonian potential of the celestial bodies today (Misner {\em et al.}, p.436, 1973).

\newpage

\subsection{Gravitation in (truncated) conical spacetimes after big bang}
\label{ConGraveq}
In CGR, the big bang can be envisaged as the watershed between two major stages of the cosmic history:
one dominated by the evolution of the vacuum state and the other by the evolution of the matter field.
We can therefore assume the big--bang time as the absolute time--unit of the theory. In the kinematic
time representation this time is denoted by $\tau_B$ and in the proper time representation by
$\widetilde{\tau}_B$ (cf. Section~\ref{CGRafterBB} of the main text).

In this subsection, we will neglect the dependence of CGR on the scale factor of vacuum dynamics
and only focus our attention on the dependence of the cosmic expansion on the average energy density
of the universe. This conceptual separation will help us to clarify the fundamental difference
between CGR and SMMC.

We represent the cosmic background of CGR's universe as a conical spacetime parameterized by
{\em polar--hyperbolic coordinates} $x=\{\tau,\varrho, \theta, \phi\}$, where $\tau$ is the
kinematic--time parameter and $\varrho, \theta, \phi$ the components of the {\em hyperbolic--Euler angle},
illustrated in Fig.1 of \S~\ref{futconegeom} of the main text. To account for the cosmic expansion,
we equip this metric with an expansion factor of the form $c(\tau)= a(\tau)\,\tau$, where $a(\tau)$
is the expansion factor of the universe. Therefore the metric matrix and its inverse shall be
written as
\begin{eqnarray}
\label{FRWmet}
&&\hspace{-10mm}g_{\mu\nu}(x) = \hbox{diag}\big[1, -c(\tau)^2, -c(\tau)^2\sinh\!\varrho^2,
- c(\tau)^2(\sinh\!\varrho\,\sin\theta)^2\big];\\
\label{invFRWmet}
&&\hspace{-10mm}g^{\mu\nu}(x) = \hbox{diag}\Big[1, -\frac{1}{c(\tau)^2},
-\frac{1}{c(\tau)^2\sinh\!\varrho^2},- \frac{1}{c(\tau)^2(\sinh\!\varrho\,\sin\theta)^2}\Big]\,,
\end{eqnarray}
from which we derive $\sqrt{-g(x)}= c(\tau)^3(\sinh \varrho)^2\sin\theta$.

This metric differs from the Robertson--Walker (RW) metric of the SMMC  in that the spacetime is not
foliated into a set of parallel 3D--hyperplanes, but into a set of spacelike hyperboloids whose
shape evolves in time. Note that for $a(\tau)=1$ the metric is flat.

The Beltrami--d'Alembert operator constructed from this metric can easily be obtained from Eqs
(\ref{hyperbdalambert}) and (\ref{unitlaplop}) by replacing $\tau$ with $c(\tau)=a(\tau)\,\tau$,
which yields
\begin{equation}
\label{hyperbdalambertc2}
D^2 f\equiv\frac{1}{\sqrt{-g}}\,\partial_\mu\Bigl[\sqrt{-g}\,g^{\mu\nu}\partial_\nu f\Bigr]=
\partial_\tau^2 f+ 3\,\bigg(\frac{1}{\tau} + \frac{\dot a}{a}\bigg)\partial_\tau f-
\frac{1}{a^2}\,\nabla^2_\Omega f,
\end{equation}
where $f$ is a function of $x$, $3\,\big(1/\tau + \dot a/a\big)\partial_\tau f \equiv
(3\,\dot c/c)\,\partial_\tau f$ works as a frictional term and
\begin{equation}
\label{unitlaplopc}
\nabla^2_\Omega f \equiv \frac{1}{\tau^2\, (\sinh\varrho)^2} \bigg\{\partial_\varrho
\big[(\sinh\varrho)^2\partial_\rho f\big]+\frac{1}{\sin\theta}\,\partial_\theta
(\sin\theta\, \partial_\theta f) +\frac{1}{(\sin\theta)^2}\,\partial^2_\phi f\bigg\}
\end{equation}
is the Laplace operator in polar--hyperbolic coordinates introduced in \S\,\ref{futconegeom}.

Since metric (\ref{FRWmet}) is diagonal, we can use Eqs (\ref{Nonzerogammas}) of Appendix
{\bf \ref{BasFormApp}} to obtain the only nonzero Christoffel symbols:
\begin{eqnarray}
\label{curvchristlist0}
&&\hspace{-18mm} \Gamma_{01}^1=\Gamma_{10}^1=\Gamma_{02}^2 =\Gamma_{20}^2 =\Gamma_{03}^3 =
\Gamma_{30}^3 =\frac{\dot c}{c};\quad\Gamma^0_{1 1} = c\,\dot c; \quad \Gamma^0_{2 2} =
c\,\dot c\, (\sinh\varrho)^2;\nonumber\\
&&\hspace{-18mm} \Gamma^1_{22} = -\sinh\varrho\cosh\varrho\,;\quad \Gamma^2_{21} =
\Gamma^2_{12} = \Gamma^3_{31} = \Gamma^3_{13} =\frac{\cosh\varrho}{\sinh\varrho}\,;
\quad\Gamma_{23}^3=\Gamma^3_{32}= \frac{\cos\theta}{\sin\theta}\,;
\nonumber\\
&&\hspace{-18mm}\Gamma^0_{3 3} = c\,\dot c\,(\sinh\varrho\,\sin\theta)^2\,;
\quad\Gamma^1_{33} =-\sinh\varrho\cosh\varrho\,(\sin\theta)^2\,;
\quad\Gamma^2_{33} = -\sin\theta\cos\theta\,;
\end{eqnarray}
to which we add for convenience the once-- and twice--index--contracted terms:
\vspace{-2mm}
\begin{eqnarray}
\label{summedGammas}
&&\hspace{-18mm} \Gamma^\rho_{0\rho} = \Gamma^\rho_{\rho\, 0}= 3\frac{\dot c}{c}\,;
\quad \Gamma^\rho_{\rho 1}= \Gamma^\rho_{1\rho} = 2\frac{\cosh\varrho}{\sinh\varrho}\,;
\quad \Gamma^\rho_{2\rho} =\Gamma^\rho_{\rho 2} =\frac{\cos\theta}{\sin\theta};
\quad \Gamma_{0\rho}^\sigma \Gamma_{0\sigma}^\rho = 3 \frac{\dot c^2}{c^2};
\nonumber\\
&&\hspace{-18mm}
\Gamma_{1\rho}^\sigma \Gamma_{1\sigma}^\rho  = 2\,\dot c^2+ 2\,\frac{\cosh\varrho^2}
{\sinh\varrho^2};\quad\Gamma_{2\rho}^\sigma \Gamma_{2\sigma}^\rho  =2\,
\dot c^2(\sinh\varrho)^2-2(\cosh\varrho)^2 +
\frac{(\cos\theta)^2}{(\sin\theta)^2};\nonumber\\
&&\hspace{-18mm} \Gamma_{3\rho}^\sigma \Gamma_{3\sigma}^\rho =
2\,\dot c^2 (\sinh\varrho)^2(\sin\theta)^2-2(\cosh\varrho)^2(\sin\theta)^2
-2(\cos\theta)^2\,.
\vspace{-2mm}
\end{eqnarray}

Using Eq (\ref{RicciTensor}) in the mixed--index form $R^\mu_\nu\!=\!g^{\mu\lambda}
\big(\partial_\rho\Gamma^\rho_{\lambda\nu} -\partial_\nu \Gamma^\rho_{\lambda\rho}+
\Gamma^\sigma_{\lambda\nu}\Gamma^\rho_{\sigma\rho}-\Gamma^\sigma_{\lambda\rho}
\Gamma^\rho_{\sigma\nu}\big)$, we obtain from Eqs (\ref{invFRWmet}),
(\ref{curvchristlist0}) and (\ref{summedGammas}) the only non--zero
components of the mixed--index Ricci tensor:
\begin{eqnarray}
\vspace{-4mm}
\label{R00term}
&& \hspace{-18mm}  R^0_0(\tau)=-3\frac{\ddot c(\tau)}{c(\tau)} =
-3\bigg[\frac{\ddot a(\tau)}{a(\tau)}
+2 \frac{\dot a(\tau)}{a(\tau)}\bigg];
\quad R^i_0(\tau) = 0\,\,(i, j =1,2,3);\\
\label{Rijterm}
&& \hspace{-18mm} R^i_j(\tau) = - \delta^i_j\bigg[\frac{\ddot c(\tau)}{c(\tau)}+
2\,\frac{\dot c(\tau)^2-1}{c(\tau)^2}\bigg]= - \delta^i_j\bigg[\frac{\ddot a(\tau)}
{a(\tau)} + 2\frac{\dot a(\tau)}{a(\tau)}+2\frac{\dot c(\tau)^2-1}{c(\tau)^2}\bigg];\\
\label{R&Rterm}
&& \hspace{-18mm} R(\tau)= -6\bigg[\frac{\ddot c(\tau)}{c(\tau)}+\frac{\dot c(\tau)^2-1}
{c(\tau)^2}\bigg]=-6\bigg[\frac{\ddot a(\tau)}{a(\tau)}+2\frac{\dot a(\tau)}{a(\tau)}
+2\frac{\dot c(\tau)^2-1}{c(\tau)^2}\bigg].
\vspace{-2mm}
\end{eqnarray}
Here are, for example, the intermediate steps that lead to $R^1_1$:
\vspace{-2mm}
\begin{eqnarray}
&& \hspace{-4mm}R^1_1 = -\frac{1}{c^2}\big[\partial_0\Gamma^0_{11} +
\partial_2\Gamma^2_{11}+\partial_3\Gamma^3_{11}-\partial_1\Gamma^\rho_{1\rho}
+ \Gamma^\sigma_{11}\Gamma^\rho_{\sigma\rho} -\Gamma^\sigma_{1\rho}
\Gamma^\rho_{1\sigma}\big] =\nonumber\\
&&\hspace{4mm}  -\frac{1}{c^2} \bigg[\partial_0 (c\,\dot c)-2\,
\partial_\varrho\bigg(\frac{\cosh\varrho}{\sinh\varrho}\Bigg)+3\,
\dot c^2  -2\,\dot c^2-2\,\frac{\cosh\varrho^2}{\sinh\varrho^2}\bigg]=
- \bigg(\frac{\ddot c}{c} + 2\,\frac{\dot c^2 -1}{c^2}\bigg).\nonumber
\vspace{-2mm}
\end{eqnarray}

Note that if $a(\tau)=1$, we have $c(\tau)=\tau$, hence $R=0$, i.e., the conical spacetime
is flat. Also note that we can have $R^0_0(\tau)=0$ with $R^i_j(\tau)\neq 0$. This mismatch
does not occur in the SMMC where $R_{00}(x)=0$ entails  $R_{\mu\nu}(x)=0$.

To clarify the physical relevance of this point, let us consider the zero--zero component of the
trace reversed gravitational equation (constructed as described in \S\,\ref{TraceRev} of
Appendix~{\bf \ref{GravGaugInvApp}}),
\begin{equation}
\label{R00Append}
R^0_0(\tau) = \kappa\,\big[\mathbb{T}^0_0(\tau)-\frac{1}{2}\mathbb{T}^\lambda_\lambda(\tau)\big] =
\frac{\kappa}{2}\,\big[\rho(\tau)+ 3\,p(\tau)\big]\,,
\end{equation}
where $\rho(\tau)$ and $p(\tau)$ are the energy density and pressure of the cosmic background.

\newpage

Eq (\ref{R00Append}) shows that $R^0_0(\tau) =0$ is possible only if $p(\tau) =- \rho(\tau)/3$,
which is negative because $\rho(\tau)$ is always positive. In which case, as explained in
\S\,\ref{RicciTensSign} of Appendix {\bf \ref{BasFormApp}}, the curvature is purely spatial.

Besides, since the zero--zero component of the gravitational tensor satisfies equation
\begin{equation}
\label{Gmunuterm}
G^0_0(\tau) = R^0_0(\tau)-\frac{1}{2}\, R(\tau)=-\frac{1}{2}\, R(\tau)= 3\,
\frac{\dot c(\tau)^2-1}{c(\tau)^2} =\kappa\, \mathbb{T}^0_0(\tau) \equiv
\kappa\,\rho(\tau),
\end{equation}
we see that the Ricci scalar is related to $\rho(\tau)$ by equation $R(\tau)=-2\,\kappa\,\rho(\tau)$,
which is consistent with the fact that the curvature of the hyperboloidal surfaces of a
the truncated conical spacetime is negative. This fact has no analog in the SMMC.

Putting $c(\tau)= a(\tau)\,\tau$ in Eq (\ref{Gmunuterm}), and identifying $H(\tau)=\dot a(\tau)/a(\tau)$
with the Hubble parameter of the cosmic background, we obtain the equation
\begin{equation}
\label{G00HubblePar}
H(\tau) = \sqrt{\frac{\kappa\,\rho(\tau)}{3}+\frac{1}{a(\tau)^2\,\tau^2}}-\frac{1}{\tau}\,.
\end{equation}

It is evident that for $\tau\rightarrow \infty$ the hyperboloids of the conical spacetime
flatten and the Hubble parameter approaches that of the cylindrical spacetime described by Eq
(\ref{RWHubblePar}).

The temporal flatness condition is condensed in equation $\ddot a(\tau)+2\,\dot a(\tau) =0$,
whose general solution is $a(\tau)= A\,(1-\tau_B/\tau)$, where $A$ and $\tau_B$ are arbitrary
positive constants. This means that the spacetime has actually the structure of a truncated
cone. We have found it natural to identify $\tau_B$ as the time of big bang.

Since $H(\tau)$ remains unvaried if $a(\tau)$ is multiplied by a constant, it is customary to
choose this constant so that the expansion factor equals 1 just today, and that the value of
$H(\tau)$ just coincides with the value of Hubble constant $H_0$ provided by astronomic
observations of nearest celestial bodies, so that $H(\tau_U) = H_0$, where $\tau_U$ is
the age of the universe. In formulas, by putting
\begin{equation}
\label{G00HubbleParU}
a(\tau) = \frac{1-\tau_B/\tau}{1-\tau_B/\tau_U}\quad\hbox{and }\,H(\tau_U) =
\sqrt{\frac{\kappa\,\rho(\tau_U)}{3}+\frac{1}{\tau^2_U}}-\frac{1}{\tau_U}=H_0\,,
\end{equation}
we obtain the best approximation to the analogous relation of the SMMC.  Backdating the time
parameter to a value $\tau < \tau_U$ we obtain instead
\begin{equation}
\label{G00HubbleParD}
H(\tau) = \sqrt{\frac{\kappa\,\rho(\tau)}{3}+\bigg(\frac{\tau_U-\tau_B}
{\tau-\tau_B}\bigg)^2\frac{1}{\tau^2_U}}-\frac{1}{\tau}\,.
\end{equation}

Considering that one of the most important discoveries of the SMMC is the ``obscure energy'',
which is estimated to be about three times greater than that of the matter field, and noting that
the energy density and the pressure of the cosmic background of CGR are related equation
$\rho(\tau)+ 3\,p(\tau)=0$, we are led quite naturally to identify the density of obscure energy with
$\rho(\tau)$ and that of the matter field as equivalent to the work done by the gradient
of pressure $p(\tau)= - \rho(\tau)/3$ between adjacent hyperboloids. The idea that the matter field
was created by a mechanism of this sort has been advanced by several authors in the last fifty years
(for instance, Brout {\em et al.}, p.3, 1978; Peacock, p.26, 1999).

We relay the discussion on this topic to \S\S\,\,\ref{CBGravTensInCHR} and \ref{EntropyTimeCourse} of the main text.

\subsection{Gravitational perturbations of the truncated conical spacetime}
\label{ConGravPerts}
By transferring the concepts introduced in \S\,\ref{CylGravPerts} to the truncated conical spacetime,
we arrive to state the perturbation $\delta R_{00}(x)$ of the $00$ component of Ricci
tensor described by Eqs (\ref{R00term})--(\ref{R&Rterm}) in the form
\vspace{-1mm}
\begin{equation}
\label{deltaR00conST}
\delta R_{00}= \frac{1}{2}\,\big(2\,D^\rho \dot h_{0 \rho}-D^2 h_{00}-\ddot h^\rho_\rho\big) =
\frac{1}{2 a^2}\,\nabla^2_\Omega\, h_{00} - 3\bigg(\frac{1}{\tau} +\frac{\dot a}{a}\bigg)\dot h_{00}+
\partial^i \dot h_{0i} -\frac{1}{2}\,\ddot h^i_i\,,
\vspace{-1mm}
\end{equation}
where expansion factor $a$ depend only on $\tau$, all component $h_{\mu\nu}$ depend on spacetime
coordinates $x$ and $\nabla^2_\Omega$ is the Laplace operator described by Eq (\ref{unitlaplopc}).
The frictional term differs from that described in Eq (\ref{CylR00Appwithc}) by the presence of
the additional term $3/\tau$, because now the truncated conical spacetime foliates into
hyperboloidal surfaces. So, as $\tau$ tends to infinity, the hyperboloids flatten and the
dependence on $\tau$ disappears.

Since the $00$ component of the Ricci tensor of the truncated conical spacetime vanishes, the
analog of Eq (\ref{CylR00Appwithc}) is missing. We can therefore  state the analog of Eq
(\ref{NewtPotInExpST}) in the form
\vspace{-1mm}
\begin{equation}
\label{NewtPotInConST}
\Delta h_{00}(x) = - 2\,\nabla^2_\Omega \Phi(x)+
6\,a(\tau)^2 \bigg[\frac{1}{\tau}+H(\tau)\bigg]\,\dot h_{00}(x),
\vspace{-1mm}
\end{equation}
where $\Phi(x)$ is the Newtonian potential, $H(\tau)$ is the Hubble parameter described in
the previous section and $a(\tau)$ is given by the first of Eq (\ref{G00HubbleParU}),
$a(\tau) = (1-\tau_B/\tau)/(1-\tau_B/\tau_U)$.

If the perturbed metric too is homogeneous and isotropic, the coordinates of the truncated
conical spacetime have the simple form $x=\{\tau, \varrho, \theta, \phi\}$, where
$\tau> \tau_B$,  and therefore the analog of Eq (\ref{ds2hmunu}) has the form
\vspace{-1mm}
\begin{eqnarray}
\label{KTConicalHarmGauge}
&&\hspace{-16mm}d\bar s^2(x) = d\tau^2\big[1\!+2\Phi(x)\big]-\nonumber\\
&&a(\tau)^2 \big[1\!-2\Phi(x)\big]\tau^2\big(d\varrho^2 +
\sinh\varrho^2\,d\theta^2+\sinh\varrho^2\,\sin\theta^2 d\phi^2\big)\,.
\vspace{-1mm}
\end{eqnarray}

\markright{R.Nobili, Conformal General Relativity - {\bf \ref{ConfInvApp}} Conformal invariance and causality}
\renewcommand\thefigure{\Alph{section}\arabic{figure}}
\setcounter{figure}{0}
\section{CONFORMAL INVARIANCE AND CAUSALITY}
\label{ConfInvApp}
Denote by ${\cal M}_n$ the $n$D Minkowski spacetime, by $x = \{x^0, x^1, x^2,..., x^{n-1}\}$
its coordinates and by $\eta_{\mu\nu} = \hbox{diag}\{1, -1,\dots, -1\}$ its metric tensor.
The largest group of coordinate transformations that preserves the causal structure of
${\cal M}_n$ is the {\em $n$--dimensional conformal group} \cite{HAAG92}, here denoted
as ${C}(1, n-1)$. The connected component of this group is formed by the infinitesimal
transformations $x^\mu \rightarrow x^\mu + \varepsilon\, u^\mu(x)$, where
$\varepsilon\rightarrow 0$ and $u_\mu(x)$ are smooth functions of $x$, that preserve the
light cones. By carrying out this transformation, the squared line--element, from $x^\mu$ t
o $x^\mu+dx^\mu$, $ds^2 =\eta_{\mu\nu} dx^\mu dx^\nu$, undergoes the change
$$
\delta ds^2 = \varepsilon\, \eta_{\mu\nu}\bigg(\frac{\partial u^\mu}
{\partial x^\rho}\,dx^\rho dx^\nu+ \frac{\partial u^\nu}
{\partial x^\rho}\,dx^\rho dx^\mu\bigg) \equiv
\varepsilon\,G_{\mu\nu}\,dx^\mu dx^\nu\,.
$$

In order for this transformation to not change the light--cone equation $ds^2(x)=0$,
we must have $G_{\mu\nu}dx^\mu dx^\nu =0$ whenever $\eta_{\mu\nu}dx^\mu dx^\nu =0$.
So $G_{\mu\nu}$ is a metric tensor, possibly depending on $x$, which has the same
light--like direction as $\eta_{\mu\nu}$, that is $G_{\mu\nu} = f(x)\,\eta_{\mu\nu}$.

By contracting $G_{\mu\nu}$ with $\eta^{\mu\nu}$, and using $\eta^{\mu\nu}$ to
lower the indices of $u^\mu$, we can eliminate $f(x)$ and obtain the following
conditions for the field of displacements $u^\mu(x)$:
\begin{equation}
\label{Gmunu2gmunu}
\frac{\partial u_\mu(x)}{\partial x^\nu}+ \frac{\partial u_\nu(x)}
{\partial x^\mu}=\frac{1}{2}\,\eta_{\mu\nu}\partial_{\rho} u^\rho(x)\,.
\end{equation}
Taylor--expanding $u_\mu(x)$ about $x=0$ yields
$$
u_\mu(x) = a^{(1)}_\mu + a^{(2)}_{\mu\nu}x^\nu +
a^{(3)}_{\mu\nu\rho}\,x^\nu x^\rho + \dots
$$
Since the homogeneous polynomial in $x$ decouple from each other, Eq (\ref{Gmunu2gmunu})
provides one separate condition for each coefficient $a^{(n)}$, which can be written
as $n$--index coefficients
$$
a^{(n)}_{\mu\nu\rho\sigma\cdots} + a^{(n)}_{\nu\mu\rho\sigma\cdots} =
\frac{1}{2}\, \eta_{\mu\nu} a^{(n)\,\lambda}_{\lambda\rho\sigma\cdots,}
$$
where $a^{(n)}_{\mu\nu\rho\sigma\cdots}$ is totally symmetric in the
last $n-1$ indices. For $n=1$ there is no restriction for $a^{(1)}$.
For $n=2$ one gets $a^{(2)}_{\mu\nu} =\omega_{\mu\nu} + c\,\eta_{\mu\nu}$,
where $\omega_{\mu\nu}=- \omega_{\nu\mu}$. For $n=3$ one
finds $a^{(3)}_{\mu\nu\rho}= \eta_{\mu\nu}\, c_\rho + \eta_{\mu\rho}\,
c_\nu - \eta_{\nu\rho}\, c_\mu$. For $n\ge3$ there are no solutions,
unless the spacetime has dimension $2$. Therefore the admissible
expression for the field of displacements is
\begin{equation}
\label{dispfield}
u^\mu(x) = a^\mu + \alpha\, x^\mu + \omega_{\nu\lambda}\,x^\lambda \eta^{\nu\mu} +
b_\lambda\big(\eta^{\lambda\mu}\, x_\nu x^\nu-2\,x^\lambda x^\mu\big)\,,
\end{equation}
where $a^\mu, \alpha, b_\lambda$ are arbitrary constants and $\omega_{\nu\lambda}$
is antisymmetric; in total $n(n+3)/2+1$ parameters. Of note, in the two--dimensional
case the spacetime is isomorphic to the Argand--Gauss plane of complex variable
$z$, so that the conformal group is isomorphic to the group of holomorphic functions
of $z$ with nowhere--zero derivative.

Eq (\ref{dispfield}) condenses the infinitesimal generators of the topologically
connected component of ${C}(1, n-1)$,  the transformations of which act on
$x^\mu$ as follows:
\begin{eqnarray}
\label{transl}
T(a):  x^\mu  & \rightarrow & x^{\mu} + a^{\mu} \quad \,\hbox{(translations)\,;} \\
\label{lorentz}
\Lambda(\omega):  x^\mu & \rightarrow &
\Lambda^\mu_\nu(\omega)\,x^{\nu} \quad \hbox{(Lorentz rotations)}\,;\\
\label{dilat} S(\alpha):  x^\mu  & \rightarrow &  e^{\alpha} x^{\mu} \qquad
\,\,\,\hbox{(dilations)}\,;\\
\label{elat}
E(b): x^\mu & \rightarrow & \frac{x^\mu - b^\mu x^2}{1-2\,bx + b^2x^2}
\quad \hbox{(elations)}\,.
\end{eqnarray}
Here $a^\mu$, $\alpha$ and the tensor $\omega\equiv\omega^{\rho\sigma} =-\omega^{\sigma\rho}$
are respectively the parameters of translations, dilation, Lorentz rotations and {\em elations};
$x^2$ and $b^2 = b_\mu b^\mu$ stand for $x_\mu x^\mu$ and $b_\mu b^\mu$, and $b x$ stands for
$b^\mu x_\mu$. $E(b)$ form an Abelian subgroup commonly known as the group of {\em special
conformal transformations}, but we call it the group {\em elations}, because this is the name
coined by Cartan in 1922 \cite{CARTAN1}.

Indicating by $P_\mu$, $M_{\mu\nu}$, $D$ and $K_\mu$ the generators of $T(a), \Lambda(\omega),
S(\alpha)$ and $E(b)$, respectively, we can easily determine their actions on $x^\mu$
\begin{eqnarray}
& & P_\mu x^\nu = -i\delta_\mu^\nu\,,\quad M_{\mu\nu}x^\lambda =
i\big(\delta_\nu^\lambda x^\mu-\delta_\mu^\lambda x^\nu\big)\,,\nonumber \\
& & D x^\mu = -i x^\mu\,, \quad K_\mu x^\nu = i\big(x^2\delta_\mu^\nu -
2x_\mu x^\nu\big) \nonumber\,,
\end{eqnarray}
where $\delta_\mu^\nu$ is the Kronecker delta. Indicating by $\partial_\mu$ the
partial derivative with respect to $x^\mu$, the actions of such generators on any
differentiable functions $f$ of $x$ are
\begin{eqnarray}
\label{dergen1} & & P_\mu f(x) = -i\partial_\mu f(x) \,;\quad  M_{\mu\nu} f(x)
= i\big(x_\mu\partial_\nu -x_\nu\partial_\mu\big)f(x)\,; \nonumber\\
\label{dergen2} & & D f(x) = -i x^\mu \partial_\mu f(x) \,;\quad K_\mu f(x) =
i\big(x^2\partial_\mu -2x_\mu x^\nu \partial_\nu\big)f(x) \,; \nonumber
\end{eqnarray}
the Lie algebra of which satisfies the following commutation relations \cite{MACKSALAM}
\begin{eqnarray}
\label{commut1}
& &[P_\mu, P_\nu]  = [K_\mu, K_\nu] =0\,;\quad [P_\mu, K_\nu] = 2i\big(g_{\mu\nu}D+M_{\mu\nu}\big) \,;\\
\label{commut2}
& &[D, P_\mu]  = i P_\mu\,; \quad [D, K_\mu] =-i K_\mu\,; \quad [D, M_{\mu\nu}] =0\,;\\
\label{commut3} & &[M_{\mu\nu}, P_\rho]  = i \big(g_{\nu\rho}P_\mu
-g_{\mu\rho}P_\nu\big) \,; \quad [M_{\mu\nu}, K_\rho]  =
i \big(g_{\nu\rho}K_\mu-g_{\mu\rho}K_\nu\big) \,;\\
\label{commut4} & &[M_{\mu\nu}, M_{\rho\sigma}]  =
i\big(g_{\mu\sigma}M_{\nu\rho} + g_{\nu\rho}M_{\mu\sigma}-
g_{\mu\rho}M_{\nu\sigma} -g_{\nu\sigma}M_{\mu\rho}\big)\,.
\end{eqnarray}

However, if we include discrete light--cone , ${C}(1, n-1)$ is somewhat larger since the partial
ordering of causal events is also preserved, for instance, by the following involution
$$
I_0: x^\mu \rightarrow - \frac{x^\mu}{x^2}\,,
$$
which we will call the {\em orthochronous inversion} with respect to event $x=0\in {\cal M}_n$ \cite{NOBILI}.
Equalities $(I_0)^2 = {\mathbf 1}$, $I_0 \Lambda(\omega)I_0 = \Lambda(\omega)$ and
$I_0S(\alpha)I_0 = S(-\alpha)$ are evident, and equation $E(b) = I_0T(b)I_0$ can easily be proved.

By a translation of the spacetime origin, we obtain the orthochronous inversion with respect
to any desired point $a\in {\cal M}_n$, which acts on $x^\mu$ as follows:
$$
I_a: x^\mu \rightarrow - \frac{x^\mu-a^\mu}{(x-a)^2}\,.
$$

If we add to Eqs (\ref{commut1})--(\ref{commut4}) the discrete transformations
\begin{equation}
\label{inverse} I_0 P_\mu I_0  = K_\mu\,; \quad\!\! I_0 K_\mu I_0  =
P_\mu\,;\quad\!\! I_0 D I_0 = -D \,; \quad\!\! I_0 M_{\mu\nu} I_0 =
M_{\mu\nu}\,,
\end{equation}
we see that $I_0$ and $P_\mu$ alone suffice to generate the connected component
of ${C}(1, n-1)$. Indeed, using Eqs  (\ref{commut1})--(\ref{commut3}) and the first
of Eqs (\ref{inverse}), we can obtain all other group generators as follows:
$$
K_\mu = I_0 P_\mu I_0\,,\quad D = \frac{i}{8}\,g^{\mu\nu}[K_\mu, P_\nu]\,,
\quad M_{\mu\nu} =\frac{i}{2}[K_\nu,P_\mu] - g_{\mu\nu}D\,.
$$

These equations show very clearly the importance of the orthochronous inversion
in the structure of conformal group ${C}(1, n-1)$. We may think of $I_0$ as an operation
of partial ordering of causally related events carried out by an observer located at $x = 0$,
which receives signals from the past and sends signals to the future; of $T(a)$ as the
operation that shifts the observer from $x = 0$ to $x = a$ in ${\cal M}_n$; and of $I_a$
as a continuous set of involutions that impart the structure of a symmetric space to the
orthocomplemented lattice formed by the causally complete regions of the spacetime
(Haag, III,\S\, 4.1; 1996).

Provided that $n$ is even, we can include, as a second discrete element of the
conformal group, parity transformation ${\cal P}: \{x^0, \vec x\} \rightarrow
\{x^0, -\vec x\}$. Time--reversal must be instead excluded, because it does not
preserve the causal order of events.

This structure marks the basic difference between GR and CGR: time reversal,
which is so familiar to GR, must be replaced in CGR by an orthochronous inversion
$I_0$ conventionally centered at some originating point $x=0$ of a conical spacetime.

\subsection{Conformal transformations of local fields}
\label{fieldtransfs} When a differential operator $g$ is applied to a differentiable
scalar function $f$ of $x$, the function changes as $gf(x) = f(gx)$, which may be
interpreted as the form taken by $f$ in the reference frame of coordinates $x'=gx$.
When a second differential operator $g'$ acts on $f(gx)$, we obtain $g'f(gx) = f(gg'x)$,
i.e., we have $g'gf(x) =  f(gg'x)$, showing that $g'$ and $g$ act on the reference
frame in reverse order.

The action of $g$ on a local quantum field $\Psi_\rho(x)$ of spin index $\rho$
has the general form $g\Psi_\rho(x)= {\cal F}_\rho^\sigma(g^{-1}, x) \Psi_\sigma(gx)$,
where ${\cal F}(g^{-1}, x)$ is a matrix obeying the composition law
$$
{\cal F}(g^{-1}_2,x)\,{\cal F}(g^{-1}_1, g_2 x)= {\cal F}(g^{-1}_2g^{-1}_1, x)\,.
$$
These equations are consistent with coordinate transformations, since the
product of two transformations $g_1, g_2$ yields
$$
g_2g_1 \Psi_\rho(x)  = \, {\cal F}^\sigma_\rho[(g_1 g_2)^{-1},
x]\Psi_\sigma(g_1g_2 x)\,,
$$
with $g_2, g_1$ always appearing in reverse order on the right--hand member.

According to these rules, the generators of the connected part of ${C}(1, n-1)$
act on an irreducible unitary representation  $\Psi_\rho(x)$ of the Poincar\'e
group that describes a field of spin index $\rho$ and length--dimension
({\em weight}) $w_\Psi$, as follows
\begin{eqnarray}
\label{dergenf1} & & [P_\mu, \Psi_\rho]  = -i\, \partial_\mu \Psi_\rho  \,; \\
\label{dergenf2} & & [K_\mu, \Psi_\rho] = i \big[x^2
\partial_\mu - 2x_\mu \big(x^\rho \partial_\rho + w_\Psi\big)\big] \Psi_\rho +
i x^\nu\big(\Sigma_{\mu\nu}\big)_\rho^\sigma\Psi_\sigma\,;\\
\label{dergenf3} & & [D, \Psi_\rho] = -i \big(x^\mu\partial_\mu +  w_\Psi\big)
\Psi_\rho \,;\\
\label{dergenf4} & & [M_{\mu\nu}, \Psi_\rho] = i \big(x_\mu \partial_\nu -
x_\nu\partial_\mu\big) \Psi_\rho - i\big(\Sigma_{\mu\nu}\big)_\rho^\sigma
\Psi_\sigma\,;
\end{eqnarray}
where $ \Sigma_{\mu\nu}$ are the spin matrices, i.e., the generators of Lorenz
rotations on the spin space.

Remember that in an $n$D spacetime, the parameters $x^\mu$ have
length--dimension $0$ and the squared line element $ds^2 = g_{\mu\nu}dx^\mu dx^\nu$
has length--dimension $2$; therefore, $g_{\mu\nu}$, $g^{\mu\nu}$ and the determinant
$g$ of matrix $[ g_{\mu\nu}]$ must have respectively length--dimensions $2$,
$-2$ and $2\,n$. Accordingly, $\partial_\mu$ and covariant gauge--fields must
have length--dimension $0$; Lagrangian densities must have length--dimension
$-n$, scalar fields $\psi$ must have length--dimension $w_{\phi}=1-n/2$ and
spinor fields $\psi$ must have length--dimension $w_\psi=(1-n)/2$.

The finite conformal transformations respectively corresponding to Eqs
(\ref{dergenf1})--(\ref{dergenf4}) are therefore:
\begin{eqnarray}
\label{conf1} & & T(a): \Psi_\rho(x)\rightarrow \Psi_\rho(x+a)\,;\\
\label{conf2} & & E(b): \Psi_\rho(x)\rightarrow {\cal E}(-b,x)^\sigma_\rho
\Psi_\sigma\bigg(\frac{x - bx^2}{1-2bx+b^2x^2}\bigg)\,;\\
\label{conf3} & & S(\alpha): \Psi_\rho(x)\rightarrow
e^{\alpha\,w_\Psi}\,\Psi_\rho(e^{\alpha} x)\,;\\
\label{conf4} & &  \Lambda(\omega): \Psi_\rho(x)\rightarrow {\cal
L}^\sigma_\rho(-\omega) \Psi_\sigma[\Lambda(\omega)\, x]\,;
\end{eqnarray}
where  ${\cal E}(-b,x), {\cal L}(-\omega)$ are suitable matrices which perform
the conformal transformations of spin components, respectively for elations and
Lorentz rotations.

As regards the orthochronous inversion, we generally have
\begin{equation}
\label{conf5}  I_0: \Psi_\rho(x)\rightarrow {\cal I}_0(x)^\sigma_\rho
\Psi_\sigma (-x/x^2)\,,
\end{equation}
where matrix ${\cal I}_0(x)$ obeys the equation
\begin{equation}
\label{IoIo}  {\cal I}_0(x){\cal I}_0(-x/x^2)=1\,.
\end{equation}
For consistency with (\ref{conf1}), (\ref{conf2}) and Eqs  $E(b) =
I_0(x)\,T(b)\,I_0(x)$, we also have
\begin{equation}
\label{ITI} {\cal E}^\sigma_\rho(-b,x)= {\cal I}_0(x)\,{\cal I}_0(x-b)\,.
\end{equation}

For the needs of a Langrangian theory, the adjoint representation of $\Psi_\alpha$ must also be
defined. It can be indicated by $\bar\Psi= \Psi^{\dag} {\cal B}$, where ${\cal B}$ is a suitable
matrix, or complex number, chosen in such a way that equation $\bar{\bar\Psi}=\Psi$ be satisfied
and that the Hamiltonian be self--adjoint. This implies ${\cal B}\,{\cal B}^\dag=1$. Therefore,
under the action of a group element $g$, the adjoint representation $\bar \Psi^\rho(x)$ is subject
to the transformation
$$
g: \bar \Psi^\rho(x)\rightarrow \bar\Psi^\sigma(gx)\,\bar{\cal F}(g^{-1},x)^\rho_\sigma\,,
$$
where $\bar{\cal F}(g^{-1}, x)= {\cal B}^\dag {\cal F}^\dag(g^{-1}, x){\cal B}$.

In standard field theory, the group of spinor transformations contains the subgroup of discrete operators formed
by parity, $\cal P$, charge conjugation, $\cal C$, and time reversal, $\cal T$. The last of these
commutes with $\cal P$, and $\cal C$ and the elicity projectors $P_\pm$ defined by equations $P_++P_-={\bf 1}$,
$\psi_R =P_+\psi$ and $\psi_L=P_-\psi$, where $R$ and $L$ stand respectively for the right--handed and the
left--handed elicities. However, passing from the Poincar\'e to the conformal group, we must exclude $\cal T$
because this violates the causal order of physical events, and replace the role of time reversal
to orthochronous inversion $I_0$ instead.

\subsection{Remarkable properties of orthochronous inversions}
\label{OnIa}
\begin{figure}[!ht]
\centering
\mbox{%
\begin{minipage}{.32\textwidth}
\hspace{-4mm}\includegraphics[scale=0.6]{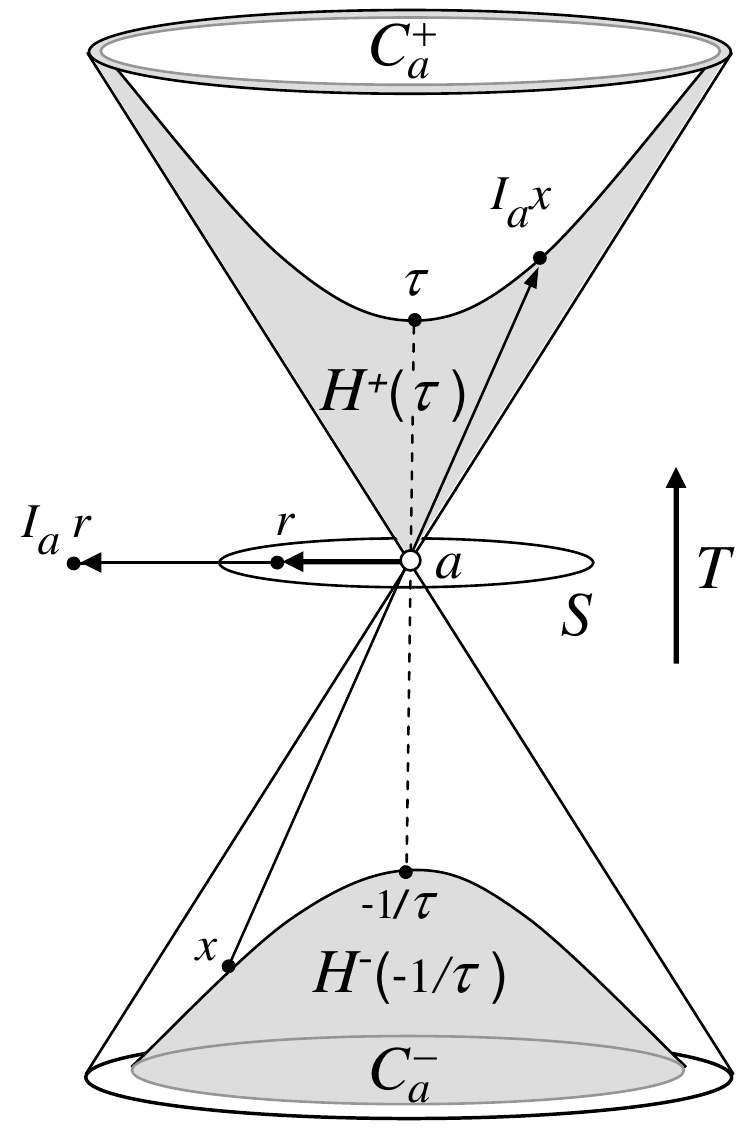} %F23
\end{minipage}%
\quad
\begin{minipage}[c]{.52\textwidth}
\caption{\small The orthochronous inversion centered at a point $a$ of an $n$D
Minkowski spacetime interchanges the events lying in the interior of the double
cone $C^+_a\cup C^-_a$. This happens in such a way that a spacetime region of the
future--cone that lies close to $a$ is mapped onto a region of the past--cone that
lies far from $a$. In particular, gray region $H^+_a(\tau)$ is mapped onto gray
region $H^-_a(-1/\tau)$ and vice versa. $T = $ direction of time axis; $S = $
$(n-1)$D unit sphere centered at $a$ and orthogonal to $T$.}
\label{ConfInvFig1}
\end{minipage}%
}
\end{figure}

Let $C^+_a$ and $C^-_a$ be respectively the future and past cones stemming from a point
$a\in{\cal M}_n$, as shown in Fig.\ref{ConfInvFig1}. Orthochronous inversion $I_a$ acts as follows:

1) It swaps $C^+_a$ with $C^-_a$, so as to preserve the direction $T$ of the time axis through $a$
and the collineation of all points lying on a straight line through $a$ within $C^+_a\cup C^-_a$.

2) It divides the events in $C^+_a\cup C^-_a$ into a two--fold foliation of
$(n-1)$--dimensional spacelike hyperboloids parameterized by the {\em kinematic time}
of origin $a$ introduced in \S\,\ref{futconegeom},
$$
\tau =\pm \sqrt{ (x^0-a^0)^2+\dots +(x^{n-1}-a^{n-1})^2}\,.
$$
In accordance with \S\,\ref{THREEWAYS}, we can respectively identify the
positive and negative parts of $\tau$ as the {\em conformal times} of
conical spacetimes $C^+_a$ and $C^-_a$.

3) It maps future--cone region $H^+_a(\tau)\subset C^+_a$, extending from
$a$ to the hyperboloid at $\tau$, onto region  $H^-_a(-1/\tau)\subset C^-_a$,
extending from the degenerated hyperboloid at $-\infty$ to the hyperboloid
at $\tau' = -1/\tau$, and vice versa (gray regions in Fig.\,\ref{ConfInvFig1}).

4) It performs the polar inversion of points $r$ internal to the
$(n-1)$--dimensional sphere $S$ of radius $1$, centered at $a$ and
orthogonal to the time axis through $a$, into points $r' = I_ar$
external to $S$, and conversely.

5) Functions of $x$ that are invariant under $I_a$ depend only on
conformal time $\tau$. Therefore, if they vanish closely near to the
origin  of the past cone, they also vanish remotely far from the
origin of the future cone, and conversely.

The latter property has the following important implication: if the
matter density of a physical system is invariant under $I_a$ and
converges to zero at $\tau=\infty$ in $C_a^+$, it converges to zero
also at $\tau=0$ in $C_a^-$, and conversely.

This is consistent with the view proposed in \S\,\ref{introduction},
according to which the history of the universe is confined to a future
cone, here identified with $C^+_0$, as the result of a spontaneous decay of
the conformal symmetry occurred at kinematic time $\tau=0$, and evolved
toward the metric symmetry of General Relativity~(GR) at $\tau=+\infty$  .

It is instead in contrast with the standard model of modern cosmology, see for
instance \cite{LIDDLE} and (Mukhanov, 2005), which, for consistency
with, GR must represent the initial state of the universe as an
infinitely dense concentration of matter spread on a spacelike surface
with constant curvature counterbalanced by a concentration of gravitational
energy.

Since $I_0$ and the group of translations suffice to generate the entire
conformal group, it is opportune to consider only systems the action of which,
${\cal A}$, is invariant under $I_0$. This is possible provided that ${\cal A}$
is the sum of two action integrals,
\begin{equation}
\label{AHplusAHminus}  {\cal A}^- = \int_{C^-_0}\!\!\!\!
\sqrt{-g(x)}\,{\cal L}(x)\,d^4x\,,\quad{\cal A}^+=
\int_{C^+_0}\!\!\!\!\sqrt{-g(x)}\,{\cal L}(x)\,d^4x\,,
\end{equation}
where $C^+_0$ and $C^-_0$ are opposite conical spacetimes, so that involution ${\cal A}^-\!\!
\xleftrightarrow{\,\, I_0 \,\,} {\cal A}^+$ be satisfied. This clearly
requires that Lagrangian density ${\cal L}(x)$ satisfies the mirroring
property
\begin{equation}
\label{IogL} \sqrt{-g(x)}\,{\cal L}(x) \xleftrightarrow{\,\, I_0 \,\,}
\sqrt{-g(I_0 x)}\, {\cal L}(I_0 x)\,,
\end{equation}
where $I_0 x = -x/x^2$.

%stemming from the common vertex at $x^0=0$

Provided the matter field is homogeneous and isotropic, and the motions
equations are only derived from ${\cal A}^+$, we can regard the systems which
satisfy these conditions as models of the universe on the large scale.

\subsection{Conformal invariance of field theories in curved spacetimes}
\label{confGrCurv}
The conformal invariance of a total action of matter and geometry in a curved spacetime
requires that the action is free from dimensional constants and invariant both under metric diffeomorphisms
and local Weyl transformations, which together form the group of conformal diffeomorphisms, as
explained in \S\,\ref{introduction}. By obvious generalization of the flat spacetime case, this is
the largest group of invariance that preserves the causal order of physical events in the curved spacetime.
If the spacetime has dimension $n>4$, in general, a Weyl transformation of the Lagrangian density of matter
and geometry, ${\cal L}= {\cal L}^M + {\cal L}^{\,G}$, does not leave $\cal{L}$ invariant, but generates an
additional expression, which is a mere surface term provided that the spacetime has dimension four
and the Ricci scalar $R$ is suitably coupled with one or more physical scalar fields $\varphi_i$
and one or more ghost scalar fields $\sigma_j$. In this case, the geometric Lagrangian density
must have the form
$$
{\cal L}^G =  (\varphi^2- \sigma^2)\,\frac{R}{12}\,,\,\,\hbox{where }\,\varphi^2 =
\sum_i\varphi^2_i\,\,\hbox{and }\, \sigma^2 = \sum_j\sigma^2_j
$$
and the vacuum expectation value of $\sum_i\varphi^2_i-\sum_j\sigma^2$ must be always negative.
This is proven in \S\,\ref{inclusCGR}. The latter condition is necessary, since otherwise the
gravitational forces would be repulsive. This point is widely discussed in \S\,\ref{introduction}
near Eq (\ref{AsigmaAvarphi}).

If the action integral is invariant under $I_0$, on account of Eq (\ref{AHplusAHminus}),
we can derive the motion equations only from the future component ${\cal A}^+$, which we
can simply denote as
$$
{\cal A} =   \int_{C^+_0}\sqrt{-g}\,\bigg[{\cal L}^M + \big(\varphi^2-\sigma^2\big)\,\frac{R}{12}\bigg] d^4x
$$
without fair of confusion.

Here we see very clearly that the implementation of the conformal invariance in a field theory defined in a curved
spacetime needs that the spacetime is a conical foliation of hyperboloidal surfaces parameterized by a time--like
parameter $\tau$, as explained in \S\,\ref{futconegeom}. This requires a hyperbolic metric tensor with the
general form $ ds^2 = \tau^2 - g_{ij}(\tau, \vec x)\,dx^i \,dx^j$, in which the gravitational field is incorporated
in the metric and depends on the matter field via the gravitational equation $\delta {\cal A}/\delta g_{\mu\nu}(x)=0$,
as described in \S\,\ref{CGR&SMEP} near Eq (\ref{CGRGRAVEQ}).

The Beltrami--d'Alembert operator associated with this metric contain frictional terms which impart a
dissipative behavior to the dynamics of scalar fields and make the potential energy terms of the Lagrangian
density evolve towards their minima. This subject is exemplified in \S\S\,\,\ref{futconegeom} and \ref{Evolvingvac}.

In quantum field theory, the conditions for conformal invariance in curved spacetime are even more selective than
in classical field theory. This is because, as discussed in \S\S\,\ref{mainproblems}, \ref{AsympConfInvinCGR} and
\ref{ShiftInvCGR}, the conformal invariance of the theory is possible only if the total one--loop term of the
effective Lagrangian is zero. But, in order for this to happen, the theory must include suitable conformal--invariant
interactions with fermion a gauge vector or axial--vector fields.  This subject is widely discussed in
Appendix {\bf \ref{PathIntApp}}.

\newpage

\markright{R.Nobili, Conformal General Relativity - {\bf \ref{BreakConfApp}} Conformal symmetry breakdown}
\section{THE BREAKDOWN OF CONFORMAL SYMMETRY}
\label{BreakConfApp}
The possibilities for the spontaneous breakdown of conformal symmetry have been studied by Fubini in 1976.
We report here his main results.

It is known that the 15--parameter Lie algebra of the conformal group $G\equiv{C}(1, 3)$, described by
Eqs (\ref{commut1})--(\ref{commut4}), is isomorphic with that of hyperbolic--rotation group $O(2,4)$ on the 6D
linear space $\{x^0, x^1, x^2, x^3, x^4, x^5\}$ of metric $(x^0)^2 + (x^5)^2 - (x^1)^2 - (x^2)^2 - (x^3)^2 - (x^4)^2$.
The spontaneous breakdown of conformal symmetry can occur only in three ways, corresponding to the
following stability subgroups of $G$:
\begin{itemize}
\item[--] $O(1,3)$: the {\em Poincar\'e group}, i.e, the 10--parameter Lie algebra generated by
$M_{\mu\nu}$ and $P_\mu$. With this choice, NG--boson VEVs are invariant under translations
and are therefore constant.

\item[--] $O(1,4)$: the {\em deSitter group} generated by the 10--parameter Lie
algebra which leaves invariant the quadric $(x^0)^2 - (x^1)^2 - (x^2)^2 - (x^3)^2 -
(x^4)^2$ \cite{MOSCHELLA}, which characterizes the class $dS_4$ of the deSitter
spacetimes as particular 4D--submanifolds, with constant positive curvature, of
the linear space $\{x^0, x^1, x^2, x^3, x^4\}$. Its generators are $M_{\mu\nu}$ and
$$
%\vspace{-4mm}
L_\mu=\frac{1}{2}\,\bigl(P_\mu - K_\mu\bigr)\,,
\vspace{-2mm}
$$
which anti--commute with orthochronous inversion $I_0$ and satisfy the commutation
relations $[L_\mu, L_\nu]=-i\,M_{\mu\nu}\,;\quad [M_{\mu\nu}, L_\rho]  =
i \big(g_{\nu\rho}L_\mu -g_{\mu\rho}L_\nu\big)$. Since vacuum state $|\Omega\rangle$
is invariant under this subgroup, the NG--field $\sigma_+(x)$ associated with the
contraction subgroup of $G$ satisfies equations
\vspace{-2mm}
\begin{equation}
\label{Lsigma-}
L_\mu\sigma_+(x)|\Omega\rangle=0\,, \quad M_{\mu\nu}\,\sigma_+(x)|\Omega\rangle\equiv
-i\big(x_\mu\partial_\nu-x_\nu\partial_\mu\big)\sigma_+(x)|\Omega\rangle=0\,;\nonumber
%\vspace{-2mm}
\end{equation}
the second of which implies that $\sigma_+(x)$ depends on $\tau^2\equiv x^2$ only.

\item[--] $O(2,3)$: the {\em anti--deSitter group} generated by the 10--parameter Lie
algebra which leaves invariant the quadric $(x^0)^2 + (x^4)^2 - (x^1)^2 - (x^2)^2
- (x^3)^2$, which characterizes the class  $AdS_4$ of the anti--deSitter spacetimes
as particular 4D--submanifolds, with constant negative curvature, of a 5D linear space.
Its generators are $M_{\mu\nu}$ and $$R_\mu=\frac{1}{2}\,\bigl(P_\mu+K_\mu\bigr)\,,$$
which commute with orthochronous inversion $I_0$ and satisfies the commutation relations
$[R_\mu, R_\nu]=i\,M_{\mu\nu}\,; \quad [M_{\mu\nu}, R_\rho]  = i (g_{\nu\rho}R_\mu
-g_{\mu\rho}R_\nu)$. Since $|\Omega\rangle$ is invariant under these transformations,
the NG--field $\sigma_-(x)$ associated with the contraction subgroup of $G$,
satisfies equations
\begin{equation}
%\vspace{-2mm}
\label{Lsigma+}
R_\mu\sigma_-(x)|\Omega\rangle=0\,, \quad M_{\mu\nu}\,\sigma_-(x)|\Omega\rangle\equiv
-i\big(x_\mu\partial_\nu-x_\nu\partial_\mu\big)\,\sigma_-(x)|\Omega\rangle=0\,,\nonumber
%\vspace{-2mm}
\end{equation}
the second of which implies that $\sigma_-(x)$  depends on $x^2=\tau^2$ only.
\end{itemize}

Comparing the results obtained for the de Sitter and anti--de Sitter groups, we note that $L_\mu$, $D$ are the
generators of the set--theoretical complement of $O(3,2)$ in $G$, and $R_\mu$, $D$ are those of the set--theoretical
complement of $O(1,4)$ in $G$. Thus, using commutation relations
$[R_\mu, D]= i\,L_\mu\,,\quad [L_\mu, D]= i\,R_\mu$, we derive
\begin{equation}
\vspace{-2mm}
\label{LRsigma}
[R_\mu, D]\,\sigma_+(\tau)|\Omega\rangle = i\,L_\mu\,\sigma_+(\tau)|\Omega\rangle=0\,
\quad [L_\mu,D]\,\sigma_-(\tau)|\Omega\rangle = i\,R_\mu\,\sigma_-(\tau)|\Omega\rangle=0\,,
\end{equation}
showing that these set--theoretical complements act respectively on $\sigma_+(\tau)|\Omega\rangle$
and  $\sigma_-(\tau)|\Omega\rangle$ as Abelian subgroups of transformations.

Using Eqs (\ref{dergenf1}) and (\ref{dergenf2}), we obtain the explicit expressions of
Eqs (\ref{LRsigma}) for $\sigma_\pm(x)$ of dimension $-1$
\begin{eqnarray}
\vspace{-2mm}
L_\mu\sigma_+(\tau)|\Omega\rangle & \equiv & -i\bigg[\frac{1+x^2}{2}\,\partial_\mu-
x_\mu(x^\nu\partial_\nu +1)\bigg]\sigma_+(\tau)|\Omega\rangle=0\,,\nonumber\\
R_\mu\sigma_-(\tau)|\Omega\rangle& \equiv & -i\bigg[\frac{1-x^2}{2}\,\partial_\mu+
x_\mu(x^\nu\partial_\nu +1)\bigg]\sigma_-(\tau)|\Omega\rangle=0\,.\nonumber
\end{eqnarray}
Contracting these equations with $x^\mu$, then putting $x^2\equiv\tau^2$ and
$x^\mu\partial_\mu \equiv \tau\partial_\tau$, we can easily verify that their
solutions are satisfied for
\begin{equation}
\label{sigmaplusminus0}
\sigma_+(\tau)= \frac{\sigma_+(0)}{1+\tau^2}\,,\quad \sigma_-(\tau)=
\frac{\sigma_-(0)}{1-\tau^2}
\end{equation}
and, which is particularly interesting, they satisfy the equations
\begin{equation}
\label{squarefs}
\Big(\partial_\tau^2 +\frac{3}{\tau}\,\partial_\tau \Big)\sigma_\pm(\tau) \pm
\lambda_{\pm}\sigma^3_\pm(\tau)=0\,,
\end{equation}
where $\lambda_{\pm} = 8/\sigma(0)^2_\pm$. Actually, Eqs (\ref{sigmaplusminus0}) are
not uniquely determined because, by applying the change of scale $\tau\rightarrow
\tau/\tau_0$,  $\tau_0 > 0$, we obtain
\begin{equation}
\label{sigmaplusminus} \sigma_+(\tau)=
\frac{\sigma_+(0)}{1+(\tau/\tau_0)^2}\,,\quad \sigma_-(\tau)=
\frac{\sigma_-(0)}{1-(\tau/\tau_0)^2}\,,\quad\mbox{where }
\lambda_\pm = \frac{8}{\tau_0^2\sigma_{\pm}(0)^2}\,.
\end{equation}

The energy spectra of these functions are respectively
$$
\int_{-\infty}^{\infty}\frac{e^{i\,\omega\tau}}{1+\tau^2/\tau_0^2}\,d\tau = 2\pi\tau_0\cosh(\tau_0\,\omega)\,;\quad
\int_{-\infty}^{\infty}\frac{e^{i\,\omega\tau}}{1-(\tau-i\epsilon)^2/\tau_0^2}\,d\tau = 2\pi\tau_0\cos(\tau_0\,\omega)\,,
$$
which are manifestly gapless and free from zero--mass poles.

\subsection{The NG bosons of the spontaneously broken conformal symmetry}
\label{Fubini2}
It is evident from Eq (\ref{squarefs}) that $\sigma_+(\tau)$ and $\sigma_-(\tau)$
can be envisaged as particular solutions of the motion equations respectively derived
from classical actions
\begin{eqnarray}
\label{ActInt+}
\hspace{-8mm}A_+ &=& \int_{\bar C^+}\sqrt{-\bar g(x)}\,\bigg\{+\frac{1}{2}\,
\bar g^{\mu\nu}(x)\big[\partial_\mu\sigma_+(x)\big]\,\partial_\nu\sigma_+(x) -
\frac{\lambda_+}{4}\,\sigma_{+}(x)^4\bigg\}\,d^4x\,,\\
\label{ActInt-}
\hspace{-8mm}A_- &=& \int_{\bar C^+}\sqrt{-\bar g(x)}\,\bigg\{-\frac{1}{2}\,
\bar g^{\mu\nu}(x)\big[\partial_\mu\sigma_-(x)\big]\,\partial_\nu\sigma_-(x)
-\frac{\lambda_-}{4}\,\sigma_{-}(x)^4\bigg\}\,d^4x\,,
\end{eqnarray}
where $\bar C^+$ denotes the flat conical spacetime equipped with metric tensor
$$
\bar g_{\mu\nu}(x)\equiv \bar g_{\mu\nu}(\tau, \vec\rho\,)=\hbox{diag}\bigl[1, -\tau^2, -\tau^2 (\sinh\varrho)^2, -\tau^2(\sinh\varrho\,\sin\theta)^2\bigr],
$$
already introduced in \S\,\ref{Evolvingvac}, and $\bar g(x)\equiv \bar g(\tau, \vec\rho\,) = -\tau^6
(\sinh\varrho\,)^4\sin\theta^2$ is the determinant of matrix $\big[\bar g_{\mu\nu}(x)\big]$. Denoting as
$d\Omega(\varrho,\theta,\phi) =(\sinh\varrho)^2\sin\theta \,d\varrho\,d\theta\,d\phi$ the 3D--volume
element of the unit hyperboloid $\Omega$, we can express the spacetime volume element $d^4x$ of integrals (\ref{ActInt+})
and (\ref{ActInt-}) as $\sqrt{-\bar g(x)}\,d^4x \equiv \tau^3 d\tau\, d\Omega(\vec \rho\,)$, where
$\vec\rho = \{\varrho,\theta,\phi\}$ are the hyperbolic--Euler--angles introduced in \S\,\ref{futconegeom}
after Eq (\ref{gmunumatrix}).

The signs of the terms in the integrals are taken in such a way that the potential energy density
is positive. Since in $A_-$ the kinetic term is negative, while in $A_+$ is positive,
we may interpret $\sigma_+(x)$ as a physical  massless  scalar field and and $\sigma_-(x)$ as a
ghost massless scalar field. If we assume that $\sigma_\pm$ should depend only on $\tau$, the integrals simplify to
\begin{eqnarray}
A_+ &=& \Omega\!\!\int_0^{\infty}\!\!\tau^3 \Big\{\!+\frac{1}{2}\,\big[\partial_\tau\sigma_+(\tau)\big]^2
-\frac{\lambda_+}{4}\,\sigma_{+}(\tau)^4\! \Big\}d\tau,\nonumber\\
A_-&=& \Omega\!\!\int_0^{\infty}\!\!\tau^3 \Big\{\!-\frac{1}{2}\big[\partial_\tau\sigma_-(\tau)\big]^2
-\frac{\lambda_-}{4}\,\sigma_{-}(\tau)^4\! \Big\}d\tau\,.\nonumber
\end{eqnarray}

The motion equations derived from these actions are exactly those indicated by Eqs (\ref{squarefs}).
Note that, while $\sigma_{+}(\tau)$ is always finite, $\sigma_{-}(\tau)$ becomes infinite
at $\tau=\tau_0$. This divergence occurs because the lower bound of the kinetic energy density is $-\infty$.

To prevent this catastrophic ending, we may combine $A_+$, $A_-$ and an interaction term of $\sigma_{+}(\tau)$
with $\sigma_{-}(\tau)$ into a conformal--invariant action $A$, in such a way that the energy of the system
remains bounded from both above and below. The simplest example is
\begin{equation}
\label{convergingA}
A =  \Omega\!\!\int_0^{\infty}\!\!\tau^3\bigg\{\frac{1}{2}\big[\,\partial_\tau\sigma_+(\tau)\big]^2
-\frac{1}{2}\big[\,\partial_\tau\sigma_-(\tau)\big]^2 -\frac{\lambda}{4}\big[\,\sigma_{+}(\tau)^2
-c^2\sigma_{-}(\tau)^2\big]^2\bigg\}\,d\tau\,,
\end{equation}
where $\lambda= \lambda_+$, $c^4 \lambda= \lambda_-$ and $0<c<1$. In this case, in fact,
the motion equations are
\begin{eqnarray}
\label{sigma+eq}
&&\partial_\tau^2 \sigma_+(\tau) + \frac{3}{\tau}\,\partial_\tau \sigma_+(\tau) + \lambda
\big[\,\sigma_{+}(\tau)^2 -c^2\sigma_{-}(\tau)^2\big]\sigma_+(\tau)=0\,,\\
\label{sigma-eq}
&&\partial_\tau^2 \sigma_-(\tau) + \frac{3}{\tau}\, \partial_\tau \sigma_-(\tau) + c^2\lambda
\big[\,\sigma_{+}(\tau)^2-c^2\sigma_{-}(\tau)^2\big]\sigma_{-}(\tau)=0\,,
\end{eqnarray}
which are the same as Eqs (\ref{deltaAvarphi}) and (\ref{deltaAsigma}), provided that
$\sigma_+(\tau)$ and $\sigma_-(\tau)$ are respectively interpreted as the VEVs of a physical
scalar field $\varphi(x)$ and of a ghost scalar field $\sigma(x)$. If we replace
$\sigma_+(\tau)$ with  $\varphi(\tau)$, $\sigma_{-}(\tau)$ with $\sigma_0\,a(\tau)$ and $c^2$
with $\mu^2/\lambda$, the solutions to Eqs (\ref{sigma+eq}) and (\ref{sigma-eq}) are like those
described in Appendix {\bf \ref{VacDynApp}}.

If the spacetime is a curved, conformal invariance would requires that
(\ref{ActInt+}) and (\ref{ActInt-}) be replaced by the actions
\begin{eqnarray}
\label{Avarphi}
& & A_+ = \int_{C^+}\bigg[+\frac{\sqrt{-g}}{2}\,
g^{\mu\nu}(\partial_\mu\sigma_+)\,\partial_\nu\sigma_+ -
\frac{\lambda_+}{4}\,\sigma_+^4 +
\frac{R}{12}\,\sigma_+^2\bigg]d^4x,\\
\label{Asigma}
& & A_- = \int_{C^+}\bigg[-\frac{\sqrt{-g}}{2}\,g^{\mu\nu}
(\partial_\mu\sigma_-)\,\partial_\nu\sigma_- - \frac{\lambda_-}{4}\,
\sigma^4_- -\frac{R}{12}\,\sigma^2_-\bigg]d^4x,
\end{eqnarray}
where $g^{\mu\nu}$ is the contravariant metric tensor of conical
curved spacetime $H^+$, $g$ the determinant of matrix $[g_{\mu\nu}]$
and $R\neq 0$ is the Ricci scalar constructed from $g_{\mu\nu}$.
The reason for the inclusion of the term in $R$ is explained in
detail in \S\,1, in \S\,\ref{inclusCGR}, near Eq (\ref{DeltaAisSurTerm}).
The term in $R$ in fact is necessary to preserve the conformal invariance
of $A_+$ and $A_-$ up to harmless surface terms, which is only possible
if $C^+$ is a 4D manifold.

But assuming Eq (\ref{Asigma}) as the action of a ghost field
$\sigma(x)$ that does not depend only from both $\tau$, would
make it impossible to suppress the propagation of free ghosts,
which is unacceptable in CGR. For this reason we must
reject the idea the equations that involve $\sigma(x)$ could
depend on $R$,

\centerline{---------------}

\centerline{------}

\newpage

\markright{R.Nobili, Conformal General Relativity - Bibliography}

\centerline{--------------------------}

\newpage
\addcontentsline{toc}{section}{SYMBOLS, UNITS AND CONVERSION TABLES}
\section*{SYMBOLS, UNITS AND CONVERSION TABLES}
\noindent  {\bf Special mathematical symbols}

\noindent $x \cong y$: $x$ is approximately equal to $y$ with more than one percent precision;

\noindent $x \approx y$: $x$ is of the order of magnitude of $y$;

\noindent $A \equiv B$: $A$ is mathematically equivalent to $B$ (different expression with same meaning);

\noindent $A \sim B$: $A$ is functionally equivalent to $B$ (for instance, in action--integral comparison).

\index{Fundamental constants}

\noindent  {\bf Fundamental constants of quantum field theory (QFT) and thermodynamics}

\noindent $c \cong 299 792 458$ m/s: speed of light;

\noindent  $h \cong 6.62607004\times 10^{-34}$ m$^2$ kg/s:  Planck constant;

\noindent  $\hbar \equiv h/2\pi \cong 1.05457180\times 10^{-34}$ m$^2$ kg/s;

\noindent 1 eV $ \cong 1.602176565\times 10^{-19}$ J (J = kg m$^2$/s$^2$);

\noindent  $K_B \cong  1.38064852 \times 10^{-23}$  J/$^{\mbox{\tiny o}}$K: Boltzmann constant
($^{\mbox{\tiny o}}$K = Kelvin).

\vspace{1mm}
\noindent  {\bf Electronvolt (eV) to metric--units (m kg s $^{\mbox{\tiny o}}$K) conversion via natural units}

\centerline{
\begin{tabular}{l r @{.} l}
1 eV  as mass ($\times \,c^{-2}$)  & $\longleftrightarrow \quad $  1&7826627$\times 10^{-36}$ kg\,;\\
1 eV$^{-1}$ as length ($\times \,\hslash\,c$)  & $\longleftrightarrow \quad $ 1&9732705$\times 10^{-7}$ m\,;\\
1 eV$^{-1}$ as time ($\times \,\hslash$)  & $\longleftrightarrow\quad$ 6&5821220$\times 10^{-16}$ s\,;\\
1 eV as temperature ($\times K_B$) & $\longleftrightarrow \quad $  1&16045220$\times 10^4\, \mbox{$^{\mbox{\tiny o}}$K}$ \,.
\end{tabular}}
%1K \cong 8.61733035\times 10{-14} GeV ; 1GeV = 1.16045220e+013 K; 1eV = 1.16045220e4 K
% T_B = 141.03 GeV = 141.03*1.16045220e13 = 1.63658574e15 K

\vspace{2mm}

\noindent From these, we derive
\vspace{-4mm}
\begin{eqnarray}
\hspace{-7mm}& & 1 \,\mbox{kg} \cong 5.6095861\times 10^{26}\mbox{GeV};\quad 1 \mbox{GeV}\cong
1.7826627\times 10^{-27}\mbox{kg}\cong 1.5192668\times 10^{24}\mbox{s}^{-1}; \nonumber\\
\hspace{-7mm}& & 1 \mbox{GeV} \cong 5.0677289\times 10^{15}\mbox{m}^{-1};\quad 1 \mbox{GeV}^{-1}
\cong 1.9732705\times 10^{-16}\mbox{m} \cong 6.5821223\times 10^{-25}\mbox{s}; \nonumber\\
\hspace{-7mm}& & 1 \mbox{m}^{-1} \cong 1.9732705\times 10^{-16}\mbox{GeV};\quad 1\mbox{s}^{-1}
\cong 6.5821223\times 10^{-25} \mbox{GeV}; \nonumber \\
\hspace{-7mm}& & 1 \mbox{m} \cong 5.0677289\times 10^{15}\mbox{GeV}^{-1};\quad 1\mbox{s}\cong
1.5192668\times 10^{24}\mbox{GeV}^{-1}; \nonumber \\
\hspace{-7mm}& & 1 \mbox{kg/m}^{3}\cong 4.3101332\times 10^{-21}\mbox{GeV}^4;\quad 1 \mbox{GeV}^4\cong
2.3201139\times 10^{20}\mbox{kg/m}^{3}; \nonumber\\
\hspace{-7mm}& & 1\, \mbox{$^{\mbox{\tiny o}}$K} \cong 8.61733035\times 10^{-14}\mbox{GeV};\quad
1 \mbox{GeV} = 1.16045220\times 10^{13}\, \mbox{$^{\mbox{\tiny o}}$K}. \nonumber
\end{eqnarray}

\vspace{-2mm}
\noindent {\bf Time parameters in kinematic--, conformal-- and proper--time coordinates}

\noindent $x=\{x^0, x^1, x^2, x^3\}$: general spacetime coordinates of the kinematic--time representation;

\noindent $\widehat{x}=\{\widehat{x\,}^0, \widehat{x\,}^1, \widehat{x\,}^2, \widehat{x\,}^3\}$:
general spacetime coordinates the conformal--time representation;

\noindent $\widetilde{x}=\{\widetilde{x\,}^0, \widetilde{x\,}^1, \widetilde{x\,}^2, \widetilde{x\,}^3\}$:
spacetime coordinates of the proper--time representation.

\noindent $\tau \equiv x^0$: kinematic time;\quad $\widehat{\tau}\equiv\widehat{x\,}^0$: conformal time;\quad
$\widetilde{\tau}\equiv\widetilde{x\,}^0$: proper time.

%\noindent  {\bf Symbols and physical constants of Conformal General Relativity (CGR)}
%
%\noindent $\varphi(x)$: Higgs field amplitude in the kinematic--time representation;
%
%\noindent $\widetilde{\varphi}(\widetilde{x})$: Higgs field amplitude in proper--time representation;

\newpage
\noindent {\bf Physical constants of CGR}

\noindent $\mu_H\cong 125.1$ GeV:  Higgs--boson mass;

\noindent $\mu=\mu_H/\sqrt{2} \cong  88.46$ GeV $\cong 1.344\times 10^{28}$s$^{-1}$: mass parameter of Higgs--field action integral;

\noindent $G \cong 6.72262488\times 10^{-39}$ GeV$^{-2}$: Newton gravitational constant
in natural units;

\noindent $M_P = 1/\sqrt{G} \cong 1.2196\times 10^{19}$ GeV: Planck mass in natural units;

\noindent $M_{rP} = M_P/\sqrt{8\,\pi} \cong 2.4328\times 10^{18}$ GeV: reduced Planck mass;

\noindent $\kappa= 8\pi G =1/M_{rP}^2\cong 1.6890\times 10^{-37}$ GeV$^{-2}$: gravitational coupling constant;

\noindent $G_F \cong  1.16637\times 10^{-5}$ GeV$^{-2}$: Fermi coupling constant;

\noindent $\lambda=\mu_H^2 G_F/\sqrt{2} \cong 0.1291$: self--coupling constant of Higgs--boson field;
%according to the Standard Model (SM) of elementary particles.

\noindent $T_B \cong 141.03$ GeV $ \cong 1.6366\times 10^{15}\, \mbox{$^{\mbox{\tiny o}}$K}$:
big--bang temperature;

\noindent $T_{BK}\cong 2.350\times 10^{-13}$GeV $\cong 2.726 \,\mbox{$^{\mbox{\tiny o}}$K}$:
temperature of cosmic--background today.

\vspace{2mm}

\noindent  {\bf Peculiar relations and constants of Conformal General Relativity (CGR)}

%\noindent {\bf Relation between scalar ghost amplitude and scale factor of dynamic vacuum}

\noindent $\sigma(\tau)$: scalar--ghost amplitude in kinematic--time representation;

\noindent $\sigma_0 = \sqrt{6/\kappa} \cong   5.959\times 10^{18}$ GeV: asymptotic
amplitude of scalar--ghost at $\tau\rightarrow \infty$;

\noindent $\sigma(0) \ll \sigma_0$ : initial value of scalar--ghost field amplitude;

\noindent $\alpha(\tau) = \sigma(\tau)/\sigma_0 \le 1$: inflation factor as a function of kinematic time.

\noindent $\alpha(0)$: initial value of inflation factor; $\alpha(\infty)=1$: final value of inflation factor;

\noindent $\alpha(\tau_B) \cong \sqrt{\alpha(0)}$: inflation factor at big bang.

\noindent $Z = 1/\alpha(0)$: inflation factor across inflation.

\noindent $\tau_B$: kinematic time of big bang;

\noindent $\widetilde{\tau}_B$: proper time of big bang;

\noindent $\tau_c \cong \tau_B$: critical kinematic time of spacetime--explosion;

\noindent $\widetilde{\tau}_c \cong\widetilde{\tau}_B \cong 0$: big--bang time in proper time units;

\noindent $\tilde U(\tilde\tau_B)=\mu_H^4/16\lambda\cong  1.18597\times 10^8$GeV$^4$  energy--density at big bang;

\vspace{2mm}

\noindent  {\bf Approximate cosmological parameters inferred from astronomical observations}

\noindent 1 Gyr $\cong 3.1557\times 10^{16}$ sec;

\noindent $H_0 \approx 67.8$ Km Mpc$^{-1}$s$^{-1}$$\cong 1.45\times 10^{-42}$GeV: Hubble constant today;

\noindent $\tilde\tau_U \approx 13.82$ Gyr $\cong 4.358\times 10^{17}$s: age of the universe in proper--time units;

\noindent $\tau_U$: age of the universe in kinematic--time units;

\noindent $\widetilde{\tau}_D\cong 0.378$ Gyr: age of photon--decoupling in proper time units;

\noindent $\tau_D$: age of photon--decoupling in kinematic--time units.

\end{document}